\documentclass[10pt]{report}
\usepackage{epsfig}

\def\refe{\par\noindent\hangindent=1.5cm}
\begin{document}

\newcommand{\del}{$\delta$}
\newcommand{\dc}{$\delta_c$}
\newcommand{\dl}{$\delta_l$}
\newcommand{\pot}{$\varphi$}
\newcommand{\res}{$\Lambda$}
\newcommand{\mres}{\Lambda}
\newcommand{\bc}{$b_c$}
\newcommand{\fc}{$F_c$}
\newcommand{\lam}{$\lambda$}
\newcommand{\luno}{$\lambda_1$}
\newcommand{\ldue}{$\lambda_2$}
\newcommand{\ltre}{$\lambda_3$}
\newcommand{\muno}{\lambda_1}
\newcommand{\mdue}{\lambda_2}
\newcommand{\mtre}{\lambda_3}

\newcommand{\be}{\begin{equation}}
\newcommand{\ee}{\end{equation}}
\newcommand{\bea}{\begin{eqnarray}}
\newcommand{\eea}{\end{eqnarray}}

\renewcommand{\r}{${\bf r}$}
\newcommand{\mr}{{\bf r}}
\newcommand{\x}{${\bf x}$}
\newcommand{\mx}{{\bf x}}
\renewcommand{\v}{${\bf v}$}
\newcommand{\mv}{{\bf v}}
\renewcommand{\u}{${\bf u}$}
\newcommand{\muv}{{\bf u}}
\newcommand{\q}{${\bf q}$}
\newcommand{\mq}{{\bf q}}
\newcommand{\s}{${\bf S}$}
\newcommand{\ms}{{\bf S}}
\renewcommand{\k}{${\bf k}$}
\newcommand{\mk}{{\bf k}}

\newcommand{\pdfd}{$P_\delta(\delta;\Lambda)$}
\newcommand{\mpdfd}{P_\delta(\delta;\Lambda)}
\newcommand{\pdfl}{$P_{\delta_l}(\delta_l;\Lambda)$}
\newcommand{\mpdfl}{P_{\delta_l}(\delta_l;\Lambda)}
\newcommand{\pdff}{$P_F(F;\Lambda)$}
\newcommand{\mpdff}{P_F(F;\Lambda)}
\newcommand{\pdfx}{$P_x(x;\Lambda)$}
\newcommand{\mpdfx}{P_x(x;\Lambda)}
\newcommand{\pdfdu}{$P_{\delta_l}^{\rm noup}(\delta_l;\Lambda)$}
\newcommand{\mpdfdu}{P_{\delta_l}^{\rm noup}(\delta_l;\Lambda)}
\newcommand{\pdfxu}{$P_x^{\rm noup}(x;\Lambda)$}
\newcommand{\mpdfxu}{P_x^{\rm noup}(x;\Lambda)}
\newcommand{\pdffu}{$P_F^{\rm noup}(F;\Lambda)$}
\newcommand{\mpdffu}{P_F^{\rm noup}(F;\Lambda)}
\newcommand{\pdfxf}{$P_x^{\rm fb}(x;\Lambda)$}
\newcommand{\mpdfxf}{P_x^{\rm fb}(x;\Lambda)}

\newcommand{\imfr}{$\Omega(<\mres)$}
\newcommand{\mimfr}{\Omega(<\mres)}
\newcommand{\imfm}{$\Omega(>M)$}
\newcommand{\mimfm}{\Omega(>M)}
\newcommand{\dmfm}{$n(M)$}
\newcommand{\dmfr}{$n(\mres)$}
\newcommand{\npk}{$n_{\rm pk}(\delta_l;\Lambda)$}
\newcommand{\mnpk}{n_{\rm pk}(\delta_l;\Lambda)}

%
%

\Huge
\begin{center}

{\bf The Cosmological Mass Function}

\bigskip
\bigskip
\bigskip
\bigskip

Pierluigi Monaco
\end{center}

\normalsize
\bigskip
\bigskip
\bigskip
\bigskip

\scriptsize
\noindent Scuola Internazionale Superiore di Studi Avanzati (SISSA), via 
Beirut 4, 34014 -- Trieste, Italy \\
Dipartimento di Astronomia, Universit\`a degli studi di Trieste, via
Tiepolo 11, 34131 -- Trieste, Italy \\
\bigskip
Present address: Institute of Astronomy, Madingley Road, Cambridge CB3 0HA,
United Kingdom
email: monaco@ast.cam.ac.uk
\normalsize

\bigskip
\bigskip
\bigskip
\bigskip
\bigskip
\bigskip
\bigskip
\bigskip
\bigskip
\bigskip

\large {\bf Abstract} 

\bigskip
\bigskip

This thesis aims to review the cosmological mass function problem,
both from the theoretical and the observational point of view, and to
present a new mass function theory, based on realistic approximations
for the dynamics of gravitational collapse. Chapter 1 gives a general
introduction on gravitational dynamics in cosmological models. Chapter
2 gives a complete review of the mass function theory. Chapters 3 and
4 present the ``dynamical'' mass function theory, based on truncated
Lagrangian dynamics and on the excursion set approach.  Chapter 5
reviews the observational state-of-the-art and the main applications
of the mass function theories described before. Finally, Chapter 6
gives conclusions and future prospects.

\bigskip

{\it Subject headings:} clusters of galaxies --- cosmology: theory ---
dark matter --- galaxies: clustering --- large-scale structure of the
universe

\normalsize
\vspace{2cm}

\begin{center} Accepted by Fundamentals of Cosmic Physics \end{center}

\tableofcontents

%
%

\setcounter{chapter}{-1}
\chapter{Introduction}

Cosmological models are designed to predict the essential features of
observed cosmological structures. Such models must then predict the
formation of structures characterized by densities much larger than
the cosmological background density; these structures can correspond
to protogalaxies, galaxies, groups and clusters of galaxies.
High-density structures are more or less isolated in many if not most
cases, so that it is possible to define a mass associated to them
(this is not the case for the large-scale network of galaxy walls and
filaments, characterized by low density contrasts). The mass
distribution of such objects, i.e. the number of objects per unit
volume and unit mass interval, is commonly called mass function or
multiplicity function; it will be referred to as MF throughout the
text.  The MF of cosmological structures is the main concern of this
thesis.

The determination of the cosmological MF is a difficult, not fully
solved problem, both from the theoretical and the observational point
of view. An analytical exact prediction of the MF is hampered, even in
the simplest cosmological models, by the fact that highly non-linear
gravitational dynamics is involved in the formation of high density
objects; it is well known that the gravitational collapse problem has
never been exactly solved, except in the case of simple symmetries
(spherical planar). Large N-body simulations can be used to determine
the MF of simulated halos; however such simulations, besides being
time-expensive and limited in resolution, provide a numerical estimate
of the final solution of the problem without directly shedding light
on the difficult problem of gravitational collapse. Approximate
analytical arguments, while being of limited validity, can provide
useful and fully-understood solutions, which can then be compared to
the results of N-body simulations.

From the observational point of view, the masses of cosmological
structures can now be estimated in a number of indirect ways: for
instance, galaxy masses up to a few optical radii can be inferred from
the velocity field of stars or from the motion of satellite galaxies,
while galaxy cluster masses can be estimated through the velocity
dispersion of galaxies, or through their X-ray emission, or through
gravitational lensing.  These method have led or will lead in the near
future to the observational determination of the MF of these
structures.

The first theoretical attempt to determine the MF was made in the
seminal paper by Press \& Schechter (1974). Their procedure was based
on an extrapolation of the linear regime of perturbation growth to the
collapse regime.  Their result completely ignores many of the
difficulties of gravitational collapse, but it is very simple, and, as
it has been shown later by means of N-body simulations, it predicts
the right number of dark-matter halos.  Between 1988 and 1992, the
theoretical MF problem witnessed a burst of effort from many authors,
who proposed many extensions or alternative procedures to overcome the
problems of the PS MF. Since 1995, the MF problem is witnessing a new
wave of interest. The problem is not still ``solved'', as remarked
before, but a number of promising approaches have been developed. On
the other hand, the PS formula still provides, according to most
authors, a simple way to effectively predict the MF for many practical
uses.

The MF problem deserves further investigation: a realistic MF theory
is the first step in the modeling of observable cosmological objects.
Reliable observational MFs, to be compared to theoretical prediction
for different cosmological models, will be available in the near
future. Besides, a careful study of the MF problem, conducted through
analytical arguments and detailed comparisons to numerical
simulations, can help to improve our understanding of the
gravitational highly non-linear regime.

This thesis is organized as follows. Chapter 1 contains a cosmological
introduction; it is focused on the problem of gravitational dynamics
in cosmology, which is addressed both qualitatively (\S 1.1) and
quantitatively (\S 1.2), through the introduction of the mathematical
tools which will be exploited in the following Chapters.  The problem
of structure formation is schematically discussed in \S 1.3.

Chapter 2 contains an extensive review of the theoretical MF problem.
The Press \& Schechter procedure is explained in \S 2.1, where its
main features, problems and merits are discussed. \S 2.2 describes the
comparison of the analytical MF with the results of numerical
simulations.  The remaining sections describe the theoretical works
which have tried to extend the validity of the original Press \&
Schechter work, or have proposed alternative procedures.

Chapters 3 and 4 describe the building of an MF theory, fully based on
realistic approximations for gravitational collapse. Chapter 3
discusses how to obtain realistic estimates of collapse times by using
dynamical approximations based on the Lagrangian formulation of fluid
dynamics.  In particular, \S 3.1 analyzes the Zel'dovich
approximation, \S 3.2 the homogeneous ellipsoidal collapse model and
\S 3.3 the Lagrangian perturbation theory. In \S 3.4 the probability
distribution function of inverse collapse times, necessary to
construct the MF, is found, while \S 3.5 gives a final discussion.

Chapter 4 describes the statistical procedures which are needed to
construct the dynamical MF. \S 4.1 describes a Press \& Schechter-like
procedure, \S 4.2 and \S 4.3 a procedure based on the excursion set
approach, while \S 4.4 describes the complications connected to the
geometry of collapsed regions in Lagrangian space. Finally, \S 4.5
gives the main conclusions on the dynamical MF.

The MF world is described in Chapter 5.  Three main topics are
discussed: observational determinations of masses, cosmological
applications of the MF theories and possible applications of the
dynamical MF theory described in Chapters 3 and 4. \S 5.1 deals with
galaxies, \S 5.2 with galaxy groups and clusters, \S 5.3 with
high-redshift objects.

Finally, Chapter 6 comments on the future prospects of the dynamical
MF theory, from the comparison to numerical simulations to the future
applications which are in progress.

%
%

\chapter[Cosmological Models]{Gravitational Dynamics In Cosmological Models}

It is now widely believed that observed cosmological structures grew
from small perturbations in a quasi-homogeneous Universe. The origin
of such perturbations is generally ascribed to quantum fluctuations
arisen during an inflation era which took place soon after the Big
Bang.  Many observational evidences strongly suggest the presence of a
gravitationally dominant dark matter component (which can be made up
of different species), responsible for matter perturbations.  Despite
many efforts, the problem of gravitational evolution of such
perturbations remains (analytically) unsolved, except in special cases
like perturbative expansions, which are accurate as long as matter
perturbations are not too large, or exact solutions in special
symmetries, like planar or spherical. These special solutions, though
of not general validity, can help to understand some important
properties of collapsed structures.

This chapter describes in some detail the cosmological models within
which the MF theory has been developed, and the process of structure
formation according to such models. \S 1.1 describes the gravitational
instability scenario of structure formation, the main families of
cosmological models not ruled out by observations, and the various
tools which help to follow the evolution of the initial
quasi-homogeneous configuration into the non-linear regime. \S 1.2
gives a summary of the mathematical formalism used in cosmology and in
the present work. \S 1.3 discusses the process of structure formation,
describing the complications of the real process and some
(over)simplifications which make the problem tractable.

This chapter aims to summarize the main arguments, to assess the
notation and to comment on those points which are relevant for further
analyses.  The arguments treated in this chapter are standard in
cosmology (except when explicitly stated); complete presentations of
these arguments can be found in many textbooks, e.g.  in Weinberg
(1972), Peebles (1980), Kolb \& Turner (1989), Padmanaban (1992),
Peebles (1993), Coles \& Lucchin (1995).

\section{Cosmological Models}

According to the ``standard'' gravitational instability scenario,
cosmological structures formed from gravitational collapse of small
fluctuations, already present at very early epochs.  Such primordial
fluctuations are often thought to have arisen during an inflationary
era as quantum fluctuations: the Universe, soon after the Big Bang
(when the temperature was about 10$^{15}$ GeV), experienced a phase of
inflation, i.e. of accelerated expansion, caused by a field (the
inflaton) whose vacuum energy happened to dominate, giving rise to
negative pressure. The stretching caused by the accelerated expansion
made the quantum fluctuations of the inflaton field to become
``classical''.  Inflation has been invoked as a mechanism able to
solve a number of other problems, giving a dynamical explanation of
the flatness problem (why is the cosmological density so similar to
the critical one), the uniformity of the CMB, and the lack of
topological defects, such as the monopoles, which can be created
during the cosmological phase transitions in the early Universe.

There is not one single theory of inflation: such theories are based
on high-energy physics theories, many of which have been
proposed. However, many inflation theories agree in predicting very
simple properties for the perturbation field. In particular, the
typical resulting Universe is (almost) flat (i.e. Einstein-de Sitter,
see \S 1.2.1), regardless of the initial geometry (which can be even
chao\-tic); this means that the actual background density is predicted
to be almost equal to the critical one. However, many particular
inflation scenarios predict matter densities less than the critical
one; global flatness is often preserved by means of a cosmological
constant term (see \S 1.2.1).  The perturbation field is usually
predicted to be Gaussian; the central limit theorem assures that, if
the situation is ``complex'' enough, Gaussian statistics are
unavoidable. However, some particular scenarios predict specific
(weak) non-Gaussian features. The power spectrum of matter
fluctuations is typically predicted to be $P(k)\propto k^n$ (see \S
1.2.7 for a precise definition), with $n$ almost equal to one (as
already proposed by Harrison (1970) and Zel'dovich (1972)), or
slightly less than one (``tilted'' power spectrum); note that
measurements based on CMB anisotropy at large scales constrain the
spectral index $n$ to be not very different from one.  Finally,
inflation can predict a stochastic background of gravitational waves
(tensor modes) to be present. These have no influence on the
subsequent evolution of perturbations, since tensor modes cannot
collapse, but would be observed on the CMB at very large scales.

Once the kinds of particles present in the Universe and the
statistical properties of primordial fluctuations are known, the
fluctuation field can be evolved up to recombination, when the
temperature dropped to small enough values to let the ionized gas
recombine, thus decoupling from radiation; this is the ``instant'' at
which the CMB is observed (more properly, recombination takes a finite
time to be completed). The evolution of perturbations up to
recombination is relatively easy to follow, as they remain small all
the time, so that linear theory can be safely used (see \S 1.2.3).
The evolution depends on the number and type of particles present in
the cosmological fluid. In this epoch the nature of the dark matter
(cold, hot, warm, mixed) becomes important: large velocity dispersions
in a hot dark matter (HDM) component smear out small-scale
perturbations, while cold dark matter (CDM) conserves fluctuations at
all cosmologically relevant scales.  Since the evolution is linear,
the power spectrum at recombination is connected to the primordial one
by means of a transfer function $T(k)$:

\be P_{rec}(k)= P_{prim}(k)\; T^2(k). \label{eq:transfer} \ee

\noindent Transfer functions for many different scenarios are 
available in the literature.

It has not been possible to decide, from cosmological observations,
which kind of matter dominates our Universe. Anyway, two conclusions
are widely accepted: purely baryonic models are nearly ruled out by
CMB, nucleosynthesis and large-scale structure constraints, especially
if the Universe is flat. Second, pure HDM models, having no
small-scale power, fail to predict the formation of small-scale
objects such as galaxies. The most widely discussed model, in the last
decade, has been the CDM model. Standard CDM is based on the
hypotheses that (i) the Universe is flat, (ii) the Hubble constant is
$H_0$=50 km/s/Mpc (see \S 1.2.1), (iii) primordial fluctuations are
Gaussian, with Harrison-Zel'dovich power spectrum. Before COBE
satellite observations of the CMB, the spectrum was normalized by
requiring that the mass variance on the scale of 8$h^{-1}$ Mpc,
$\sigma_8$, were equal to one, as suggested by observations of galaxy
catalogs (Davis \& Peebles 1983); in either case, the variance in the
number of galaxies can be different from the underlying mass variance,
by a factor $b=1/\sigma_8$ ({\it bias factor}; see Kaiser 1984, or
Dekel \& Rees 1987). The {\bf biased CDM} models assumed a certain
value for the bias parameter.

After COBE, all models are normalized on the CMB fluctuations at large
scales (typically neglecting the possible contribution of
gravitational waves); the COBE-normalized standard CDM model gives too
much power at small scales, and so is excluded by observations.
However, it remains a good ``template'' theory to understand
large-scale structure: most current and popular models are just
variations or extensions of the CDM one. The main families of models
currently proposed in the literature are:

\begin{itemize}
\item
{\bf open CDM:} CDM in an open Universe;
\item
{\bf $\Lambda$ CDM:} CDM in a flat Universe with cosmological constant;
\item
{\bf tilted CDM:} CDM with tilted primordial power spectrum ($n\neq 1$);
\item
{\bf C+HDM:} two dark matter components, a hot (20-30\%) and a cold
one.
\end{itemize}

It is necessary to stress, before going on, that the general scenario
just described is not the only one proposed. An important class of
alternative scenarios is that of topological defects. These objects
can form during particular cosmological phase transitions which take
place after the inflation era: a spontaneous symmetry breaking causes
different points of space to get to equivalent but different vacuum
states; topological defects form at the interface of zones dominated
by different vacuum configurations. Their topology can range from
domain walls to strings, monopoles and global textures; while domain
walls and monopoles are ruled out by current observations, the
validity of the cosmic strings and textures scenarios will be decided
by future CMB observations, especially by the COBRAS/SAMBA experiment.
Such topological defects act as seeds of structure formation; they
can, for instance, provide small-scale power in the HDM scenario.

After recombination, fluctuations continue growing, until they turn
non-linear. At this stage structure formation takes place: collapsed
clumps form, which eventually reach an equilibrium configuration
(virialization), passing through a number of complicated transient
configurations. As a matter of fact, large cosmological structures
(e.g. galaxy clusters or superclusters) are likely to be in their
transient phase, while smaller structures (e.g. galaxies) have
probably reached an equilibrium state, at least in their inner parts.
The non-linear evolution of perturbations can be safely followed by
means of the Newtonian approximation of gravity, as long as non-linear
perturbations are well inside the horizon, peculiar velocities are
non-relativistic and the wavelengths considered are much larger than
the event horizon of every compact object present. Newtonian gravity
is non-local ({\it action at distance}), as every particle is
influenced in its motion by all the other particles in the Universe.
Non-locality makes the Newtonian problem very difficult to solve: for
instance, it is not possible in an N-body simulation to follow the
motion of a particle without following at the same time the motion of
all the other particles.

Different methods have been developed to follow the evolution of
perturbations in the non-linear regime. These are qualitatively
described below; the relevant equations will be given in the next
section.

\begin{itemize}
\item 
{\bf N-body simulations:} a direct method to follow the non-linear
evolution of perturbations is that of numerically solving the
equations of motion for given initial conditions. This can be done by
means of large N-body simulations, where a perturbed lattice of
particles approximates the continuum of matter in the Universe. N-body
simulations have the great merit of giving good numerical
approximations of the exact solution, but have three main problems:
they are very time-expensive, limited in resolution range, and give a
``blind'' solution, which does not directly shed light on the various
dynamical effects responsible for the behavior of the system. Anyway,
a deep comprehension of gravitational collapse can come by comparing
various analytical arguments to N-body simulations; as a matter of
fact, cosmological N-body simulations have become an invaluable tool
to describe and understand the non-linear evolution of perturbations.

\item 
{\bf Linear theory:} on large scales, much larger than the typical
correlation length of the matter field, perturbations are still small
compared to the background. It is then a reasonably good approximation
to describe them as if they were in linear regime, i.e. neglecting the
non-linear transfer of power from small to large scales. Linear theory
is presented in full detail in Peebles (1980). As a crude argument,
one could try to predict the onset of non-linearity by extrapolating
linear theory up density contrasts of order one; this has been shown
to be a poor description of the true non-linear evolution of
structures (see, e.g., Bond et al. 1991; Coles, Melott \& Shandarin
1993).

\item 
{\bf Spherical collapse:} this is one of the few cases in which
analytical solutions can be found (Gunn \& Gott 1972; Peebles
1980). In particular, the collapse of a top-hat perturbation (constant
overdensity inside a sphere) has often been used to describe the
collapse of isolated structures, such as galaxies or galaxy clusters.
Note that spherical collapse is local: the initial overdensity is the
only quantity needed to evolve the system.  Mathematically, collapse
corresponds to the formation of a singularity (infinite density at the
center). In practice, real perturbations are not perfectly
homogeneous, and they do not reach a singularity; it is typically
assumed that collapse to a singularity corresponds to the
virialization of the structure. All these ideas will be criticized in
\S 1.3.

\item 
{\bf Ellipsoidal collapse:} the collapse of a homogeneous ellipsoid is
described by a set of coupled ordinary differential equations; no
analytical solution is known for this, but the evolution equations can
be easily integrated numerically (White \& Silk 1979; Peebles 1980;
Monaco 1995; Bond \& Myers 1996a). At variance with the spherical
model (which is contained as a limiting case), the ellipsoidal model
takes the surrounding Universe into account by means of a global tidal
force. The ellipsoidal model has sometimes been used to describe the
collapse of structure; it will be described in detail in \S 3.2.

\item 
{\bf Peaks:} an idea in embryo in Doroshkevich (1970), and then
developed by Peacock \& Heavens (1985), Bardeen et al. (1986) and many
others, is that structures form in the peaks of the initial density
field, smoothed on a relevant scale.  Objects can then be modeled as
evolving from peaks, whose statistical properties are known. However,
the peak paradigm has been criticized both from the theoretical point
of view (see, e.g., Shandarin \& Zel'dovich 1989), and from the
numerical point of view (Katz, Quinn \& Gelb 1993, van de Weigaert \&
Babul 1994; but see Bond \& Myers 1996b for a different view). The
main criticism is that the peak paradigm neglects the important role
of tides, which are able to fragment peaks or make them merge with
neighboring peaks.

\item 
{\bf Eulerian perturbation theory:} linear theory is just the first
term of a perturbative solution of the evolution equations for the
matter field. Further-order terms give some information on the
non-linearly evolved matter field, such as the skewness or the
kurtosis\footnote{
Skewness and kurtosis of a distribution are suitably normalized
measures of the third and fourth moment of the distribution itself;
the skewness indicates the degree of asymmetry of a distribution, the
kurtosis the sharpening or the broadening with respect to the
Gaussian one.}
of the density contrast (see, e.g., Bouchet 1996).  Unfortunately,
Eulerian perturbation theory can describe just weak deviations from
linearity: it breaks down (all the contributions are of the same
order) as soon as the density contrast becomes comparable to one.

\item 
{\bf Zel'dovich approximation:} Zel'dovich (1970) proposed a model, in
which particles maintain their peculiar velocity (in comoving
coordinates!) and make straight paths through space. Then, structure
formation is a process similar to caustic formation in geometrical
optics. The main features of Zel'dovich approximation will be
described in \S 1.2.6 and \S 3.1; it is to be noted that it has become
a very simple and powerful tool to understand gravitational evolution
for density contrasts of order unity, the so-called {\it mildly
non-linear regime}.

\item
{\bf Lagrangian perturbation theory:} the Zel'dovich approximation is
the first term of a perturbative series; the perturbed quantity is not
the density, as in Eulerian perturbation theory, but the displacement
of the particles from the initial position. This change has
far-reaching consequences: Lagrangian perturbation theory retains its
validity in the mildly non-linear regime, extending in some important
cases up to first collapse. It will be deeply analyzed in
\S1.2.6 and \S 3.3.

\item
{\bf Other approximations:} a number of other approximate schemes to
describe the mildly non-linear evolution of perturbations have been
proposed; they have been reviewed by Sahni \& Coles (1996). Among
them, the {\it adhesion model} (Gurbatov, Saichev \& Shandarin 1986)
is based on the idea of sticking particles as their trajectories,
found by means of Zel'dovich approximation, try to intersect, to mimic
the formation of bound gravitational structures; the {\it frozen-flow
approximation} (Matarrese et al. 1992) holds the peculiar velocity
field constant in space, as if particles moved without inertia; the
{\it linear potential approximation} (Bagla \& Padmanaban 1994,
Brainerd, Scherrer \& Villumsen 1994) makes particles move in a
constant gravitational potential.

\item
{\bf BBGKY hierarchy:} all the approximations listed above attempt to
describe the linear or mildly non-linear regime of gravitational
clustering. The highly non-linear regime, which describes structure
evolution toward a relaxed equilibrium state (if reached), is much
more difficult to describe. An attempt in this direction was made by
means of the so called BBGKY hierarchy of equations (see, e.g.,
Peebles 1980), a statistical approach which has been successful in
describing plasma physics.  However, due to the long-range character
of gravitational forces, the BBGKY infinite hierarchy never closes;
some equilibrium ans\"atze have been used in literature, but the
matter is still debated.  In any case, all the proposed closure
ans\"atze apply to the strong clustering regime, where the two-point
correlation function is much larger than one; this happens only in the
inner parts of the formed structures, so these theories are not
relevant when attempting to describe the main global properties of
collapsed structures.
\end{itemize}

\section{Theoretical cosmology: a summary}

This section aims to sum up the main cosmological equations which are
relevant for subsequent discussions, to set up the notation and
terminology (which often varies from author to author), and to comment
on some points which will be relevant in the following.  A complete
presentation can be found in the textbooks cited at the beginning
of this chapter.

\subsection{Background cosmology}

It is assumed that our observable Universe is a perturbed
Friedmann-Robertson-Walker (hereafter FRW) one (with a possible
cosmological constant term); perturbations are assumed to be small at
recombination, at a redshift $z\sim 1500$.

FRW background cosmologies are justified by the observation that the
Universe is isotropic at large scales; if the metric is analytical,
this assumption, together with the cosmological principle, which
states that there are not preferred observers in the Universe,
suffices in demonstrating that the Universe is homogeneous (Weinberg
1972). A different view can be found in Pietronero, Montuori \&
Sylos-Labini (1996): if the matter distribution is non-analytical,
like a fractal or a multifractal, the observed isotropy does not imply
homogeneity.  Many observations, especially those of CMB, seem to
confirm large-scale homogeneity, even though the final demonstration
is still matter of debate (see, e.g., Peebles 1993, Davis 1996,
Pietronero et al. 1996).

The FRW metric can be obtained from Einstein's equations by assuming
the largest degree of spatial symmetry (invariance for translations
and rotations). A relevant quantity is the scale factor $a(t)$, which
describes how distances scale with time, as a consequence of Hubble
expansion. It is connected to the cosmological redshift by:

\be 1+z(t)=a(t_0)/a(t), \label{eq:redshift} \ee

\noindent 
where $t_0$ is the present time and the redshift $z$ is measured for
an object in our past light cone at the cosmological epoch $t$.  If
$\bar{\rho}$ is the background energy density and \res\ is a
cosmological constant term, the evolution equation for the scale
factor is the following:

\be \dot{a}^2 = \frac{8}{3}\pi G\bar{\rho} a^2 + \frac{1}{3} \Lambda a^2 -
R^{-2}, \label{eq:frw_general} \ee

\noindent 
where $R$ is the curvature of the Universe (it is infinite if the
Universe is flat, imaginary if the Universe is open), and the dot
denotes the time derivative. The normalization of the scale factor is
arbitrary; it is then normalized as:

\be a(t_0)=1. \label{eq:a_normal} \ee

To determine the evolution of $a(t)$, it is necessary to give an
equation of state for the matter or radiation present in the
Universe. After the recombination, the Universe is dominated by
non-relativistic pressureless matter, so that:

\be p=0, \label{pressure_less} \ee

\noindent and

\be \bar{\rho}(t) \propto a^{-3}. \label{eq:state_matter} \ee

The Hubble parameter is defined as: 

\be H(t)=\dot{a}/a; \label{eq:hubble_def} \ee 

\noindent 
the Hubble constant is then $H(t_0)=H_0$. If $\Lambda=R^{-1}=0$, the
Universe is flat, and is called Einstein-de Sitter; the background
density in this case is called critical density:

\be \bar{\rho}_{cr}(t) = \frac{3H^2}{8\pi G}. \label{eq:critical_density} \ee 

\noindent It is then convenient to define, for any FRW cosmology, the
following density parameter:

\be \Omega(t) = \bar{\rho}/\bar{\rho}_{cr}. \label{eq:omega_def} \ee

\noindent 
$\Omega=1$ denotes the Einstein-de Sitter cosmology; in this case it
remains constant in time. If $\Omega<1$, and the cosmological constant
is null or not large enough, then the Universe is open; if $\Omega>1$
(and there is no negative cosmological constant), the Universe is
closed.  The cosmological density parameter $\Omega_0$ is:

\be \Omega_0 = \Omega(t_0). \label{eq:omegazero_def} \ee

The three constants $H_0$, $\Omega_0$ and $\Lambda$, together with the
normalization of the scale factor, define uniquely the FRW background
model.  The Hubble constant is parameterized as usual:

\be H_0=h\ 100\ {\rm km\ s}^{-1}\ {\rm Mpc}^{-1}, \label{eq:hnot_value} \ee

\noindent 
where $h$ contains our ignorance on the real value of $H_0$ ($0.5 \leq
h \leq 0.8$).

In the following, only three families of FRW cosmological models will
be considered (the others are not of cosmological interest):

\begin{itemize} 
\item [1)] flat models with no cosmological constant: $\Omega=1$, 
$\Lambda=0$;
\item [2)] open models with no cosmological constant: $\Omega<1$,
$\Lambda=0$;
\item [3)] flat models with positive cosmological constant: $\Omega<1$,
$\Lambda\neq 0$, $\Omega+\Lambda/3H_0^2 = 1$.
\end{itemize}

In model (1), the FRW equation becomes:

\be \frac{\dot{a}^2}{a^2}=\frac{8}{3}\pi G\bar{\rho}, \label{eq:frw_flat} \ee

\noindent whose solution is:

\be a(t)=(t/t_0)^{2/3}. \label{eq:a_ev_flat} \ee

In model (2):

\be \frac{\dot{a}^2}{a^2}=\frac{8}{3}\pi G\bar{\rho}\left(1+
\left(\Omega_0^{-1} -1\right) a\right), \label{eq:frw_open} \ee

\noindent 
and the $a(t)$ evolution can be expressed through the following
parametric representation:

\bea a(\eta)&=&\frac{\Omega_0}{2(1-\Omega_0)}({\rm cosh} \eta -1) \\
     t(\eta)&=&\frac{\Omega_0}{2H_0(1-\Omega_0)^{3/2}}({\rm sinh}\eta -\eta).
\nonumber \label{eq:a_ev_open} \eea

In model (3):

\be \frac{\dot{a}^2}{a^2}=\frac{8}{3}\pi G\bar{\rho}\left(1+
\left(\Omega_0^{-1} -1\right) a^3\right), \label{eq:frw_cosm} \ee

\be a(t) = \left(\Omega_0^{-1}-1\right)^{-1/3}{\rm sinh}^{2/3}
\left(\frac{3}{2}\sqrt{\frac{\Lambda}{3}} t\right). \label{eq:a_ev_cosmo} \ee

\subsection{Perturbations}

The evolution of perturbations will be followed by approximating
cosmological matter as a perfect pressureless fluid.  This choice is
reasonable as long as the scales considered are much larger than the
mean interparticle distance. For dark matter perturbations, pressure
can be neglected if dark matter is cold (CDM); even for HDM, free
streaming which survives after recombination can be neglected at least
at the scales of galaxy groups and clusters.  Finally, as mentioned
above, Newtonian gravity will be used.

It is convenient to subtract the effect of Hubble expansion from the
evolution of perturbations. If \r\ is the physical coordinate, 
the comoving coordinates \x\ is defined as:

\be  \mr = a \mx. \label{eq:comoving_coo} \ee

\noindent In this way, a point comoving with the background has fixed
comoving coordinates \x. The peculiar velocity is defined as:

\be \mv = \dot{\mr}-H\mr = a \dot{\mx}. \label{eq:pecv_def} \ee

\noindent The density contrast is defined as:

\be \delta(\mx)=(\rho(\mx)-\bar{\rho})/\bar{\rho}. \label{eq:delta_def} \ee

\noindent 
This dimensionless quantity is bound to be larger than $-1$, while it
can grow up to infinite (very large) positive values.

In Newtonian cosmology, it is not possible to construct a meaningful
gravitational potential whose Laplacian gives the background density:
Newtonian gravity cannot provide a self-consistent cosmological
model. Nonetheless, it is possible to construct a peculiar potential
for matter fluctuations:

\be \nabla^2_\mx\Phi = 4\pi G (\rho-\bar{\rho}) a^2. \label{eq:poisson_old} \ee

\noindent 
This Poisson equation, which connects the peculiar potential to the
matter distribution, is one of the equations for the evolution of cold
matter perturbations. The subscript \x\ in the $\nabla$ operator indicates
that the differentiation is performed with respect to the comoving
coordinate.

The gradient of the peculiar potential gives the gravitational force
acting on fluid elements. The Euler equation of motion for a generic
fluid element is:

\be \frac{\partial \mv}{\partial t} + \frac{1}{a}
\left(\mv \cdot \nabla_\mx \right) \mv
+ \frac{\dot{a}}{a} \mv = - \frac{\nabla_\mx\Phi}{a}. \label{eq:euler_old} \ee

\noindent 
The last evolution equation is the continuity one:

\be \frac{\partial \delta}{\partial t} + \frac{1}{a} \nabla_\mx \cdot
(1+\delta) \mv = 0. \label{eq:continuity_old} \ee

\subsection{Eulerian linear theory}

When the density contrast is much smaller than one, $\delta\ll 1$, and
peculiar velocities are small enough to satisfy $(vt/d)^2\ll \delta$,
where $t$ is the cosmological time and $d$ is the coherence length of
the matter field, the system of equations (\ref{eq:poisson_old} --
\ref{eq:continuity_old}) can be linearized, leading to the equation:

\be \frac{\partial^2 \delta}{\partial t^2} + 2\frac{\dot{a}}{a}
\frac{\partial\delta}{\partial t}=4\pi G \bar{\rho}\delta
\label{eq:linear_evol}\ee

\noindent 
This second-order equation gives two solutions, which represent a
growing and a decaying mode. The growing mode, denoted by $b(t)$, is
the one of cosmological interest, as it is the one responsible for the
growth of small perturbations; it will always be assumed that the
decaying mode has already faded away.  In this case, the velocity
field is connected to the density contrast as follows:

\bea \nabla_x\cdot\mv &=& -a\frac{\partial\delta}{\partial t} \nonumber\\
\nabla_x \times \mv &=& 0 \label{eq:delta_pecv_rel} \eea

\noindent
in the sense that any difference from this relation fades away with
the decaying mode.  An interesting consequence is that any primordial
vorticity is damped out by linear evolution.

In the following, the solutions of Eq. (\ref{eq:linear_evol}) for the
growing modes, relative to the three background models of \S 1.2.1,
are reported:

$\Omega=1$:

\be b(t)=a(t). \label{eq:flat_growingmode} \ee

$\Omega< 1$, $\Lambda=0$: it is useful to use the time variable

\be \tau = (1-\Omega(t))^{-1/2} = \sqrt{(a(\Omega_0^{-1}-1))^{-1}+1}.
\label{eq:tau_def} \ee

\noindent Then:

\be b(\tau) = \frac{5}{2(\Omega_0^{-1}-1)}\left( 1+3(\tau^2-1) \left(
1+\frac{\tau}{2} \ln \left( \frac{\tau-1}{\tau+1} \right)\right)\right).
\label{eq:open_growingmode} \ee

\noindent 
Note that this $b(t)$ function saturates to the value
$5/2(\Omega_0^{-1}-1)$ at large times.

$\Omega<1$, $\Lambda\neq 0$, flat: it is useful to use the time
variable

\be h = {\rm coth}(3t\sqrt{\Lambda/3}/2). \label{eq:htime_def} \ee

\noindent Then,

\be b(\tau) = h \int_h^\infty (x^2(x^2-1)^{1/3})^{-1}dx . 
\label{eq:cosm_growingmode} \ee

\noindent
Growing modes are normalized so as to give $b(t)\simeq a(t)$ at early
times, and $a(t_0)=1$.

In the MF theory, collapse time estimates are often based on an
extrapolations of the linear regime to density contrasts of order one.
It is then convenient to define the quantity:

\be \delta_l\equiv \delta(t_i)/b(t_i). \label{eq:deltal_def} \ee

\noindent 
This is the initial density contrast linearly extrapolated to the
time at which $b(t)=1$, which, in an Einstein-de Sitter background, is
the present time; it will be used in next chapters.

Using the growing mode $b(t)$ as time variable, it is possible to
write the equations of motion (\ref{eq:poisson_old} --
\ref{eq:continuity_old}) in a more compact way. Defining the peculiar
velocity \u\ as

\be \muv = d\mx/db = \mv/a\dot{b}, \label{eq:u_def} \ee

\noindent the Lagrangian (convective) derivative $d/db$ as 

\be \frac{d}{db} = \frac{\partial}{\partial b} + \muv \cdot \nabla_\mx,
\label{eq:lag_deriv} \ee 

\noindent and the rescaled peculiar gravitational potential \pot\ as 

\be \varphi = 2a\Phi/3bH_0^2\Omega_0, \label{eq:rescpecpot_def} \ee 

\noindent 
the following system of equations is obtained:

\bea \frac{d\muv}{db} + \frac{3}{2b}\frac{\Omega}{f^2(\Omega)}\muv &=& 
    -\frac{3}{2b}\frac{\Omega}{f^2(\Omega)}\nabla_\mx \varphi 
    \label{eq:euler_new}\\
   \frac{d\delta}{db} + (1+\delta) \nabla_\mx \cdot \muv &=& 0 
	\label{eq:continuity_new} \\
   \nabla^2_\mx \varphi &=& \frac{\delta}{b}. \label{eq:poisson_new} \eea

\noindent 
The function $f(\Omega)=\dot{b}/Hb\simeq\Omega^{0.6}$ is defined in
Peebles (1980); note that $\Omega/f^2(\Omega)\simeq\Omega^{-0.2}$ is
weakly time-dependent.

\subsection{Spherical Collapse}

Spherical symmetry is one of the few cases in which gravitational
collapse can be solved exactly (Gunn \& Gott 1972; Peebles 1980).  In
fact, as a consequence of Birkhoff's theorem, a spherical perturbation
evolves as a FRW Universe with density equal to the mean density
inside the perturbation.

The simplest spherical perturbation is the top-hat one, i.e. a
constant overdensity $\delta$ inside a sphere of radius $R$; to avoid
a feedback reaction on the background model, the overdensity has to be
surrounded by a spherical underdense shell, such to make the total
perturbation vanish. The evolution of the radius of the perturbation
is then given by a Friedmann equation.

The evolution of a spherical perturbation depends only on its initial
overdensity. In an Einstein-de Sitter background, any spherical 
overdensity
reaches a singularity (collapse) at a final time:

\be t_c=\frac{3\pi}{2}\left(\frac{5}{3}\delta(t_i)\right)^{-3/2} t_i.
\label{eq:spherical_coll} \ee

\noindent
By that time its linear density contrast reaches the value: 

\be \delta_l(t_c)=\delta_c=\frac{3}{5}\left(\frac{3\pi}{2}\right)^{3/2}
\simeq 1.69. \label{eq:delta_c_sph}\ee

\noindent
In an open Universe not any overdensity is going to collapse: the
initial density contrast has to be such that the total density inside
the perturbation overcomes the critical density. This can be
quantified (not exactly but very accurately) as follows: the growing
mode saturates at $b(t)=5/2(\Omega_0^{-1}-1)$, so that a perturbation
ought to satisfy $\delta_l>1.69\cdot 2(\Omega_0^{-1}-1)/5$ to be able
collapse.

Of course, collapse to a singularity is not what really happens in
reality. It is typically supposed that the structure reaches virial
equilibrium at that time. In this case, arguments based on the virial
theorem and on energy conservation show that the structure reaches a
radius equal to half its maximum expansion radius, and a density
contrast of about 178. In the subsequent evolution the radius and the
physical density of the virialized structure remains constant, and its
density contrast grows with time, as the background density decays.
Similarly, structures which collapse before are denser than the ones
which collapse later.

Spherical collapse is not a realistic description of the formation of
real structures; however, it has been shown (see Bernardeau 1994b for
a rigorous proof) that high peaks ($> 2\sigma$) follow spherical
collapse, at least in the first phases of their evolution. However, I
will show in Chapters 3 and 4 that a small systematic departure from
spherical collapse can change the statistical properties of collapse
times.

Spherical collapse can describe the evolution of underdensities. A
spherical underdensity is not able to collapse (unless the Universe is
closed!), but behaves as an open Universe, always expanding unless its
borders collide with neighboring regions. At variance with
overdensities, underdensities tend to be more spherical as they
evolve, so that the spherical model provides a very good approximation
for their evolution.

\subsection{Lagrangian formulation}

The fluid equations described in \S 1.2.2 are based on the so-called
{\it Eulerian} formulation: the dynamical variables are defined in the
(comoving) real space \x. In other words, the dynamical variables are
measured by observers fixed in the comoving space (i.e. comoving with
the background). Alternatively, one can decide to describe the system
by means of observers comoving with the perturbed matter; this is the
{\it Lagrangian} formulation of fluid dynamics (see, e.g., Shandarin
\& Zel'dovich 1989). Note that the freedom of choosing among different
equivalent formulations corresponds to the (actually much wider) gauge
freedom of general relativity.

Let $\mq=\mx(t_i)$ be the position of a mass element (contained in a
vanishingly small volume centered on \q) at an initial time $t_i$; it
is defined as the {\it Lagrangian coordinate}. Note that the
Lagrangian coordinate can then be seen as a ``label'' which uniquely
identifies the mass elements. The trajectory $\mx(\mq,t)$ of the
element \q\ can be written in terms of a displacement \s\ from the
initial, Lagrangian to the final, Eulerian position:

\be \mx(\mq,t) = \mq + \ms(\mq,t). \label{eq:mapping} \ee

\noindent 
The problem of the evolution of a self-gravitating fluid can be
reformulated in terms of equations for the displacement field
$\ms(\mq,t)$.

A natural limit of the Lagrangian formulation (at least within an
analytical approach) lies in the following fact.  Eq.
(\ref{eq:mapping}) is a mapping from the space \q\ to the space \x.
In the dynamical evolution of cold matter, it can happen that two mass
elements get to the same point.  This event is called {\it orbit
crossing} (hereafter OC), or shell crossing (this expression is
taken from spherical collapse, where it describes different shells
of material crossing each other). OC has a number of consequences:

\begin{itemize}
\item 
The $\mq\rightarrow\mx$ mapping is not biunivocal from OC on, as
different (at least three) mass elements \q\ get to the same Eulerian
position \x. The situation after OC is called {\it multi-streaming},
and the places where it happens are called {\it multi-stream regions}.
\item 
The OC instant can mathematically be defined as the instant at which
the Jacobian determinant of the $\mq\rightarrow\mx$ mapping vanishes:

\be J(\mq,t)=\det\left(\frac{\partial x_a}{\partial q_b}\right)=0.
\label{eq:jacobian} \ee

\item 
Because of continuity, at OC the density formally goes to infinity;
this is called {\it caustic formation}. A complete description of the
geometry and topology of gravitational caustics, together with a
classification taken by Arnol'd's catastrophe theory and an
interesting connection to geometrical optics, can be found in
Shandarin \& Zel'dovich (1989).
\item 
While before OC the Lagrangian formulation can be used to get powerful
approximations for the mildly non-linear regime, after OC the
analytical approximations break down.
\item 
If a (subdominant) dissipative, cold component, e.g. a baryonic
component, is present, then shock waves in it will form at OC; this is
caused by the fact that the acoustic speed is very low at that moment.
\item 
If the initial field is perturbed at small scales, then OC will take
place soon after recombination, and multi-stream regions will
dominate.
\end{itemize}

The last point would suggest that the Lagrangian approach is, as it
stands, not very useful. However, as linear theory is supposed to
apply at very large scales, where small-scale nonlinearities can be
neglected, so the Lagrangian scheme can be applied to smoothed
versions of the initial field, in the hypothesis that small-scale
multi-streaming does not influence much the dynamics of larger scales.
This is probably the strongest hypothesis which is at the basis of
many dynamical approximations based on Lagrangian formulation.

The equations for the displacement field \s\ have been found by
several authors: Buchert (1989), Bouchet et al. (1995) and Catelan
(1995). As a matter of fact, all of these authors give equivalent but
formally quite different versions of the same system of equations; I
report the equations by Catelan (1995):

\be [(1+{\bf \nabla_\mq}\cdot\ms)\delta_{ab} - S_{a,b} + S^C_{a,b}] 
\frac{d^2 S_{b,a}}{d\tau^2} = \alpha(\tau)[J-1] \label{eq:lag1} \ee

\be \varepsilon_{abc}[(1+{\bf \nabla_\mq}\cdot\ms)\delta_{bd} - S_{b,d} 
+ S^C_{b,d}]\frac{dS_{c,d}}{d\tau} = 0\; . \label{eq:lag2} \ee

\noindent 
In these equations, the time variable $\tau$ defined in Eq.
(\ref{eq:tau_def}) has been used; the function $\alpha(\tau)$ is equal
to $6/(\tau^2+k)$ ($k=-1$, 0 or 1 for open, flat and closed models).
The quantities $\delta_{ab}$ and $\varepsilon_{abc}$ are,
respectively, the Kronecker tensor and the Levi-Civita antisymmetric
pseudotensor.  The tensor $S_{a,b}$ is commonly called {\it
deformation tensor}, $J$ is again the Jacobian determinant defined in
Eq. (\ref{eq:jacobian}), and $S^C_{a,b}$ is the cofactor matrix of
$S_{a,b}$; comma denotes partial derivative with respect to the
Lagrangian coordinate $q_a$. The first set of equations determines the
evolution of the \s\ field, while the second one is an irrotationality
condition in Eulerian space, which does not hold if the initial
displacement field has non-vanishing vorticity.

In this framework, the only dynamical variable is the displacement
field \s. The quantities defined in \S 1.2.2 can be expressed as
function of \s:

\bea 
{\bf v}(\mq,t)  & = & a(t)d\ms(\mq,t)/dt \label{eq:quantities_from_s}\\
1+\delta(\mq,t) & = & J(\mq,t)^{-1} [1+\delta(\mq,t_i)]\; , \nonumber
\eea

\noindent Here $t_i$ denotes again an initial time.

A defect of the evolution equations for \s\ is that it is difficult to
understand the behavior of the system by a qualitative analysis. An
alternative ``mixed'' formulation can be found by means of the system
of equations (\ref{eq:euler_new}--\ref{eq:poisson_new}).  In this
system time derivatives are Lagrangian, in the sense that the time
differential operator follows the motion of the mass element
considered. However, Eulerian-space derivatives are still present.  To
eliminate them, one can consider the space derivatives of the velocity
field \u\ as separate dynamical variables:

\be u_{a,b} = \theta\delta_{ab}/3 + \sigma_{ab} + \omega_{ab}.
\label{eq:shear_etc} \ee

\noindent 
The expansion $\theta$ is the divergence of the velocity field, and
gives the global expansion of the mass element centered on the point
considered; the shear $\sigma_{ab}$ is its traceless symmetrical part,
and gives the relative deformation of the principal axes; the
vorticity $\omega_{ab}$, its traceless antisymmetric part,
represents a global rotation.  By differentiating the Euler equation
with respect to the Eulerian space coordinate, one can obtain
``Lagrangian'' evolution equations (i.e. with non space derivatives)
for the density contrast, expansion (the Raychaduri equation), shear
and vorticity (see, e.g., Ellis 1971). It will be shown in \S 3.2 that
the evolution equation for the shear couples with the so-called tidal
tensor:

\be E_{ab} = \varphi_{,ab} - \nabla^2_\mx \varphi/3 \label{eq:tides} \ee

To close such a system of equations, it is necessary to find an
evolution equation for the tidal tensor. An attempt in this direction,
based on general relativity, can be found in Matarrese, Pantano \&
Saez (1993,1994) and Bertschinger \& Jain (1994), but its application
to Newtonian gravity has been criticized, for instance, by Kofman \&
Pogosyan (1995) and Matarrese (1996).  As a conclusion, such mixed
Eulerian--Lagrangian formulation, which is useful to understand many
characteristics of the gravitational problem, is not useful to
construct a self-consistent formulation of gravitational dynamics.

\subsection{Lagrangian perturbations and other approximations}

The Lagrangian system, Eq. (\ref{eq:lag1}) and (\ref{eq:lag2}), can be
perturbatively solved for small displacements. This has been done by
Buchert (1989), Moutarde et al. (1991), Bouchet et al. (1992), Buchert
(1992), Buchert \& Ehlers (1993), Lachi\`eze-Rey (1993a,b), Buchert
(1994), Bouchet et al. (1995) and Catelan (1995); see also Bouchet
(1996) and Buchert (1996) for reviews. The first order solution, for
suitable initial conditions (as given by the linear growing mode), is
the well-known Zel'dovich (1970) approximation:

\be \mx(\mq,t) = \mq - b(t)\nabla_\mq \varphi(\mq). \label{eq:zeldovich} \ee

\noindent 
This is equivalent to say that particle ``velocities'' remain constant in
``time'':

\be \frac{d\muv}{dt} = 0 \label{eq:zeldovich_bis} \ee

\noindent 
(although particles do accelerate in physical coordinates).  This is
not the place to list all the characteristics, merits and limits of
the Zel'dovich approximation; see Shandarin \& Zel'dovich (1989) for a
review. It is sufficient to recall that the ``truncated'' version of
this approximation, obtained by applying it to filtered versions of
the initial field, is very effective in describing the mildly
non-linear behavior of N-body density field (see, e.g., Coles, Melott
\& Shandarin 1993). After OC, the Zel'dovich approximation breaks down:
particles maintain their \u\ velocity, so that the collapsed
``pancakes'' are soon washed away by particle motions; in the real
problem collapsed regions typically remain bound.

The perturbative series has been calculated up to third order; here it
will be assumed that the decaying mode is not present, the initial
velocity field is irrotational and initial peculiar velocity and
acceleration are parallel (see Buchert 1989, 1992 and Buchert \&
Ehlers 1993 for a discussion).  It turns out that, at any order, the
terms can be written as a finite sum of terms which are separable in
space and time (Ehlers \& Buchert 1997). At the third order, the
perturbative term is divided into three separable modes. Then:

\bea \lefteqn{\ms(\mq,t) = b_1(t) \ms^{(1)}(\mq) +
b_2(t)\ms^{(2)}(\mq)+b_{3a}(t)
\ms^{(3a)}(\mq)} \label{eq:pert_series} \\ 
&& + b_{3b}(t)\ms^{(3b)}(\mq) + b_{3c}(t)
\ms^{(3c)}(\mq) + \ldots \nonumber\eea

First and second order solutions are irrotational in Lagrangian space,
i.e. the matrices $S^{(1)}_{a,b}$ and $S^{(2)}_{a,b}$ are
symmetric. The third-order contribution is made up of two irrotational
modes, 3a and 3b, a purely rotational one, as $S^{(3c)}_{a,b}$ is
antisymmetric (the reason for the presence of this rotational
component is that the $\mq\rightarrow \mx$ transformation is
non-Galileian; see the discussions in Buchert 1994 and Catelan 1995).

The solution of the time equations, which will be called time
functions, are accurately represented by the following expressions
(valid for the growing modes):

\bea
b_1 & = & -b(t)\nonumber\\ 
b_2 & = & -\frac{3}{14} b_1^2 \nonumber \\
b_{3a} & = & \frac{1}{9} b_1^3\label{eq:time_functions} \\
b_{3b} & = & -\frac{5}{42} b_1^3 \nonumber \\
b_{3c} & = & \frac{1}{14} b_1^3. \nonumber 
\eea

\noindent 
These expressions are exact for an Einstein-de Sitter Universe, and
very accurate for Universes with $\Omega\sim 1$ (see Bouchet et
al. 1995; Catelan \& Theuns 1996b).  It is then apparent that the use
of the growing mode as time variable allows us to factorize out the
explicit dependence of the displacement field on the background
cosmology.

It is convenient to express the $\ms^{(i)}$ corrections in terms of
scalar or vector potentials:

\bea
\ms^{(i)}  & = & {\bf \nabla} \varphi^{(i)},\ \ n=1,2,3a,3b 
\label{eq:potentials}\\
\ms^{(3c)} & = & {\bf \nabla} \times \varphi^{(3c)}.
\nonumber  \eea

\noindent 
Defining the principal and mixed invariants of one or two tensors as
follows:

\bea
\mu_1(A_{ab}) & = & {\rm tr}(A_{ab})=A_{aa}\nonumber \\
\mu_2(A_{ab},B_{ab}) & = & \frac{1}{2}(A_{aa}B_{bb}-A_{ab}B_{ab}) 
\label{eq:invariants}\\
\mu_2(A_{ab}) & = & \mu_2(A_{ab},A_{ab}) \nonumber\\
\mu_3(A_{ab}) & = & \det(A_{ab}), \nonumber 
\eea

\noindent 
the perturbative potentials can be written as the solutions of the
following Poisson equations:

\bea
\varphi^{(1)} & = & \varphi \nonumber\\
\nabla^2\varphi^{(2)} & = & 2\mu_2(\varphi^{(1)}_{,ab}) \nonumber\\
\nabla^2\varphi^{(3a)} & = & 3\mu_3(\varphi^{(1)}_{,ab}) 
\label{eq:space_functions} \\
\nabla^2\varphi^{(3b)} & = & 2\mu_2(\varphi^{(1)}_{,ab},\varphi^{(2)}_{,ab})
\nonumber \\
\nabla^2\varphi^{(3c)}_a & = & \varepsilon_{abc}\varphi^{(1)}_{,cd}
\varphi^{(2)}_{,db}. \nonumber
\eea

\noindent 
The first equality is just a consequence of initial conditions.  One
can recognize in the first perturbative term the Zel'dovich
approximation, Eq. (\ref{eq:zeldovich}).

It has been shown that the second-order term improves the predictive
power of the truncated Zel'dovich approximation in the mildly
non-linear regime, while the third-order correction does not seem to
introduce any significant improvement (Melott, Buchert \& Wei\ss\
1994; Buchert, Melott \& Wei\ss\ 1994).

The main problem with Zel'dovich and perturbative Lagrangian
approximations is that they break down after OC.  Many authors have
then tried to develop approximations which avoid OC or make particles
oscillate around pancakes. The main ones are the following (see Sahni
\& Coles 1996 for a complete review).

\begin{itemize}
\item 
{\bf Adhesion model} (Gurbatov, Saichev \& Shandarin 1989; see also
Shandarin \& Zel'dovich 1989): a vanishing artificial viscosity is
introduced in the Zel'dovich equation of motion, Eq.
(\ref{eq:zeldovich_bis}):
\be \frac{d\muv}{db} = \nu\nabla_\mx^2\mv. \label{eq:adhesion} \ee
As a consequence, particles move as in Zel'dovich approximation up to
caustic formation, but remain stuck as they try to cross each other.
Adhesion is able to describe the large-scale skeleton of structures.
Notably, it does not need any truncation, as no small-scale
multi-stream regions form. The evolution of the velocity field is
governed by a Burgers equation, a well known equation in mathematical
physics, and it is possible to find analytical solutions for it.
\item 
{\bf Frozen flow approximation} (Matarrese et al. 1992): the velocity
field, in the Eulerian space, is supposed to be constant in time,
\be \frac{\partial\muv}{\partial b} = 0; \label{eq:frozen_flow} \ee
this at variance with the Zel'dovich approximation, where the
Lagrangian time derivative of the velocity field vanishes. In
practice, particles move free of inertia, as tracers of a constant
velocity field. A consequence of this approximation is that OC never
occurs: particles slow down as they approach the pancakes, reaching
them only asymptotically.
\item 
{\bf Linear potential approximation} (or frozen potential; Bagla \&
Padmanabhan 1994; Brainerd, Scherrer \& Villumsen 1994): the peculiar
potential remains constant, as in linear theory:
\be \frac{\partial\Phi}{\partial b} = 0. \label{eq:frozen_pot} \ee
Then particles do cross each other, but, instead of going away from
the pancake, they oscillate around it.  Unfortunately, in this
approximation it is not possible to find analytical solutions; it is
useful to construct numerical codes which are faster then usual N-body
ones, as the potential is not recalculated at every step.
\end{itemize}

\subsection{Statistics}

The initial configuration of the matter field is usually given at
recombination; in standard cosmological scenarios, no perturbation can
grow non-linear before that moment. The initial perturbed density
contrast field is given as a random field, whose statistical
properties depend on the outcomes of inflation and on the kind(s) of
dark matter particles which are present. As mentioned in \S 1.1, this
random field is usually assumed to be Gaussian.

For a rigorous definition of random fields, see, e.g., the textbook by
Adler (1981). I give here a qualitative definition: a random field is
a field which can take random values at any point; the probability
distribution function (hereafter PDFs) of the values of the field at
any point is given, together with all the N-point joint PDFs. The
random fields considered here are homogeneous (the 1-point PDFs are
equal at all the points), isotropic, ergodic (ensemble averages are
equivalent to space averages over large-enough volumes; see Adler 1981
for a rigorous definition); moreover, they will be supposed to be
sufficiently well-behaved (continuous, many times differentiable). A
Gaussian random field is characterized by the fact that all the joint
N-point PDFs are multivariate Gaussians.

A complete description of the statistical properties of a Gaussian
field is given by its 2-point correlation function, or by its Fourier
transform, the power spectrum. If $\delta(\mx)$ is a Gaussian random
field, and

\be \tilde{\delta}(\mk) = \frac{1}{(2\pi)^{3/2}} \int d\mx \delta(\mx) 
{\rm e}^{-i\mk\cdot\mx} \label{eq:delta_kappa} \ee

\noindent is its Fourier transform, the power spectrum is defined as:

\be P(k) = \langle\tilde{\delta}(\mk)\tilde{\delta}(\mk')\rangle 
\delta_D(\mk+\mk'), \label{eq:power_spectrum} \ee

\noindent 
where $\delta_D$ is a Dirac delta function (note that, because
$\delta(\mx)$ is real, $\tilde{\delta}(-\mk)= \tilde{\delta}^*(\mk)$,
where $^*$ denotes complex conjugation). The power spectrum gives
information only on the modulus of the transformed density field; with
Gaussian statistics, phases are randomly distributed between 0 and
2$\pi$.  Specifying a non-Gaussian distribution is equivalent to
specifying a non-flat distribution for the phases.

In conclusion, if initial conditions are Gaussian, the primordial
power spectrum and the transfer functions are sufficient to specify
the subsequent dynamical evolution. The statistic of the evolved
density contrast field is however not Gaussian; to understand this,
note that the density contrast is bounded to be larger than $-1$, but
can grow as large as infinite. It has been shown that the 1-point PDF
of the evolved density field is nearly lognormal (its logarithm is
Gaussian), at least for some kind of initial conditions (Coles \&
Jones 1991; Bernardeau \& Kofman 1995).

Transfer functions for different scenarios have been published by many
authors; for instance, Bardeen et al. (1996) give good analytical
approximations for the CDM and HDM transfer functions.  However, it is
often useful, when testing particular dynamical procedures, to
consider simple power spectra, as power laws:

\be P(k) \propto k^n. \label{eq:power_law} \ee

\noindent 
$n$ is called {\it spectral index} (it is not connected in general to
the primordial one).  As most ``realistic'' spectra are gently curved
(in a $\log k$ -- $\log P(k)$ space), such power laws can be used to
approximate limited parts of the spectrum. In this case, an effective
spectral index can be defined as the logarithmic derivative of $P(k)$
at one point. For instance, the effective spectral index at very small
scales, if some cold dark matter is present, is $-3$, while at the
scales relevant for galaxy formation is about $-2$, rising to $-1$ or
more at galaxy clusters and superclusters scale. Note that if $n\geq
-3$ matter clustering proceeds from small to large scales
(hierarchical, bottom-up clustering), while for steeper spectra, as
the HDM one, large scale structure forms before small scale one
(top-down scenario).

As mentioned before, the two-point correlation function of the matter
field, $\xi(x)$, is simply the Fourier transform of the power
spectrum:

\be \xi(x) = \frac{1}{(2\pi)^{3/2}} \int d\mk P(k)
{\rm e}^{i\mk\cdot\mx}. \label{eq:corfun_def} \ee

\noindent 
Another important quantity, the density contrast variance (often
called {\it mass variance}) is connected to the power spectrum.  For
realistic power spectra, which have power at all scales, the mass
variance is usually infinite; however, if the field is smoothed by
means of a window function $W(\mx;R)$, which filters out all the
perturbations on scales smaller than $R$, the mass variance is:

\be \mres \equiv \sigma^2 = \int d\mk P(k) \tilde{W}(\mk;R).
\label{eq:mass_variance} \ee

\noindent 
Here $\tilde{W}(\mk;R)$ denotes the Fourier transform of the filter
function. The 1-point PDF of the density contrast, smoothed on the
scale $R$, is then:

\be \mpdfd = \frac{1}{\sqrt{2\pi\mres}}\exp\left(-\frac{\delta^2}{2\mres}
\right) d\delta. \label{eq:gaussian_onepoint} \ee

\noindent
If the power spectrum is a power law, as in Eq. (\ref{eq:power_law}),
then the mass variance scales with the filter radius as:

\be \mres \propto R^{-(n+3)}. \label{eq:var_radius} \ee

As a final remark, initial conditions can be given also in terms of
the initial peculiar potential, defined as in Eq.
(\ref{eq:rescpecpot_def}), which, being connected to the density
contrast through a Poisson equation (Eq. \ref{eq:poisson_new}), shares
with it the same statistical properties. In this case the effective
spectral index for the potential is connected to that of the density
contrast through the relation $n_\Phi = n_\delta -4$.

\section{The formation of collapsed structures}

Once the background geometry is fixed, the kinds of cosmologically
relevant particles are known, and the statistical properties (power
spectrum, phase distribution) of the matter field at recombination are
known, the subsequent evolution of the matter field is in principle
determined. In practice, the exact behavior of collapsed objects is
so complicated that it is hard to get precise predictions on
astrophysical objects such as galaxies.  The necessary introduction of
gas dynamics further complicates the problem, through a chain of
events which are very difficult to understand and model, from star
formation to feedback due to supernovae explosions, to the formation
of supermassive black holes. When such events take place, very small
scales (fractions of 1 pc) can directly or indirectly interact with
cosmological scales (several Mpc); then, to get any reliable
prediction, something like 10 orders of magnitude in scale have to be
taken into account at once.

In the following, the various phases of structure formation are
outlined.  As a matter of fact, structure formation is a complex
mixture of hierarchical, bottom-up clustering and top-down, pancake
fragmentation. In fact, it has been recently realized that, while
hierarchical clustering is definitely the most promising picture for
describing structure formation, large scales evolve approximately in a
top-down fashion, forming pancakes which fragment to give rise to
smaller structures.

\begin{itemize} 
\item 
{\bf Linear evolution:} at very large scales, or just after
recombination, perturbations are very small. They quietly evolve
according to linear theory; however, higher-order corrections, given
by Eulerian perturbation theory, soon become appreciable.

\item 
{\bf Mildly non-linear regime:} when perturbations reach density
contrasts of order one, their evolution is characterized by the
formation of low-contrast, sheet-like structures, commonly called
pancakes or walls, which give rise to filaments and knots at their
intersections.  Such structures form a network which fill the space,
giving rise to a sponge-like (or swiss cheese-like) topology. This
topological situation is recognizable in current redshift surveys:
large voids are surrounded by structures like the great wall, which
are characterized by density contrasts of order one; rich galaxy
clusters would correspond to the already collapsed knots.

\item 
{\bf First collapse:} as observed in \S 1.2.4, the mildly non-linear
evolution ends with caustic formation. The local geometry of this
collapse is usually pancake-like; however, the most massive objects,
which more or less correspond to the knots or to the peaks of the
initial density field, soon collapse along all the directions, giving
rise to spheroidal structures. The global geometry of the collapsed
regions is more complicated: pancakes and filaments can be still
collapsing while the knots have already collapsed along all
directions. Nonetheless, the sponge-like topology is just a transient:
all matter flows from walls to filaments and then to knots, which are
the only stable structures.

\item 
{\bf (Violent) relaxation:} when matter collapses to form a bound
structure, the mass elements which fall into the structure soon lose
memory of their previous dynamical status and tend towards some kind of
equipartition. This happens as a consequence of the collective
interaction of the falling mass elements with the whole gravitational
potential of the forming structure. This event, which is characterized
by short time scales, is called violent relaxation (Lynden-Bell 1967),
and is probably responsible for the relaxed state of clusters. It
leads to the equipartition of velocities, not of energies.

\item 
{\bf Substructure erasure, virialization:} clumps falling into a
collapsed and (violently) relaxed structure can maintain their
identity for a while. This transient ends up when the structure
reaches a global (thermo)dynamical equilibrium, which corresponds to
virialization.  This is caused by particle-particle interactions
(two-body relaxation); this kind of relaxation, which leads to
equipartition of energy, is slower than violent relaxation.  Note that
non-virialized matter continues to fall into the relaxing structure,
at least as long as the filamentary network is still present;
moreover, the high-density cores of small clumps can maintain their
identity inside almost virialized structures.  Observed galaxy
structures, such as groups and clusters, have probably undergone
violent relaxation (loose groups may not!), and are evolving toward a
fully virialized state, which has not yet been reached in many if not
most cases.

\item 
{\bf Gas heating and cooling:} gas does not affect the evolution and
collapse of dark matter, provided its density is much smaller than the
dark matter density. At OC, the gas gets shock-heated to very high
temperatures, then radiatively cools down.  In some structures, like
galaxy clusters, this hot gas (10$^8$ K) is visible in the X-ray band.
To form stars or other compact, observable objects, the gas, or at
least a component of its, has to cool to low temperatures. This takes
place if the density of the gas cloud is large enough for the cooling
(Compton cooling with the CMB, bremsstrahlung, molecular hydrogen
lines etc.) mechanisms to be effective.  Cooling time-scales can
explain why galaxies have a maximum size: gas cannot efficiently cool
in very large mass clumps, whose dynamical times are shorter than
cooling times.

\item 
{\bf Star formation, supernovae feedback:} if the gas is able to cool
to low enough temperatures, molecular hydrogen forms, and the gas
fragments into small clumps which give birth to new stars.  Depending
on the shape of the initial mass function of stars, some or many OB
stars can form; such stars soon become type II supernovae.  Notably,
this step of structure formation is one of the hardest to model.  The
interaction of gas with such energetic objects can strongly influence
the cooling history of gas: supernovae can sweep the gas away from
small halos, or create a background ultraviolet radiation able to
maintain the gas ionized.  Further particulars of these processes are
discussed in \S 5.1.2 and
\S 5.3.1.

\item 
{\bf QSO activity:} another important event, which couples very small
with large scale, is the formation of a supermassive black hole (SBH),
giving rise to QSO activity. In large dark matter clumps, gas can
dissipate its angular momentum, acquired during the mildly non-linear
evolution, and then collapse into a rather compact configuration. Such
an object can go through a chain of events, which soon ends up in the
formation of a SBH, with mass of order 10$^6$ -- 10$^9$ $M_\odot$.
Further in-fall of gas make the SBH grow in mass and emit a huge amount
of radiation at all wavelengths (QSO activity).  Such radiation can
strongly interact with the surrounding gas, influencing the properties
of the host galaxy or large-scale environment. SBH formation and
evolution is another step of structure formation which is quite
difficult to model.

\item
{\bf Further dynamical and chemical evolution of galaxies:} once
galaxies are formed, their dynamics determine the optical morphology:
if a differentially rotating disk forms, the galaxy will be observed
as a spiral, while if coherent rotation gets lost and a spheroid
forms, the galaxy will be observed as an elliptical.  Further
interactions with the surrounding environment can strongly influence
galactic internal dynamics (see, e.g., Giuricin et al. 1993).  On the
other hand, the chemical abundances of stars and intergalactic gas
evolve, due to supernovae explosions and galactic winds. The end
products of such events are present-day galaxies (see, e.g., the
review by Matteucci 1996).
\end{itemize}

Cosmological models are trying to go through this chain of events, to
predict the properties of galaxies, galaxy groups and clusters, QSOs
and other observable structures. Unfortunately, many of the steps
listed just above are not fully understood; future generations of
models will hopefully be able to get precise predictions on
astrophysical objects.

Nonetheless, as long as dark matter is concerned, gas dynamical events
can safely be neglected, in the hypothesis that baryons consitute a
sub-dominant matter component. As shown in this chapter, even this
drastically simplified problem is not easy to solve. In the following
chapters I will present the approaches proposed to address the MF
problem.  All these approaches are based on some approximation to
gravitational dynamics, of the kinds reported in \S 1.2. The simplest
and most used approximation in this context is a combination of linear
theory and spherical collapse: a linearly-extrapolated perturbation
which reaches a density contrast of order one is going to collapse;
this is correct for a spherical top-hat perturbation, in which case
the threshold density contrast is 1.69.

With such an approximation, the formation history of an object is much
simpler than what actually happens; in particular:

\begin{itemize}
\item 
as the top-hat perturbation collapses to a point at a certain instant,
collapse is well defined; all the geometrical complications of real
collapse are absent;
\item 
virialization is assumed to take place just after collapse; all
transients connected to the erasure of substructures are absent;
\item 
further accretion of material is supposed to be spherical, while
accretion takes place preferentially through filaments.
\end{itemize}

After many gravitational transients, collapsed and relaxed clumps
become more or less spheroidal; this happens when some kind of
equilibrium is reached and further matter infall is negligible, at
least in the inner parts of the clump. In this situation, a spherical
symmetry appears a reasonable approximation; however, this is not
caused by a global sphericity of collapse, but is a result of
gravitational equilibrium.

When the collapse of a realistic matter field, with no special
symmetry imposed, is considered, all the difficulties discussed above
appear.  In particular, the definition of collapse is not so
straightforward as in the spherical case. A reasonable choice is to
define collapse as the OC: in fact, caustics form at that instant, and
dynamics becomes ``complicated'' and ``interesting''.  However, it can
be argued that caustics correspond to the collapse along one axis,
while it would be better to wait for the collapse along all the three
axes. In this apparently reasonable objection two different things are
confused, namely the local and global geometry of collapse, which can
be quite different from each other. As an example, a matter element
collapsing onto a spherical peak (but not a top-hat) experiences a
{\it spindle} collapse, in which the two axes perpendicular to the
symmetry axis (the one pointing toward the center of the peak)
collapse, while the element is infinitely elongated along the symmetry
axis.  Nonetheless, matter accretes with perfect spherical symmetry on
the peak.

On the other hand, OC is a necessary condition for many events to
take place: infinite density, infinite (negative) expansion and
infinite shear take place (this is shown in \S 3.1), while violent
relaxation, virialization, gas heating and cooling and other
astrophysical events need OC to be triggered.  Finally, while
single-stream dynamics can be described by means of analytical tools,
multi-stream dynamics is very difficult to follow, at least at the
present state-of-the-art. Then, it is convenient to define collapse as
OC, with the confidence that the structures described in this way
constitute those {\it high-density environments} within which
astrophysical structures form.

%
%

\chapter[Mass function theory]{A not brief history of the mass function theory}

There is general consensus in setting the birth date of the MF theory
in 1974, when the seminal paper of Press \& Schechter (hereafter PS)
was published (surprisingly, the same PS formula can be found in
Doroshkevich 1967). That paper proposed a heuristic procedure, based
on linear theory, to obtain the distribution of the masses of
collapsed clumps.  That work inspired the fit for the galaxy
luminosity function proposed by Schechter (1976), but received a
limited attention for more than a decade. A real explosion of
attention to the MF theory started in 1988, when the first large
N-body simulations started to reveal a surprising adherence of their
results with the PS formula.  Many authors tried to extend the PS
procedure in many directions, or proposed different, alternative or
complementary procedures. These last years are witnessing a new wave
of interest, which has not yet been exhausted.

The chapter is organized as follows.  The PS procedure is described
and commented in \S 2.1.  Comparisons to N-body simulations are
reviewed in \S 2.2.  \S 2.3 reviews the works based on the excursion
set formalism, \S 2.4 the works based on the peak hypothesis, and \S
2.5 outlines different attempts to model an MF based on more realistic
dynamical arguments.

\section{The Press \& Schechter Mass Function}

It is surprising, in reading the PS paper, to find out that
their procedure was developed for a cosmological framework explicitly
different from the ``standard'' one presented in Section 1.2.
Inflation had not been proposed in 1974, so the perturbation spectrum
was a somehow {\it ad hoc} element in the theory.  Press and Schechter
aimed to analyze the case in which Poisson-distributed small-mass
seeds are responsible for matter perturbations. As a matter of fact
their analysis, which relies on the linear evolution of perturbations
(\S1.2.3), is applicable (at the same heuristic level) to the standard
cosmological scenario; their Poisson-distribution case would then
correspond to a white noise ($P(k) \propto k^0$) power spectrum, their
``minimum variance'' case (seeds on a perturbed lattice) to
$P(k)\propto k$. Efstathiou, Fall \& Hogan (1979) generalized the PS
procedure to general (flat) cosmological models with scale-free
(power-law) spectra. The PS procedure is presented here with the same
notation that will be followed throughout the paper.

Let's consider an Einstein-de Sitter Universe filled with cold
(pressureless) gas, with Gaussian initial perturbations; different
cosmologies will be analyzed in Chapters 3 and 4. In this case, the
linear growing mode (\S 1.2.3) is equal to the scale factor $b(t)
\equiv a(t)\propto t^{2/3}$.  The density contrast, linearly
extrapolated to the present (defined by $a(t_0)=1$), is denoted, as in
\S 1.2, by $\delta_l = \delta_i/a_i$ (the subscript $i$ denotes an
initial time).  Let \dl\ be smoothed at a scale $R$ with a filter
$W(\mx,R)$. It will be shown below that the shape of the window
function has a great importance; here $W$ will be supposed to be
a top-hat in real space:

\be W(\mx,R)\propto\theta(1-x/R), \label{eq:tophat} \ee 

\noindent 
where $\theta$ is the Heavyside step function. Let's call $\Lambda(R)
\equiv\sigma_{\delta_l}^2(R)$ (\S 1.2.7) the variance of the \dl\ field, 
when smoothed on the scale $R$. Both \res\ and $R$ can be used as {\it
resolution} variables; in the following the \res\ variable will be
mainly used.

It is assumed that a collapsed structure forms whenever \dl\ reaches a
given threshold \dc, of order one.  As seen in \S 1.2.4, spherical
collapse gives $\delta_c\simeq 1.69$; this is supposed to correspond
to the formation of a virialized dark-matter halo.  The fraction of
collapsed mass, at any resolution (\res\ value) smaller than \res, can
then be calculated as the probability of \dl\ being larger than the
threshold \dc:

\be \mimfr = \int_{\delta_c}^{\infty} P_{\delta_l} (\delta_l;
    \Lambda) d\delta_l. \label{eq:ps_integral} \ee

\noindent 
To express Eq. (\ref{eq:ps_integral}) in the mass variable M, it is
necessary to determine the mass associated to \res, or to the scale
$R(\mres)$. In the top-hat smoothing case, it is quite reasonable to
set:

\be M= 4\pi\bar{\varrho} R^3(\mres) /3. \label{eq:tophat_mass_res} \ee

\noindent 
A point collapsed at a scale $R$ can either be part of a structure of
size $R$ or, more generally, of a larger structure. Then, the number
density of structures of mass between $M$ and $M+dM$ --- the MF \dmfm\
--- is related to the derivative of the integral MF, \imfr\, through
the following ``golden rule'' (as named by Cavaliere, Colafrancesco \&
Scaramella 1991):

\be M n(M) dM = \bar{\varrho} \left|\frac{d\Omega}{dM}\right| dM = 
\bar{\varrho} n(\mres) \left| \frac{d\Lambda}{dM}\right| dM. 
\label{eq:ps_plain} \ee

Eq. (\ref{eq:ps_plain}) is the PS formula for the MF. It is possible
to recognize two relevant terms in the right-hand-side of that
equation: a ``dynamical'' and ``statistical'' term, $n(\mres)=
|d\Omega/d \mres|$,

\be n(\mres)d\mres=\frac{1}{2}\frac{\delta_c}{\sqrt{2\pi\mres^3}} 
\exp \left( -\frac{\delta_c^2}{2\mres} \right) d\mres, 
\label{eq:ps_ndires} \ee

\noindent which contains information on the collapse dynamics (through
the parameter \dc) and on the statistics of the initial field, and a
``cosmological'' term, $|d\mres/dM|$, which contains information on the
power spectrum.  If the power spectrum is a power-law, $P(k)\propto
k^n$, then $\mres \propto R^{-(n+3)} \propto M^{-(n+3)/3}$ (Eq. 
\ref{eq:var_radius}), and the PS MF can be analytically expressed as 
follows:

\be n(M)dM = \frac{\bar{\varrho}}{M_*^2} 
    \left(\frac{1}{\sqrt{2\pi}} \right) 
    \left( \frac{n+3}{6} \right) 
    \left( \frac{M}{M_*} \right) ^{\frac{n+3}{6}-2} 
    \exp \left( - \frac{1}{2}\left(\frac{M}{M_*} \right) ^{\frac{n+3}{3}} 
      \right) dM.
    \label{eq:ps_powerlaw} \ee

\noindent 
$M_*$ is the mass corresponding to the scale at which \res\ is equal
to \dc$^2$.  Eq. (\ref{eq:ps_powerlaw}) has the following
characteristic shape, shown in Fig. 2.1 (for $n=-2$ and 1): a power
law at small masses, with slope $(n+3)/6-2$ (which is $\sim 2$ when
$n\sim -2\div -1$), and a modified exponential cutoff at masses larger
than the typical mass $M_*$.  In Fig. 2.1 the effect of a change in
the \dc\ parameter from 1.69 to 1.5 is also shown: the large-mass part
is pushed to larger masses, while the small-mass part is slightly
lowered.  If the spectrum is not a power-law, as it happens in
realistic cases, Eq. (\ref{eq:ps_plain}) can be used to explicitly
calculate the MF. However, typical power spectra are very gently
curved, and can be well approximated by a power-law in restricted
ranges of scales. In this cases, the power-law expression given in Eq.
(\ref{eq:ps_powerlaw}) can still be used.

\begin{figure}
\begin{center}
\hbox{
\epsfig{file=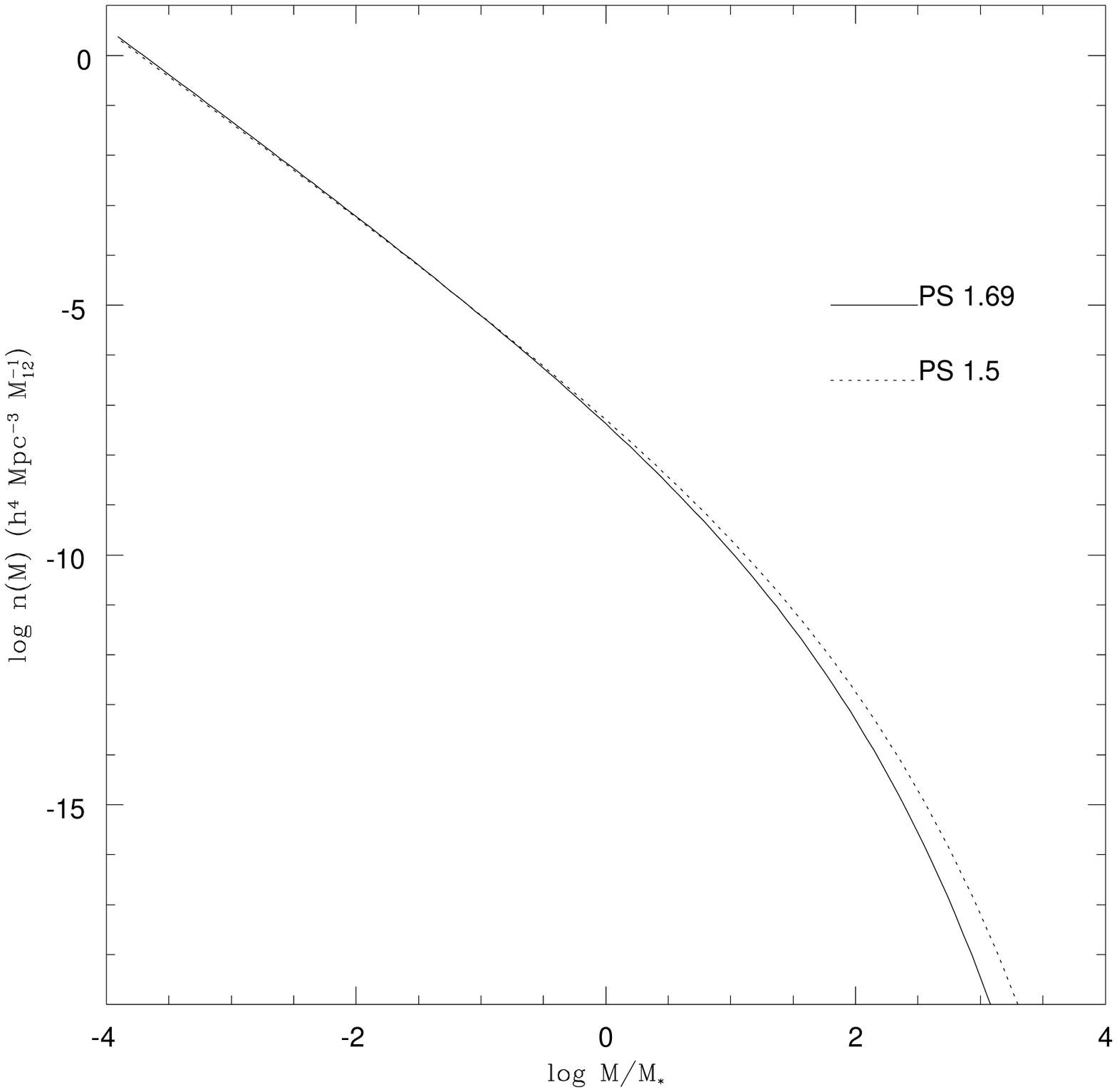,width=7cm}
\epsfig{file=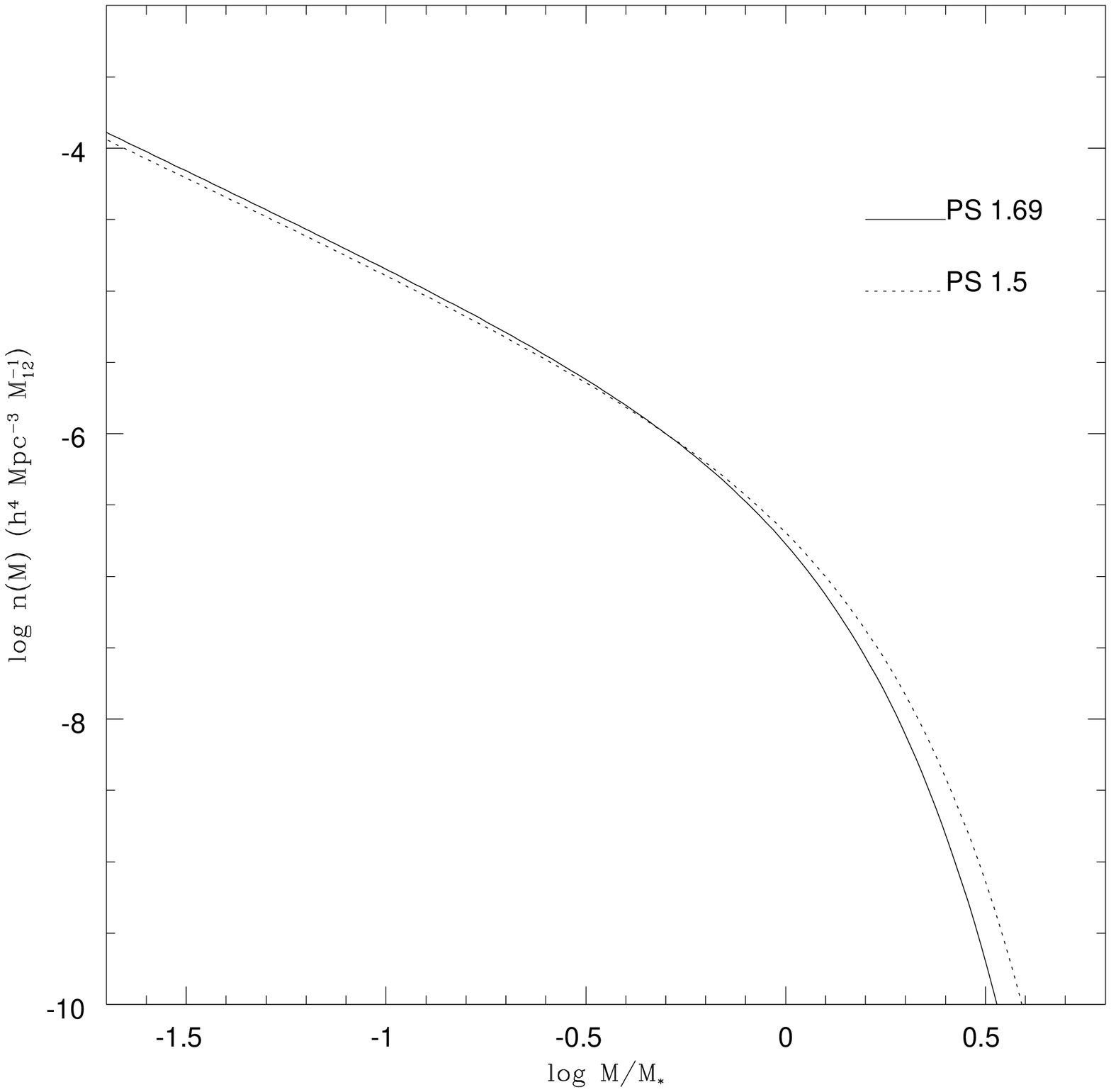,width=7cm}}
\caption{PS mass function: (a) $n=-2$, (b) $n=1$.}
\end{center}
\end{figure}

The PS theory has a number of problems, that can be summarized as
follows:

\begin{itemize}
\item 
{\bf Statistical problems:} in the limit of vanishing smoothing radii,
or of infinite variance, the fraction of collapsed mass, Eq.
(\ref{eq:ps_integral}), asymptotes to 1/2. This is a signature of
linear theory: only initially overdense regions, which constitute half
of the mass, are able to collapse. Nonetheless, underdense regions can
be included in larger overdense ones, or, more generally,
non-collapsed regions have a finite probability of being included in
larger collapsed ones; this is commonly called {\it cloud-in-cloud
problem}.  PS argued that the missing mass would accrete on the formed
structures, doubling their mass without changing the shape of the MF;
however, they did not give a true demonstration of that. Then, they
multiplied their MF by a ``fudge factor'' 2. Other authors (see
Lucchin 1989) used to multiply the MF by a factor $(1+f)$, with $f$
denoting the fraction of mass accreted by the already formed
structures.

\item
{\bf Dynamical problems:} the heuristic derivation of the PS MF
bypasses all the complications related to the highly non-linear
dynamics of gravitational collapse, described in \S 1.3. Spherical
collapse helps in determining the \dc\ parameter and in identifying
collapsed structures with virialized halos. However, the PS procedure
completely ignores important dynamical elements, such as the role of
tides and the transient filamentary geometry of collapsed structures.
Moreover, supposing that every structure virializes just after
collapse is a crude simplification: when a region collapses, all its
substructure is supposed by PS to be erased at once, while in
realistic cases the erasure of substructures is connected to the
two-body interaction of already collapsed clumps, an important piece
of gravitational dynamics which is completely missed by the PS
procedure.

\item
{\bf Geometrical problems:} to estimate the mass function \dmfm\ from
the fraction of collapsed mass at a given scale, \imfr, it is
necessary to relate the mass of the formed structure to the resolution
\res. The reasonable relation given in Eq.  (\ref{eq:tophat_mass_res})
can be considered only as an order-of-magnitude estimate. In practice,
the true geometry of the collapsed regions in the Lagrangian space
(i.e. as mapped in the initial configuration) can be quite complex,
especially at intermediate and small masses; in this case a different
and more sophisticate mass assignment ought to be developed, so that
geometry is taken into account. For instance, if structures are
supposed to form in the peaks of the initial field, a different and
more geometrical way to count collapsed structures could be based on
peak abundances.
\end{itemize}

Despite all of its problems, the PS procedure proved successful, as
compared to N-body simulations, and a good starting point for all the
subsequent works on the subject.

\section{N-body simulations}

Press and Schechter were the first to performed N-body simulations to
test the validity of their formula. They found some encouraging
agreement, but their simulations were limited to 1000 bodies, a very
small number to reach any firm conclusion.  Efstathiou, Fall \& Hogan
(1979) performed similar simulations, with the same number of point
masses, obtaining the same conclusions as PS.

Later, Efstathiou et al. (1988) compared the results of larger (32$^3$
P3M, scale-free power spectra) N-body simulations to the PS formula:
their dynamical range in mass was large enough to test the knee of the
MF.  The surprising result was that the PS formula nicely fitted their
abundances of simulated halos (as found by means of a percolation
friend-of-friend algorithm).  Further comparisons with N-body
simulations were performed by Efstathiou \& Rees (1988), Narayan \&
White (1988), Carlberg \& Couchman (1989), Carlberg (1990), Bond et
al. (1991), Brainerd \& Villumsen (1992), White, Efstathiou \& Frenk
(1993), Ma \& Bertschinger (1994), Jain \& Bertschinger (1994), Gelb
\& Bertschinger (1994), Katz, Quinn, Bertschinger \& Gelb (1994),
Lacey \& Cole (1994), Efstathiou (1995), Klypin \& Rhee (1994), Klypin
et al. (1995), Bond \& Myers (1996b).  Most authors reported the PS
formula to fit well their N-body results; nonetheless, all the authors
agree in stating the validity of the PS formula to be only
statistical, i.e.  the existence of the single halos is not well
predicted by the linear overdensity criterion of PS (see in particular
Bond et al. 1991).

There are however some exceptions to this general agreement: Brainerd
{\&} Villumsen (1992) reported their MF, based on a CDM spectrum, to
be very similar to a power-law with slope $-2$, different from the PS
formula both at small and at large masses.  Jain \& Bertschinger
(1994), Gelb \& Bertschinger (1994) and Ma \& Bertschinger (1994)
noted that, to make the PS formula agree with their simulations (based
on CDM or CHDM spectra), it is necessary to lower the value of the
\dc\ parameter as redshift increases.  The same thing was found by
Klypin et al. (1995), but was interpreted as an artifact of their
clump-finding algorithm. Recent simulations seem to confirm this trend
(Governato, private communication).

Lacey \& Cole (1994) extended the comparison to N-body simulations to
the predictions for merging histories of dark-matter halos (see \S
2.3); they found again a good agreement between theory and
simulations. This fact is noteworthy, as merging histories contain
much more detailed information about hierarchical collapse.

It is opportune to comment on two important technical points about
such comparisons.  First, the \dc\ parameters used by different
authors as ``best fit'' values range from the 1.33 of Efstathiou \&
Rees (1988) to the 1.9 found (in a special case) by Lacey \& Cole
(1994). The precise value of the \dc\ parameter is influenced by the
shape of the filter used to calculate the PS, Gaussian filters
requiring lower \dc\ values. Recent simulations tend to give
$\delta_c\simeq 1.5$ (e.g. Klypin et al. 1995) or $\delta_c=1.69$
(e.g. Lacey \& Cole 1994).  If \dc\ changes with time, a value 1.5
could be good at high redshifts, lowering to 1.7 at low redshifts.

Second, the numerical MF deeply depends on the way halos are picked up
from simulations. Typical algorithms, such as the friend-of-friend or
DENMAX, are parametric, i.e. they contain free parameters. For
instance, the frequently used friends-of-friends algorithm defines as
structures those clumps whose particles are separated among them by
distances smaller than a percolation parameter $b$ times the mean
interparticle distance. A heuristic argument, based on spherical
collapse, suggests a value of 0.2 for $b$ (with this percolation
parameter the mean density contrast of halos is about 180, which is
the expected density contrast of a virialized top-hat perturbation).
Obviously, the use of different $b$ parameters leads to different MFs.
In practice, what is obtained in this case is not ``the'' MF, but the
``friend-of-friend, $b$=0.2'' MF.  Then, the numerical MF contains
some hidden parameters, which, together with the \dc\ parameter (and
the mass associated to the filter in the Lacey \& Cole (1994) paper),
makes such comparisons more similar to parametric fits, rather than to
comparisons of a theory to a numerical experiment.

\section{Excursion set approach}

The terminology ``excursion set approach'' was introduced by Bond et
al.  (1991), to indicate that the MF determination is based on the
statistics of those regions in which the linear density contrast \dl\
is larger than a threshold \dc\ (such regions are called excursion sets
in the theory of stochastic processes; see, e.g., Adler 1981).  The PS
procedure is clearly included in this approach.  This section presents
those works which are based on the excursion set approach.

\subsection{Extensions of the PS formalism}

As mentioned before, the original PS work was developed within a model
in which structures grow from small seeds, either Poisson distributed
or set on a perturbed lattice. The extension of the PS work to more
general and ``standard'' cosmological settings was due to Efstathiou,
Fall \& Hogan (1979), who limited their analysis to power-law spectra
and an Einstein-de Sitter background.  The first application of PS
to a CDM spectrum was made by Schaeffer \& Silk (1985). They
``discovered' the PS MF (the procedure is described without any
reference to the PS paper), complete with its unjustified fudge factor
2, and criticized it in some interesting points; in particular, they
tried to model, on purely geometrical grounds, objects which do not
collapse spherically, like pancakes and filaments.

Starting from 1988, many authors extended the PS approach in many
directions, trying to understand why it appeared to work, in spite of
its heuristic and not fully satisfactory derivation.  The situation in
those years is reviewed in Lucchin (1989).  Just one year before,
Kashlinsky (1987) tried to determine the 2-point correlation function
of structures, collapsed according to the PS prescription.
Martinez-Gonzalez \& Sanz (1988a) performed similar calculations with
a different method, while Martinez-Gonzalez \& Sanz (1988b) attempted
to determine the luminosity function of galaxies of different
morphological types by means of an approach which was intermediate
between the PS and the peak one, described in \S 2.4.  Lucchin \&
Matarrese (1988) formulated a PS MF for the case of non-Gaussian
perturbations.  Note that, non-Gaussian perturbations can be both
primordial and arise as a natural result of gravitational dynamics:
the results of Lucchin \& Matarrese (1988) indirectly introduced some
dynamics in the MF theory.  Lilje (1992) and Lahav et al. (1991)
extended the PS result to open Universes and to flat Universes with a
cosmological constant.  Zhan (1990) changed he PS procedure by taking
into account the correction to the background density given by the
initial density contrast; as a matter of fact, such a correction is
negligible if the initial time is small.  Schaeffer \& Silk (1988a)
used the PS procedure to justify the presence of some small-scale
power in the HDM cosmology; in fact, the use of Gaussian smoothing
causes some large-scale power to be spread toward small scales.  Again
Schaeffer \& Silk (1988a,b), and Occhionero \& Scaramella (1988)
applied the PS formula to get many cosmological predictions on
collapsed structures.  Later on, the PS MF became a standard tool in
cosmology, some of its applications will be described in Chapter 5.

\subsection{The cloud-in-cloud problem}

As mentioned in \S 2.1, a region which is not predicted to collapse at
a scale can be included in a larger-scale collapsing one; this is not
taken into account by the PS formula. Although the name cloud-in-cloud
was given by Bardeen et al. (1986), the first one to recognize and
solve this problem was Epstein (1983, 1984). Schaeffer \& Silk (1988a)
addressed the problem, reaching the correct conclusion that, if Gaussian
smoothing is used, the cloud-in-cloud problem is not severe at large
masses.  The cloud-in-cloud problem was again recognized and solved
independently by Peacock \& Heavens (1990) and by Bond et al. (1991)
(see also the discussion in Efstathiou 1990). Their conclusions were
in agreement with those of Epstein (1983).

\begin{figure}
\begin{center}
\epsfig{file=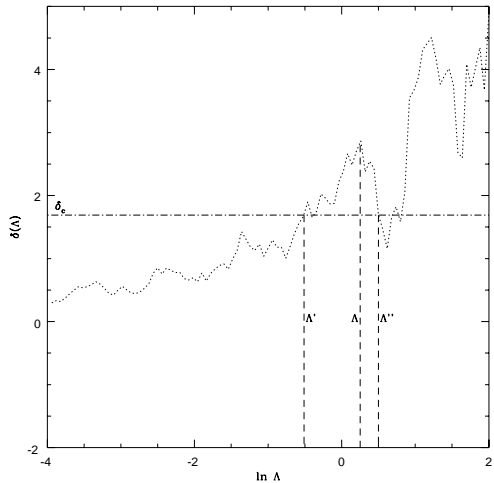,width=7cm}
\caption{The absorbing barrier problem.}
\end{center}
\end{figure}

The cloud-in-cloud problem has origin in the following inconsistency
of the original PS procedure. A collapse prediction is given to any
point (of the Lagrangian space; see \S 1.2.5) {\it for any resolution}
\res; in other words, a whole trajectory in the $\delta_l$--\res\
plane is given to any point, as in Fig. 2.2. Such trajectories start
from 0 at \res=0 (vanishing resolution implies complete smoothing, and
then vanishing density contrast), then wander around the plane,
eventually upcrossing or even downcrossing the threshold line
$\delta_l=\delta_c$.  When a trajectory lies above the threshold, the
point is assumed to be part of a collapsed region of radius $R'\geq
R(\mres)$; it is clear that the size $R'(\mres')$ of the formed
structure is connected to the point $\mres'$ of first upcrossing of
the trajectory. On the other hand, when a trajectory downcrosses the
barrier at a resolution $\mres''$, the point is interpreted as not
included in any region of size $>R''(\mres'')$ ($R$ decreases with
increasing \res), which is clearly in contradiction with what stated
above.

To solve the cloud-in-cloud problem, a point whose trajectory has
experienced its first upcrossing of the threshold line has to be
considered as collapsed at that scale, regardless of its subsequent
downcrossings. This can be done as follows: an absorbing barrier is
put in correspondence with the threshold line, so as to eliminate any
downcrossing event (Bond et al. 1991). Alternatively, a non-collapse
condition can be formulated as follows: a point is not collapsed at
\res\ if its density contrast {\it at any} $\mres'<\mres$ is below the
threshold (Peacock \& Heavens 1990).

The mathematical nature of the problem, and the resulting MF, strongly
depend on the shape of the filter.  For general filters, trajectories
are strongly correlated in \res, and then all the N-point correlations
of the process at different resolutions have to be known to solve the
problem. However, if the smoothing filter sharply cuts the density
field in the Fourier space, then independent modes are added as the
resolution changes, and the resulting trajectories are Gaussian random
walks. Such kind of filter is commonly called {\it sharp} $k$-{\it
space filter}; it will be referred to as SKS filter throughout the
text.

In the SKS case, the problem is suitably solved within the diffusion
framework proposed by Bond et al. (1991). Random walks, which are a
particular type of Markov processes (in stochastic language, they are
Wiener processes), are characterized by independent increments: if
$d\delta_l(\mres)$ is the increment of the process at \res, then

\be \langle d\delta_l(\mres)\; d\delta_l(\mres') \rangle = 
\delta_D(\mres-\mres'), \label{eq:random_walk} \ee

\noindent 
where $\delta_D$ is the Dirac delta function. The PDF of the
$\delta_l$ values can then be found as the solution of a Fokker-Planck
equation:

\be \frac{\partial}{\partial \mres} \mpdfl = \frac{1}{2} \frac{\partial^2}
{\partial \delta_l^2} \mpdfl. \label{eq:fokker_planck} \ee

If an absorbing barrier at \dc is inserted, then this equation has to
be solved with the constraint that the solution \pdfdu\ always
vanishes at \del=\dc. This solution is found and discussed in \S 4.2.
The integral MF can be written as:

\be \mimfr=1-\int_{-\infty}^{\delta_c} \mpdfdu d\delta_l, 
\label{eq:omega_cic} \ee

\noindent 
and this turns out to be exactly equal to the PS formula, including
the fudge factor 2. Notably, Epstein (1983) arrived to this same
result by considering a cosmological model analogous to that of the
original PS work, namely Poisson distributed seeds in a flat Universe.
His procedure, while formally different, is equivalent to the
diffusion problem with SKS smoothing.

Non-SKS filters change the results considerably; Fig. 4.10 (\S 4.3)
shows the deep difference between SKS and Gaussian trajectories.  To
find \pdfdu, all the N-point correlations of the $\delta(\mres)$
process have to be considered, and the problem becomes mathematically
intractable.  Qualitatively, the stability of the trajectories has the
consequence that, if a trajectory is just below the barrier, it cannot
easily jump above it as a result of noise. As a consequence, the PS
formula {\it without} the factor 2 is recovered at the large mass end
(as suggested by Schaeffer \& Silk 1988a), while the small mass part
is characterized by different, spectrum and filter dependent slopes.
Anyway, both Peacock \& Heavens (1990) and Bond et al. (1991) have
proposed reasonable and successful procedures to approximate the MF;
the procedure by Peacock and Heavens will be described in detail in \S
4.3.

Recently, in a paper by Jedamzik (1996), it was claimed that the
solution of the cloud-in-cloud given by Bond et al. (1991), in the SKS
case, is wrong.  Jedamzik analyzed the problem by means of a different
formalism, which is, as a matter of fact, very similar to that
proposed by Peacock \& Heavens (1990). However, because of a technical
mistake, Jedamzik's claim is not correct; this was shown by Yano,
Nagashima \& Gouda (1996).

A possible inconsistency of the diffusion model is the following: the
linear density contrast \dl, which is of order one, can easily become
smaller than $-1$, which would imply negative mass densities.  To
avoid this, Porciani et al. (1996) put either a reflecting or an
absorbing barrier at $\delta_l=-1$; their result was an increase of
large mass objects and a dramatic cutoff at small masses, at a second
mass scale, different from $M_*$. In practice, such a second barrier
was put to understand the possible effects of the introduction of
non-Gaussianity (which avoids $\delta_l<-1$) in the diffusion
formalism.

Another direct way to solve the cloud-in-cloud model is to construct
Monte Carlo realizations of the initial density field, and then apply
some multi-scale algorithm of clump identification. Two models of this
kind have been proposed, the ``block model'' by Cole \& Kaiser (1988),
and that of Rodriguez \& Thomas (1996). Such models have the advantage
of a satisfactory treatment of the geometry of collapsed regions in
Lagrangian space, but are somewhat limited in resolution and do not
provide analytical formulae for the MF. Anyway, they are dramatically
faster than an N-body simulation.

\subsection{Merging histories}

Merging histories are an important piece of information in the
formation history of dark matter objects. They are a natural outcome
of MF theories: a mass point, found in an object of mass $M$ at a
given time, will be found in another, more massive object at a
subsequent time; from the conditional probabilities connected with
such events it is possible to construct the statistics of accretion
and merging histories of collapsed structures.  This was attempted
first by Carlberg (1990), whose results are in contradiction with more
recent works outlined in the following. Bower (1991) constructed
merging histories by using the PS formalism, obtaining the same
results as that obtained by means of the diffusion formalism, which
are now reported.

Bond et al. (1991; see also Efstathiou 1990) proposed to use an
extension of the diffusion formalism, with two absorbing barriers, to
determine the merging histories. Following this suggestion, Lacey \&
Cole (1993) developed an analytical framework: the position of the
absorbing barrier changes with redshift according to (in an
Einstein-de Sitter background)

\be \delta_c(z)=\delta_c(z=0)(1+z). \label{eq:lacey_dc} \ee

\noindent
Then, the upcrossing history of a SKS trajectory above two barriers at
different positions is interpreted as a point mass belonging to two
structures of different mass at different times. The conditional
probability of such events is just:

\be P(\delta_l=\delta_c(z'),\mres'|\delta_l>\delta_c(z),\mres) d\mres'=
\frac{(\delta_c(z')-\delta_c(z))}{(2\pi)^{1/2} (\mres'-\mres)^{3/2}}
\exp\left[-\frac{(\delta_c(z')-\delta_c(z))^2}{2(\mres'-\mres)}\right]
d\mres', \label{eq:bower} \ee

\noindent with $\mres'>\mres$, and $\delta_c(z')>\delta_c(z)$.

To describe the accretion history of an object, it suffices to lower
the barrier in a continuous way, and follow the position in \res\ of
the first upcrossing point: if this performs a discontinuous jump
(which happens when the trajectory goes down and then up again), the
object containing the mass point considered suffers a discontinuous
merging with a structure of comparable size, while if the point moves
continuously the object is just accreting material.  It is clear that,
if the trajectory is a random walk, the first upcrossing point will
always perform discrete jumps; continuous accretion will be recognized
only if an (arbitrary) minimum resolution step is fixed.

Lacey \& Cole (1993) also proposed a Monte Carlo approach to simulate
ensembles of formation histories, based on realizing a large number of
random walks. Their Monte Carlo method for simulating merging
histories is commonly used to model the formation of virialized
galactic halos, in which gas dynamical is inserted ``by hand''. Such
Monte Carlo models of galaxy formation will be discussed in \S 5.1.2.
As a matter of fact, they found a weak inconsistency in their
formalism (a probability density going slightly negative), probably
caused by the simplicistic mass assignment, Eq.
(\ref{eq:tophat_mass_res}).  As mentioned in \S 2.2, the same authors
(Lacey \& Cole 1994) compared their results to N-body simulations,
finding an overall satisfactory agreement.

Finally, Sheth (1995,1996) obtained a complete analytical description
for the merging histories of objects formed from a Poisson
distribution of seed masses, the problem analyzed by the original PS
paper. The resulting MF has turned up to be the same as the
distribution function proposed by Saslaw \& Hamilton (1984).

\subsection{Space correlations}

If spherical collapse is considered, and if top-hat smoothing is
consistently used, then the density contrast in a point is just the
mean density contrast over a sphere, so that if that point is
predicted to collapse, all the points contained in the sphere are
going to collapse. Then, the collapse condition of a point is not
simply that of having a density contrast, smoothed on the scale
$R(\mres)$, larger than the threshold \dc, or, in other words, of
being {\it at the center} of a collapsing region, but that of being at
a distance smaller than $R$ from a point whose smoothed density contrast
is larger than \dc, or, in other words, of being {\it embedded} in a
collapsing region. In Monaco (1996a,b) I have called such reasoning
{\it global interpretation} of the collapse time.

This point was already contained in the surprising paper by Epstein
(1983), who considered points embedded in both spherical and
non-spherical regions.  More recently, Blanchard, Valls-Gabaud \&
Mamon (1992) and Yano et al.  (1996) proposed the same argument. The
former authors concluded that the resulting MF changes at the large
mass end, but is just slightly flatter than the PS one at small
masses.  The latter authors addressed the problem by explicitly
introducing the two-point correlation function into the collapse
condition (they used the formalism proposed by Jedamzik 1996). They
concluded, in agreement with Blanchard et al. (1992), that this new
element does influence the large mass part of the MF, while the small
mass part is very weakly affected.

More recently, Betancort-Rijo \& Lopez-Corredoira (1996) explicitly
tried to determine the distribution of sizes of excursion sets, thus
giving an important contribution to the treatment of the geometry of
collapsed regions in Lagrangian space; however, their procedure is
valid only for high ($>2\sigma$) thresholds.  In agreement with the
authors cited above, they adopted the global interpretation of
collapse times, by adding to the excursion sets a whole strip of width
$R$, thus containing all the points which are going to be included
into collapsed regions. They again obtained a mass function with many
more massive objects at large masses, probably too massive to agree
with N-body simulations.  They concluded that the global
interpretation cannot be applied to typical points of the Lagrangian
space, while it can probably be applied to peaks. Anyway, their
results can be used to obtain an analytical description of the block
model proposed by Cole \& Kaiser (1988).

\section{Peak approach}

An idea which traces back to Doroshkevich (1970) is to suppose that
structures form in the peaks of the initial density field.  This idea
became a standard paradigm in the framework of biased galaxy
scenarios, discussed in \S 1.1: Kaiser (1984) noted that high-level
peaks show an enhanced correlation with respect to the underlying
matter field, a fact which provided an explanation for the large
correlation length of clusters with respect to galaxies, and gave
freedom to tune the normalization of the CDM model to reproduce the
large-scale distribution of galaxies.  Peacock \& Heavens (1985) and
Bardeen et al. (1986) calculated a number of statistical expectation
values for the peaks of a Gaussian random field, as the number density
of peaks of given height. This quantity seemed suitable to determine a
peak MF, but two important problems, recognized by Bardeen et
al. (1986) (who did not attempt to determine an MF from the peak
number density) hampered such a determination: (i) it was not clear
which mass had to be assigned to a peak; (ii) the peak number density
was based on the initial field smoothed at a single scale, so the peak
MF suffered from the same cloud-in-cloud problem (the {\it
peak-in-peak} problem) as the PS one.

The excursion set and peak approaches are somewhat complementary.  In
fact, excursion sets are effective in determining the total fraction
of collapsed mass, and then the global normalization, but are not
accurate in deciding how the collapsed mass fragments into clumps,
i.e. to count the number of objects formed.  On the contrary, the peak
approach clearly determines the number of formed objects, but does not
determine the mass to be associated with the structures, and hence the
global normalization.

A peak MF can be formulated as follows. Let's denote by \npk\ the
number density of peaks with linear density contrast \dl\ (per unit
\dl\ interval); the initial field is assumed to be smoothed
at a resolution \res. Then, if $M_{\rm pk}(\delta_l,\mres)$ is the
mass associated to the peak, and if the $M$ variable is used in place
of \res\ (the uncertainty in the $\mres\rightarrow M$ relation can
be absorbed in the $M_{\rm pk}$ definition) the fraction of collapsed
mass can be written as:

\be \mimfm = \frac{1}{\bar{\rho}} \int_{\delta_c}^\infty d\delta_l
n_{\rm pk}(\delta_l,M) M_{\rm pk}(\delta_l,M), \label{eq:intmf_peak} \ee

\noindent 
where \dc\ is a density threshold for the peak. By using the same
``golden rule'' as in the PS approach (Eq. \ref{eq:ps_plain}), one
obtains:

\be n(M)dM = \left| \frac{d[n_{\rm pk}(\delta_c;M)M_{\rm pk}(\delta_c,M)]}
{dM} \right| dM.\label{eq:peak_plain} \ee

As a matter of fact, there is not a general agreement on the actual
validity of the peak paradigm. From the theoretical point of view,
structures are {\it not} predicted to form in the peaks of the initial
field; for instance, according to Zel'dovich approximation, structures
form in the peaks of the largest eigenvalue of the deformation
tensor. Then the peak paradigm can not be valid in general, except for
the highest peaks.  Some numerical simulations (Katz, Quinn \& Gelb
1993; van de Weygaert \& Babul 1994) have shown that galactic-size
peaks often disrupt or merge with larger structures, as a result of
tidal interactions with external structures. Manrique \&
Salvador-Sol\'e have argued that such results are due to the lack of
correction for the peak-in-peak problem. On the other hand, Bond \&
Myers (1996b) have found their peak-patch structures (see \S 2.4.2),
which account for the peak-in-peak problem, to represent well N-body
structures.

\subsection{The peak mass}

The first to use the peak MF, with $M_{\rm pk}$ given by the mass
inside the filter (and independent of \dc), were Carlberg \& Couchman
(1989), who compared it with their N-body simulations, finding a good
agreement; however, the peak MF is not much different from the PS one
in the range tested by those simulations. Besides, Bond (1989)
proposed an analogous formula to model the MF in a biased CDM
scenario.

Alternative, more motivated expressions for the mass of peaks were
proposed by a number of authors. Peacock \& Heavens (1985) modeled
the peak as a triaxial ellipsoid, and estimated its mass by means of
the volume of the ellipsoid within $\delta_l>0$.  Following a
suggestion by Bardeen et al. (1986), Hoffman (1988) and Ryden (1988)
determined the mass of the peak as that contained within the radius at
which the mean density profile equals its dispersion.  Colafrancesco,
Lucchin \& Matarrese (1989) also modeled the peak as an ellipsoid,
and estimated its mass by means of the volume inside the ellipsoid
surface with density larger that a given threshold. Finally, Peacock
\& Heavens (1990) estimated the mass of a (Gaussian-smoothed) peak as
that contained by a sphere which produces the same variance as that
obtained by a top-hat smoothing. They also noted that a peak mass
ought to satisfy the constraint of a global normalization of the MF.

All these reasonable definitions give rise to different MFs. To decide
which mass assignment is correct, two things ought to be done: (i) the
peak-in-peak problem ought to be solved, and then (ii) the mass
assignments ought to be compared to N-body simulations.  Attempts
in this direction are reported in next subsection.

\subsection{The peak-in-peak problem}

A point which is a peak on a given scale can be part of a larger-scale
(and larger mass) peak; as in the diffusion formalism, real structures
can be connected to those peaks whose density contrast just upcross a
given threshold when \res\ reaches a given value (and then are not
included in a larger peak). A first heuristic solution of the
peak-in-peak problem was given by Peacock \& Heavens (1990), with a
method very similar to that used to solve the cloud-in-cloud problem
with non-SKS filters.  Appel \& Jones (1990) calculated the MF by
means of the fraction of peaks whose height became smaller than the
threshold when the smoothing scale increases (and then \res\
decreases), an event which is analogous to the upcrossing in the
diffusion formalism.  This nearly solved the peak-in-peak problem, but
still some small mass peaks were found inside large mass ones. This can
be explained as follows: Appel \& Jones used Gaussian filtering, so
the $\delta(\mres)$ trajectories corresponding to the peak points were
quite stable (see \S 4.3); as a consequence, downcrossing events were
quite rare and could happen only at large resolution. They assigned to
their peaks the mass contained by the filter at the scale at which the
peak disappeared; note how this definition coincides with the global
interpretation of the collapse time. Moreover, they did not check the
overall normalization of their MF.

Following and extending the suggestions by Appel \& Jones (1990),
Manrique \& Salvador-Sol\'e (1995,1996) constructed a {\it confluent
system} formalism to describe the formation history of peaks. As in
Appel \& Jones, they considered those peaks which go below the
threshold at a given scale, but added the condition of not being
included in any larger-scale peak.  To their peaks they associated a
mass proportional to that of Appel \& Jones; the proportionality
constant was then fixed to guarantee the overall normalization.  They
constructed $\delta$--\res\ trajectories for the peak points
(correctly following their motion in Lagrangian space as \res\
changed), which could be interpreted as merging histories for the
peaks. A relevant point is that such merging histories make a clear
difference between continuous accretion (the peak remains well
defined) and discontinuous merging (some peaks are destroyed, a larger
peak forms); this is at variance with Lacey \& Cole (1993), who could
only distinguish between merging with small and large clumps. In my
opinion, this difference is due essentially to the use of different
filters (Gaussian for Manrique a\& Salvador-Sol\'e, SKS for Lacey \&
Cole), rather than to a difference between the peak and excursion set
approaches; moreover, as Manrique and Salvador-Sol\'e have recognized,
accretion is in practice merging with small-mass halos, and a physical
difference between the two events can be seen only in the effects that
merging has on the internal dynamics of structure.  As a final remark,
Manrique \& Salvador-Sol\'e (1995) found their MF to be very similar
to the usual PS one, but only if quite a large threshold (as large as
6 or more) is used.

Finally, Bond \& Myers (1996a,b,c) developed a Monte Carlo extension
of the peak formalism, called peak-patch formalism. They identified
structures by considering peaks of an initial field, filtered on a
hierarchy of scales. They identified the patch, which is going to
collapse with the peak, as the matter contained by a homogeneous
ellipsoid which can collapse along its three axes. The peaks were then
moved according to Zel'dovich approximation. This formalism has a
number of merits, as it correctly takes into account a number of
important dynamical event, as the effect of the shear on collapse
dynamics (through the ellipsoidal model, described in \S 3.2);
moreover, it has been found (Bond \& Myers 1996b) to reproduce
correctly the structures present in N-body simulations, but this time
not only from a statistical point of view.  This method is then able
to generate Monte Carlo object catalogs in a faster way than N-body
simulations, but is too complex for an analytical or semi-analytical
treatment.

\section{Dynamical models}

Both the excursion set and the peak approaches agree in identifying
collapsed structures as those regions whose linear density contrast
exceeds some threshold. But linear theory is not suitable to follow
the complicated behavior of collapsing matter; spherical collapse, on
the other hand, neglects important elements such as the role of tides.
Such simplifications have been seen in \S 1.3 to lead to oversimplified
and misleading arguments.  A number of authors have tried to insert
elements of realistic dynamics in the theoretical MF. Such attempts
are reviewed in the following.

\subsection{PS-like approaches}

Some authors have inserted elements of realistic dynamics in the MF
problem by extending the original PS approach or the diffusion or the
peak one.  The already cited paper by Lucchin \& Matarrese (1988) can
be considered as an attempt in this direction: they inserted
non-Gaussianity of dynamical origin in the PS procedure, with the
result of moving the exponential cutoff toward larger masses. Porciani
et al. (1996) justified their introduction of a reflecting (or
absorbing) barrier at $\delta_l=-1$ as a trick to introduce some
non-Gaussianity of dynamical origin in the diffusion formalism:
indeed, the negative density problem would be avoided if the true
non-Gaussianity of the evolved field were properly taken into
account. Their result was again an increase of large-mass objects,
together with an interesting low-mass cutoff.

The peak-patch formalism by Bond \& Myers (1996a), described above, is
also characterized by a more realistic description of the dynamical
evolution of peaks, even though structures are always identified
through the peaks of the linear field.  Other determinations of the MF
were proposed by Henriksen \& Lachi\`eze-Rey (1990), where collapsed
regions were identified by means of correlated velocity structures,
and by Newman \& Wasserman (1990) and Bernardeau \& Schaeffer (1991),
who related (in quite different ways) the MF to the correlation
properties of the matter field.

Finally, Monaco (1995) constructed a MF in a PS-like approach, based
on realistic collapse time estimates, found by means of extensions of
the Zel'dovich approximation, and by the use of the homogeneous
ellipsoid collapse mode. Again, dynamics led to the prediction of more
large-mass objects.  This work will be described in detail in Chapters
3 and 4.

\subsection{Time scales}

One of the problems with the PS approach, reported by Cavaliere et al.
(1991), is that it supposes matter clumps to instantaneously pass from
non-collapsed to collapsed, and to be immediately incorporated in a
larger clump. In other words, PS seems to imply vanishing time scales
for the formation and destruction of clumps. As a matter of fact, PS
simply does not contain any information on such timescales: the change
of the MF with time is a combination of {\it creation} of new clumps,
{\it destruction} of old clumps and {\it accretion} of mass onto
existing clumps (see the discussion reported in \S 2.4.2: accretion is
in practice merging with very small halos).  Such terms cannot be
disentangled by means of the PS approach alone, without further
assumptions: for instance, the ``static'' procedure proposed by Lacey
\& Cole (1993), which is based only on statistics, cannot provide a
precise definition of formation time (it is arbitrarily, though
reasonably, defined as the time taken by a clump to double its mass).

Cavaliere et al. (1991) proposed a {\it dynamical} procedure, based on
creation and destruction time scales, to model the MF. The evolution
equation for $n(M,t)$ (without the accretion term) is given by:

\be \frac{\partial n(M,t)}{\partial t} = \frac{n(M,t)}{\tau_+} - 
\frac{n(M,t)}{\tau_-}. \label{eq:ccs_evol} \ee

\noindent 
At variance with the approaches listed above (with the exception of
the formalism proposed by Jedamzik 1996 and used by Yano et al. 1996),
the definition of the mass function is implicit, and an evolution
equation is given for it. Note also that this evolution equation is of
the kind of a non-conservation equation: $\partial n/\partial
t=S(n(M))$, where $S$ is a source term.  They proposed the following
expressions for the time scales:

\bea \tau_+ & = & 2 t_c (M/M_c)^{-\Theta}/\Theta \nonumber \\
     \tau_- & = & 2 t_c /\Theta . \label{eq:time_scales} \eea

\noindent 
Here $M_c= M_c(z)$ is a typical collapsed mass scale, and $t_c=t_c(z)
=\dot{M_c}/M$ is a typical collapse time; both quantities are redshift
dependent, and can be estimated, to within a factor of order one, by
means of linear theory (as a consequence of self-similarity, if the
spectrum is scale-free or gently curved).  The formation time scale
exponentially cuts off at masses much larger than the typical mass,
while the destruction time scale is mass independent. The typical
behavior of such MF is a power-law at small masses, with slope
$-2+\Theta/2$, and a modified exponential cutoff at large masses,
$n(M)\propto \exp(-(M/M_c) ^\Theta)$.  If $\Theta=(n+3)/3$ is chosen,
then the typical behavior of the PS mass function is recovered.
Finally, accretion of mass at a rate $\dot{M}$ can be inserted in the
evolution Eq. (\ref{eq:ccs_evol}) by adding a term of the form
$\partial (\dot{M}n(M,t))/\partial M$ in the left-hand-side of Eq.
(\ref{eq:ccs_evol}).

Blain \& Longair (1993a,b) and Sasaki (1994) used a similar approach,
obtaining identical results. They imposed that the destruction time
scale has no characteristic time, then demonstrated that such time
scale is mass-independent, and explicitly derived the formation time
scale, by imposing Eq. (\ref{eq:ccs_evol}) to give the PS MF as a
solution.  In particular, Blain \& Longair (1993) found a numerical
solution, while Sasaki (1994) derived an analytical expression for the
formation time scale.

\subsection{Kinetic approach}

A completely different approach to the MF was proposed by Silk (1978)
and Silk \& White (1978). Aggregation (and fragmentation) of collapsed
clumps of similar size can be described by means of an aggregation
kinetic equation (Smoluchowski 1916; Ernst 1986), of the kind:

$$ \frac{\partial n(M,t)}{\partial t} = \frac{1}{2} 
\int_0^M K(M,M'-M,t) n(M',t)n(M-M',t)dM' $$ 
\be - n(M,t)\int_0^\infty K(M,M',t)n(M',t)dM'. \label{eq:smoluchowsky} \ee

\noindent 
The kernel $K(M,M',t)$ gives the probability that two mass clumps of
mass $M$ and $M'$ coagulate into one single clump of mass $M+M'$.  A
similar equation holds for the fragmentation of clumps.  The first
term in the right-hand-side of Eq. (\ref{eq:smoluchowsky}) gives the
number of objects of mass $M$ formed by coagulation of two clumps of
mass $M'$ and $M-M'$, while the second term gives the number of
objects lost by coagulation. The similarity of this equation to Eq.
(\ref{eq:ccs_evol}) is apparent: both are evolution equations for the
MF, and in both cases the definition of clump is implicit; an
important difference is that Smoluchowsky equation is non-linear in
$n(M,t)$.

The behavior of the MF given by Smoluchowsky equation has been
reviewed elsewhere (see, e.g., Lucchin 1989; Cavaliere, Colafrancesco
\& Menci 1991b; Cavaliere, Menci \& Tozzi 1994); here I list some main
points.  As a general remark, such MFs quickly lose their dependence on
initial conditions, their final shape depending only on the $K$
kernel.  Binary aggregations can evolve in two ways, through
geometrical collisions or focused, resonant interactions (Lucchin
1989; Cavaliere, Colafrancesco \& Menci 1992). The first mode is
prevalent in environments of moderate density, and can cause a
flattening of the MF, due to the coagulation of small clumps into
large ones (Cavaliere, Colafrancesco \& Menci 1992; Cavaliere \& Menci
1993). It will be shown in \S 5.1 that these merging events can
explain both the flatness of the luminosity function and the abundance
of blue galaxies at high redshifts.

The second mode is triggered in environments characterized by high
densities and velocity dispersions comparable to the internal
dispersions of subclumps, and leads to a {\it merging runaway},
i.e. to the formation of a single large clump whose mass is comparable
to that of the whole system; this could correspond to the formation of
cD galaxies through cannibalism of smaller galaxies, or to the erasure
of substructure in clusters (Cavaliere \& Menci 1991; Menci,
Colafrancesco \& Biferale 1993; see also Kontorovich, Kats \&
Krivitisky 1992). Menci \& Valdarnini (1994) have shown, by means of
N-body simulations, that both coagulation modes can be active in
large-scale galaxy filaments.

Recently, Shaviv \& Shaviv (1993, 1995) have analyzed the Smoluchowsky
equation (always in the gravitational context) in a different way; at
variance with Cavaliere and coworkers, they found their MF to depend
on initial conditions.  Another application of Smoluchowsky equation
in a cosmological context is due to Edge et al. (1990), to explain the
evolution of the X-ray luminosity function of galaxy clusters.

As a general remark, such kinetic approaches can describe those
aggregation events which take place between already collapsed clumps.
The direct hierarchical (first) collapse of structures remains well
described by a diffusion formalism, like the one proposed by Bond et
al. (1991). Cavaliere \& Menci (1994) proposed a formalism, based on
Cayley trees, to unify diffusion and aggregations. In practice,
aggregation events can be inserted in a diffusion formalism by means
of {\it branching} events, i.e. the splitting of a trajectory in two;
such branching event are controlled by a Smoluchowsky-like equation.
This formulation of the gravitational clustering problem has
interesting connections with the works by Sheth (1995;1996).

\subsection{Adhesion model}

The first determination of a mass function, based on a self-consistent
realistic dynamical approximation, is probably due to the works on the
adhesion model (see \S 1.2.6). Within this model collapsed structures
can clearly be identified with shocks in the collapsing medium;
further restrictions on (local!) collapse geometry can select
subclasses of collapsed regions, e.g. knots, if one wants to avoid to
count filamentary structures.  Some authors compared the predictions
of clump formation, as given by the adhesion model, to N-body
simulations, in 1D (Doroshkevich \& Kotok 1990; Williams et al. 1991)
and 2D (Nusser \& Dekel 1990; Kofman et al. 1992), finding
satisfactory agreement. 

Vergassola et al. (1994), attempted an analytical estimate of the MF
with adhesion, by using and extending a number of theorems
demonstrated by Sinai (see references in the cited paper).  They were
able to find the asymptotic dependences of the MF: it behaves
exactly like PS at large masses (the exact position of the typical
mass is not determined), but has a different slope at small
masses. However, they could not find the exact normalization of the
MF. Cavaliere, Menci \& Tozzi (1996) applied their Cayley trees
formalism to the adhesion model, and found that some branching (see
the previous subsection) is present in such model; moreover, they
found that the PS behavior is recovered at small masses if only
isotropic shocks (knots) are considered.

\subsection{Lagrangian perturbations}

In \S 1.2.6 the powerful Lagrangian perturbative formalism was
presented.  As will be shown later, this formalism can give reliable
dynamical prediction up to OC; as a consequence, it can be used to
predict the OC instant, which is defined as the collapse time. Then,
Lagrangian perturbations can be used to construct an MF fully based on
realistic dynamics. At variance with adhesion theory, this dynamical
approximation is of truncated type, i.e. small-scale power has to be
filtered out to avoid small-scale multi-stream regions. Moreover,
Lagrangian perturbations show interesting connections with the
homogeneous ellipsoid collapse model, which can also be used to give
reliable collapse time estimates, as in Monaco (1995). Such topics
have been addressed in Monaco (1996a,b), and will presented in full detail
in the next two chapters.

\chapter[Dynamics]{Dynamics and the Mass Function}

In chapter 1 a number of dynamical approximations and simplified
models of gravitational dynamics were presented. Such models were
based on two equivalent formulations of the gravitational problem,
namely the Eulerian and the Lagrangian ones. Clearly, Eulerian-based
approximations, such as Eulerian perturbation theory, frozen flow or
linear potential approximations, are suitable to describe what happens
to the matter field in the real, Eulerian space, no matter where the
mass contained in a point comes from. On the other hand,
Lagrangian-based approximations, such as Lagrangian perturbation
theory, or adhesion model, can describe the fate of mass elements
which come from a given point, giving both its deformation and its
trajectory. The Lagrangian formulation appears then suitable to
analyze a problem such as the MF one, where the deformation (and the
eventual collapse) of a mass element, which starts from a given
Lagrangian position, is wanted. In this sense, the MF is an
intrinsically Lagrangian quantity.

In this chapter and in the next a new MF theory, based on Lagrangian
dynamics, will be presented. This chapter addresses the problem,
purely dynamical in nature, of the determination of the collapse time
of a generic mass element in a continuous and differentiable
(smoothed) density field. \S 3.1 presents some attempts to estimate
collapse times by means of Zel'dovich approximation; the determinant
role of tides is also discussed. \S 3.2 Introduces the homogeneous
ellipsoid collapse model. \S 3.3 presents collapse time estimates
based on Lagrangian perturbation theory up to third order; the
connection of such theory with ellipsoidal collapse is analyzed, then
collapse times of generic mass elements in Gaussian fields are
calculated by means of Monte Carlo methods. In \S 3.4 the statistical
distributions of the inverse collapse times is determined; this is the
main outcome of this chapter, and will be used in next chapter to
determine an MF.  Finally, some important points are discussed in \S
3.5: the ``local'' and ``non-local'' nature of the dynamical
approximations used is discussed, and the ``punctual'' interpretation
of the collapse time estimates is stressed.  All the topics presented
in this chapter are contained in Monaco (1995;1996a), and discussed in
Monaco (1994;1996c-f).

\section{Zel'dovich approximation}

\subsection{The role of tides}

It has been shown in Chapter 2 that most MF theories proposed in the
literature are based at best on spherical collapse. It is then
interesting to understand which kind of dynamical interactions are
missed by this collapse model, and then which is the simplest way to
take them into account.

Spherical top-hat collapse is a truly local dynamical approximation:
the fate of a spherical perturbation is determined just by its initial
overdensity. In other words, the dynamical role of the whole Universe
outside the perturbation (according to Birkhoff's theorem) is assumed to
be negligible.  As mentioned in \S 1.2.5, it is possible to construct
a mixed Eulerian-Lagrangian system from the evolution equations of
fluid elements, Eqs. (\ref{eq:euler_new} -- \ref{eq:poisson_new}), by
decomposing the tensor of (Eulerian) space derivatives of the peculiar
velocity \u\ into an expansion scalar $\theta$, a shear tensor
$\sigma_{ab}$ and a vorticity tensor $\omega_{ab}$
(Eq. \ref{eq:shear_etc}).  In this way, the following evolution
equation for the density contrast can be obtained (see, e.g., Ellis
1971; here the growing mode $b(t)$ is used as time variable):

\be \frac{d^2\delta}{db^2}+4\pi G\bar{\rho}\frac{b}{\dot{b}^2}
\frac{d\delta}{db}  = \frac{4}{3} \left(\frac{d\delta}{db}\right)^2 
\frac{1}{(1+\delta)} + (1+\delta) \left(4\pi G\bar{\rho}
\frac{b}{\dot{b}^2} \delta + 2\sigma^2-2\omega^2\right) 
\label{eq:delta_evol} \ee

\noindent 
Here $\sigma^2=\sigma_{ab}\sigma_{ab}/2$ and $\omega^2=\omega_{ab}
\omega_{ab}/2$ (note that $\sigma^2$ in this context is not the
mass variance).  The fact that shear and vorticity enter through
positive definite quantities, with different signs, has two important
consequences: (i) shear always accelerates the collapse of a mass
element (see Bertschinger \& Jain 1994), (ii) vorticity always
decelerate it.  According to linear theory (\S 1.2.3), any vortical
mode is severely damped in the early gravitational evolution, and this
remains true during the mildly non-linear regime, up to OC (at OC
vorticity couples with the growing mode; see Buchert 1992). Then it is
reasonable to assume vanishing vorticity in the present framework. On
the other hand, the shear does not vanish in general: it provides the
link between the mass element and the rest of the Universe.

The evolution equation for the shear reads as follows:

\be \frac{{d\sigma}_{ab}}{db} + \frac{2}{3}\vartheta\sigma_{ab} + 
\sigma_{ac}\sigma_{cb} + 4\pi G\bar{\rho}\frac{b}{\dot{b}^2}\sigma_{ab} - 
\frac{2}{3}\sigma^2 \delta_{ab} = - 4\pi G\bar{\rho}\frac{b}{\dot{b}^2} 
E_{ab}. \label{eq:shear_evol}  \ee

\noindent 
The tensor $E_{ab}$, already defined in Eq. (\ref{eq:tides}),
represents the tidal interactions between the mass element and the rest
of the Universe. Then, tides are the relevant dynamical interaction
neglected by spherical collapse.

An important remark has to be made at this point. The equations just
presented describe the behavior of a vanishing mass element in a
smooth density field; any conclusion, including the calculation of
collapse times, is relative to mass elements; in Monaco (1996a) this
has been called {\it punctual} interpretation of collapse times. This
is at variance with spherical top-hat collapse, according to which the
whole spherical region which surrounds a given point, whose
overdensity is large enough, is going to collapse ({\it global}
interpretation; see \S 2.3.4). To understand the difference between
punctual and global dynamical descriptions, the example already
presented in \S 1.3 can be used: a mass element accreting on a
perfectly spherical peak, which is not a top-hat, experiences spindle
collapse (two axes collapse, one axis is shrunk to infinity). For such
mass element, the tidal interaction with the peak is determinant, even
though the global symmetry of the collapse is spherical.

\subsection{ZEL collapse times}

It is useful to find which is the simplest way to introduce tides in
the evolution of a mass element. It has been shown in \S 1.2.6 that
the simplest, realistic approximation of gravitational evolution in
the (mildly) non-linear regime is the Zel'dovich approximation; in the
following it will be referred to as ZEL.  With ZEL, the
Lagrangian-to-Eulerian mapping is written as follows:

\be \mx(\mq,t) = \mq - b(t)\nabla_\mq \varphi(\mq). \label{eq:zel_bis} \ee

\noindent 
The evolution of all the kinematic and dynamical quantities relative
to the mass element can be found by means of the ZEL deformation
tensor, $S^{(1)}_{a,b}=\varphi_{,ab}$, and Eqs.
(\ref{eq:quantities_from_s}).  The ZEL deformation tensor is obviously
symmetric, and can then be diagonalized. If \luno, \ldue and \ltre are
its three eigenvalues, ordered according to:

\be \lambda_1\geq\lambda_2\geq\lambda_3 \label{eq:lambda_order} \ee

\noindent 
(note that, because of Poisson equation, \luno+\ldue+\ltre=$\delta_l$),
then the Jacobian determinant $J$ of the $\mq\rightarrow\mx$
transformation, the density contrast $\delta$, the expansion $\theta$
and the shear $\sigma_{ab}$ evolve as follows:

\bea J(\mq,t) & = & (1-b(t)\lambda_1)(1-b(t)\lambda_2)(1-b(t)\lambda_3)
\label{eq:varie_zel_evol}\\
\theta(\mq,t) & = & - \frac{\lambda_1}{1-b(t)\lambda_1} - 
\frac{\lambda_2}{1-b(t)\lambda_2}-\frac{\lambda_3}{1-b(t)\lambda_3}\nonumber\\
\sigma_{ab}(\mq,t)  & = & {\rm diag}\left(\frac{\lambda_1}{1-b(t)\lambda_1}-
\frac{\theta}{3}, \frac{\lambda_2}{1-b(t)\lambda_2}-\frac{\theta}{3},
\frac{\lambda_3}{1-b(t)\lambda_3}-\frac{\theta}{3}\right)\nonumber\\
\delta(\mq,t)  & = & ((1-b(t)\lambda_1)(1-b(t)\lambda_2)
(1-b(t)\lambda_3))^{-1} -1.
\nonumber \eea

In the equation for $\delta$, the initial density contrast has been
neglected with respect to one. Note also that the density reported
here is given by the continuity equation. As a matter of fact, it is
possible to find another expression for the density (sometimes called
{\it dynamical} density, in order to distinguish it from the {\it
continuity} density given above), by combining Poisson equation with
the equality $\muv = -\nabla_\mq\varphi$; valid for ZEL.  It is:
$\delta_d = - b(t)\theta$ (see, e.g., Shandarin, Doroshkevich \&
Zel'dovich 1983). This is a symptom of the fact that ZEL is not a
consistent solution of the dynamical problem; on the other hand, the
two densities are very similar at early times, and their difference
remains finite at caustics, so it is not important for the present
purposes to use one or the other density.

When $b(t)=1/\lambda_1$, caustic formation takes place: the Jacobian
determinant vanishes, and all the other quantities go to infinity.  It
is then quite reasonable, from the point of view of the mass element,
to define such instant as the collapse time. As already mentioned in
\S 1.3, hereafter collapse will always be defined as the OC event.
Note also that, after OC, the ZEL evolution along the second and third
axes is not really meaningful, as ZEL does not work after OC.

It is then possible to give collapse time estimates for any mass
element.  Initial conditions are ``locally'' given by the three
$\lambda$ eigenvalues, but the evolution is not physically local, as
initial conditions contain non-local information about tides.
It is useful to define the following variables:

\bea x & = & \lambda_1 - \lambda_2 \label{eq:x_and_y} \\
     y & = & \lambda_2 - \lambda_3, \nonumber \eea

\noindent 
and to use the growing mode $b(t)$ as time variable.  Moreover, it is
possible to consider regions with linear initial density contrast
$\delta_l=1$ or $-1$, as all the other cases can be obtained by a
simple rescaling of $b$. The ZEL collapse time is finally given by:

\be b_c^{ZEL} = \frac{3}{\delta_l+2x+y} \label{eq:zel_bc} \ee

\begin{figure}
\begin{center}\hbox{
\epsfig{file=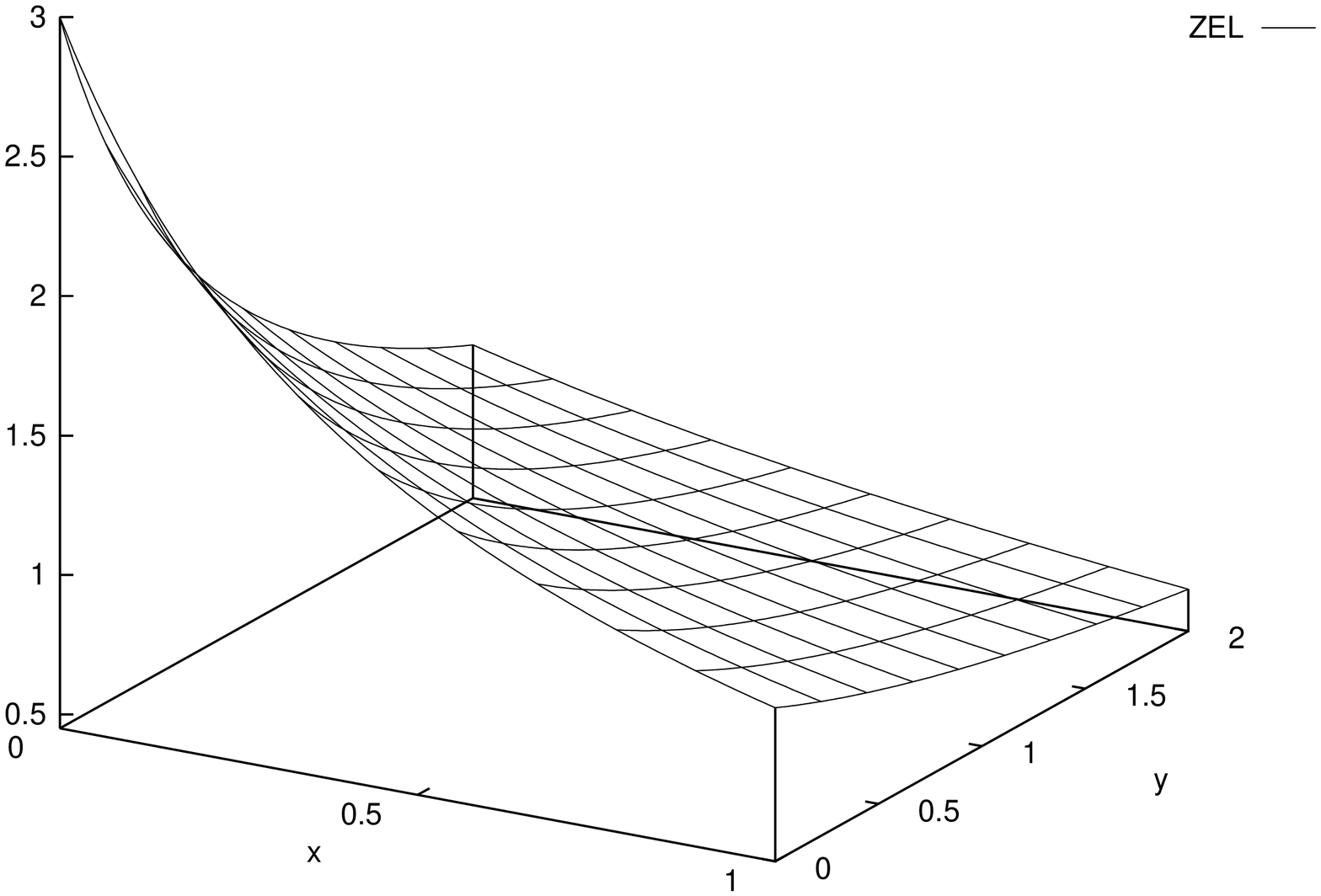,width=6cm,angle=-90}
\epsfig{file=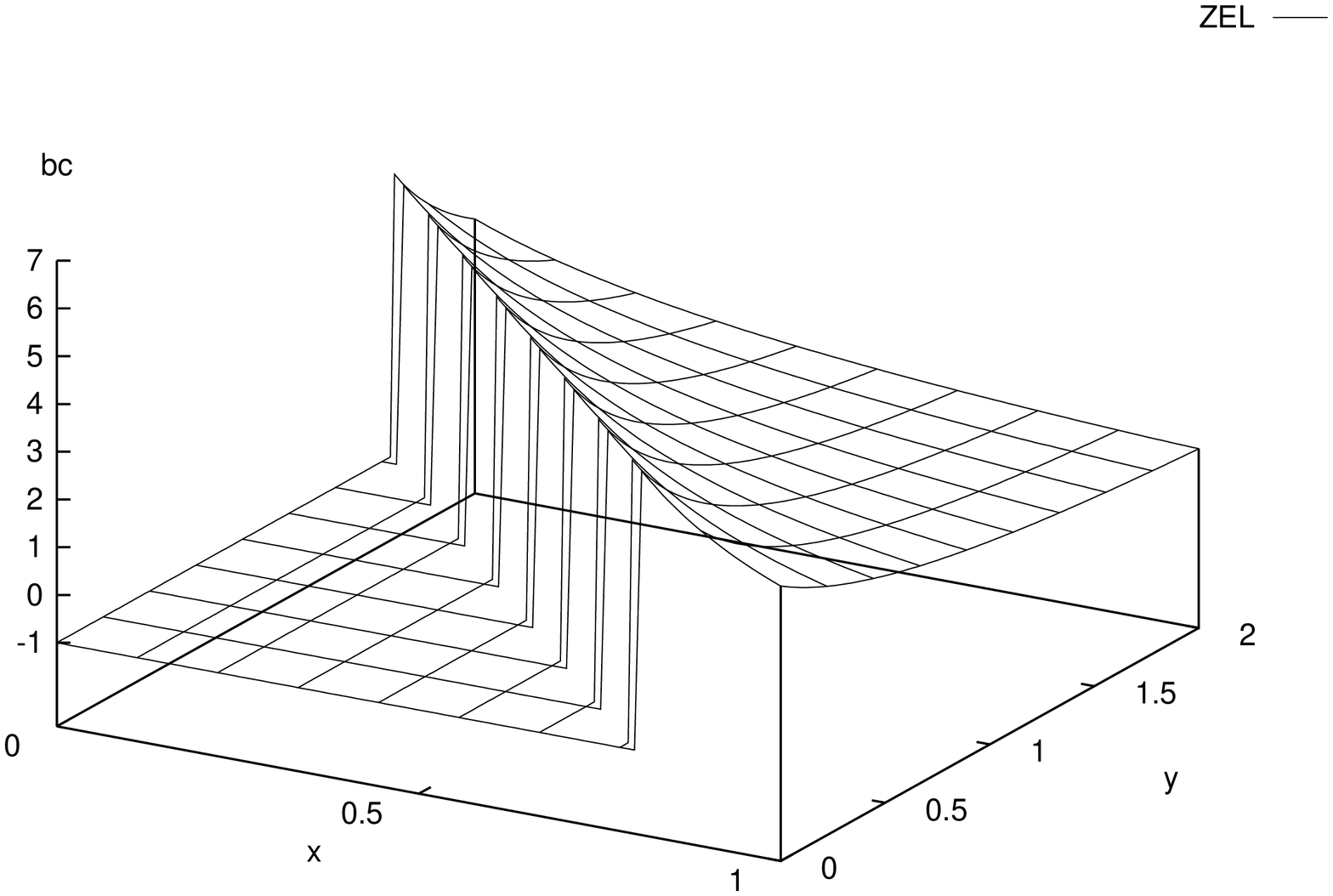,width=6cm,angle=-90}}
\caption{Collapse times with ZEL.}
\end{center}
\end{figure}

Fig. 3.1 shows the collapse time curves $b_c(x,y)$ for $\delta_l=1$
and $-1$. A problem is soon apparent: in the spherical case, when
$x=y=0$, the collapse time is $3/\delta_l$, instead of the exact
$1.69/\delta_l$ value. This discrepancy of nearly a factor of two is
easy to understand: ZEL is an exact solution (before OC of course) in
one dimension (see, e.g., Shandarin \& Zel'dovich 1989), and is then
able to describe the collapse of pancake-like structures, while it
severely underestimates the collapse rate in spherical symmetry.

A way to overcome this problem is to try some simple {\it ansatze} for
the ``true'' shape of the collapse time curve. In practice, a truly
realistic collapse time curve will depend not only on the $\lambda$
eigenvalues, but also on the values of the density (or potential)
field in all the points of (Lagrangian) space. However, as long as ZEL
gives a first-order description of gravitational collapse, it is
possible to think to a first-order collapse time which requires the
same initial conditions as ZEL but reduces to the correct spherical
value in the relevant limit.  In next subsection it will be shown that
the homogeneous ellipsoidal model provides such a collapse time; in
the meantime it is useful to consider simple variations of the ZEL
collapse time.

\begin{figure}
\begin{center}
\hbox{
\epsfig{file=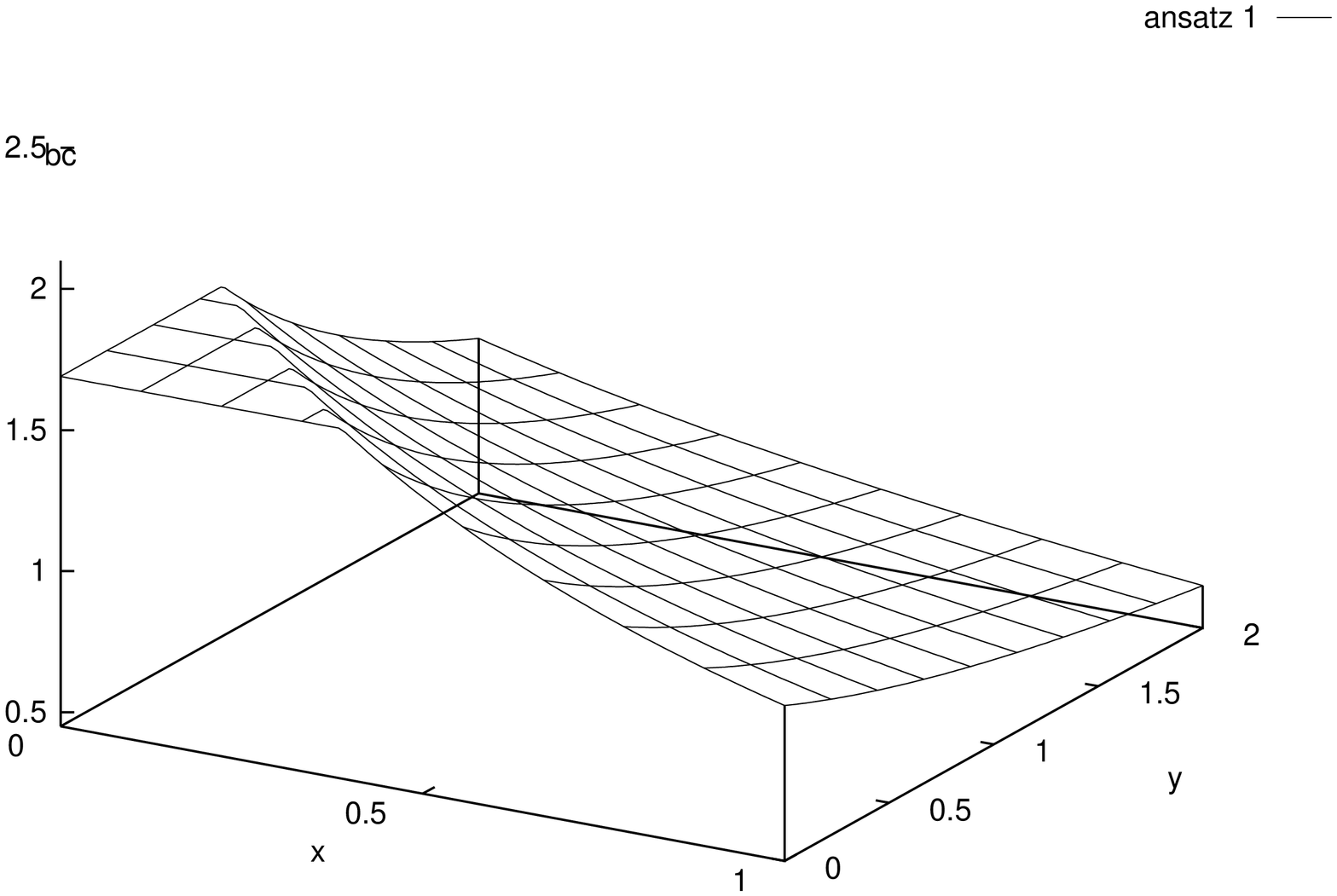,width=6cm,angle=-90}
\epsfig{file=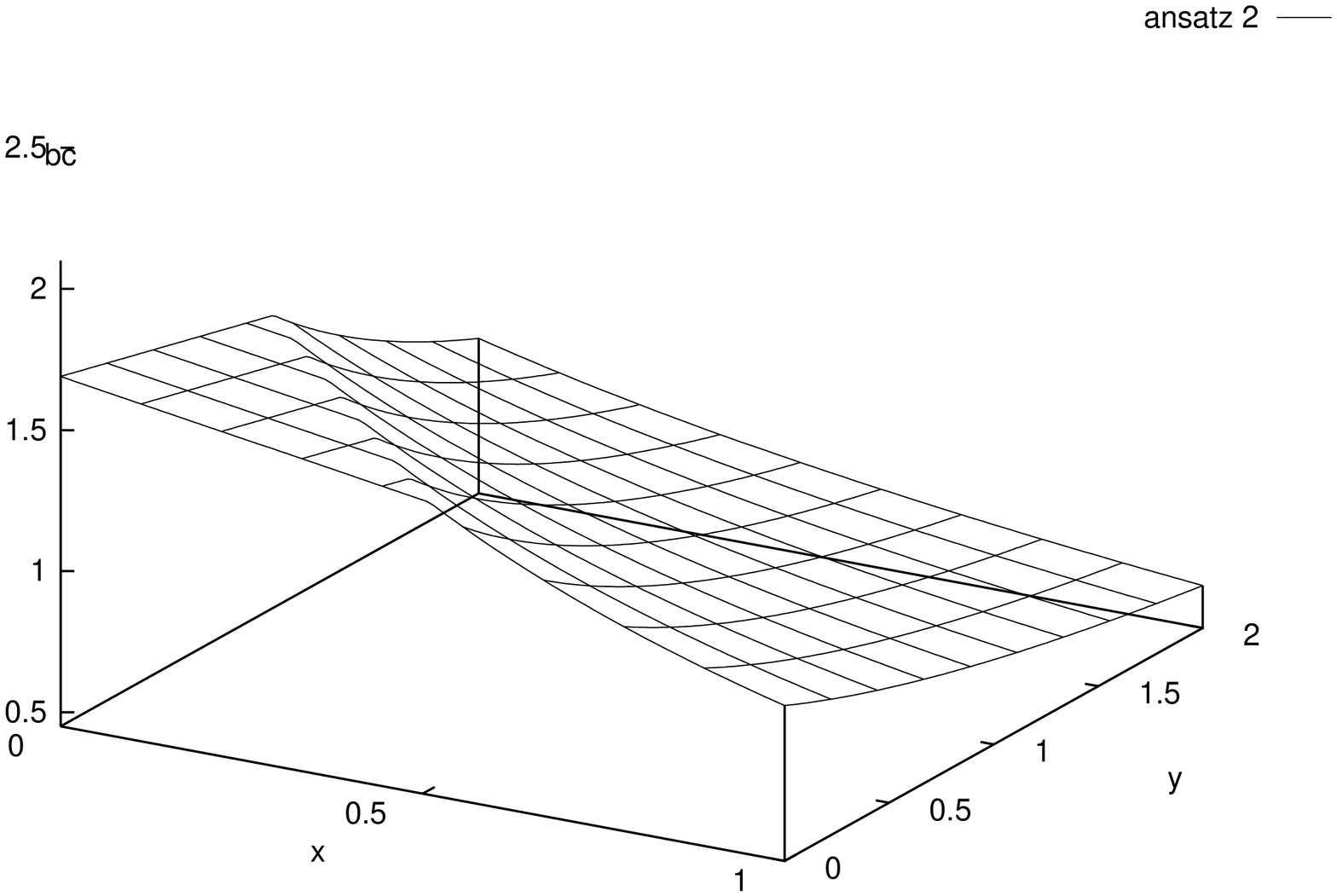,width=6cm,angle=-90}
}
\caption{Collapse times with the two {\it ansatze}.}
\end{center}
\end{figure}

According to Eq. (\ref{eq:delta_evol}), spherical (no shear) collapse
is the slowest, so a first way to change the ZEL prediction is to
force it not to assume values larger than the spherical one:

\be b_c^{an1}=\min \{ b_c^{ZEL}, 1.69/\delta_l\}. \label{eq:ansatza} \ee

\noindent 
This $b_c$ curve is shown in Fig. 3.2a for $\delta_l=1$; it has a
plateau, of height 1.69, at small $x$ and $y$ values.  On the other
hand, it is unlikely that all quasi-spherical collapses have exactly
the same $b_c$ value; a systematic trend of lower $b_c$ values at
nonvanishing $x$ and $y$ values is more realistic.  Such a $b_c$ curve
can be modeled as the intersection of the ZEL prediction with a
slightly inclined plane which reaches 1.69 at $x=y=0$:

\be b_c^{an2}= \min \{ b_c^{ZEL}, 1.69/\delta_l-\epsilon(2x+y) \}. 
\label{eq:ansatzb} \ee

\noindent
This curve, with $\epsilon=0.2$, is shown in Fig. 3.2b.  Monaco (1995)
contains a more complete discussion of such {\it ansatze}.  A further
possibility, examined in Monaco (1995), is to insert ``by hand'' in
Eq.  (\ref{eq:delta_evol}) the shear as given by ZEL (Eq.
\ref{eq:varie_zel_evol}), and then solve the equation numerically; the
resulting collapse time is very similar to that of Fig. 3.2a, but the
transition from spherical to ZEL regime is smooth.

An interesting conclusion, which will be found in next chapter, is
that the reasonable systematic trend shown in Fig. 3.2b does influence
the large-mass part of the mass function, moving it toward large
masses, even though spherical collapse is recovered for very large
overdensities (which are characterized by small $x$ and $y$
values). In other words, the fact that large, rare fluctuations
asymptotically follow spherical collapse does not guarantee that the
``spherical'' PS MF (with \dc=1.69) is recovered at large masses.

\section{Ellipsoidal collapse}

The convenience in using the homogeneous ellipsoid collapse model
resides in the fact that it can easily be solved by means of a
numerical integration of a system of three second-order ordinary
differential equations.  One of the advantages of spherical symmetry
is that, because of Birkhoff's theorem, it is possible to introduce in a
background metric a perturbation without influence the rest of the
Universe, provided any positive perturbation is compensated for by an
(outer) negative one, such to make the total mass perturbation
vanish. This is necessary to ensure the self-consistency of the
problem: the background has to evolve as if it were unperturbed.  This
reasoning is not valid any more when introducing a triaxial
perturbation in an unperturbed background: this is going to influence
the background, through non-linear feedback effects.  To use
ellipsoidal collapse in a cosmological context, the correct strategy
is not to try to insert an ellipsoid in a uniform background, but to
extract an ellipsoid from a perturbed FRW Universe.

\subsection{Homogeneous ellipsoids from a perturbed Universe}

A homogeneous triaxial ellipsoid is characterized by its mean
overdensity and its axial ratios; it can experience a global
expansion, a deformation or a global rotation. It can be recognized
that its properties are analogous to that of a mass element. As a
matter of fact, it is possible to write down evolution equations for a
Newtonian homogeneous ellipsoid (see Peebles 1980, \S 20), which turn
out to be formally equal to the ``mixed'' Eulerian-Lagrangian
evolution equations for a mass element; for instance, Eqs.
(\ref{eq:delta_evol}) and (\ref{eq:shear_evol}) are recovered.  The
reason why this happens can be understood as follows: a homogeneous
ellipsoid possesses a ``minimal'' geometric complexity which make its
structure analogous to that of a mass element. The fundamental
difference between a homogeneous ellipsoid and a generic mass element
is in the role of the potential, a quadratic form in the first case
and a whole random (Gaussian) field in the other.

Following Bond \& Myers (1996a), to extract an ellipsoid from a
perturbed potential field in a point $\mq_0$, it suffices to expand
the potential around that point in a Taylor series:

\be \varphi(\mq) = \varphi(\mq_0) + \frac{\partial \varphi}{\partial q_i}
(\mq_0) q_i + \frac{1}{2} \frac{\partial^2 \varphi}{\partial q_i \partial q_j}
(\mq_0) q_i q_j + \ldots \label{eq:taylor_expan} \ee

\noindent 
The first term is an unimportant constant; the second term produces a
bulk motion of the mass element, but does not influence the internal
properties of the ellipsoid. The third, quadratic term is the first
one which is relevant for internal dynamics; it is then possible to
approximate the potential as a quadratic form. The next step is to
split the potential into an internal and an external term:

\be \varphi = \varphi_{int} + \varphi_{ext} \label{eq:split} \ee

\noindent 
(this corresponds to the ``extraction'' of the ellipsoid).  The second
term, divergenceless, is supposed to give external tides.  It can be
held constant in the evolution of the ellipsoid: it is accurately
constant in the linear and quasi-linear regime, while it becomes
negligible, with respect to the internal potential, in the collapse
phase (see Bond \& Myers 1996a).  The first term, the internal
potential, is written as the potential of a homogeneous ellipsoid.

Notably, initial conditions are just given by the initial values of
the second derivatives of the potential; in its principal frame, they
are just the three \lam\ eigenvalues. Then, as anticipated before, the
homogeneous ellipsoid collapse model provides a way to determine
collapse times which require the same initial conditions as ZEL, and
obviously reduce to spherical collapse in the relevant limit.

\subsection{Collapse times}

The dynamical variables of ellipsoidal collapse are the three axes
$a_i(t)$ of the ellipsoid; they are normalized as the scale factor:
$a_i(t)=a(t)$ if the ellipsoid is a sphere with null density contrast.
Their evolution equations are:

$$ \frac{d^2a_i}{da^2} - (2a(1+(\Omega_0^{-1}-1)a))^{-1}
\frac{da_i}{da} + (2a^2(1+(\Omega_0^{-1}-1)a))^{-1}a_i$$ 
\be\times \left[ \frac{1}{3} +\frac{\delta}{3} + \frac{b'_i}{2} \delta + 
\lambda'_{vi} \right]=0 \label{eq:ellips_open} \ee

\noindent in the open case (the Einstein-de Sitter case can be obtained 
by setting $\Omega_0=1$), while in the flat case with cosmological
constant they are:

$$ \frac{d^2a_i}{da^2} - \frac{1-2(\Omega_0^{-1}-1)a^3}{2a(1+(\Omega_0^{-1}
-1)a^3)}\frac{da_i}{da} + (2a^2(1+(\Omega_0^{-1}-1)a))^{-1}a_i $$ 
\be\times \left[ \frac{1}{3} + \frac{\delta}{3} + \frac{b'_i}{2} \delta + 
\lambda'_{vi} \right]=0. \label{eq:ellips_cosm} \ee

\noindent 
Note that this time the scale factor $a(t)$ has been used as time
variable.  The density contrast $\delta$ is:

\be \delta = \frac{a^3}{a_1a_2a_3} - 1,\label{eq:ellips_delta} \ee

\noindent 
while the quantities $b'_i$ and $\lambda'_{vi}$ are defined as:

\be b'_i=\frac{2}{3} [a_i a_j a_k R_D(a_i^2,a_j^2,a_k^2)-1]\;\;\;\;\;
i\neq j\neq k \label{eq:bprime} \ee

\noindent (where the $R_D$ is the Carlson's elliptical integral

\be R_D(x,y,z) = \frac{3}{2}\int_0^\infty \frac{d\tau}{(\tau+x)^{1/2}
(\tau+y)^{1/2}(\tau+z)^{3/2}}, \label{eq:carlson} \ee

\noindent 
which can be calculated by means of the routine given by Press \&
Teukolsky (1990)) and

\be \lambda'_{vi} = -\frac{a}{a_0} \left(\frac{\delta}{3} - 
a_0 \lambda_i \right). \label{eq:lambdavi} \ee

Initial conditions can be set by imposing that the $a_i$ evolve
according to ZEL at early times:

\be a_i \simeq a(1-a\lambda_i) \label{eq:ellips_incond} \ee
\be \frac{d a_i}{da} \simeq \frac{1}{a} (a_i(a)-a^2 \lambda_i)\; . \ee

These three coupled second-order ordinary differential equations, Eqs.
(\ref {eq:ellips_open}) or (\ref{eq:ellips_cosm}), can be integrated
by means of standard routines, as the Runge-Kutta one given in Press
et al. (1992).  The numerical integration has to be pushed to the
singularity, when at least one axis vanishes (and the density
diverges).  To do this, it is useful to use logarithmic variables, to
have more controlled variations from quasi-homogeneity to collapse.
Moreover, the integration can be divided into two parts: the first is
stopped at decoupling, defined as the instant at which the density
starts to increase, while in the second part the density is used as
time variable, and the integration is pushed up to large density
values (Monaco 1996a). The overall precision of the numerical
integration is better than 1\% for the spherical collapse, but becomes
about 8\% for pancake-like collapses. It is to be noted that, in any
case, the first collapsing axis is that corresponding to \luno, the
largest \lam\ eigenvalue.

\begin{figure}
\begin{center}
\hbox{
\epsfig{file=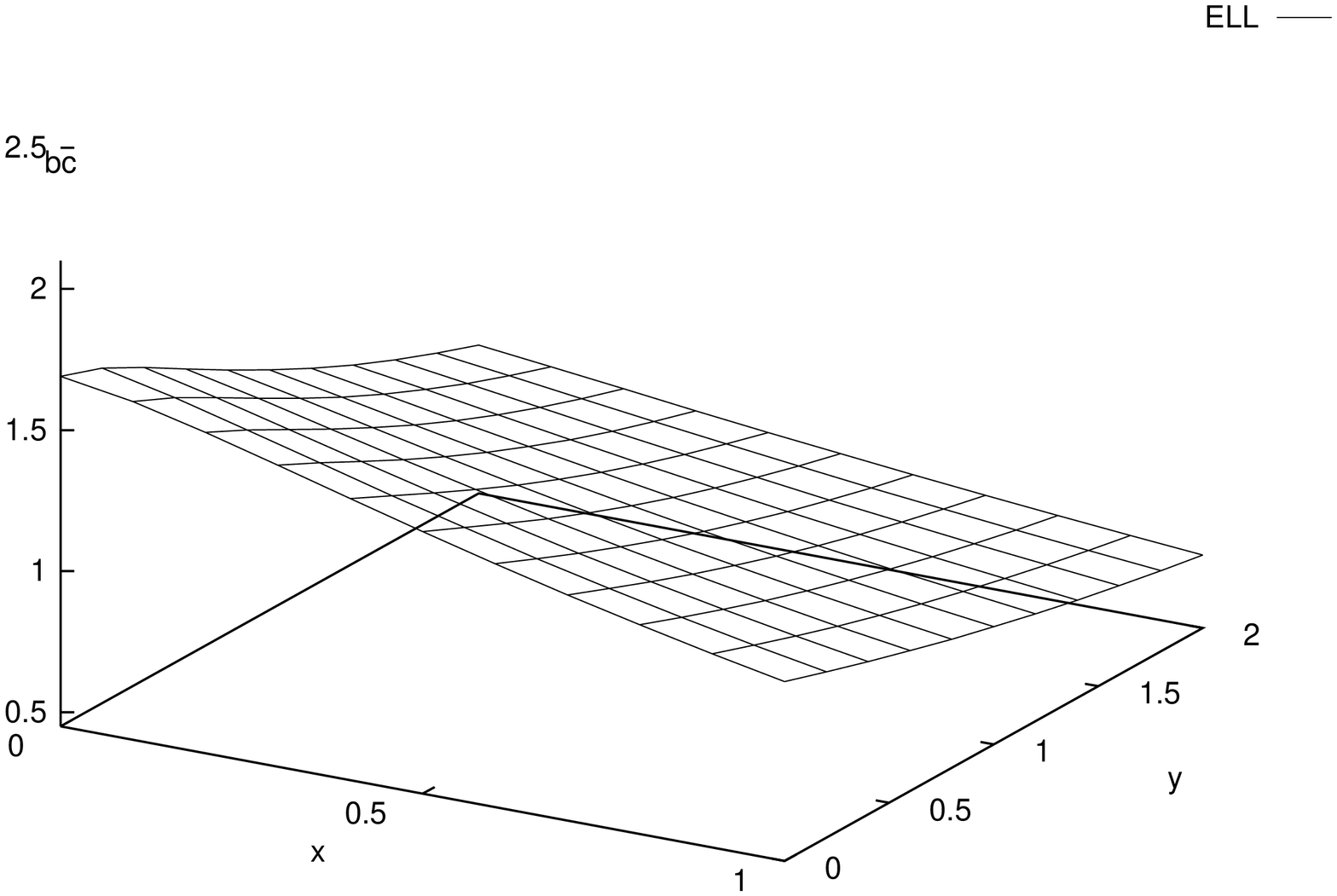,width=6cm,angle=-90}
\epsfig{file=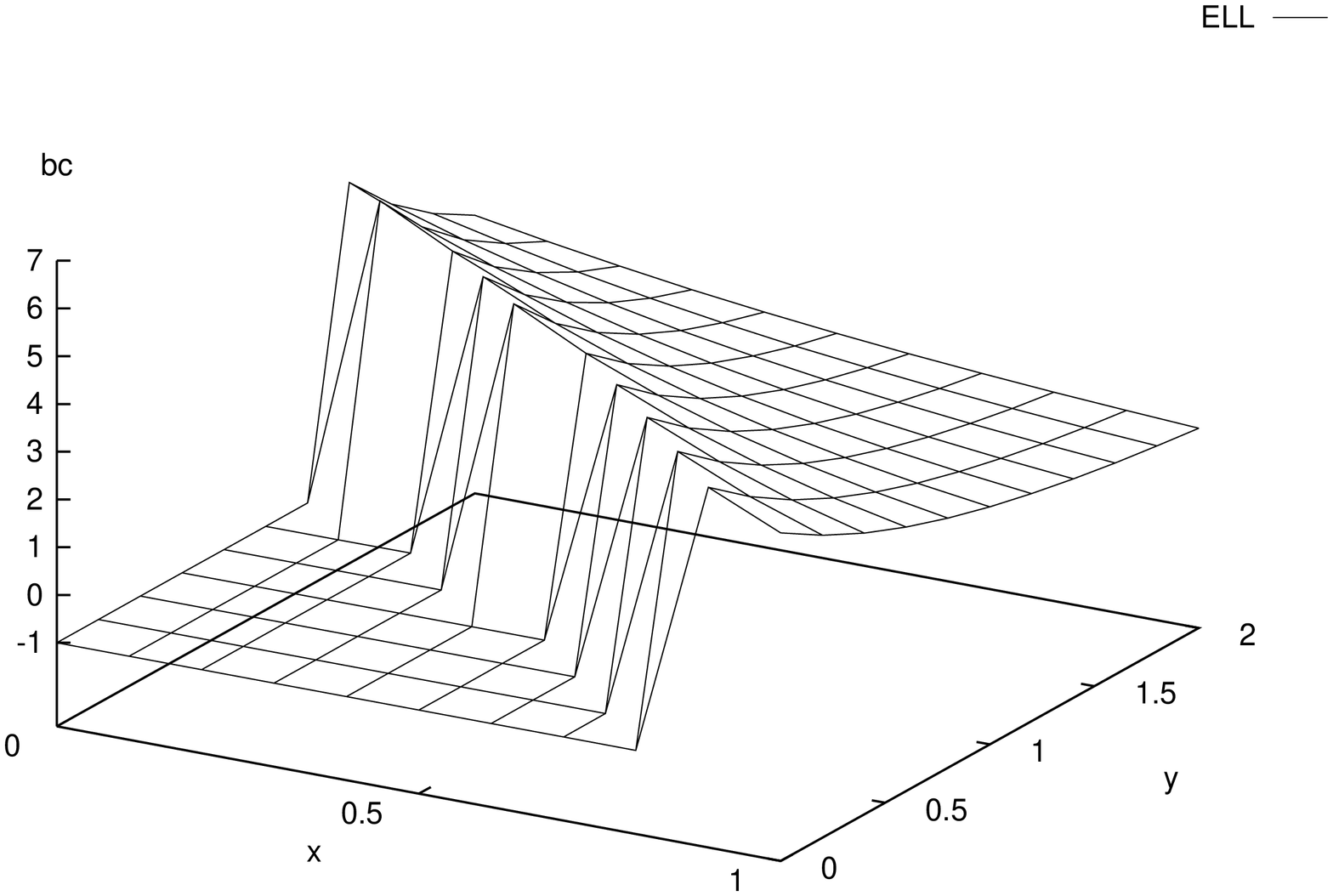,width=6cm,angle=-90}
}
\caption{Collapse times with ELL.}
\end{center}
\end{figure}

Fig. 3.3 shows the collapse ``times'' $b_c$ of initially overdense
(\dl=1) and underdense (\dl=$-$1) ellipsoids in an Einstein-de Sitter
Universe (in this case $b_c=a_c$). Spherical collapse is obviously
recovered at $x=y=0$, while quasi-spherical collapses reasonably show
a systematic departure from spherical collapse, as in Fig. 3.2a.
The large-shear behavior is similar but not identical to that
predicted by ZEL; at variance with what happens with quasi-spherical
collapses, in this range ZEL tends to underestimate the collapse time.
Fig. 3.4 shows the $b_c$ curve for ellipsoids in an open Universe;
this time \dl=3 has been chosen, to allow all the ellipsoids to
collapse. As expected, this curve is nearly identical to the one shown
in Fig. 3.3a, apart from an obvious rescaling. Notably, numerical
calculations of collapsing ellipsoids in open Universes are affected
by larger errors than that quoted above if the ellipsoid takes a long
time to collapse.

\begin{figure}
\begin{center}
\epsfig{file=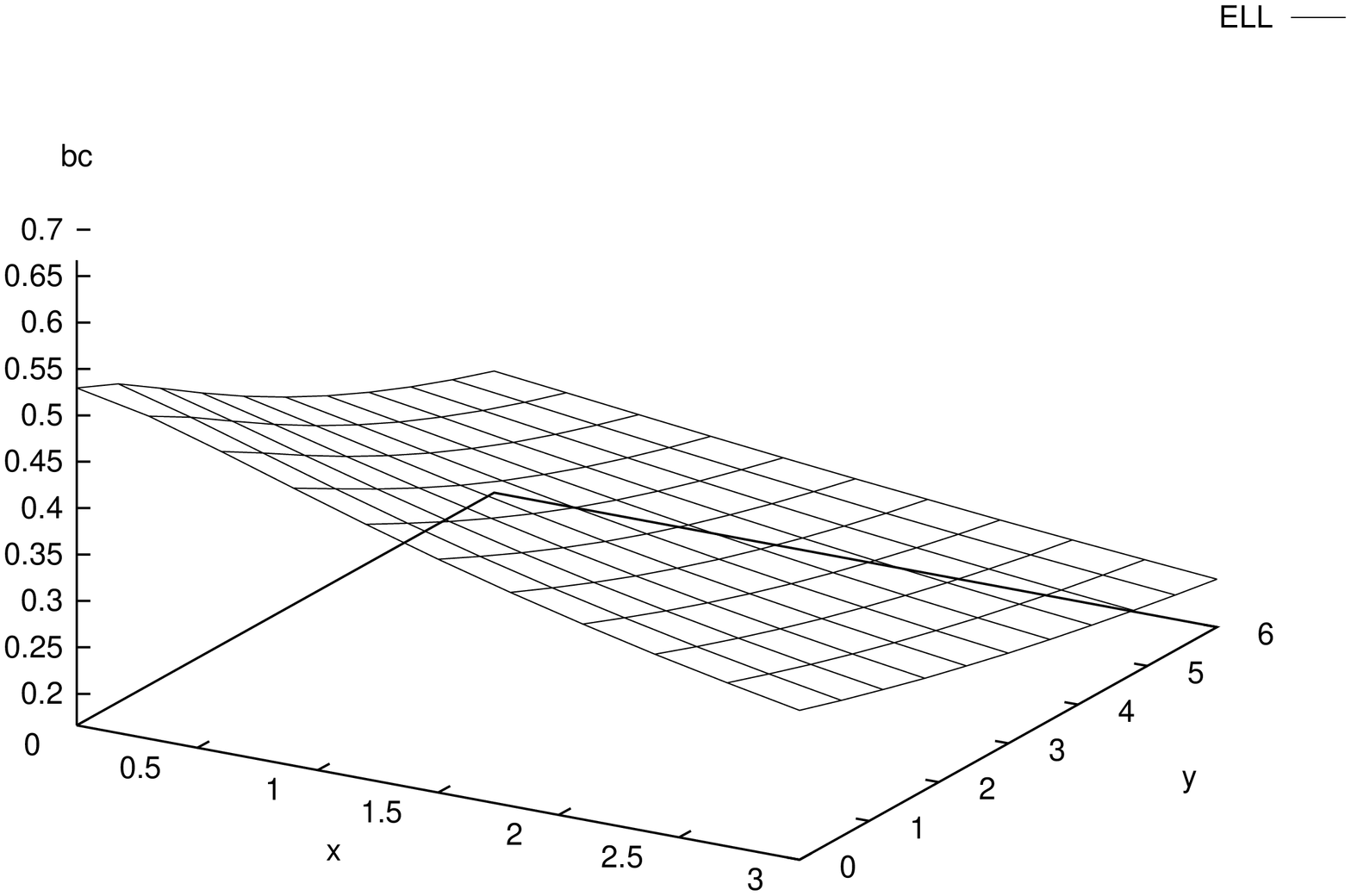,width=6cm,angle=-90}
\caption{Collapse times with ELL, open Universe.}
\end{center}
\end{figure}

The homogeneous ellipsoid collapse model has been used in the
cosmological context, besides Bond \& Myers (1996a,b,c), by White \&
Silk (1979), Peebles (1980), Barrow \& Silk (1981), Hoffman (1986),
and, more recently, by Bartlemann, Ehlers \& Shneider (1993), who used
it to estimate collapse times, by Eisenstein \& Loeb (1995a), who
modeled a collapsing structure by means of an ellipsoid, in order to
calculate its angular momentum acquired in the mildly non-linear
regime, and by Audit \& Alimi (1996). A relevant difference of such
papers with respect to what is proposed here (with the exception of
the last one cited) is that homogeneous ellipsoids are usually assumed
to model {\it extended} distributions of mass, and not vanishing mass
elements. A consequence of this is that the collapse can be followed
even after first collapse: to model a partial virialization, the
collapse of an axis is ``frozen'' when the axis becomes smaller than
a fixed fraction of the scale factor, and the evolution of the other
axes is followed up to the collapse of the third axis (if it takes
place), which is defined as the collapse time. With this collapse
definition, the shear always {\it slows down} the collapse (see also
Peebles 1990).

\section{Lagrangian perturbations}

The powerful formalism of (truncated) Lagrangian perturbation theory,
presented in \S 1.2.6, provides a precious tool for determining
collapse times of generic mass elements. The first order term, the
Zel'dovich approximation, has already been analyzed in \S 3.1; it
provides a first determination of collapse times, but depends on a
limited amount of information (the \lam\ eigenvalues). This is, on one
hand, a practical advantage, as will be seen in next chapter, but it
is, on the other hand, a physical limitation, as actual collapse times
require much more information (the whole density or potential field).
Higher-order perturbative terms add both dynamical precision and
non-local information, at the expenses of an increased complexity in
the calculations.

A relevant difference between usual applications of truncated
Lagrangian perturbation theory and what is proposed here is that
Lagrangian perturbations are commonly used to predict the evolution of
density fields in the mildly non-linear regime, up to mass variances
slightly larger than one. On the other hand, Lagrangian perturbations
will be used in the following to predict the OC instant; at that time,
the various contributions are all of the same order, so the
convergence of the series toward a solution is not guaranteed by
construction. It is then useful to test the convergence of the series
in a simple case, such as the homogeneous ellipsoid collapse; this
test will reveal some interesting connections between the two schemes.

\subsection{Lagrangian perturbations and ellipsoidal collapse}

Let's consider the quadratic potential,

\be \varphi(\mq) = \frac{1}{2}(\lambda_1 q_1^2 + \lambda_2 q_2^2 +
\lambda_3 q_3^2). \label{eq:ell_pot} \ee

\noindent 
The \lam\ coefficients are again the eigenvalues of the second
derivatives tensor of the initial potential; they are ordered as in
Eq. (\ref{eq:lambda_order}).  To find the displacement field up to
third order, it is necessary to solve the Poisson equations
(\ref{eq:space_functions}). This is not difficult, as the \pot\
potential has a very simple form, but the perturbative potentials can
be found in a direct way by using the so called ``local forms'' given
by Buchert \& Ehlers (1993), Buchert (1994) and Catelan (1995).

Local forms are solutions of the Poisson equations which are valid
only for restricted classes of initial conditions; they have the 
very interesting property of depending only on the values of
the potential and its derivatives {\it in the point considered}.
As an example, the second order displacement can be written as:

\be \ms^{(2)} = [{\bf \nabla}\varphi(\nabla^2\varphi)
- ({\bf \nabla}\varphi\cdot{\bf \nabla}){\bf \nabla}\varphi] 
+ {\bf R}^{(2)} = \ms^{(2L)} + {\bf R}^{(2)}. \label{eq:local_form} \ee

\noindent
The vector $\ms^{(2L)}$ is the local form for $\ms^{(2)}$; it
possesses the right divergence, but is not irrotational in general, so
that a divergenceless, purely rotational vector ${\bf R}^{(2)}$ must
be added to it to have the right irrotational solution; this vector
contains all the non-local information missed by the local form.
Analogous expressions can be found for the local parts of the
third-order modes (see the references cited above, and Monaco 1996a).
If the initial conditions fulfill particular conditions, given in
Buchert \& Ehlers (1993) and Buchert (1994), then the local forms are
irrotational; in this case, they are the solutions of the Poisson
equations (\ref{eq:space_functions}).

In the simple case of ellipsoidal potential, it is easy to show that
the local forms provide the correct solutions (see appendix B of
Monaco 1996a). This is easy to understand, as ellipsoidal collapse
requires only the \lam\ eigenvalues as initial conditions. Moreover,
while the general local deformation tensors depend on derivatives of
the initial potential of order equal to the Lagrangian order plus one,
the ellipsoidal deformation tensor can only include second
derivatives.  Then, the contributions to the deformation tensor of a
homogeneous ellipsoid are:

\bea 
S^{(2E)}_{a,b} & = & \varphi_{,ab}\varphi_{,cc} - \varphi_{,ac}
\varphi_{,bc} \nonumber\\
S^{(3aE)}_{a,b} & = & \varphi_{,ac}\varphi_{,bc}^C \label{eq:ellips_forms} \\
S^{(3bE)}_{a,b} & = & \frac{1}{2}[S^{(2E)}_{a,b}\varphi_{,cc} - S^{(2E)}_{b,c}
\varphi_{,ac} + \varphi_{,ab}S^{(2E)}_{c,c} - \varphi_{,bc}S^{(2E)}_{a,c}]
\nonumber\\ 
S^{(3cE)}_{a,b} & = & 0; \nonumber \eea 

\noindent 
$\varphi_{,ab}^C$ is the cofactor matrix of $\varphi_{,ab}$.  Such
ellipsoidal terms can be seen as a truncation of local forms, when all
the more-than-second derivatives are neglected. In other words,
ellipsoidal collapse can be seen as a truncation of Lagrangian
perturbation theory, which makes the evolution of a mass element
depend only on the \lam\ eigenvalues.

It is then possible to find the collapse times by solving the equation
$J(\mq,b_c)=0$; the linear growing mode is again used as time
variable, to parameterize out, with very great accuracy, any
dependence on the background cosmology (see \S 1.2.6). It is easy to
show that all the contributions to the deformation tensor are diagonal
in the same frame, and that in that frame their diagonal components
are:

\bea
\varphi_{,11} &=& \lambda_1\nonumber\\
\varphi^{(2)}_{,11} &=& \lambda_1(\lambda_2+\lambda_3)
\label{eq:_ell_pert_eig}\\
\varphi^{(3a)}_{,11} &=& \lambda_1\lambda_2\lambda_3\nonumber\\
\varphi^{(3b)}_{,11} &=& \lambda_1\lambda_2\lambda_3+\lambda_1\delta_l
                         (\lambda_2+\lambda_3)/2.\nonumber \eea

\noindent 
The other diagonal components can be obtained by rotating the indexes.
The $J=0$ equation is then:

\be 1-\lambda_i b_c - \frac{3}{14} \lambda_i (\delta_l-\lambda_i)
b_c^2 - \left( \frac{\mu_3}{126} + \frac{5}{84}\lambda_i
\delta_l(\delta_l-\lambda_i)\right)b_c^3 = 0. \label{eq:ell_collapse} \ee

\noindent 
$\mu_3$ (given in Eq. \ref{eq:invariants}) is equal to \luno\ldue\ltre.

It is interesting to analyze the spherical case, \luno=\ldue=\ltre, to
understand whether Lagrangian perturbations help in improving the
discrepancy of nearly a factor of two between ZEL and the the exact
solution.  If Eq. (\ref{eq:ell_collapse}) is solved in the spherical
case up to first, second and third order, the following solutions are
obtained:

\bea b_c^{(1)} &=& 3/\delta_l \nonumber\\ b_c^{(2)} &=& 2.27/\delta_l
\label{eq:spher_pert_coll}\\ b_c^{(3)} &=& 2.05/\delta_l. \nonumber\eea

\noindent 
For \dl=1, the difference from 1.69 is reduced from 1.31 to .36.
Then, Lagrangian perturbations, in the spherical case, show a (not
very fast) convergence toward the correct solution; this is
encouraging, as spherical symmetry is the hardest to deal with for
this approximation scheme.

The first-order solution of Eq. (\ref{eq:ell_collapse}) is simply the
ZEL approximation discussed above, $b_c^{(1)} = 1/\lambda_1$. The
second-order part of Eq. (\ref{eq:ell_collapse}) gives two solutions;
one of these turns out to be the correct one:

\be b_c^{(2)} = \frac{7\lambda_1 - \sqrt{7 \lambda_1(\lambda_1+6\delta_l)}}
{3\lambda_1(\lambda_1-\delta_l)}. \label{eq:ell_second} \ee

\noindent 
This solution is real as long as $\delta_l\geq -\lambda_1/6$, i.e.
only for initial overdensities or small underdensities. The other
solution (with a $+$ in front of the square root) either gives larger
$b_c$ values, or negative ones, or gives false solutions which can
make even a spherical void collapse. This bad behavior of second-order
perturbations in underdensities had already been noted by Sahni \&
Shandarin (1996).

It is then necessary to use third order perturbations to correctly
handle underdensities. The third-order collapse time is given by the
smallest non-negative solution of Eq. (\ref{eq:ell_collapse}).  It is
possible to write down analytical expressions for the solutions; they
are given in appendix B of Monaco (1996a). However, in most cases it
is possible but not straightforward to choose analytically the right
solution, and the best way to address the problem is by means of a
computer.

\begin{figure}
\begin{center}\hbox{
\epsfig{file=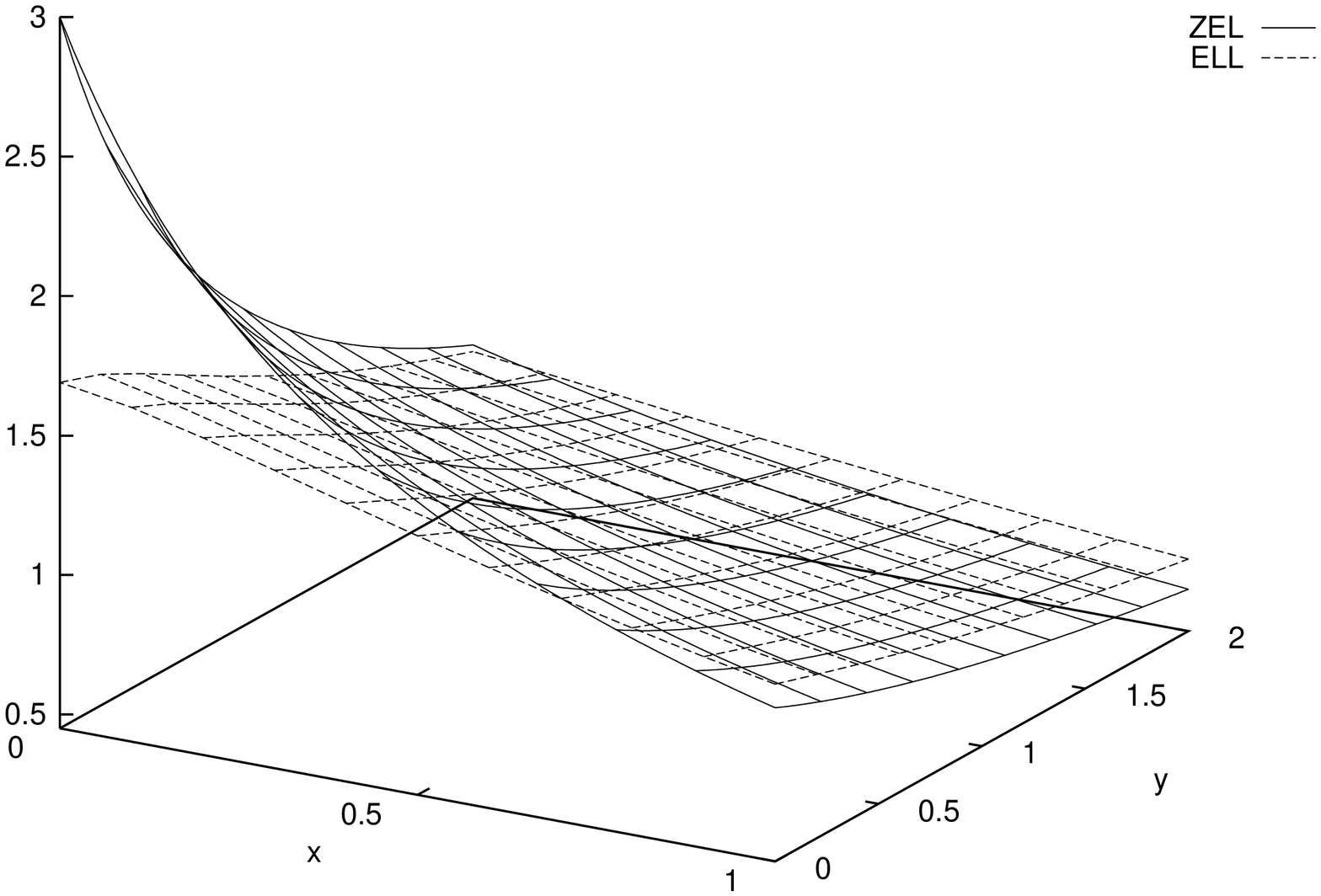,width=6cm,angle=-90}
\epsfig{file=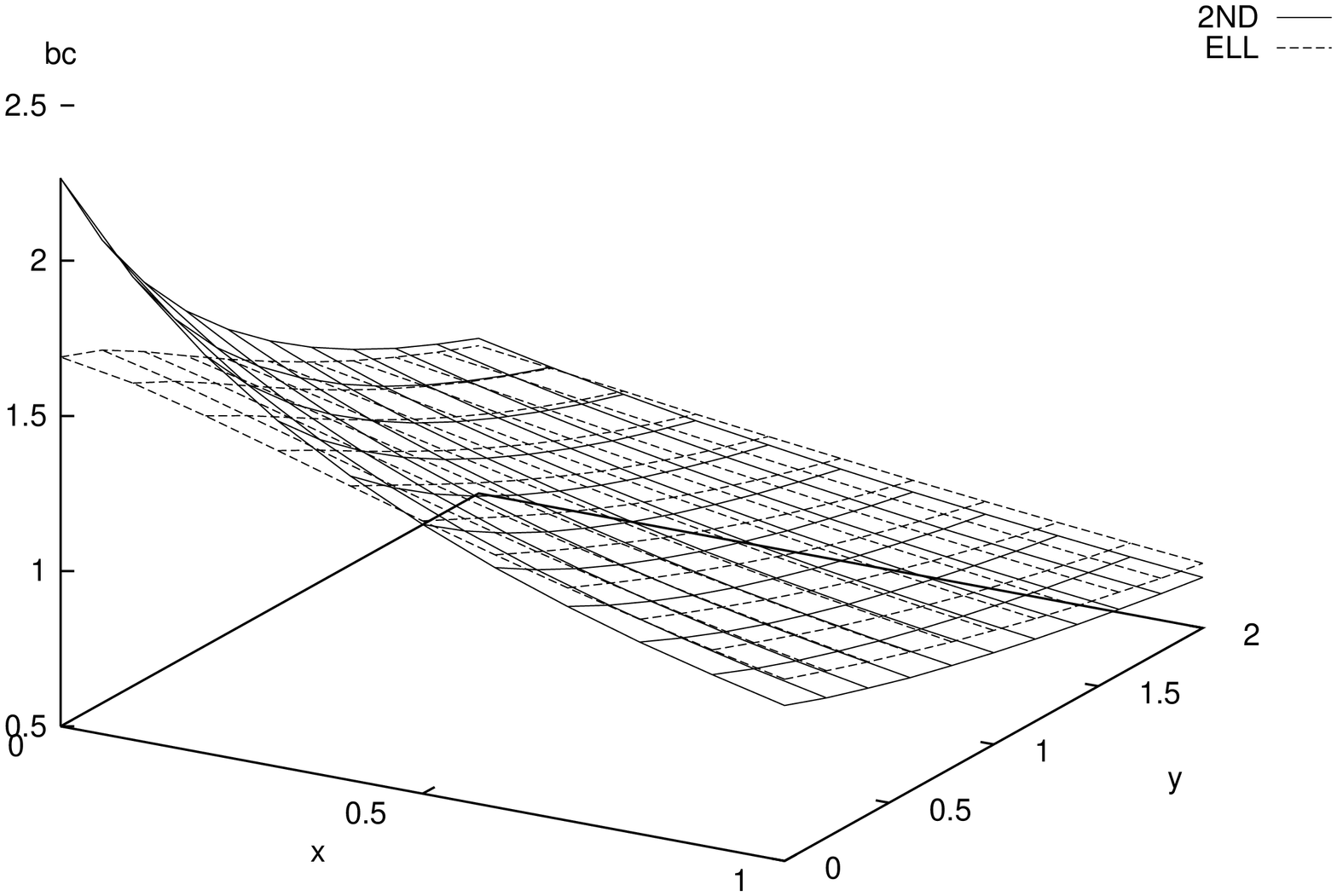,width=6cm,angle=-90}}
\epsfig{file=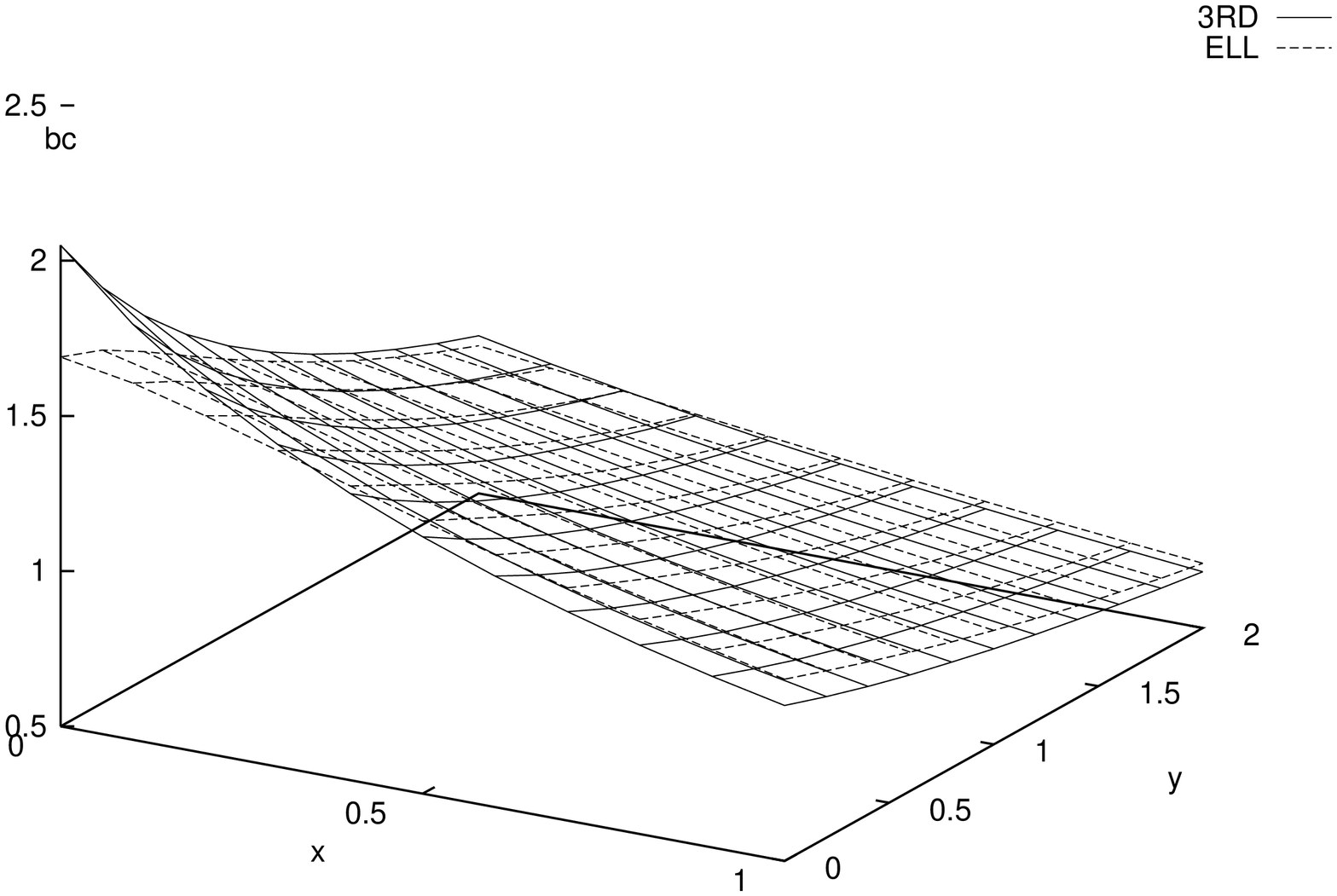,width=6cm,angle=-90}
\caption{Collapse times for overdense ellipsoids.}
\end{center}
\end{figure}

A common feature of second- and third-order solutions is that they
correctly make the 1-axis collapse before the others; this can be
explicitly checked for the second-order collapse time by
differentiating Eq. (\ref{eq:ell_second}) with respect to \luno, and
verifying that such derivative is always negative, so that the largest
axis always gives the smallest collapse time. An analogous calculation
can be performed for the third-order solution in the case of
quasi-spherical collapses, while for general collapses the thing can
be verified with simple computer calculations.  This is by itself an
indication of convergence, as the first-order prediction gives the
most important contribution to the collapse.

Fig. 3.5 shows the comparison between the numerical solution of
ellipsoidal collapse, as given in Fig. 3.3, and the various Lagrangian
predictions, for \dl=1; Fig. 3.6 shows the same for \dl=$-$1.  Such
calculations are based on an Einstein-de Sitter background.  The
following conclusions can be drawn:

\begin{itemize}
\item[(i)] in the overdensity case, the predictions at increasing
Lagrangian orders converge to the numerical value, within the
numerical errors quoted in \S 3.2.2;
\item[(ii)] in the initial underdensity case, only the odd Lagrangian orders 
show a convergence toward the numerical solution, while the
second-order prediction gives completely meaningless solutions (the
false solution which makes spherical voids collapse is shown in
Fig. 3.6);
\item[(iii)] the convergence is very fast for large shears; in this
case (for initially overdense ellipsoids), the third-order solution does
not much improve the agreement with the numerical solution;
\item[(iv)] initially underdense ellipsoids can collapse if the
shear is large enough; in this case the third-order prediction is
always sufficiently accurate.
\end{itemize}

Calculations performed with other cosmologies fully confirm these
conclusions, as expected.

\begin{figure}
\begin{center}\hbox{
\epsfig{file=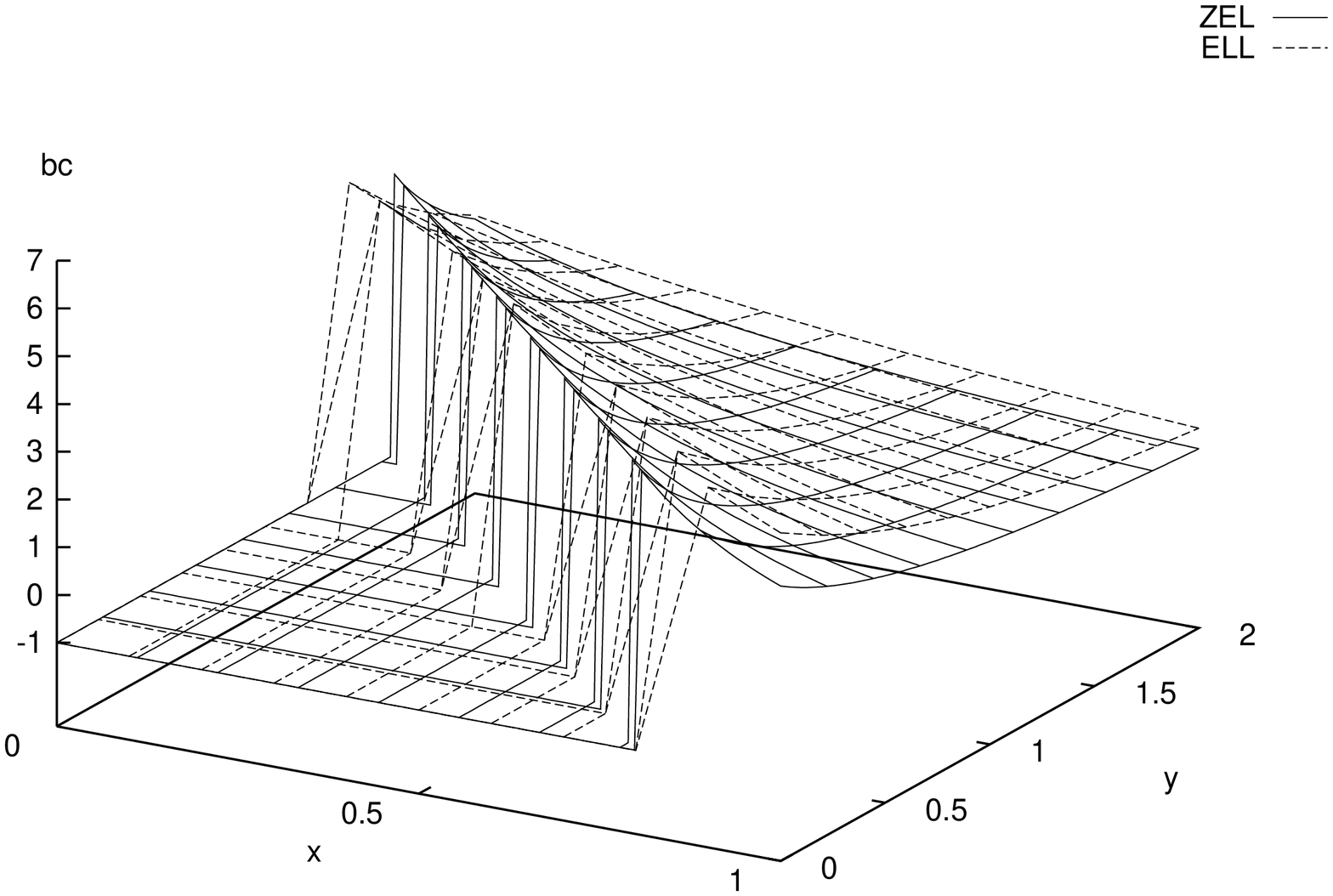,width=6cm,angle=-90}
\epsfig{file=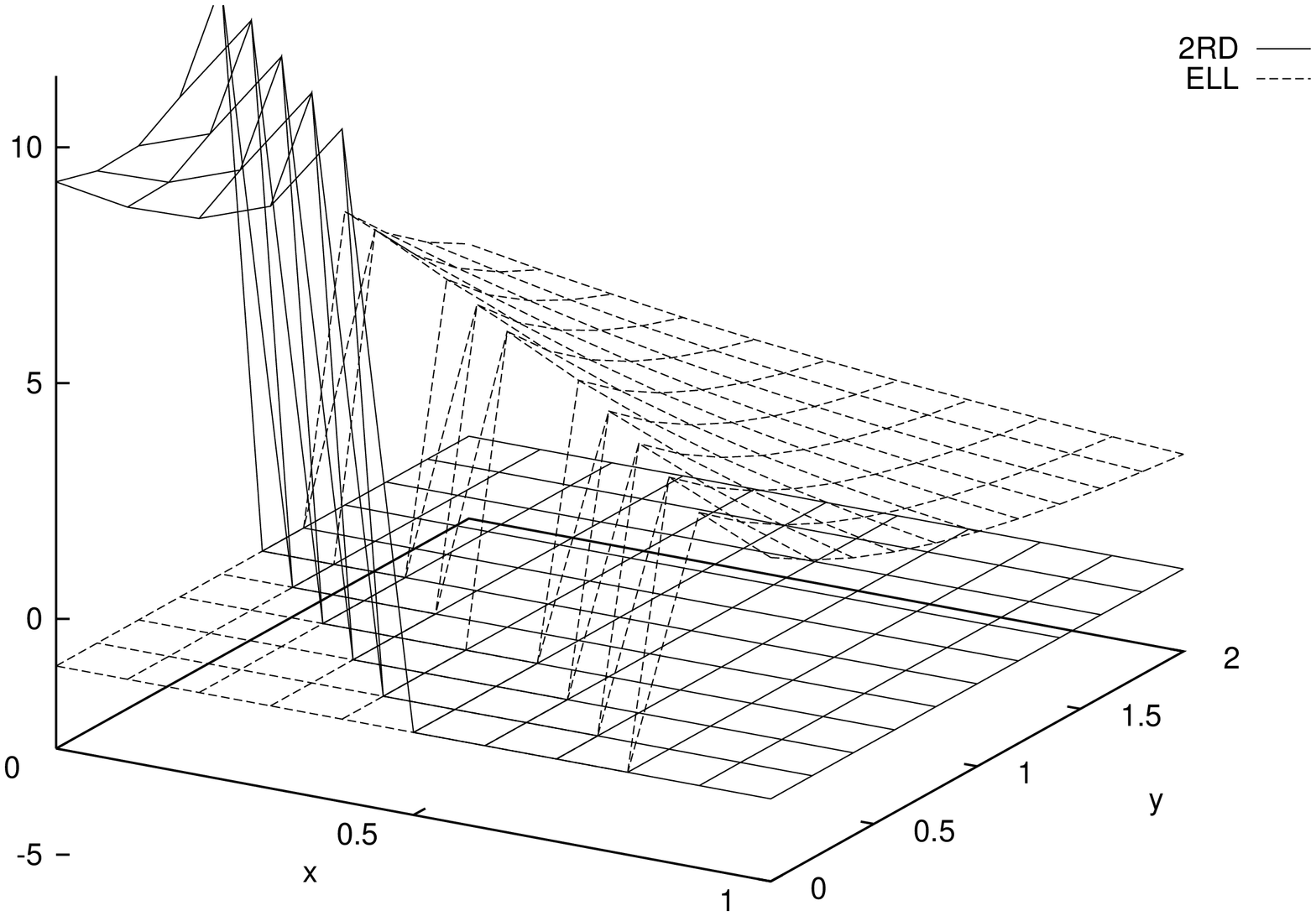,width=6cm,angle=-90}}
\epsfig{file=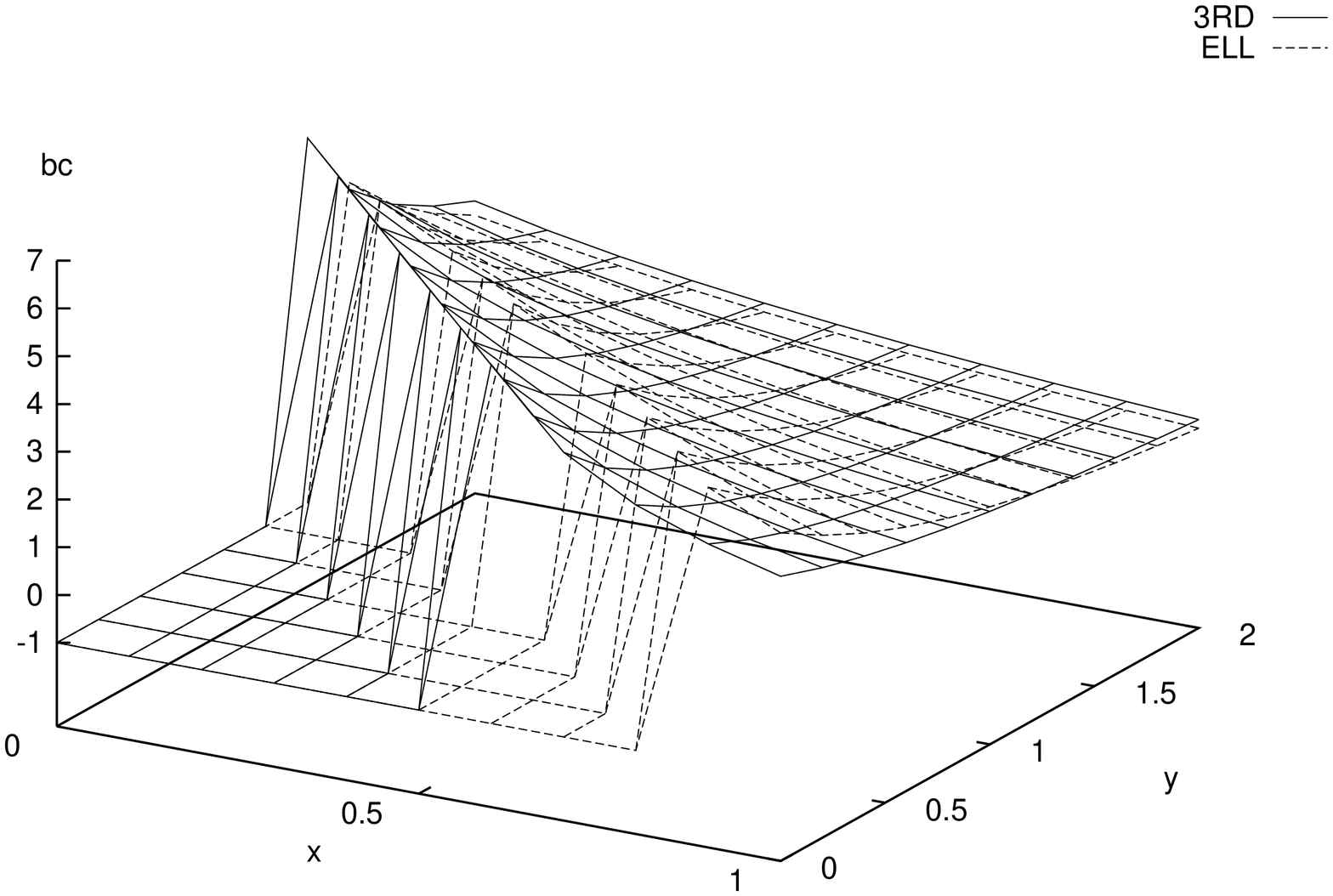,width=6cm,angle=-90}
\caption{Collapse times for underdense ellipsoids.}
\end{center}
\end{figure}

In the large-shear range, Lagrangian predictions are probably more
accurate than the numerical integration, but quasi-spherical collapse
times are significantly overestimated. The following formula corrects
for this:

\be b_c^{(nC)} = b_c^{(n)} - \Delta\exp(-ax-by)\; , \label{eq:ell_corr} \ee

\noindent 
where $n$=2,3, and the three coefficients take on the following
values:

\bea      & {\rm 2nd\ ord}&\ \ {\rm 3rd\ ord} \nonumber\\
\Delta  = &\ 0.580        & {\rm or} \ \ 0.364 \label{eq:ell_corr_coef}\\
     a  = &\ 5.4          & {\rm or} \ \ 6.5 \nonumber\\
     b  = &\ 2.3          & {\rm or} \ \ 2.8\; . \nonumber \eea

\noindent 
This corrections are applied only when $\delta_l>0$; no correction is
needed when $\delta_l \leq 0$.

As a conclusion, the connection between Lagrangian perturbations and
ellipsoidal collapse can be summarized as follows:

\begin{itemize}
\item 
Ellipsoidal collapse can be considered as a particular truncation of
the Lagrangian series, when all the more-than-second derivatives of
the initial potential are neglected in the local forms for the
deformation tensor.
\item 
With a small correction for quasi-spherical collapses, Lagrangian
perturbations can be used to follow ellipsoidal collapse in a very
fast and accurate way.
\end{itemize}

\subsection{General Gaussian fields}

When Lagrangian perturbation theory is considered in all its
complexity, a purely analytical determination of collapse times \bc\
of generic mass elements in Gaussian fields is prohibitive even at
second order. In fact, \bc\ is the smallest non-negative solution of
the equation $J=\det(\delta_{ab} +S_{a,b})=0$. The $S^{(1)}_{a,b}$ and
$S^{(2)}_{a,b}$ matrices are symmetric, but in general not diagonal in
the same frame.  Moreover, it is possible to obtain mean values for
the matrix elements of $S^{(2)}_{a,b}$, but it is discouragingly
difficult to obtain the joint PDF of all the matrix elements of both
perturbative matrices, a quantity which would be necessary in order to
analytically obtain the 1-point PDF of collapse times.

It is definitely more convenient to find collapse times by simulating
Monte Carlo realizations of Gaussian fields, with given power spectra,
in cubic grids with periodic boundary conditions.  Such simulations do
not need to be very extended: even a small, 8$^3$ grid can give a
sufficient degree of non-locality to allow a satisfactory
determination of collapse-time statistics, at least at the 1-point
level.  In fact, calculations based on 16$^3$ and 32$^3$ grids have
given indistinguishable results.  The use of cubic grids, with
multiple-of-two sides, allows the use of Fast Fourier Transform (FFT)
techniques to solve the Poisson equations for the perturbative
potentials (Eqs. \ref{eq:space_functions}).

Potential fields \pot\ have been simulated, in place of the usual
density fields; in the Fourier space, such fields are just connected
by a multiplicative $k^2$ term.  Power spectra $P_\varphi(k)\propto
k^{n-4}$, with $n$ from $-2$ to 1, have been considered ($n$ is the
usual spectral index for the matter power spectrum; see \S 1.2.7).
The fields have been normalized so as to give a total unity mass
variance: $\mres=\sigma^2_\delta=1$.  Analogously to what is usually
done to set up initial conditions for N-body simulations, potential
fields have been directly simulated in the Fourier space, by assigning
a Raileigh-distributed modulus and a random phase to any mode.
Perturbative potentials have been found by solving the corresponding
Poisson equations in Fourier space, i.e. by dividing the transform of
the source term by $k^2$. Third-order terms have been found to be
quite sensitive to numerical errors, a fact which was already reported
by Buchert, Melott \& Wei\ss\ (1994); as a consequence, third-order
calculations are not expected to be very accurate. This was noted
especially in the case of the transverse 3c term; however, it has been
verified that this term can be safely neglected in the calculation;
this is again in agreement with Buchert et al. (1994).  The
interpretation of this fact is simple: the 3c term, being purely
rotational, is not expected to influence much the density of the mass
element.

The following collapse times \bc\ have been calculated for every
point: spherical (hereafter SPH), Zel'dovich (ZEL), second-order (2ND)
third-order (3RD) and ellipsoidal (ELL) ones.  SPH, ZEL and ELL
collapse times have been calculated analytically: SPH is simply
1.69/\dl, ZEL is 1/\ltre, while ELL has been calculated by finding the
smallest positive root of Eq. (\ref{eq:ell_collapse}), and by
correcting for quasi spherical collapses as in Eq.
(\ref{eq:ell_corr}).  2ND and 3RD collapse times have been found by
looking for the first instant at which $J<0$, then using conventional
root-finding algorithms to find the accurate value. As a matter of
fact, it is possible that the Jacobian determinant $J$ returns
positive just after having become negative; a very small number of
such events, with an incidence of a few times 10$^{-5}$, were missed
by the algorithm used in the calculations.

\begin{figure}
\begin{center}\hbox{
\psfig{figure=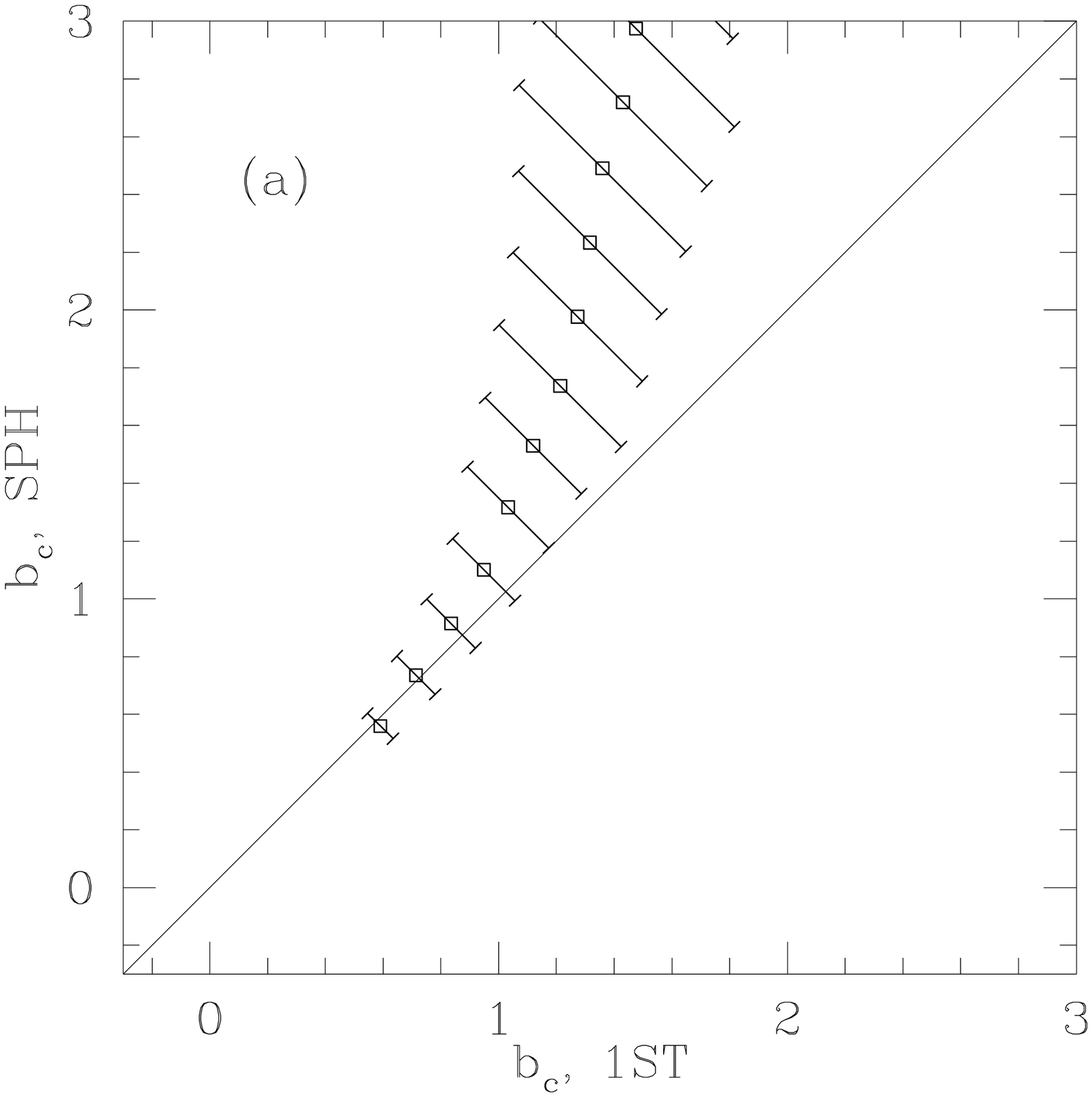,width=5cm}
\psfig{figure=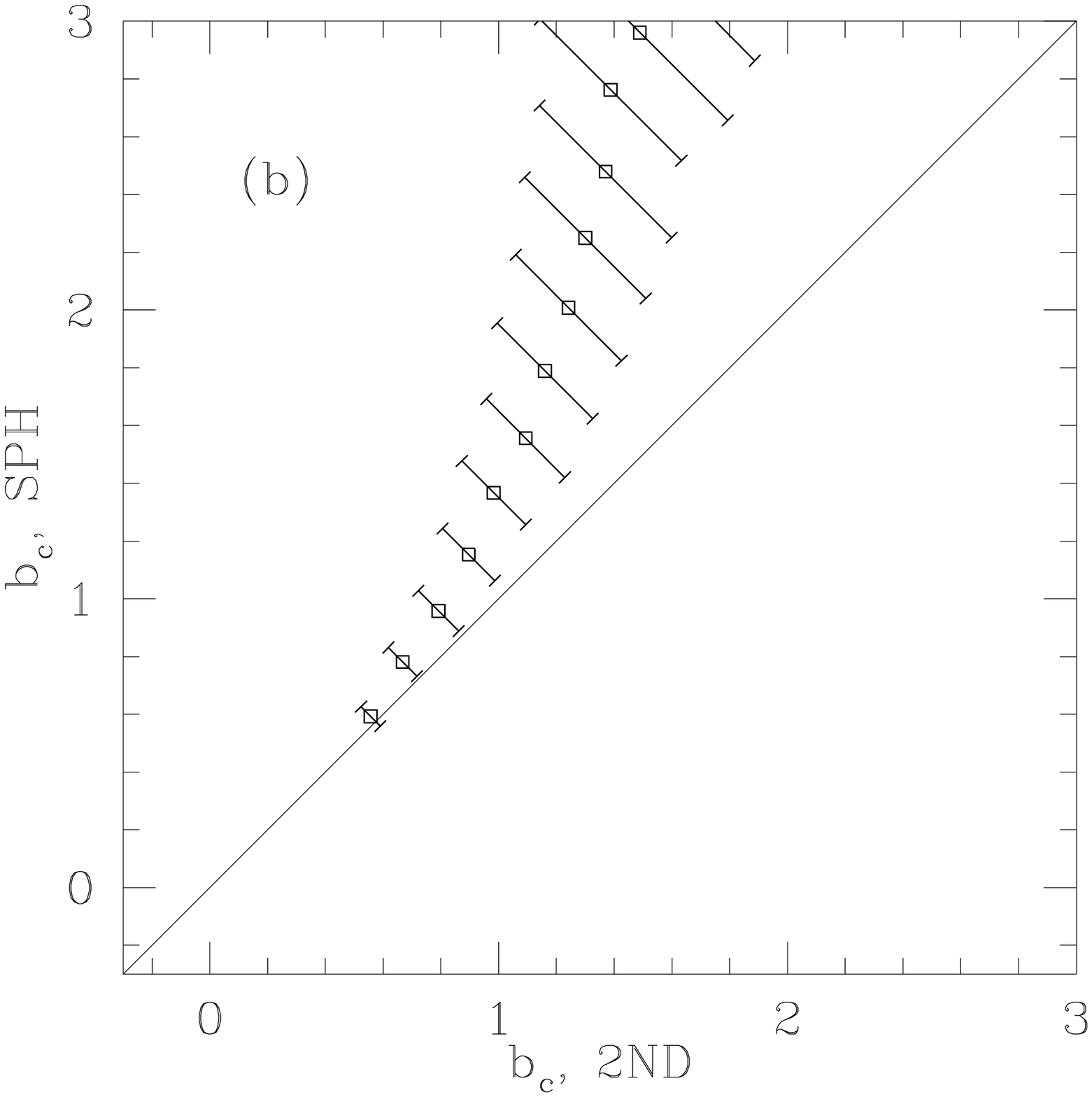,width=5cm}
\psfig{figure=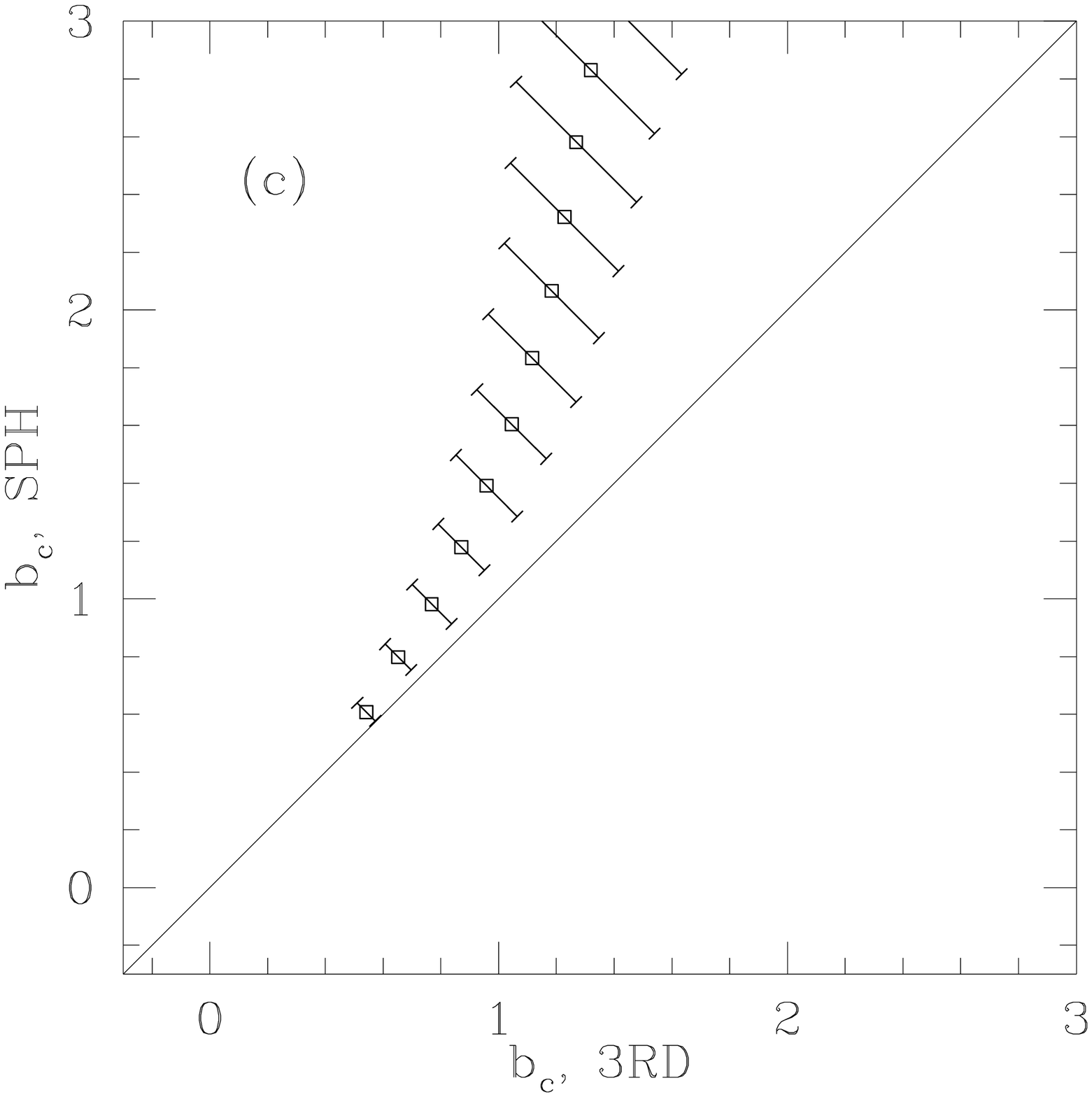,width=5cm}
}
\hbox{
\psfig{figure=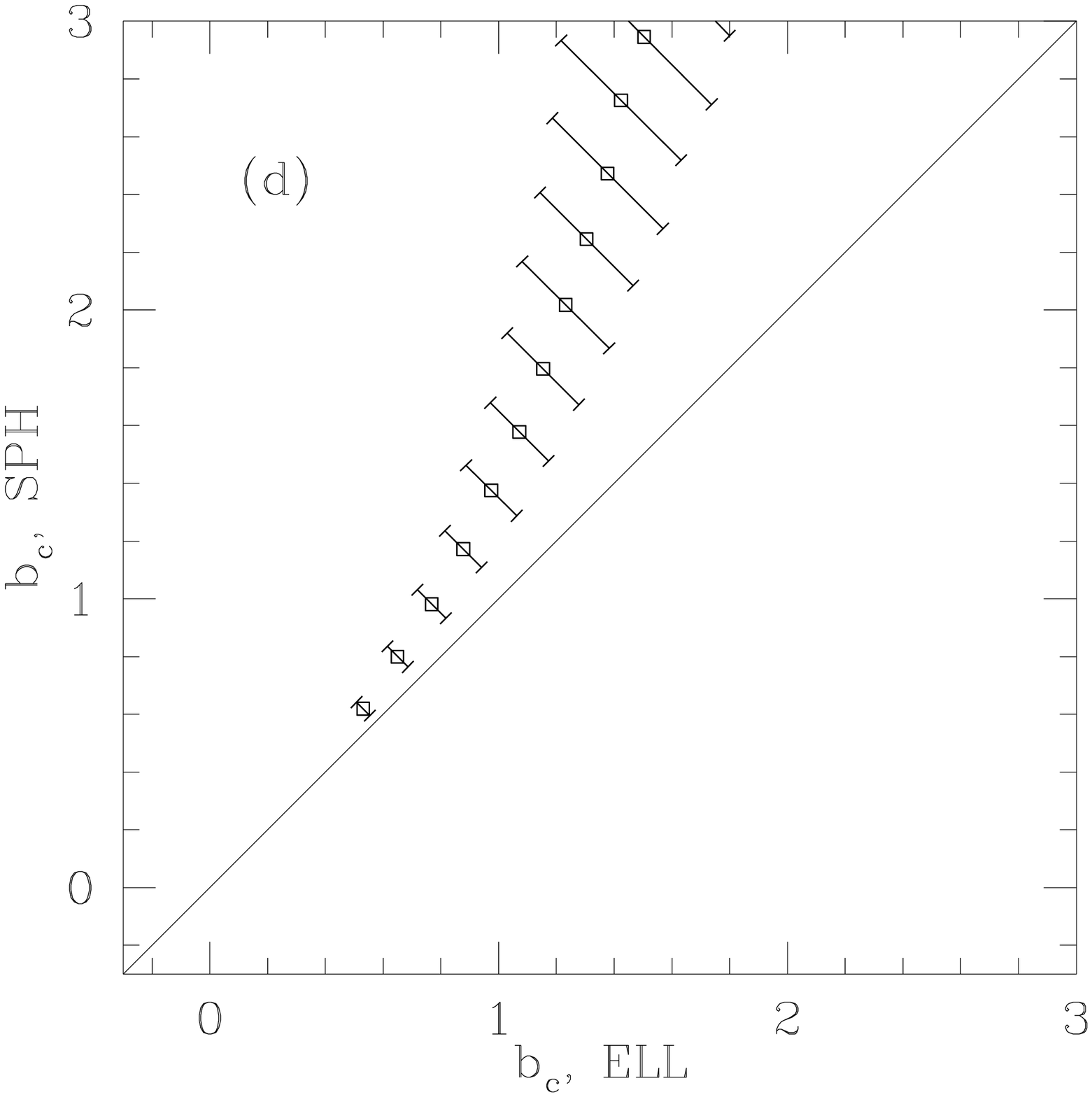,width=5cm}
\psfig{figure=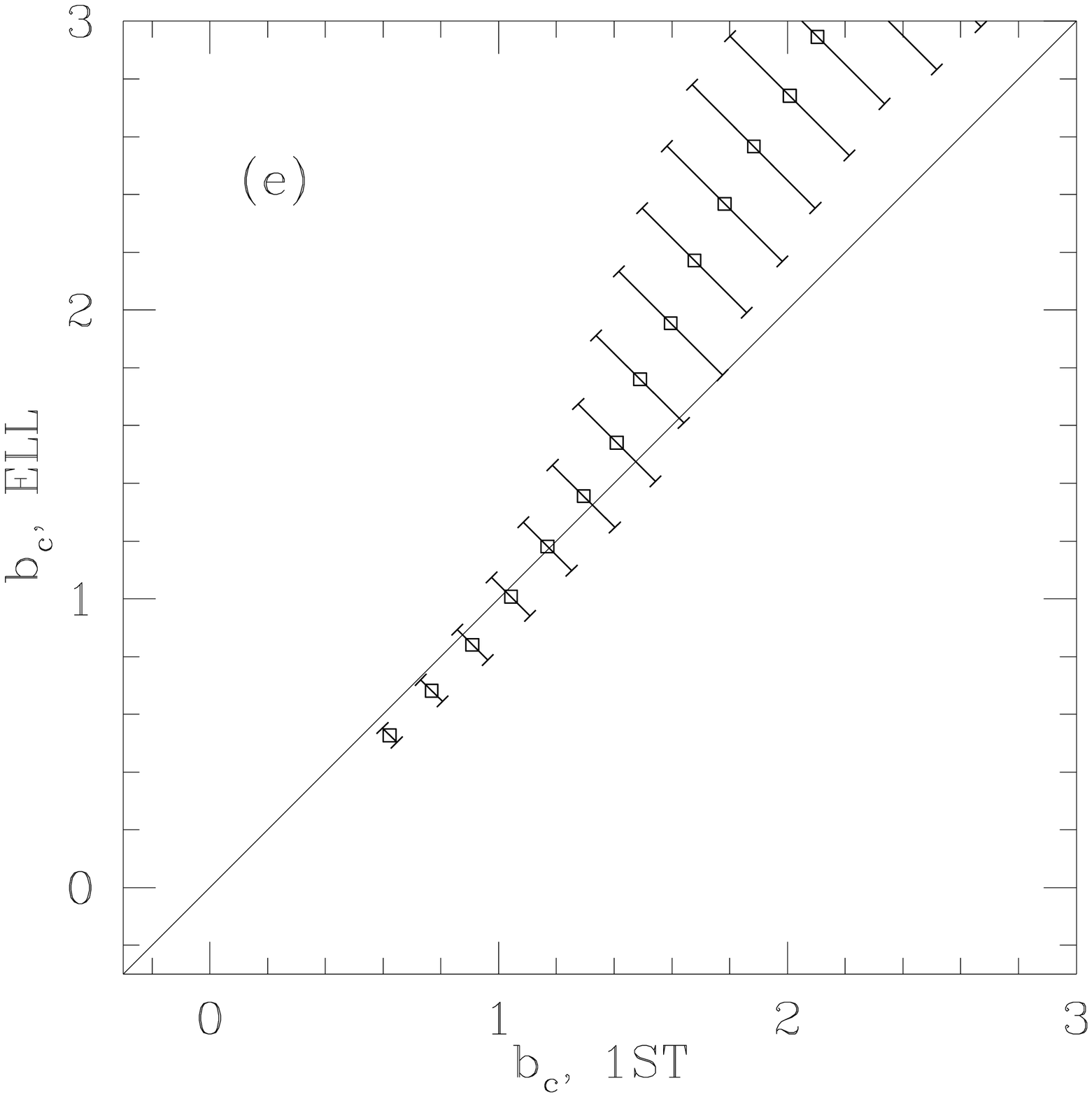,width=5cm}
\psfig{figure=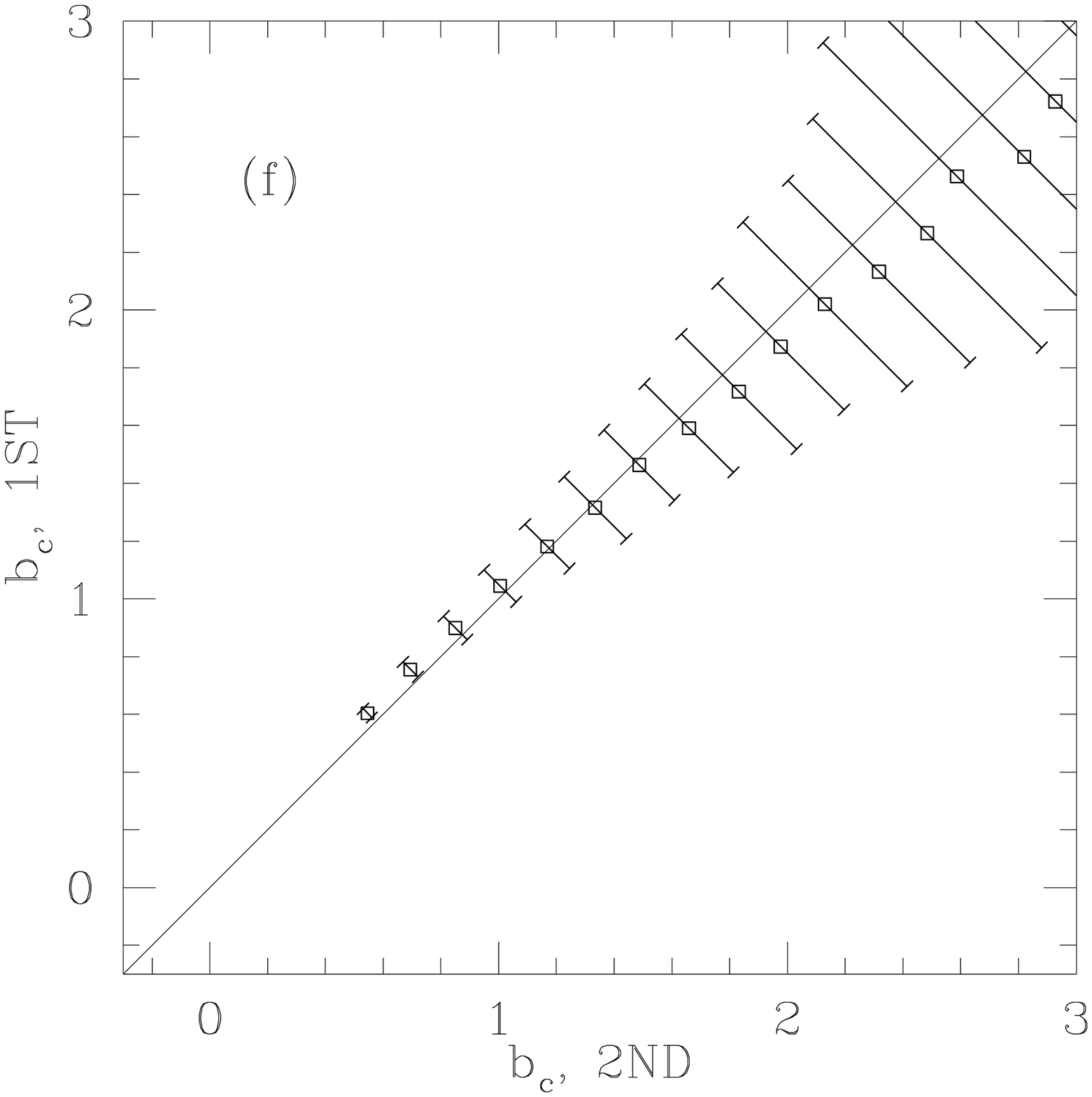,width=5cm}
}
\hbox{
\psfig{figure=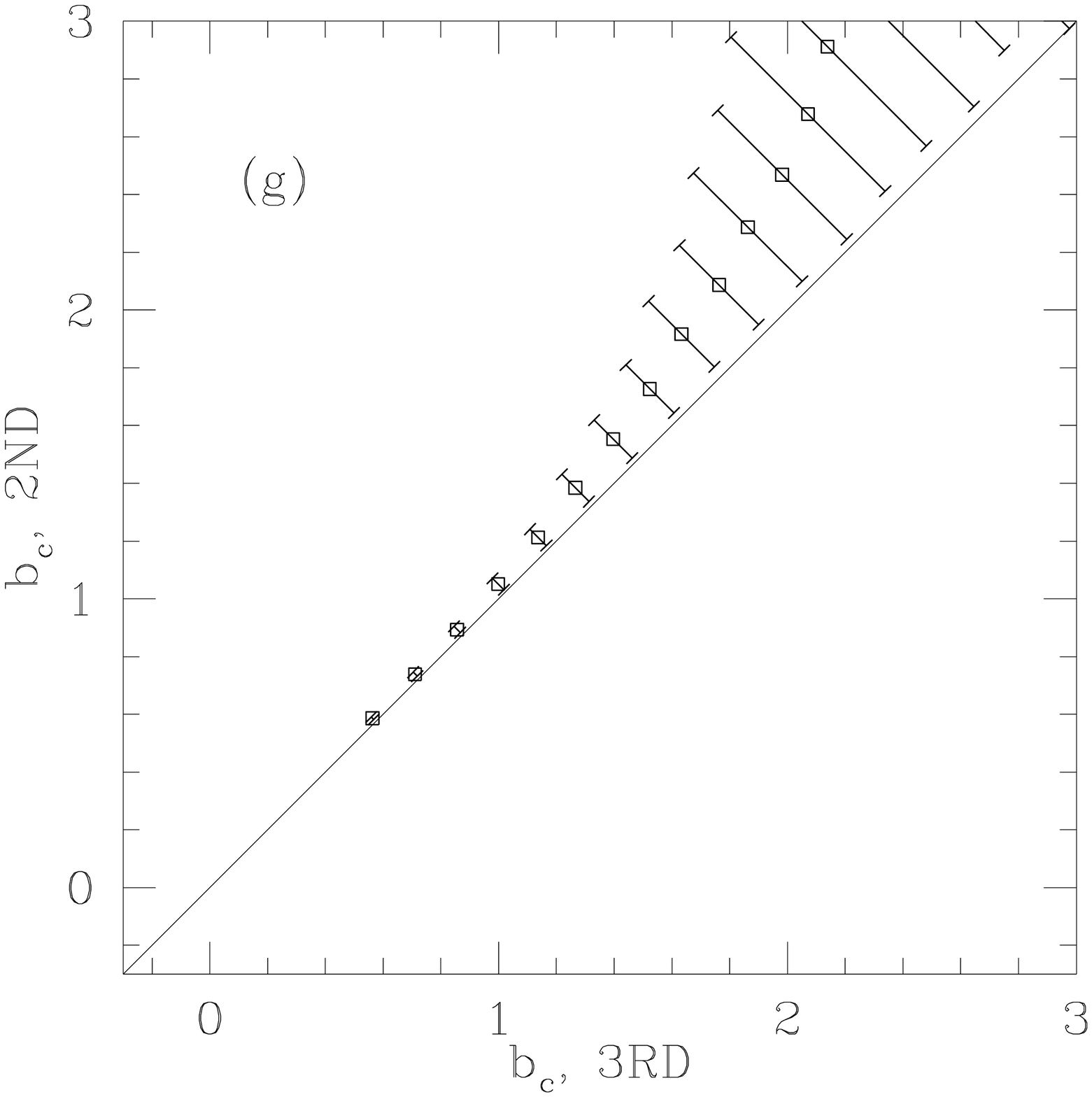,width=5cm}
\psfig{figure=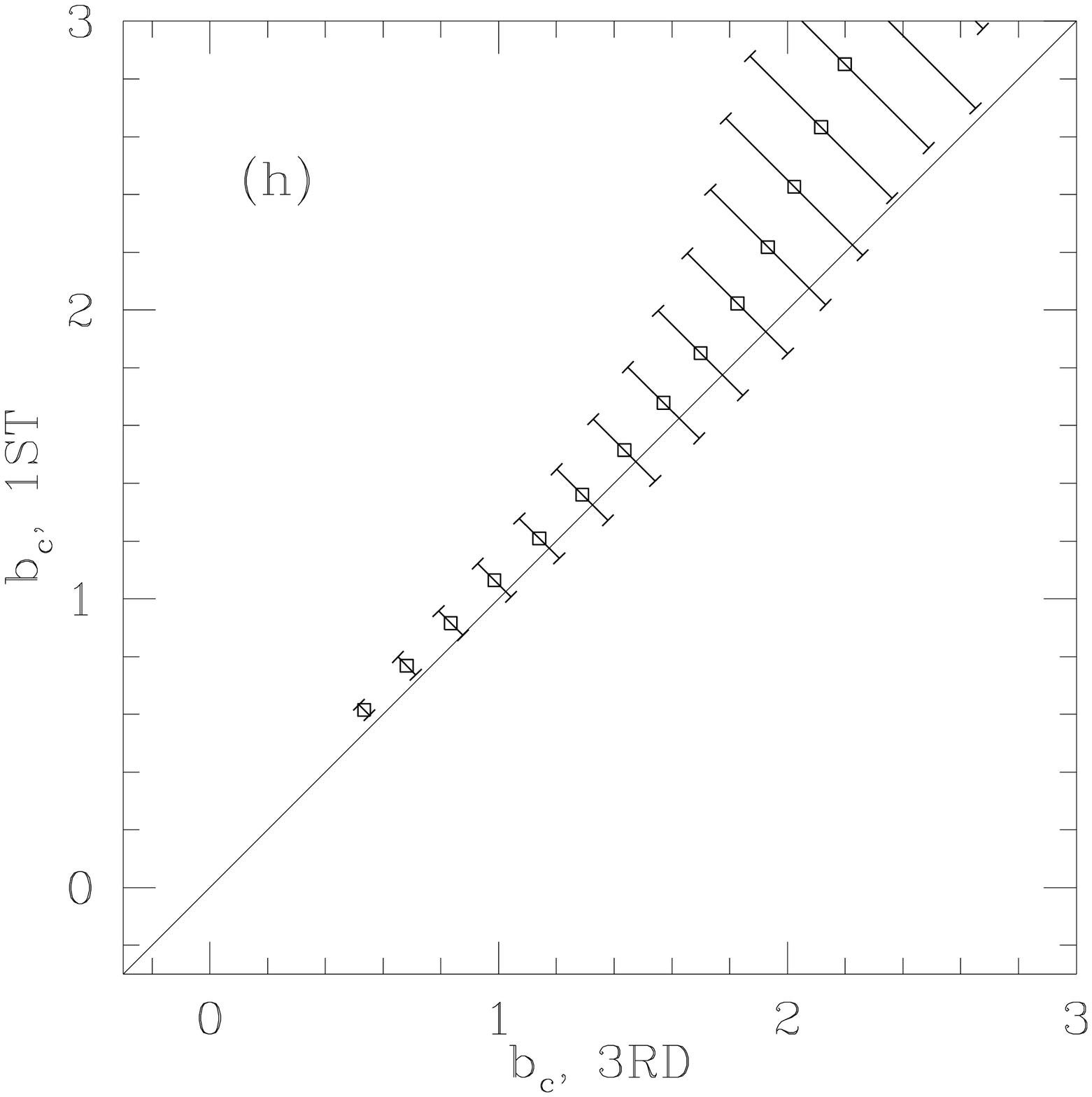,width=5cm}
\psfig{figure=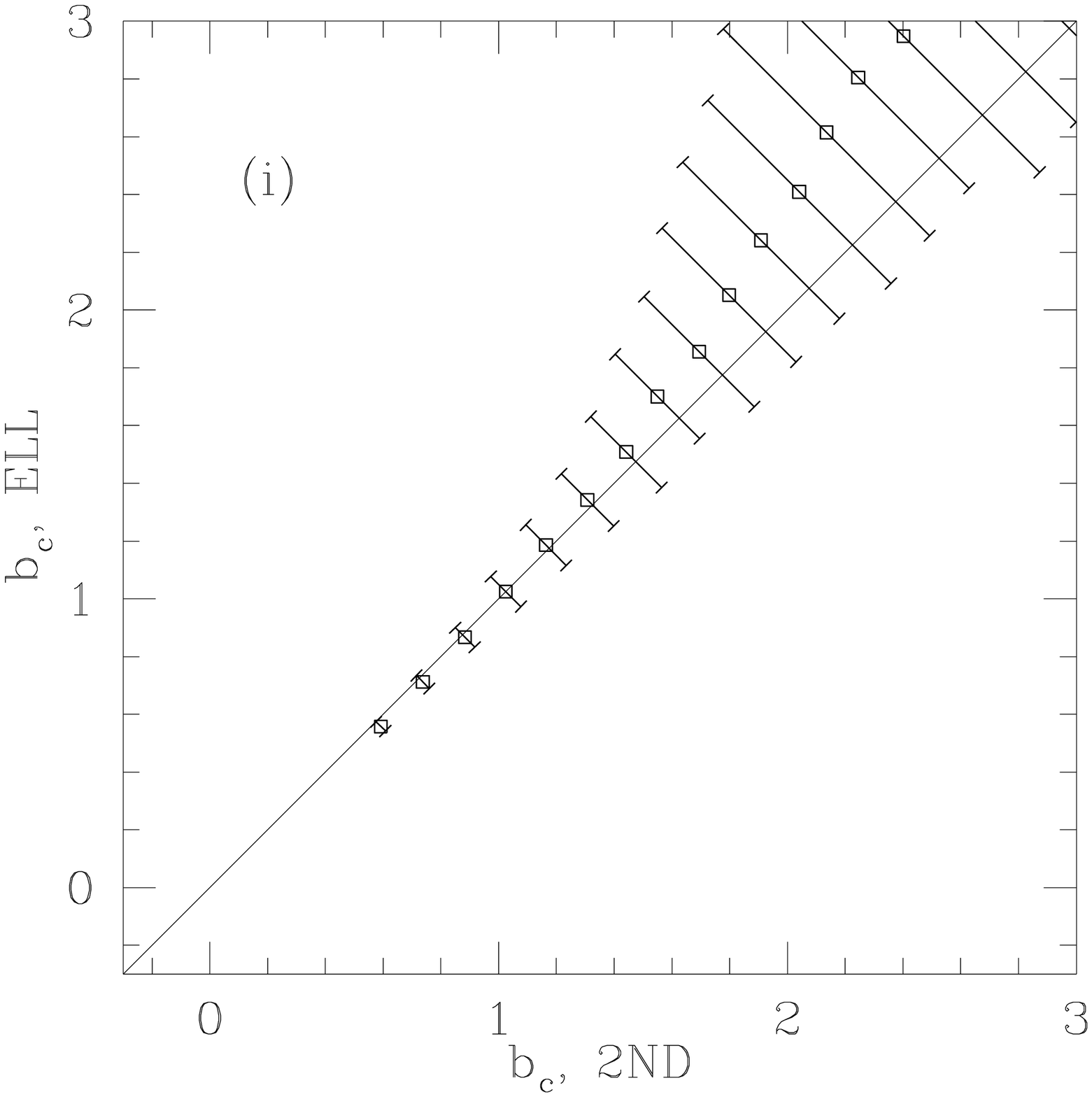,width=5cm}
}
\psfig{figure=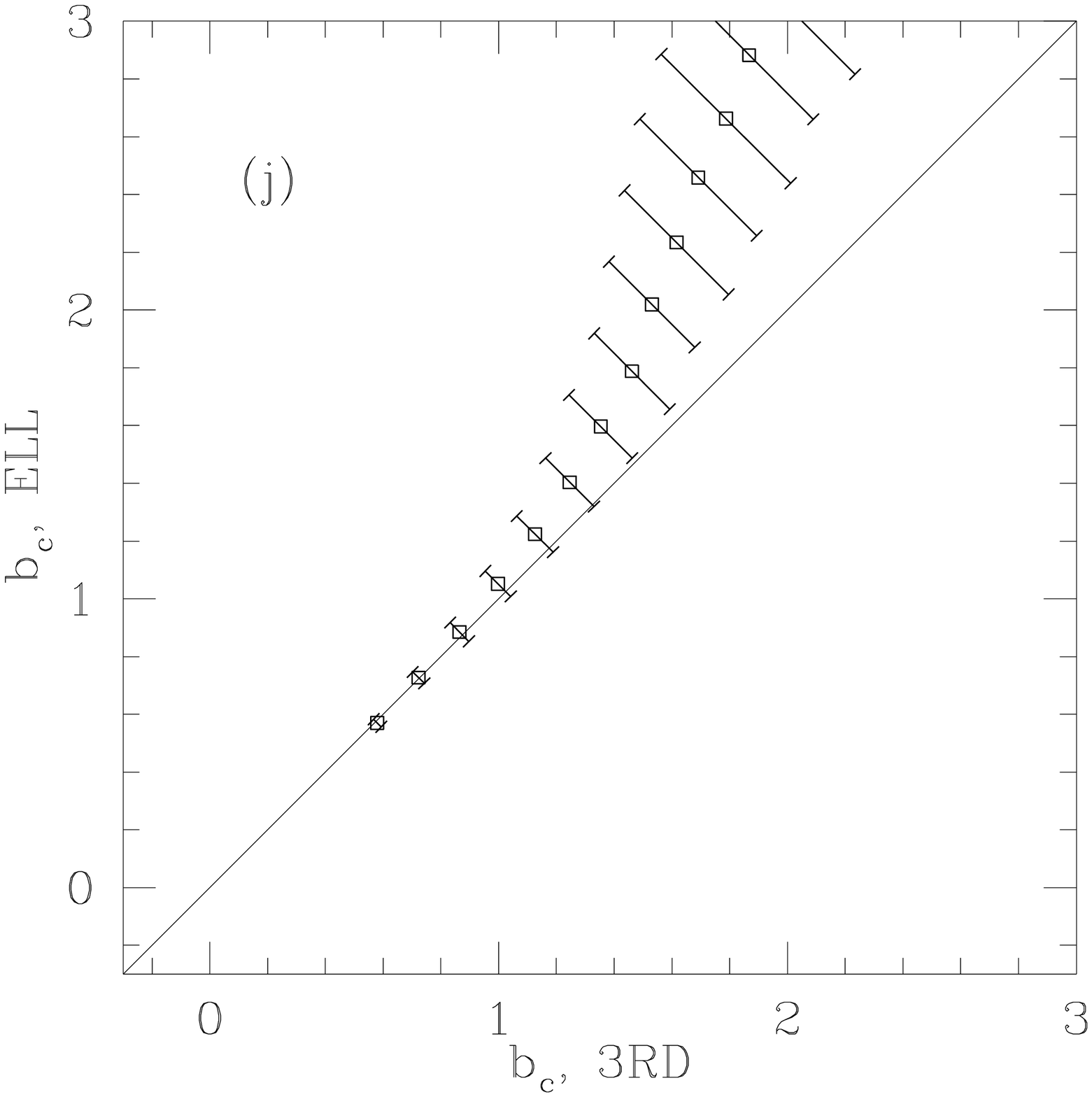,width=5cm}
\caption{Scattergrams of the various collapse times.}
\end{center}
\end{figure}

\begin{figure}
\begin{center}
\hbox{\epsfig{file=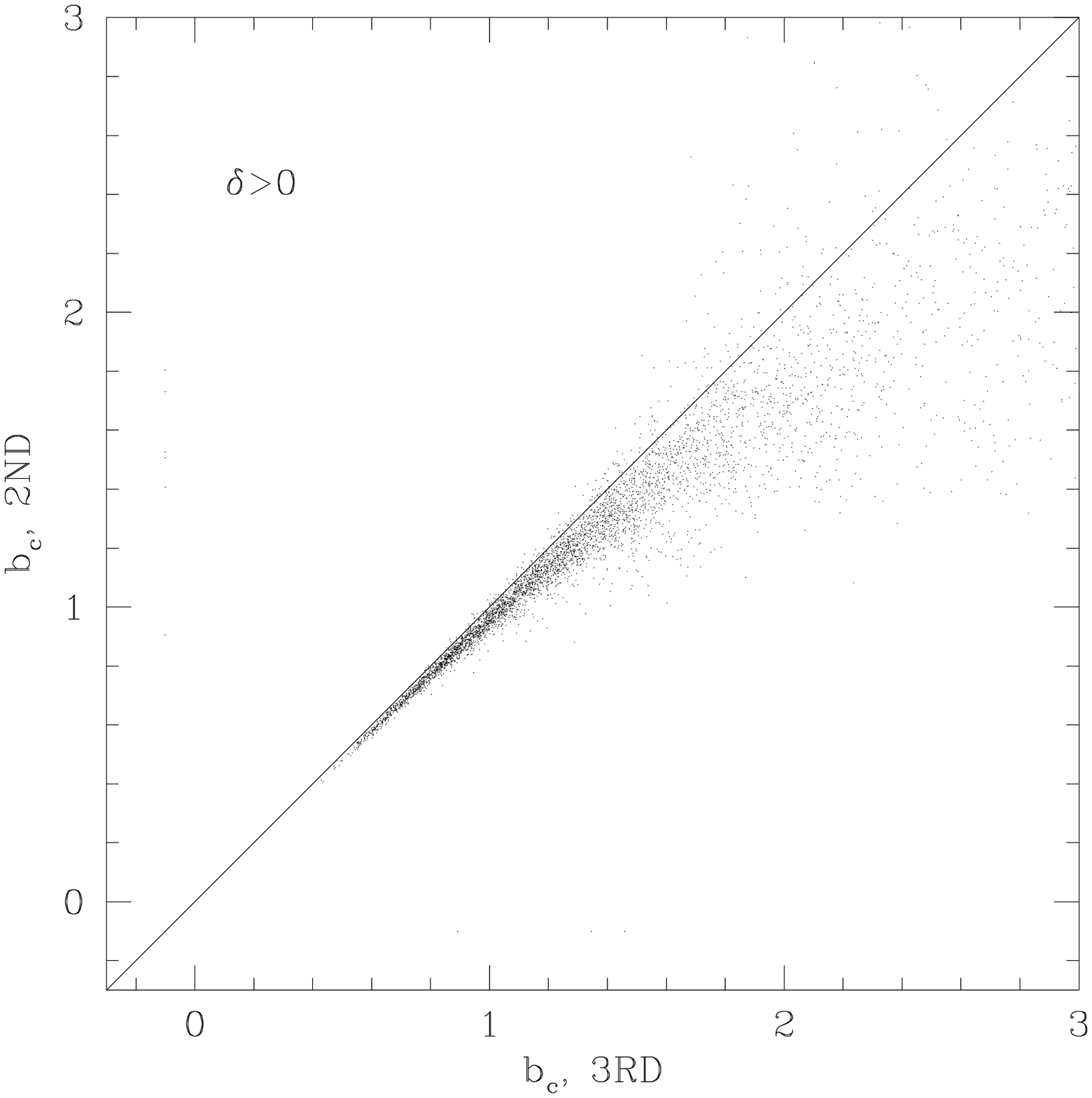,width=7cm}
\epsfig{file=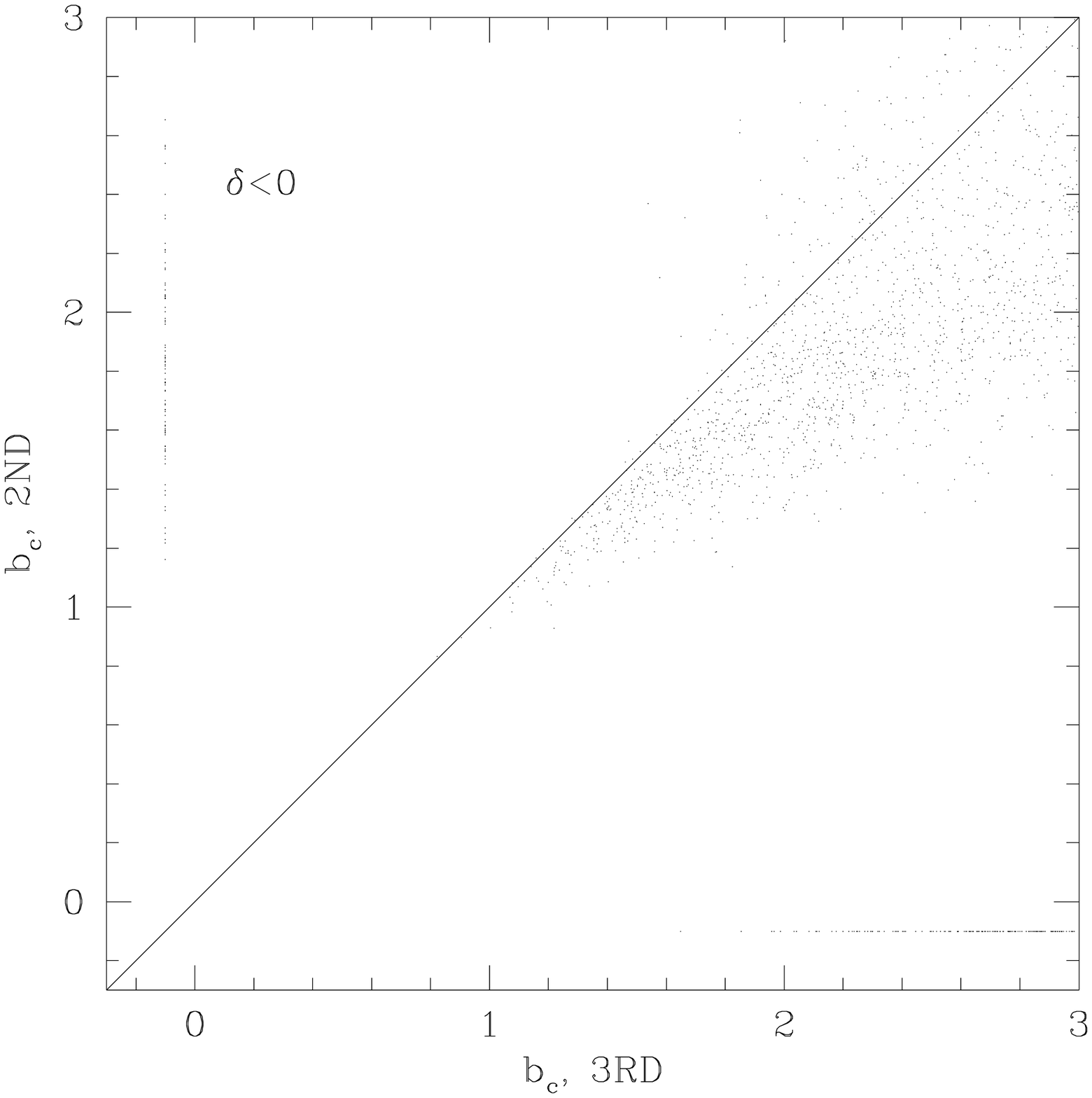,width=7cm}}
\caption{2ND--3RD, overdense and underdense points are separated.}
\end{center}
\end{figure}

Ten realizations have been performed for each power spectrum, in order
to eliminate the ``cosmic variance'' due to limited volume effects.
Nearly 10$^5$ points have been extracted from every set of
realizations.  The behavior of the various collapse times can be
understood by analyzing their mutual scattergrams; these are reported
in Figs. 3.7a-j for the $n=-2$ case (these figures focus on the
interesting $b_c<3$ part of the scattergrams).  Non-collapsing points
have been assigned a small negative \bc\ value of $-0.1$, in order to
be visualized together with the other points.  In this way, mass
elements which do not collapse according to both the two predictions
of the scattergrams lie in the lower left corner, while mass elements
which are predicted to collapse according to a prediction but not to
the other are visible as horizontal or vertical rows of points; such
points will be referred to as {\it discordant} in the following.
Figs. 3.7 also reports, superimposed on the scattergrams, the means
and dispersions of non-discordant points around the bisector line.

The following conclusions can be drawn by analyzing Figs. 3.7a-j:

\begin{enumerate}

\item 
All the results listed below are essentially independent of $n$; this
is exact for SPH, ZEL and ELL (the PDF of the \lam\ eigenvalues just
depends on the mass variance; see Eq. \ref{eq:dorosh}).  Some weak
differences in 2ND and 3RD will be quantified in next section.  For
this reason only the $n=-2$ case is shown here; the $n=1$ case is
shown in Monaco (1996a).

\item 
As expected, SPH correlates with the other predictions only for the
fastest collapsing points, and it badly overestimates collapse times
in general cases; moreover, many points (those with $\delta_l<0$!) are
incorrectly not predicted to collapse. Again, realistic (punctual)
collapses are always faster than the spherical one.  As a conclusion,
spherical collapse is not suitable, even statistically, for describing
gravitational collapse.

\item 
The ZEL -- 2ND and the 2ND -- 3RD correlations at small collapse times
are increasingly good; this demonstrates the convergence of the
Lagrangian series in predicting the collapse time of the fast
collapsing points. The ZEL -- 3RD correlation is similar to that of
ZEL -- 2ND.

\item 
The discordant points in the 2ND -- 3RD scattergram are either some
weak underdensities, for which 2ND does not find any solution, as in
the ellipsoidal case, (and then $b_c^{(2ND)}$=0), or strong
underdensities which are predicted to collapse by 2ND (and then
$b_c^{(3RD)}$=0). This is shown in Fig. 3.8, where the same
scattergram is shown separately for initially overdense and underdense
points. The same features are recognizable in the other 2ND
scattergrams. 

\item
As a consequence, third order is necessary for reliably calculating
collapse times.

\item 
3RD accelerates the collapse of the points with $b_c>1$ with respect
to all the other predictions. Given the uncertainties connected with
third-order calculations, this feature is believed only at a
qualitative level.

\item 
Both 2ND and 3RD predict the collapse of more points than ZEL.
Anyway, it is not clear whether this feature, which regards points
completely outside the convergence range of the Lagrangian series, has
any meaning.  Probably Lagrangian perturbations are not a good means
for determining the fraction of collapsed mass in a smooth Universe.

\item 
ELL shows an encouraging correlation with 2ND and an even better one
with 3RD, when $b_c$ is small.  In other words, when $b_c$ is small
the Lagrangian series converges to a solution which tightly correlates
with ELL. This has two implications, namely that the Lagrangian series
probably converges to the right solution, and that ELL can be used to
approximate 3RD collapse time. Finally, ELL tends to overestimate
collapse time for $b_c>1$, slightly with respect to 2ND and strongly
with respect to 3RD. This means that the non-locality not contained in
the ELL estimate speeds up the collapse.

\end{enumerate}

The conclusions just presented are strictly valid in the Einstein-de
Sitter case, but retain their validity in more general cosmologies, as
long as the weak lack of proportionality of $b_n(t)$ with $b(t)^n$ is
neglected.  However, in the open case collapse times larger than
$5/2(\Omega_0^{-1}-1)$ imply no collapse, as the growing mode
saturates at that value.

\begin{figure}
\begin{center}
\hbox{
\epsfig{file=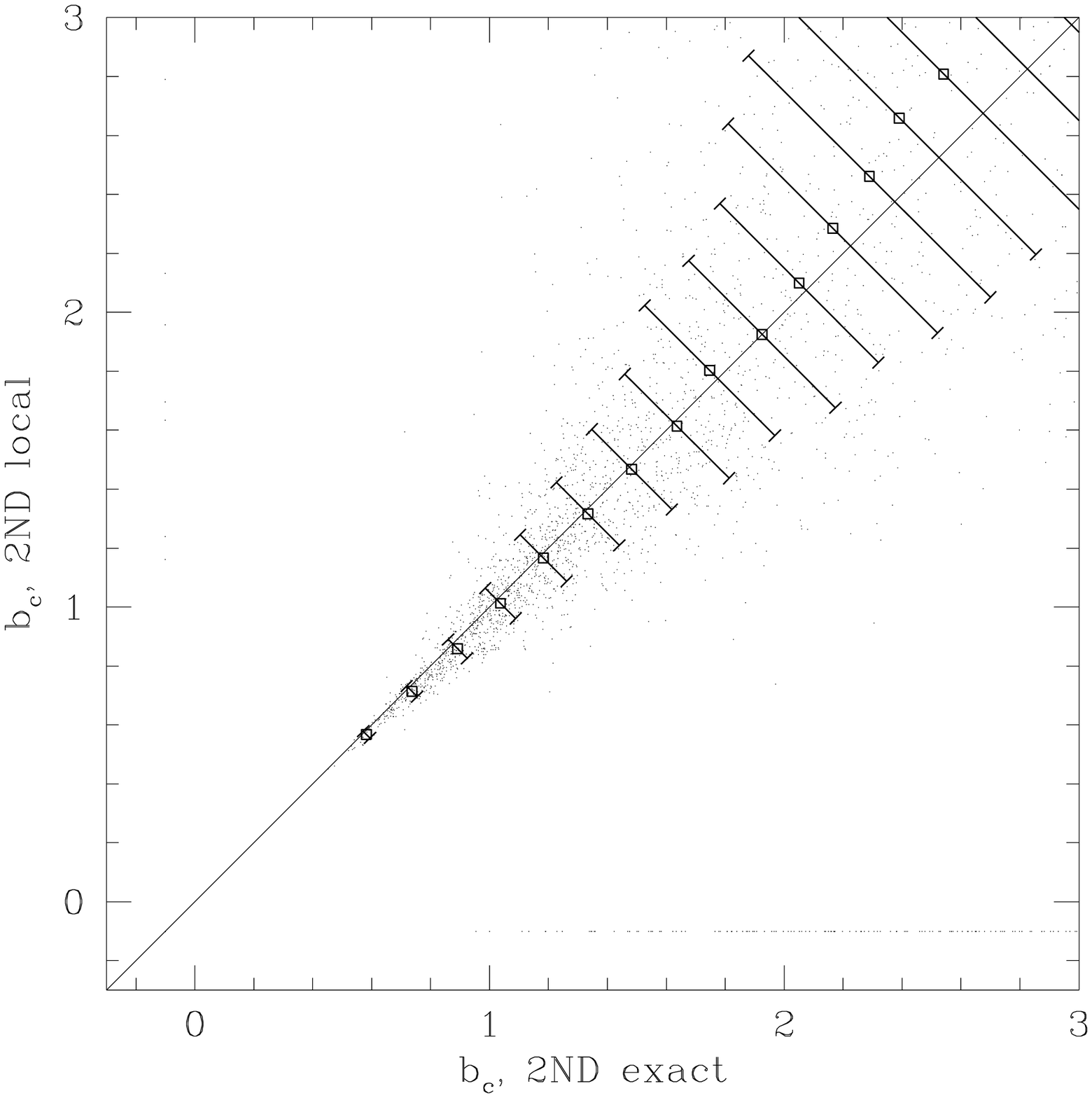,width=7cm}
\epsfig{file=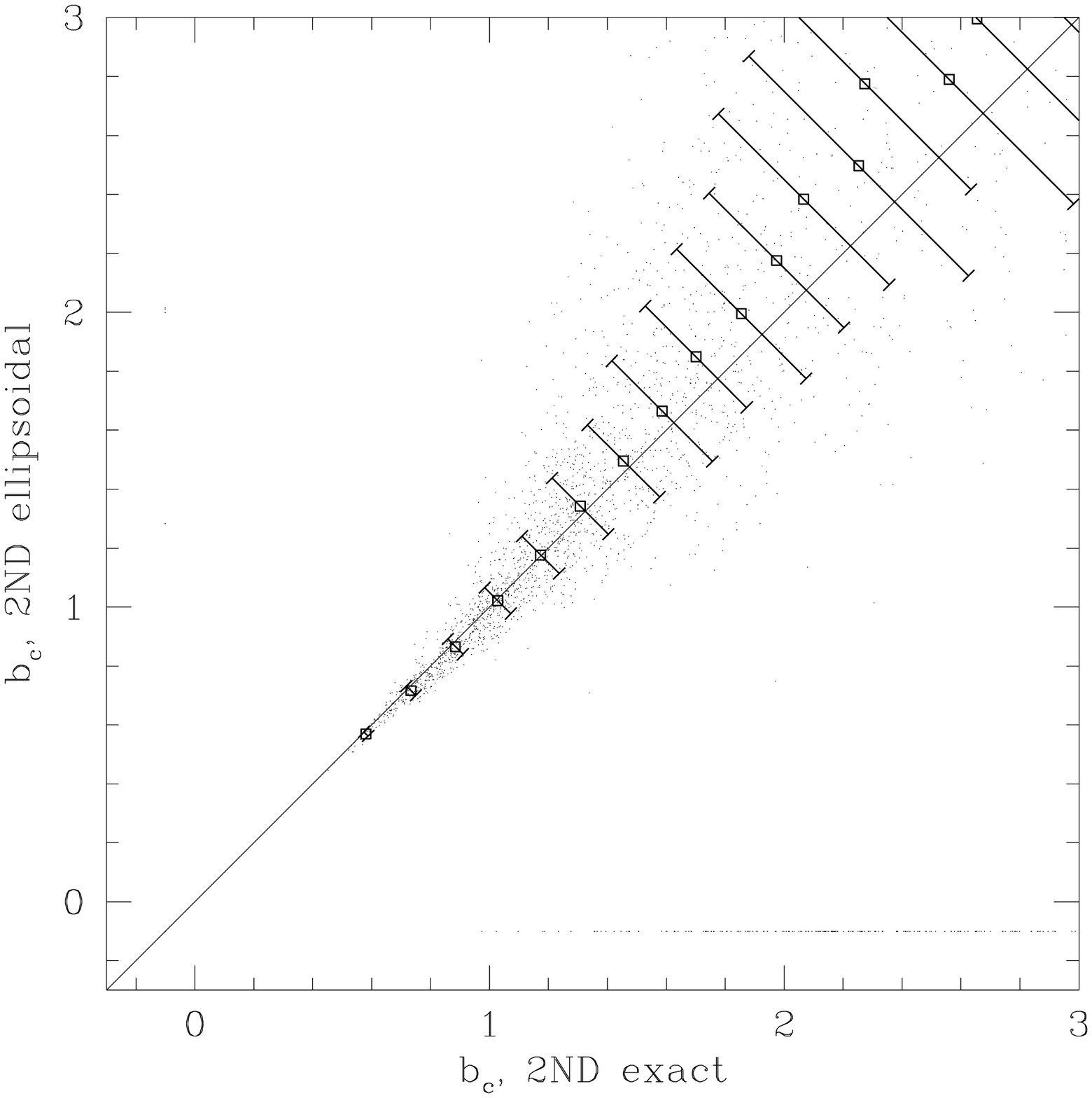,width=7cm}}
\caption{Check on the 2ND local and ellipsoidal forms.}
\end{center}
\end{figure}

As a final remark, the calculations presented here can be used to test
the efficiency of the local truncations of the Lagrangian series,
already discussed in the previous subsection. In particular, they can
be used to decide whether the full local forms, given by Buchert \&
Ehlers (1993), Buchert (1994) and Catelan (1995), better reproduce
collapse times with respect to the ellipsoidal truncations presented
here.  This test has been performed for second-order perturbations:
Fig. 3.9 shows the scattergrams of 2ND collapse prediction with
respect to its 2nd-order local and ellipsoidal parts (see Monaco 1996a
for further details). It can be seen that the full local part only
causes a modest increase of precision with respect to the ellipsoidal
truncation. It is then concluded that the ellipsoidal truncations
presented above can be preferred to the local ones, as they are much
simpler to deal with, at the expenses of only a modest loss of
accuracy.

\section{Inverse collapse times}

Collapse times present the disadvantage that the passage from
collapse to non-collapse takes place through infinity. It is more
convenient to define the quantity $F(\mq;\mres)$ as the inverse
collapse time of the point \q, when the initial field is smoothed at a
resolution (mass variance) \res:

\be F(\mq;\mres)=\frac{1}{b_c(\mq;\mres)} \label{eq:inverse_coll} \ee

\noindent 
$F$ is the basic dynamical quantity needed to construct the MF. In the
SPH case $F$ is simply proportional to the linear density contrast,
$F=\delta_l/1.69$; in the Zel'dovich (ZEL) case $F$ is just equal to
\luno.

The quantity $F$ is obviously a non-Gaussian process, and it is not a
simple non-linear function of a Gaussian field (such as, for instance,
a lognormal distribution): it is a complicated non-linear and non-local
functional of the whole initial Gaussian perturbation field. As shown
in Chapter 2, its 1-point PDF is the minimal amount of statistical
information needed to construct the MF; this point will be further
discussed in \S 4.4.

\begin{figure}
\begin{center}
\epsfig{file=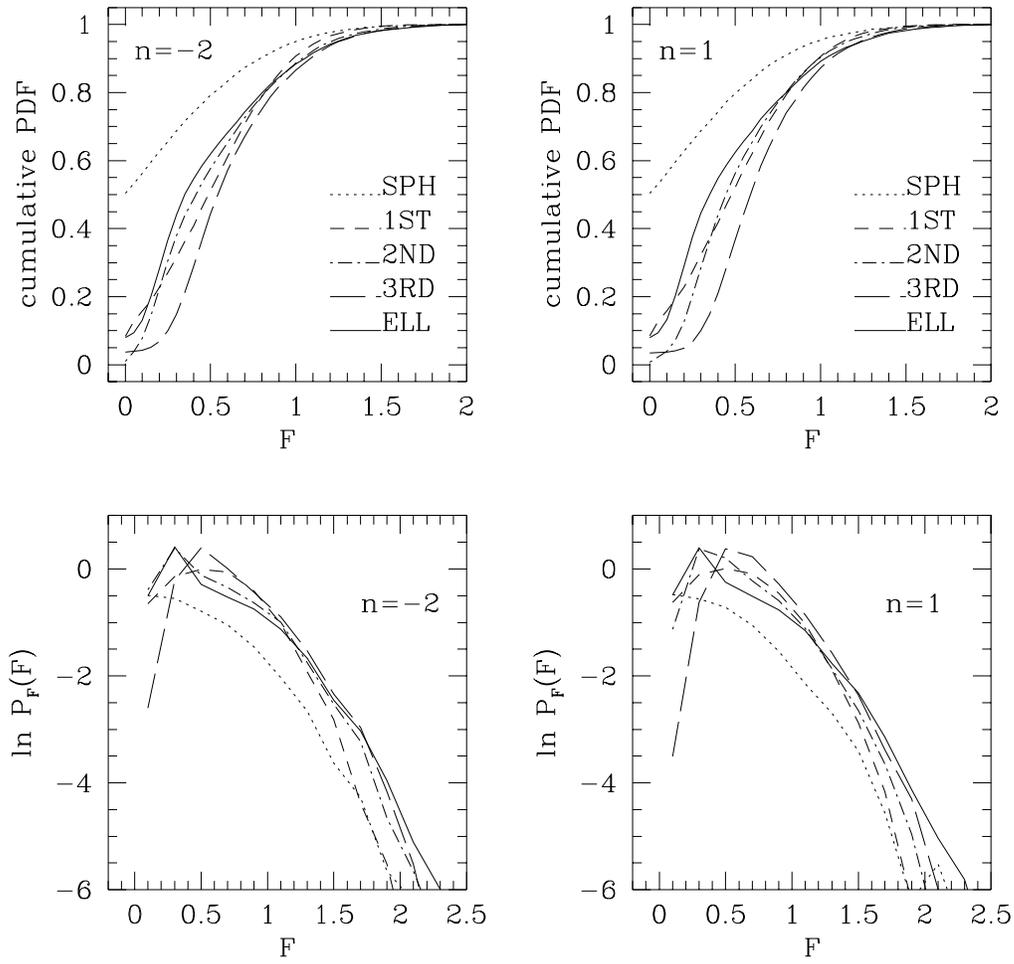,width=14cm}
\caption{Cumlative and differential PDFs of $F$.}
\end{center}
\end{figure}

The 1-point PDFs for the $F$ processes, relative to different
dynamical predictions, can be estimated by means of the Monte Carlo
realizations described in last section. Fig. 3.10 shows such PDFs for
two different spectral indexes, namely $n=-2$ and 1. Both cumulative
and differential curves (the latters in logarithm scale) are shown:
cumulative curves, being binning-free, are less noisy and directly
show the total fraction of mass which lies in a certain range of $F$,
while differential curves better represent the behavior of the various
PDFs, especially at large $F$ values. The following conclusions can be
drawn:

\begin{enumerate}
\item 
The SPH curve is quite different from all the others, even in the high
$F$ tail, which corresponds to fast collapsing points. As will be seen
again in next Chapter, a systematic deviation of collapse times from
the spherical value strongly influences the statistics of the rare
event tail, even though spherical collapse is asymptotically recovered
for the rarest points.

\item 
In the range $F\geq1$ the ZEL, 2ND and 3RD curves show a monotonic
shift toward large $F$ values, which is larger when passing from ZEL
to 2ND than from 2ND to 3RD; this can be interpreted as convergence
toward a solution. This is not true for $F<1$; the bad behavior of 2ND
in initial underdensities is one probable cause.

\item 
ELL and 3RD nearly coincide down to $F=1.2$, which corresponds to
$b_c=0.83$, and overall have a similar behavior; ELL slightly makes
more mass collapse at large $F$ values, because 3RD slightly
underestimates quasi-spherical collapses.  The main differences come
out in the range where the convergence of the Lagrangian series is not
guaranteed, but both ZEL, 2ND and 3RD have larger medians than ELL; so
ELL probably underestimates the collapses around $F=0.5$ or $b_c=2$.
It is to be stressed again that the behavior of Lagrangian dynamical
predictions in this range is not considered very robust.

\item 
The $n$ dependence can be appreciated in Fig. 3.10.  As expected, SPH,
ZEL and ELL are independent of $n$, while the $n$ dependence of 2ND
and 3RD is weak. Moreover, the difference between ELL and 3RD is
smaller for smaller $n$, i.e. when more large-scale power is present,
while we would expect the opposite if the difference of 3RD from ELL
were due to non-locality induced from large scales. In the following,
such a weak $n$-dependence will be neglected.

\item 
The range $F>1$, where the Lagrangian series converges and is in
agreement with ELL, corresponds to more than 10 \% of mass. 20 \% of
mass is found within $F>0.8$, where ELL and 3RD are still not very
different, while half of the mass is contained within $F\sim0.5$,
below which the various PDFs are different even qualitatively.  Then,
the convergence of the Lagrangian series and its agreement with ELL
influences a significant quantity of mass, which corresponds to more
than the large-mass tail of the MF, while the non-robustness of the
low-$F$ part is going to affect the power-law, small-mass part of the
MF.
\end{enumerate}

\begin{figure}
\begin{center}
\epsfig{file=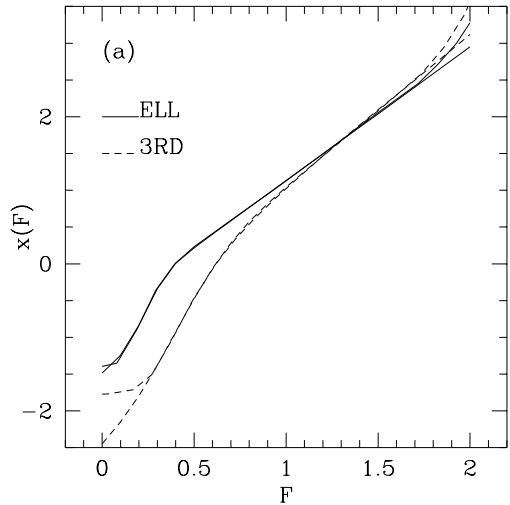,width=7cm}
\caption{$F\rightarrow x$ transformation.}
\end{center}
\end{figure}

In the following, the ELL and 3RD predictions will be considered, and
the PDFs will be mediated over four the spectral indexes, namely
$n=-2$, $-1$, 0 and 1. To obtain an analytical fit for the PDFs, it is
useful to find a transformation such as to ``Gaussianize'' the
distributions, i.e. such as to give:

\be P_F(F;\mres)dF = P_G(x(F);\mres)dx(F) = \frac{1}{\sqrt{2\pi\mres}}
{\rm e}^{-x(F)^2/2\mres} dx(F). \label{eq:Gaussianize} \ee

\noindent 
In practice, the transformed $x$ quantity is such to have its 1-point
PDF equal to that of the linear density contrast. (This does not mean
that $x$ is a Gaussian process: its N-point PDFs will not be that of
\dl). The $x(F)$ transformation, for the \res=1 case shown in
Fig. 3.10, can be found by solving the following ordinary differential
equation:

\be \frac{dx(F)}{dF} = \sqrt{2\pi} P_F(F) \exp\left(\frac{x(F)^2}{2}
\right). \label{eq:Gauss_transfom} \ee

\noindent 
This equation has been solved with an ordinary Runge-Kutta
algorithm. Initial conditions have been set by imposing $x=0$ at the
median of the $F$ distribution, and then integrating toward positive
and negative $x$ values. The solutions for ELL and 3RD are shown in
Fig. 3.11. At large $F$ values the transformation curves are simple
lines, which implies that the $P_F(F)$ curves are Gaussians (note that
the weak rises of the transformation curves at $F\sim 2$ are artifacts
of the numerical integration). Such curves are fit by the following
functions:

\bea x(F)_{ELL}&=& -0.69 + 1.82F - 0.4 ({\rm erf}(-7.5 F + 1.75) + 1)
\label{eq:Gauss_fit}\\
     x(F)_{3RD}&=& -1.02 + 2.07F - 0.75({\rm erf}( -3  F + 1.18) + 1).
\nonumber \eea

\noindent 
These analytical fits are shown in Fig. 3.11; note that the not
interesting $F<0.3$ part is not well reproduced by the fits.  The
first two terms carefully fit the linear parts of the $x(F)$ curves,
while the error functions reproduce the drop at small $F$ values.

\begin{figure}
\begin{center}
\epsfig{file=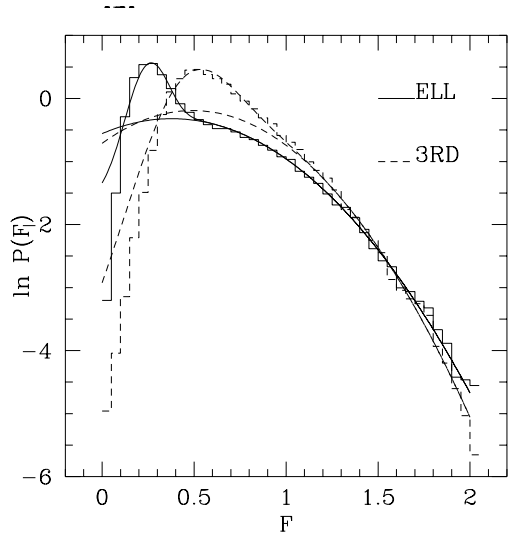,width=7cm}
\caption{$F$-PDFs for  ELL and 3RD, with Gaussian and complete fits.}
\end{center}
\end{figure}

Fig. 3.12 shows the ELL and 3RD PDFs, together with their linear and
complete fits. The linear fits reproduce well the large-mass parts,
which are then demonstrated to be accurately Gaussian, but do not
reproduce the pronounced peak present at $F\sim 0.5$. Such a peak is
well reproduced by the complete fits. Note also that the ELL PDF is
accurately Gaussian down to $F=0.5$, while the 3RD prediction starts
to deviate from the Gaussian curve at $F=1$.

The $x(F)$ and \pdff\ curves given above are relative to \res=1. Their
expression for any \res can be obtained by exploiting the following
scaling relations:

\bea P_F(F,\mres) &=& \sqrt{\mres}\, P_F(F/\sqrt{\mres},1) \nonumber\\
     x(F,\mres) &=& \sqrt{\mres}\, x(F/\sqrt{\mres},1).
\label{eq:rescaling} \eea 

\noindent 
As a final technical remark, it has been shown by Bernardeau (1994b)
that the shape of the filter does not influence the statistical
properties of the filtered process, as long as filtering is performed
in Lagrangian space.

\section{Discussion}

The MF problem can be formulated in terms of a non-Gaussian process
$F$, representing the inverse collapse time of a generic mass
element. If linear theory with a threshold (or spherical collapse) is
considered, $F$ is proportional to the linear density contrast, and is
then Gaussian. More realistic estimates of collapse times lead to $F$
processes which have a different statistical behavior, even in the
rare event tail.

As a matter of fact, the $F$ process is a function of the resolution;
this implies that, analogously to what happened in the spherical case,
a whole $F(\mres)$ trajectory is given to any point.  It is then
natural to think about applying to the $F$ process $F$ the absorbing
barrier formalism, described in \S 2.3.2, or another of the approaches
described in Chapter 2.  However, the complicated non-Gaussian
character of the $F$ process can considerably complicate the
calculations with respect to the linear theory case. In particular, an
analytical approach based on the peaks of the $F$ process (such as the
confluent formalism of Manrique \& Salvador-Sol\'e (1995)) is
essentially hopeless: an analytical description of the peaks of a
random field can be obtained only for Gaussian or closely related
fields (see Adler 1981), or for asymptotically high peaks (Catelan,
Lucchin \& Matarrese 1988).  On the other hand, the simpler excursion
set approach requires knowledge of only the 1-point PDF of the $F$
process, which has been obtained above. Section 4.2 will be devoted to
showing that, in the SKS smoothing case, the diffusion formalism can
be applied to the $F$ process.

On the other hand, the use of the excursion set formalism has to be
carefully justified in the present dynamical context. As a matter of
fact, the collapse times described above predict the collapse of a
large fraction of the mass (see Fig. 3.10), 92\% for ELL (just like
the Zel'dovich approximation) and even more for 3RD, even though the
exact amount is not considered as a robust prediction; this is at
variance with linear theory, which predicted the collapse of only 50\%
of mass. As a consequence, an MF calculated by means of a PS-like
approach, i.e. by determining the fraction of mass which is predicted
to collapse at a given scale, will be nearly normalized: the ``fudge
factor'' would be only, $1/0.92 \simeq 1.09$, or even more similar to
1 in the 3RD case.  On the other hand, the inconsistency connected to
the interpretation of downcrossing events remain: when a $F(\mres)$
trajectory first upcrosses a barrier, the point is interpreted as
collapsed at that resolution; if the trajectory downcrosses the
barrier, the point is interpreted as not collapsed at a larger
resolution, i.e. at a smaller scale. As a point which belongs to a
multi-stream region at a scale cannot be in single-stream regime at a
smaller scale, such downcrossing events are not considered as
physical, and then, to avoid them, trajectories are absorbed by the
barrier. The thing can be seen from another point of view: collapse
predictions are just approximate, and when different collapse
predictions are in contradiction the one at the smallest resolution is
believed.

Another important point regards the punctual nature of the dynamical
predictions used. Any collapse prediction, though
resolution-dependent, is just relative to a point of vanishing volume
and mass, not to an extended region of filter-length size.  Then,
there is no need to introduce spatial correlations in the diffusion
problem, as it happened within the global interpretation of the
collapse time (see \S 2.2.4).  On the other hand, if the underlying
potential field is smoothed at a scale $R$, the excursion sets will
have a typical size of order $R$, which is the relevant characteristic
scale, but the exact relation between $R$ and the size of collapsed
objects in Lagrangian space will be determined by the same spatial
correlations which have been avoided in the diffusion formalism. In
other words, the punctual interpretation allows us to neglect spatial
correlations in the diffusion problem, but the same correlations come
in play within the geometrical problem of estimating the size, and
then the mass, of the collapsed regions.  This point will be deepened
in next \S 4.4.

Two further technical points have to be noted. First, as $F$ is the
inverse of a time, the position of the absorbing barrier has a precise
meaning: it has to be put at the (inverse) instant at which the MF is
wanted. In other words, there is no free $\delta_c$ parameter in this
MF theory; this is due to the improved dynamical description of
collapse.  Second, the shape of the smoothing filter ought to be
chosen to optimize the dynamical prediction; in this sense, Gaussian
smoothing is usually suggested (Melott, Pellman \& Shandarin 1994;
Wei\ss, Gottl\"ober \& Buchert 1996), but such preference has been
found by testing density fields in the mildly non-linear regime, while
in principle collapse predictions could be optimized by different
filters.

Finally, I want to make some remarks on the concept of locality in
gravitational dynamics, which has recently been stressed again by
Kofman \& Pogosyan (1995). From the mathematical point of view, a
continuous system is said to evolve locally if its evolution equations
are ordinary differential equations, with no spatial derivatives (and
if there is no explicit coupling between different points). Indeed,
the absence of spatial derivatives in the evolution system makes any
point evolve independently by all the other points, once initial and
boundary conditions have been given. From a physical point of view, a
system can be said to evolve locally if mass elements are not
influenced in their evolution by the surrounding matter. Linear theory
and spherical collapse predict local evolution, as the fate of any
mass element is decided only by its initial density. On the other
hand, ZEL is able to predict the trajectory and deformation of every
mass element once the \lam\ eigenvalues relative to the point are
given, so that the evolution is local from the mathematical point of
view; but the \lam\ eigenvalues contain non-local information about
tides, so that the ZEL evolution of mass elements is physically
non-local. Such non-locality is due to the fact that the density and
potential fields are connected by a non-local Poisson equation.

It is interesting to note that the same reasoning can be applied to
Lagrangian perturbation theory at any order: initial conditions
contain non-local information through the perturbative potentials,
which are connected to the initial potential through complicated
non-linear and non-local Poisson equations, but, once the perturbative
potentials are known in a given point, the fate of that point can be
locally followed, neglecting the fate of all the other points. Then,
Lagrangian perturbation theory describes a dynamical evolution which
is non-local from the physical point of view, but local, and then easy
to compute, from the mathematical point of view.  As a further
example, N-body simulations follow the dynamical problem in its full
non-locality; as a consequence, all the particle trajectories have to
be followed at once, and this makes such calculations dramatically
slower than the ``local'' Lagrangian perturbation theory.

\chapter[Statistics]{Statistics and the Mass Function}

Chapter 3 was dedicated to finding realistic estimates of collapse
times of generic mass elements.  The problem of translating such
information into an expression for the MF is of purely statistical
nature. It was shown in Chapter 2 that such a statistical problem, in
the simple case in which the inverse collapse time $F$ is proportional
to the initial density contrast, has received much attention in the
scientific literature. Two main approaches were identified, namely the
excursion set and the peak ones; the first approach was shown to be
easier to manage than the second one, at the expense of a simplified
treatment of the geometry of collapsed regions in Lagrangian space,
while the peak approach, whose validity relies on the validity of the
peak hypothesis, better takes into account geometry, at the expense
of an increased complexity of the formalism, especially when trying to
include a proper treatment of the peak-in-peak problem.

It is not trivial to extend these approaches to the case in which $F$
is a complicated non-Gaussian process. In particular, the peak approach
seems to be far too complex to be used in this context; even in the
Gaussian case, it is not easy to calculate simple expectation values,
such as the mean number of peaks of given height or their mean shape,
and such quantities have never been calculated for non-Gaussian
fields, except those which are simple functions of Gaussian fields
(see Adler 1981), or in the case of asymptotically high peaks
(Catelan, Lucchin \& Matarrese 1988). The reason for that is that the
peak constraint (the condition which has to be satisfied by a point to
be a peak) requires knowledge of spatial correlations.  On the other
hand, the simpler excursion set approach only requires knowledge of
the 1-point PDF of the process considered, provided the punctual
interpretation of collapse times applies.  Anyway, the extension of
the excursion set formalism to a general collapse prediction requires
much care.

In \S 4.1 the MF will be determined by means of a simplified PS-like,
single-scale approach, which consists of estimating the probability of
having initial conditions that make the mass element collapse.  In \S
4.2 it will be demonstrated that it is possible to extend the
diffusion formalism of Bond et al. (1991) to the $F(\mres)$
trajectories, if SKS filters are used. To this aim, a number of
concepts from stochastic calculus will have to be used.  Then it will
be shown that the diffusion problem can be recast in term of a Wiener
process (a random walk) absorbed by a moving barrier; such problem
will be numerically solved and some accurate analytical approximate
formulas, valid at large masses, will be given. \S 4.3 will be
dedicated to the absorbing barrier problem in the case of Gaussian
smoothing: a useful analytical approximation, proposed by Peacock \&
Heavens (1990), can be used to calculate the first upcrossing rate.
The complex geometry of collapsed regions in Lagrangian space will be
shown, in \S 4.4, to have a relevant role in the passage from the
resolution to the mass variable. Final remarks are presented in
\S 4.5. All the topics presented in this chapter are contained in
Monaco (1995;1996b), and discussed in Monaco (1994;1996c-f).

\section{PS-like approach}

A first determination of the MF can be obtained by applying the same
statistical approach as in the original PS paper: the fraction of
collapsed mass is obtained by integrating, at a given fixed scale,
over all initial conditions which make a mass element collapse before
a given time.  This determination obviously suffers from the same
cloud-in-cloud problem as the PS one; however, as noted in \S 3.5,
most mass is predicted to finally collapse by realistic collapse time
estimates, so a PS-like MF is nearly normalized, by more than 90\%. It
is then natural to suspect that an MF obtained by means of the
absorbing barrier formalism can be not very different from the PS-like
one, as only a minor part of the mass, lying in the strongest
underdensities, has to be redistributed; this mass is not expected to
influence the MF in any interesting mass range. In \S 4.2 and 4.3 it
will be shown under which conditions the PS-like and the absorbing
barrier MFs are very similar; here I just anticipate that the
simplified PS-like approach suffices in finding the main features of
the dynamical MF.

Integration over initial conditions requires that initial conditions
are specified, and that their joint PDF is known. In the PS case,
where linear theory and spherical collapse were used, initial
conditions were simply provided by the initial density contrast.  In
the most general case, initial conditions are given by the value of the
density (or, equivalently, of the potential) at every point, so that a
direct integration is hard to perform. An intermediate case is
provided by the Zel'dovich (ZEL) approximation, and by the other
approximations which require the same initial conditions, as the {\it
ansatze} proposed in \S 3.1 and ellipsoidal (ELL) collapse. In this
case, the joint PDF of initial conditions, the \lam\ eigenvalues, is
known (Doroshkevich 1970):

$$ P_\lambda(\muno,\mdue,\mtre)d\muno d\mdue d\mtre = 
\frac{675\sqrt{5}}{8\pi\mres^3} \exp\left( -\frac{3}{\mres} \mu_1^2+
\frac{15}{2\mres}\mu_2\right) $$
\be \times (\muno-\mdue)(\muno-\mtre)(\mdue-\mtre)d\muno d\mdue d\mtre.
\label{eq:dorosh} \ee

\noindent 
\res\ is again the mass variance; $\mu_1=\muno+ \mdue+ \mtre$ and
$\mu_2=\muno\mdue+ \muno\mtre+ \mdue\mtre$ are principal invariants of
the ZEL deformation tensor, as defined in Eq. (\ref{eq:invariants}).
It is convenient to express this PDF in terms of the linear density
\dl\ and the $x$ and $y$ variables defined in \S 3.1, Eq.  
(\ref{eq:x_and_y}); in this case the joint PDF is factorized into a
Gaussian for \dl\ and a joint PDF for $x$ and $y$:

$$ P(\delta_l,x,y) d\delta_ldxdy= \frac{1}{\sqrt{2\pi\mres}}\exp\left( 
-\frac{\delta_l^2} {2\mres} \right)d\delta_l\frac{225}{4} \sqrt{\frac{5}{2\pi}}
\frac{1}{\mres^{5/2}} \exp \left(-\frac{5}{2\mres} (x^2+xy+y^2)\right) $$
\be \times xy(x+y)dxdy=\mpdfl d\delta_l \times P_{x,y}(x,y;\mres)dxdy. 
\label{eq:dorosh_due} \ee

\begin{figure}
\begin{center}
\epsfig{file=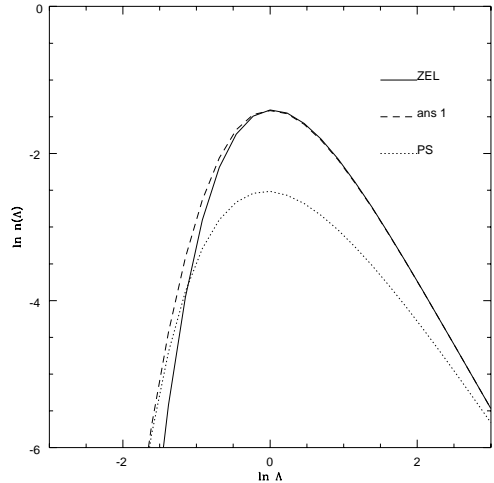,width=7cm}
\caption{PS-like $n(\mres)$ curves for ZEL and ans\"atz 1, compared to the PS 
one.}
\end{center}
\end{figure}

The fraction of collapsed mass can then be obtained as follows:

\be \mimfr = \int_0^\infty \!dx \!\int_0^\infty \!dy P_{x,y}(x,y;\mres)
\!\int_{\delta_c(x,y)}^\infty \! d\delta_l\, \mpdfl, \label{eq:integralold} \ee

\noindent 
where the function $\delta_c(x,y)$, which substitutes the \dc\
parameter of PS, is defined as the solution of the equation:

\be b_c(\delta_c,x,y)=b(t_0), \label{eq:deltac_def} \ee

\noindent
where $b(t_0)$ is the instant at which the MF is wanted (it will
usually be set equal to one). By writing the function $\delta_c$ as
$\delta_0-f(x,y)$, where $\delta_0$ is the spherical value 1.69 and
the positive function $f(x,y)$, vanishing at the origin, gives the
effect of the shear, it is possible to write the \dmfr\ function as:

\be n(\mres)=n_{PS}(\mres)\times {\cal I}(\mres), \label{eq:correction} \ee

\noindent 
where $n_{PS}(\mres)$ is the PS curve, Eq. (\ref{eq:ps_ndires}), and
${\cal I}(\mres)$ is a correction term:

$$ {\cal I}(\mres) = \frac{1}{\mres} \int_0^\infty \!dx 
\!\int_0^\infty \!dy\, P_{x,y}(x,y) \exp \left( -\frac{1}{2\mres}f^2(x,y)+
\frac{\delta_0}{\mres} f(x,y) \right)$$
\be \times\left( 
1-\frac{1}{\delta_0} \left(f-x\frac{\partial f}{\partial x} -y 
\frac{\partial f}{\partial y} \right)\right). \label{eq:i_factor} \ee

\dmfr\ curves have been calculated for ZEL, ELL, and for the two {\it
ansatze} presented in \S 3.1; details of the calculations are reported
by Monaco (1995). Fig. 4.1 presents the \dmfr\ curve for ZEL and for
the first {\it ans\"atz}, Eq. (\ref{eq:ansatza}), in which spherical
collapse is recovered when ZEL predicts a slower collapse. The
canonical PS \dmfr curve is shown for comparison; the PS curve has not
been multiplied by the fudge factor of two at this stage, in order to
compare the results of the PS-like procedures in the different cases,
with no guarantee of normalization (in Monaco 1995 all the curves were
multiplied by two).  It can be seen that the ZEL curve underestimates
the number of large-mass\footnote{I freely use the word mass in this
context to indicate the large-mass (small \res) or small-mass (large
\res) part of the MF} objects, and gives more intermediate- and
small-mass objects, also thanks to its better normalization. The {\it
ans\"atz} curve reproduces the PS one at large masses, and reduces to
the ZEL one at small masses, as expected.

\begin{figure}
\begin{center}
\epsfig{file=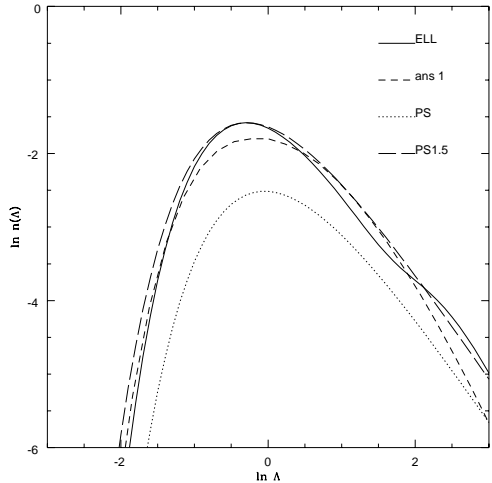,width=7cm}
\caption{PS-like $n(\mres)$ curves for ELL and ans\"atz 2, compared to 
the PS one and with a PS with \dc=1.5 and the fudge factor 2; see text.}  
\end{center}
\end{figure}

Fig. 4.2 shows the ELL prediction, in comparison with the second {\it
ans\"atz}, Eq. (\ref{eq:ansatzb}) (with $\epsilon=0.2$), the canonical
PS curve (without the fudge factor 2, as before) and a PS curve with
\dc=1.5, representative of the typical outcome of N-body simulations
(and then with the factor 2; see \S 2.2). Both the {\it ans\"atz}
curve and ELL one predict an overabundance of large-mass clumps with
respect to the canonical PS curve: it is again demonstrated that a
systematic displacement of collapse times from the spherical value
influences the large-mass tail of the MF, even though spherical
collapse is asymptotically recovered. In particular, the ELL curve is
quite similar to the PS 1.5 curve. It is however to be stressed that
this similarity, while encouraging, has to be taken with care, as it
is not clear how the collapsed objects predicted by this theory are
related to the N-body clumps.  As a technical remark, this ELL curve
has been computed by using the full $b_c$ curves found in \S 3.2; in
Monaco (1995) only the overdense curve was considered, and the ZEL
behavior was forced at large $x$ and $y$ values.

The following conclusions, which will be fully confirmed below, can be
drawn:

\begin{enumerate}
\item 
realistic collapse predictions cause an enhanced production of
large-mass clumps; the resulting MF is similar to a PS one with
$\delta_c=1.5$;
\item 
the fact that spherical collapse is asymptotically recovered for the
strongest overdensities does not guarantee that the ``spherical''
canonical PS MF, with \dc=1.69, is recovered at large masses.
\end{enumerate}

\begin{figure}
\begin{center}
\epsfig{file=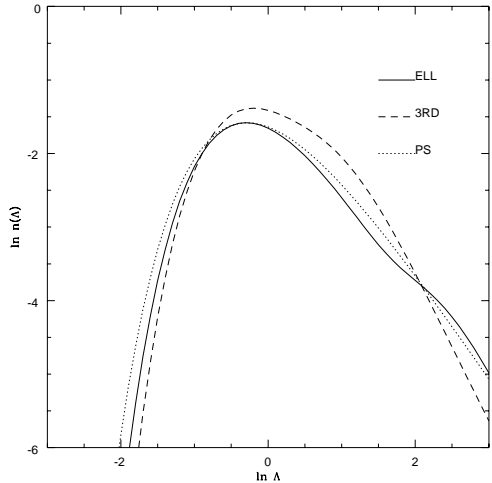,width=7cm}
\caption{PS-like $n(\mres)$ curves for ELL and 3RD, compared to 
a PS with \dc=1.5 and the fudge factor 2; see text.}
\end{center}
\end{figure}

The procedure described above is difficult to apply to Lagrangian
perturbation theory at second and third order, because it is hard to
find the joint PDF of the relevant initial conditions, i.e. of the
matrix elements of the perturbative contributions to the deformation
tensor.  On the other hand, the determination of the 1-point PDF of
the (inverse) collapse times, performed in \S 3.3 though Monte Carlo
calculations, implies in practice an integration of the joint PDF of
initial conditions over all the variables minus one. It is then
possible to calculate a PS-like MF in the following way:

\be \mimfr = \int_{F_c}^\infty dF\, P_F(F;\mres). \label{eq:pslike} \ee

\noindent
$F_c$ is the inverse of the time at which the MF is wanted (again set
to one for simplicity). Fig. 4.3 shows the ELL and 3RD PS-like curves,
again together with a PS curve with \dc=1.5 and the fudge factor 2.
As expected, the 3RD curve is similar to the ELL one at large and
intermediate masses, but has a different slope at small masses, where
the dynamical predictions are considered not robust.  It is also
apparent that ELL predicts slightly more large-mass clumps the 3RD;
this is caused by the fact that 3RD slightly overestimates
quasi-spherical collapse times, as shown in \S 3.3. Finally, both ELL
and 3RD are similar to the PS 1.5 at large and intermediate masses.

\section[SKS smoothing]{The diffusion formalism: sharp $k$-space smoothing}

In the excursion set approach, and within the punctual interpretation
of the collapse time, the cloud-in-cloud problem can be solved with
the absorbing barrier formalism proposed by Bond et al. (1991) (and
implicitly by Epstein 1983 and Peacock \& Heavens 1990). As explained
in \S 2.3.2, these authors showed that, if sharp $k$-space (SKS)
filtering of the initial field is used, the $\delta_l(\mres)$
trajectories are random walks, and the problem can be reformulated as
the diffusion of a Wiener process, absorbed by a barrier at \dc.  It
is not trivial to understand whether this formalism can be extended to
the case of the complicated non-Gaussian process $F(\mres)$, i.e.
whether the $F(\mres)$ process is a diffusion process; if this were
the case, the first upcrossing rate of $F$ could be found by solving a
1D Fokker-Planck equation. It will be numerically shown in this
section that this is the case.

\subsection{A Fokker-Planck equation for $F$}

It is necessary to introduce, at a qualitative level, a number of
concepts from stochastic calculus, some of which were already
mentioned in \S 2.3.2. All the definitions and theorems presented in
this sections can be found in many textbooks; I make reference to
Arnold (1973), an introductory textbook on stochastic calculus, and
Risken (1989), a textbook on the Fokker-Planck equation.  A general
process $\xi(\mres)$ is said to possess the {\it Markov property} if
its history at any given resolution $\mres'>\mres$ is determined just
by its value at {\res}, and is independent of its past history. In
more precise terms, its increments at different resolutions do not
correlate:

\be \langle d\xi(\mres)d\xi(\mres') \rangle \propto \delta_D(\mres-\mres').
\label{eq:markov} \ee

\noindent 
Note that this definition, and all the ones given below, apply also if
$\xi$ is a vector process; in this case, the proportionality constant
can include correlations of the various components of the process at
fixed \res.  A Markov process is called {\it Wiener process} if it is
Gaussian and if the proportionality constant in Eq. (\ref{eq:markov})
is 1, as in Eq.  (\ref{eq:random_walk}). The trajectories of a Wiener
process are called random walks (this definition follows the use of
Bond et al. (1991), but in Risken (1989) this name is reserved to
processes with a viscosity term). The PDF of a Markov process, $P_\xi
(\xi,\mres)$, obeys a Fokker-Planck (hereafter FP) equation, provided
some regularity conditions are satisfied (see Arnold 1973 and Monaco
1996b). In this case the process is called {\it diffusion process};
note that some authors reserve this name only to the Wiener process.
A FP equation has the form:

\be \frac{\partial P_\xi(\xi,\mres)}{\partial \mres} = \left[ - \frac{\partial}
{\partial \xi}D^{(1)}(\xi,\mres) + \frac{1}{2} \frac{\partial^2}
{\partial \xi^2} D^{(2)}(\xi,\mres) \right] P_\xi(\xi,\mres).
\label{eq:general_fp} \ee

\noindent 
The $D$ functions are called drift and diffusion coefficients; for the
Wiener process, $D^{(1)}=0$ and $D^{(2)}=1$.

The $F_\mq(\mres)$ process (the subscript \q\ indicating the Lagrangian
point of which the collapse time is calculated) is a non-linear and
non-local functional of the initial Gaussian potential
$\varphi(\mq;\mres)$:

\be F(\mq,\mres) = {\cal F}_\mq[\varphi(\mq',\mres)]. \label{eq:functional} \ee

\noindent 
It is important to note that ${\cal F}$ is deterministic and does not
act directly on the \res\ variable, but just on the \pot\ field {\it
at fixed resolution}.  The potential \pot\ can be considered as an
infinite-dimensional Wiener process, any dimension corresponding to
the value of the potential in a point. Then, the $F$ process is a
non-linear function of an infinite-dimensional Wiener process. It is
useful to consider discrete spaces $\{\mq_i\}$, as, e.g., in N-body
simulations; any of the considerations which follow will be valid in a
continuum limit. In this case, the \pot\ process is a vector Wiener
process of finite (although large) length, and the functional becomes
an ordinary function:

\bea 
F(\mq,\mres) & \rightarrow & F(\mq_i,\mres) = F_i(\mres) \nonumber\\
\varphi(\mq,\mres) & \rightarrow &\varphi(\mq_i,\mres) = \varphi_i(\mres) 
\label{eq:disc} \\
{\cal F}_\mq[\varphi(\mq',\mres)] & \rightarrow & {\cal F}_{\mq_i}
(\{\varphi(\mq_j,\mres)\}) = {\cal F}_i(\{\varphi_j(\mres)\})\; . \nonumber
\eea

Then, the MF problem can be recast in terms of a multi-dimensional
diffusion process: in fact, as \res\ changes, the point
$\{\varphi_i\}$, representing the potential configuration, performs a
multi-dimensional random walk. It then suffices to absorb such a
random walk with a barrier defined by the equation:

\be {\cal F}_i(\{\varphi_j(\mres)\}) = F_c. \label{eq:multi_barrier} \ee

\noindent 
This formulation does not seem easily manageable, but highlights the
Markov nature of the problem.  In the Zel'dovich case, the situation is
simpler: initial conditions are given by the six independent matrix
elements of the first-order deformation tensor $S^{(1)}_{a,b}
=\varphi_{,ab}$, which obviously are six correlated Gaussian Wiener
processes. The diffusion process is then a six-dimensional random
walk, which is absorbed by the barrier defined by (recall that
$F=\lambda_1$ in this case):

\be \lambda_1(S^{(1)}_{1,1}(\mres),\ldots,S^{(1)}_{3,3}(\mres))=F_c. 
\label{eq:zel_barrier} \ee

On the other hand, the multi-dimensional random walk, in six or in a
very large number of dimensions, can be mapped into an $F(\mres)$
trajectory; as long as the ${\cal F}$ functional is well-behaved
enough, there will be a biunivocal correspondence between any
upcrossing of the \pot\ process above the complicated multi-dimensional
barrier and any upcrossing of the $F$ process above the simple \fc\
barrier. Then, if the $F$ process were a Markov (a diffusion) process,
the problem could be equivalently formulated in terms of the 1D $F$
process, with a great gain terms of mathematical simplicity.

\begin{figure}
\begin{center}
\epsfig{file=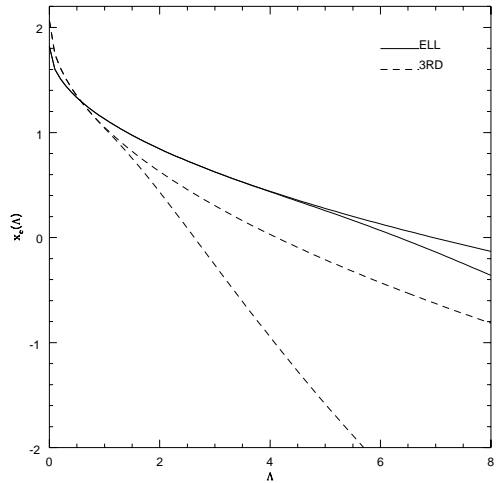,width=7cm}
\caption{Moving barriers, linear and complete.}
\end{center}
\end{figure}

The evolution equations for the $F_i$ processes are obtained by means
of chain-rule differentiation:

\bea dF_i &=& \frac{\partial{\cal F}_i}{\partial \varphi_j} (\{\varphi\})
d\varphi_j \label{eq:whole_system} \\ d\varphi_i &=& f_{ij}(\mres) d{\rm W}_j. 
\nonumber \eea

\noindent 
$dW_j$ are $N$ independent Wiener processes, and the $f_{ij}$
coefficients in the equation for $\varphi_i$ are such to reproduce the
correct variances and correlations.  This is a non-linear Langevin
system, and this simple fact, together with some regularity conditions
(the drift and diffusion coefficients have to be defined as suitable
averages of the transition probability; see Arnold 1973) suffices in
demonstrating that the whole $\{F_i,\varphi_j\}$ system is a diffusion
process, and then admits a FP equation. Then, the MF problem can be
reformulated in terms of the complicated multidimensional
$\{F_i,\varphi_j\}$ process with the simple $F=F_c$ absorbing barrier.

The fact that each of the $F_i$ processes is a component of a
diffusion process does not guarantee that it is a diffusion process by
itself. On the other hand, it is not possible to find a FP equation
for $F_i$ by just integrating out all the other variables from the FP
equation for the whole system, as the drift and diffusion coefficients
are not constant, and then can not be taken out from the integration.
As a consequence, there is no immediate reason to believe that a
solution of the 1D problem, obtained by treating $F$ as a diffusion
process, would give the correct solution of the multimensional problem
for the statistics of the first upcrossings of $F$. Nonetheless, this
solution of the 1D problem can be considered as an useful {\it
ans\"atz} to the true solution, to be compared {\it a posteriori} with
a numerical solution of the multidimensional problem. I will show in
the following how to obtain the solution of the 1D problem, and how
this solution perfectly describes the actual first upcrossing rate of
$F$, found by means of the Monte Carlo simulations described in \S
3.3. Then, it will be concluded that the diffusion formalism, in the
SKS case, can be extended to the $F$ process. This is enough for the
present purposes, but leaves the question open on the possible Markov
nature of the $F$ process.

To solve the 1D problem, it is necessary to construct a FP equation
whose solution is \pdff. Through Eq. (\ref{eq:Gauss_fit}), given in \S
3.4, it is possible to transform $F$ into a process $x$, whose PDF is
a Gaussian with null mean and variance \res. The FP equation for $x$
is obviously that of a Wiener process:

\be \frac{\partial P_x(x,\mres)}{\partial \mres} = \frac{1}{2}
\frac{\partial^2}{\partial x^2} P_x(x,\mres).  \label{eq:wiener_fp} \ee

\noindent 
It is possible from this equation and from the $x(F)$ transformation
to obtain the FP equation for $F$; this is done in Appendix A of
Monaco (1996b). However, it is much more convenient to work in terms
of the Wiener $x$ process. In this case, it can be seen from
Eqs. (\ref{eq:Gauss_fit}) and (\ref{eq:rescaling}) that the fixed
barrier at \fc\ is transformed into a moving barrier for $x$.

In conclusion, if SKS smoothing is used, the fraction of collapsed
mass, calculated by assuming that $F$ is a diffusion process, can be
found by solving the problem of the diffusion of a Wiener process with
a moving absorbing barrier.

\subsection{The moving barrier problem}

The solution of the fixed barrier problem was found by Chandrasekar
(1943): consider Eq. (\ref{eq:wiener_fp}), subject to the constraint
$P_x(x_c;\mres)=0$ at any \res, and to the initial condition
$P_x(x,0)= \delta_D(x)$.  If a negative image is put in a symmetric
position with respect to the barrier, i.e. if the initial condition is
changed to $P_x(x,0)=\delta_D(x)-\delta_D(2x_c-x)$, then it is easy to
show that the solution \pdfxf\ (fb denoting fixed barrier) is:

\be \mpdfxf dx = \frac{1}{\sqrt{2\pi\mres}} \left[ \exp \left( 
-\frac{x^2}{2\mres} \right) - \exp \left( -\frac{(2x-x_c)^2}{2\mres} 
\right) \right] dx. \label{eq:chandra} \ee

\noindent 
This solution manifestly satisfies the constraint at any \res;
moreover, it makes sense only for $x<x_c$, turning negative beyond the
barrier.  Finally, note that no meaningful solution exists for
$x_c<0$, as the trajectories are all absorbed from the start.  

\begin{figure}
\begin{center}
\epsfig{file=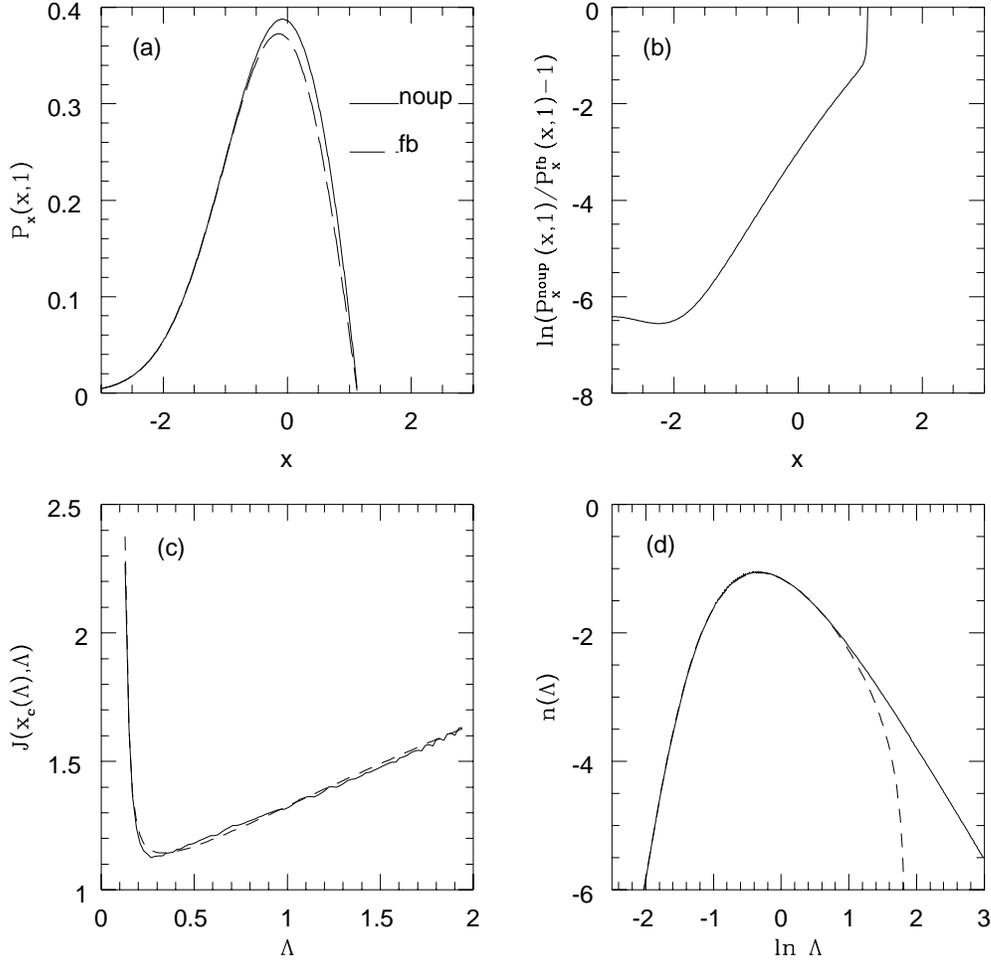,width=14cm}
\caption{Moving barrier problem: ELL prediction, linear barrier.}
\end{center}
\end{figure}

Such procedure does not apply to the moving barrier case.  To show
this, it is useful to trace the problem back to the $F$ variable: the
barrier is then fixed, but the process has a non-vanishing (and
possibly time-dependent) drift coefficient, which causes the mean
value of the process to change with \res.  In this case, any negative
image, put specularly with respect to the barrier, ought to move
specularly with respect to the positive solution, i.e. with a drift of
opposite sign, in order to ensure the barrier constraint to be
satisfied at any \res.  Then, the image would not obey the same FP
equation as the positive solution, but another equation with drift of
opposite sign.

\begin{figure}
\begin{center}
\epsfig{file=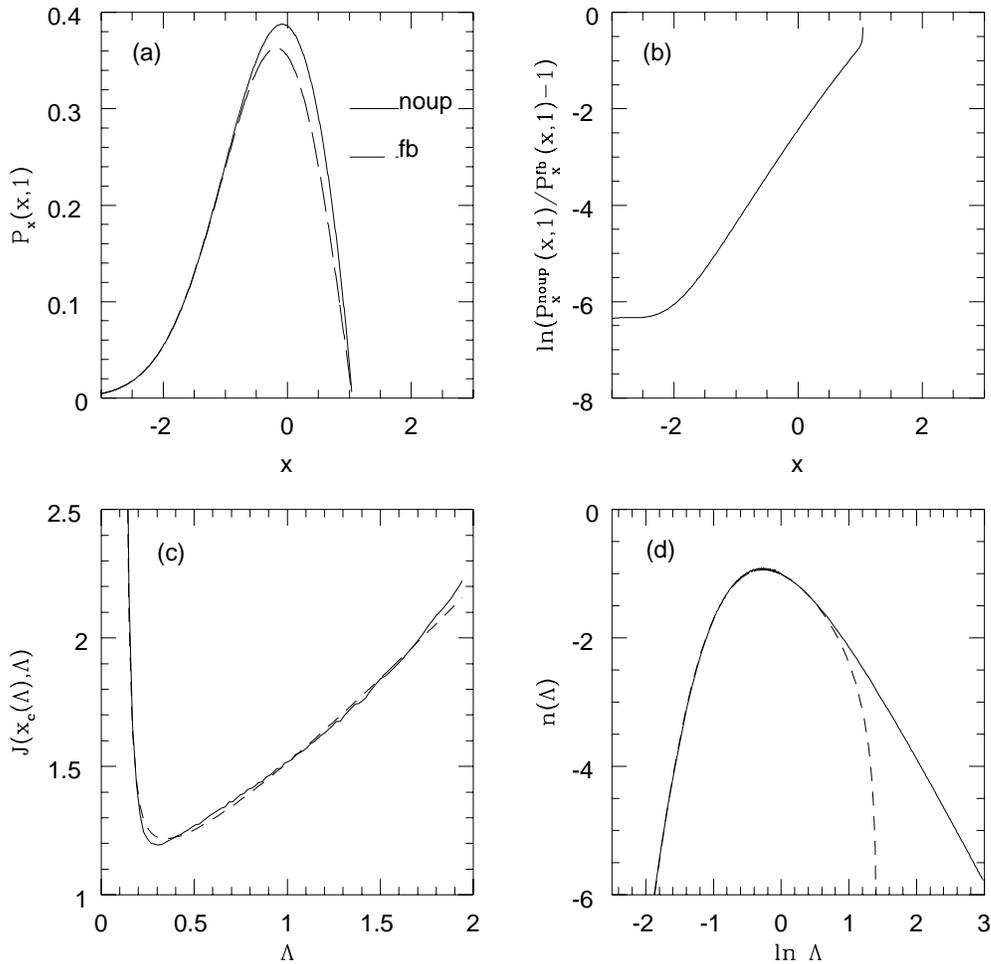,width=14cm}
\caption{Moving barrier problem: 3RD prediction, linear barrier.}
\end{center}
\end{figure}

No analytical solution has been found for the moving barrier problem;
some attempts are reported in Appendix B of Monaco (1996b). It is
however possible to numerically integrate the FP equation with a
standard Cranck-Nicholson method, as that described in Press et
al. (1992). It consists of a finite-interval integration; intervals
have been chosen as $\Delta x = 7.5\; 10^{-3}$, $\Delta\mres = 5\;
10^{-5}$, so that the critical parameter $\alpha=\Delta\mres/ 2(\Delta
x)^2$, which determines the stability of the calculation, is 0.444;
this is small enough to give precise results. Initial conditions have
been assigned by assuming an unperturbed Gaussian for the PDF, at \res\
values small enough that no trajectory has yet met the barrier.  The
numerical solution of the moving barrier problem will be denoted as
\pdfxu.

The absorbing barrier $x_c(\mres)$ can be found by setting $F=F_c$ in
the transformation Eq. (\ref{eq:Gauss_fit}), generalized to any \res\
by means of the scaling relation (\ref{eq:rescaling}). For the ELL and
3RD predictions, they are:

\bea x_c(\mres) &=& 1.82 F_c - 0.69 \sqrt{\mres} - \label{eq:barriers}\\&& 
0.4 \sqrt{\mres} ({\rm erf}(-7.5 F_c/\sqrt{\mres} + 1.75) + 1)\ \ {\rm (ELL)}
\nonumber\\ 
     x_c(\mres) &=& 2.07 F_c - 1.82 \sqrt{\mres} - \nonumber\\&& 0.75 
\sqrt{\mres} ({\rm erf}( -3  F_c/\sqrt{\mres} + 1.18) + 1)\ \ {\rm (3RD)}. 
\nonumber \eea

\noindent
It was recognized in \S 3.4 that the $x(F)$ transformation is
accurately linear for moderate and large $F$ values. In this case, the
terms in the first lines of Eq. (\ref{eq:barriers}) are obtained.
Such barriers are called {\it linear barriers}; note however that they
are not linear in \res! Fig. 4.4 shows the $x_c(\mres)$ barriers: for
the ELL prediction, linear and complete barriers remain very similar
up to large resolutions, while in the 3RD case they become
significantly different after \res=1.5. Then, linear barriers provide
a good approximation for the large- and intermediate-mass range.

\begin{figure}
\begin{center}
\epsfig{file=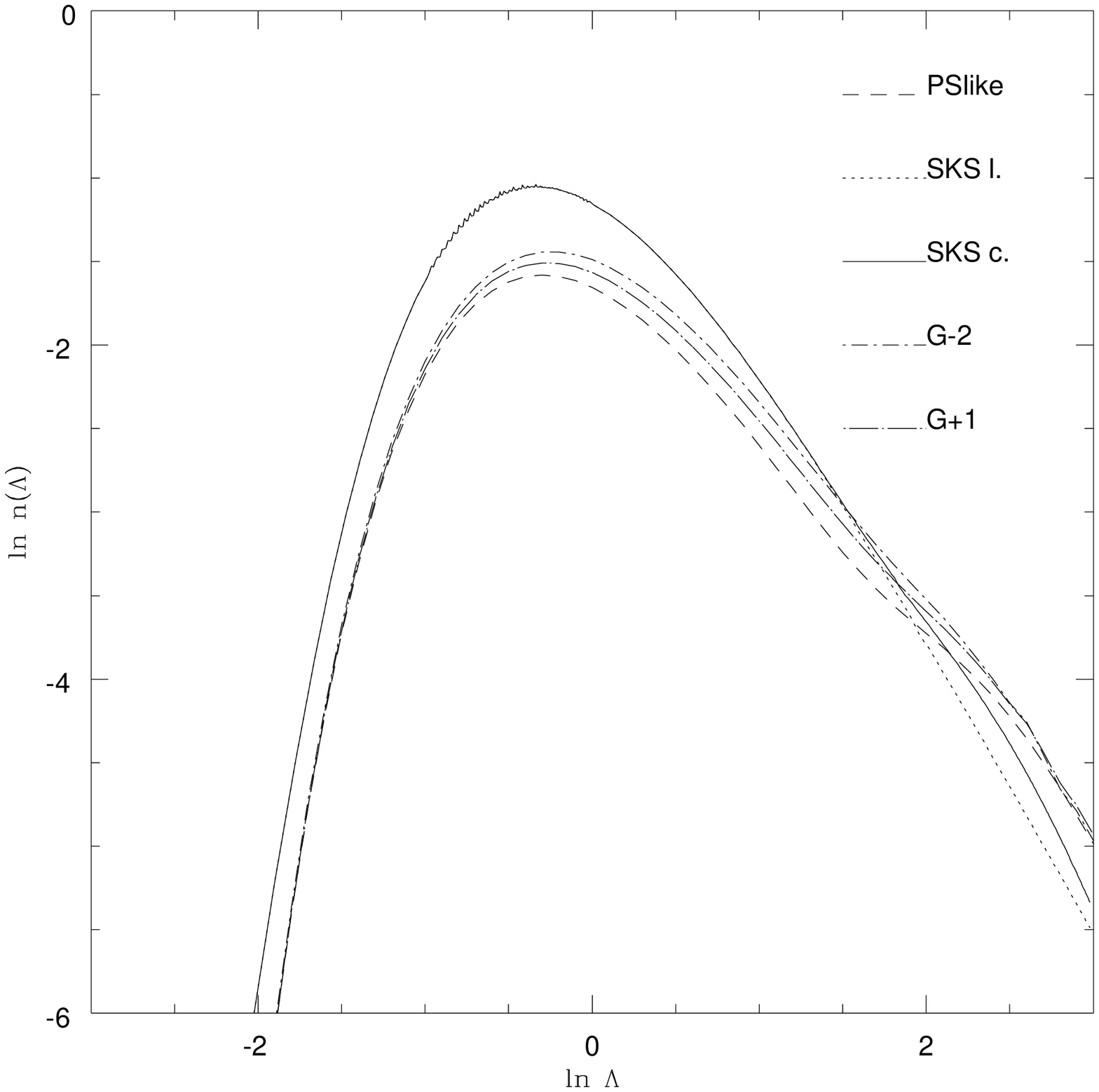,width=7cm}
\caption{$n(\mres)$ curves for the ELL prediction.}
\end{center}
\end{figure}

In the linear barrier case, the numerical solution has been found to
be related to the fixed barrier solution, found by inserting the
moving barrier $x_c(\mres)$ into Eq. (\ref{eq:chandra}):

\be \mpdfxu/\mpdfxf = {\cal J}(x,\mres) = 1+\exp(f(\mres)x+g(\mres)).
\label{eq:fb_correction} \ee

\noindent
Figs. 4.5a and 4.6a show, for ELL and 3RD, the numerical and fixed
barrier solutions at \res=1. Figs. 4.5b and 4.6b show the quantity
$\ln(\mpdfxu/$ $\mpdfxf -1)$, which is accurately linear up to errors
due to the numerical precision.  To find the $f(\mres)$ and $g(\mres)$
functions, it has to be noted that the corrected function, ${\cal J}
P_x^{\rm fb}$, has to be precise enough to correctly reproduce not
only the numerical curve but also its second derivative in $x$ and its
first derivative in \res, in order to satisfy the FP equation. Putting
this constraint, the following functions have been found:

\bea f(\mres) & = & -0.5  + 2.41\mres^{-1.08}\label{eq:linear_fit_ell}\\
     g(\mres) & = &  2.23 - 4.90\mres^{-1} \nonumber \eea

\noindent for the linear ELL transformation, and

\bea f(\mres) & = & -0.3  + 2.05\mres^{-1.13}\label{eq:linear_fit_trd}\\
     g(\mres) & = &  1.15 - 4.25\mres^{-1} \nonumber \eea

\noindent 
for the linear 3RD transformation. Figs. 4.5c and 4.6c show the
numerical ${\cal J}$ function and their analytical fits, which appear
quite satisfactory.

The integral MF is:

\be \mimfr = 1 - \int_{-\infty}^{x_c(\mres)}\mpdfxu dx. \label{eq:sks_imf}\ee

\noindent
It is then easy to show that the \dmfr\ curve can be written again as:

\be n(\mres)=n_{PS}(\mres)\times {\cal I}(\mres), \label{eq:correct_bis} \ee

\noindent where the correction function ${\cal I}$ is:

\be {\cal I}(\mres) =   \frac{x_c(\mres)}{\delta_c}\exp\left(-
\frac{2(x_c(\mres)-\delta_c)+(x_c(\mres)/\delta_c-1)^2}{2\mres}\right)
{\cal J}(\mres). \label{eq:correc_factor_bis} \ee

\noindent 
Figs. 4.5d and 4.6d show the numerical curves and their analytical
approximations, which appear to be quite satisfactory at large and
intermediate masses. The sudden fall of the analytical approximations,
which takes place at moderately large resolutions, is caused by the
fact that the barrier turns negative, and then the fixed barrier
solution cannot be used any more; this fact has no relevant meaning.

\begin{figure}
\begin{center}
\epsfig{file=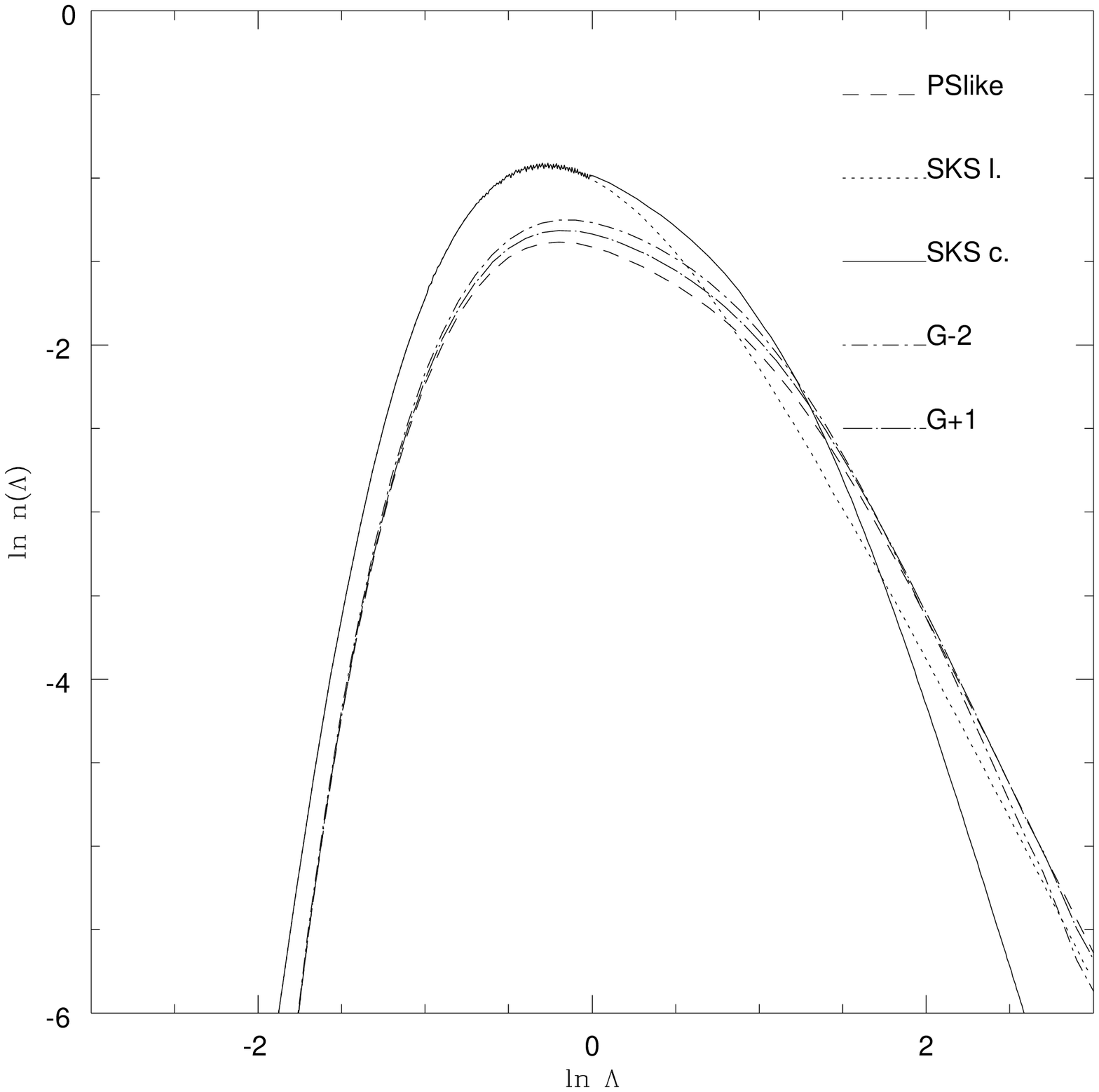,width=7cm}
\caption{$n(\mres)$ curves for the 3RD prediction.}
\end{center}
\end{figure}

The numerical \dmfr\ curves obtained by using the full barriers are
shown in Fig. 4.7 (ELL) and 4.8 (3RD), in comparison with the
(numerical) linear barrier solutions and with the PS-like curves of \S
4.1. In the ELL case, the two curves start to be different only at
small masses, while in the 3RD case this happens at intermediate
masses. In both cases the SKS curve gives, at large masses, roughly a
factor 2 more objects than the PS-like one, just like in the canonical
PS case, even though the ``fudge factor'' in the present case is not
more than 1.09.  Then, the justification of the factor 2 by means of
the diffusion formalism, in the PS case, appears just like a fortunate
coincidence.

\begin{figure}
\begin{center}
\epsfig{file=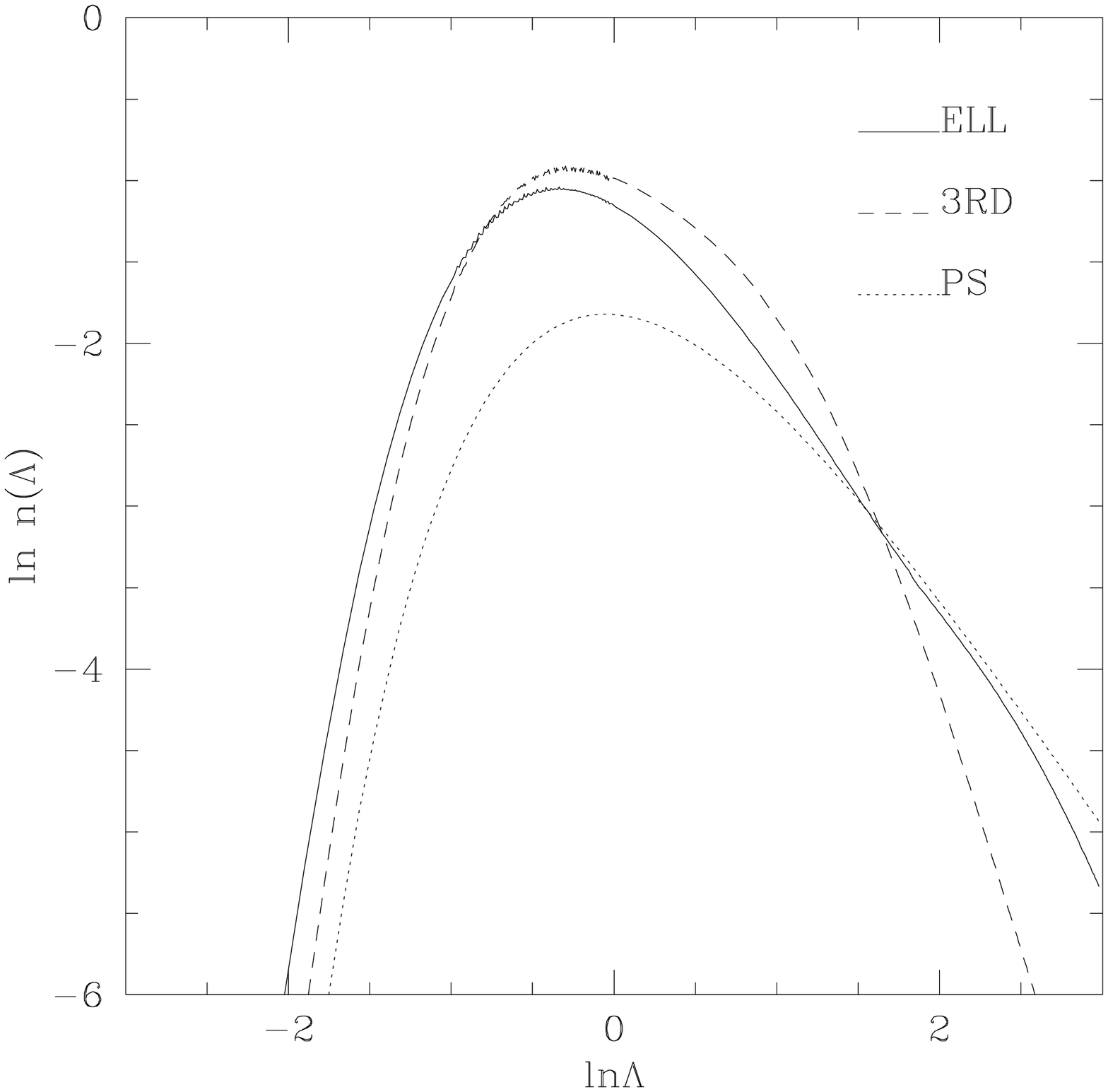,width=7cm}
\caption{$n(\mres)$ curves with SKS filtering.}
\end{center}
\end{figure}

Fig. 4.9 finally shows the (complete) ELL and 3RD curves, in
comparison with the corresponding canonical PS solution of the
diffusion problem ({\it with} the factor 2). The following conclusions
can be drawn:

\begin{enumerate}
\item both ELL and 3RD predict more large-mass objects  than the
canonical PS prediction;
\item both ELL and 3RD predict nearly a factor 3 more intermediate-mass
objects;
\item both ELL and, especially, 3RD predict steeper \dmfr\ curves at 
small masses (which means {\it flatter} MFs at small masses); the
exact slope depends sensitively on the uncertain particulars of the
moving barrier at large resolutions.
\end{enumerate}

The first conclusion is in agreement with what had been found in \S
4.1; the second one is worrisome, as the PS curve agrees well with
N-body simulations; this will be solved in next section. Finally, the
third conclusion shows how the small-mass part of the MF sensibly
depends on the uncertain dynamics of slowly-collapsing mass clumps.
Anyway, the trend of a decrease of small-mass objects, with a
corresponding bump at intermediate masses, goes in the same direction
as the findings of Porciani et al. (1996), of a low-mass cutoff
influenced by non-Gaussianity. While their effect is much more
dramatic, it is confirmed that a non-Gaussian $F$-distribution can
somehow introduce a new small-mass characteristic scale, which takes
place when most mass has collapsed (to be more precise, when the
non-linear barrier sets in).

\begin{figure}
\begin{center}
\hbox{\epsfig{file=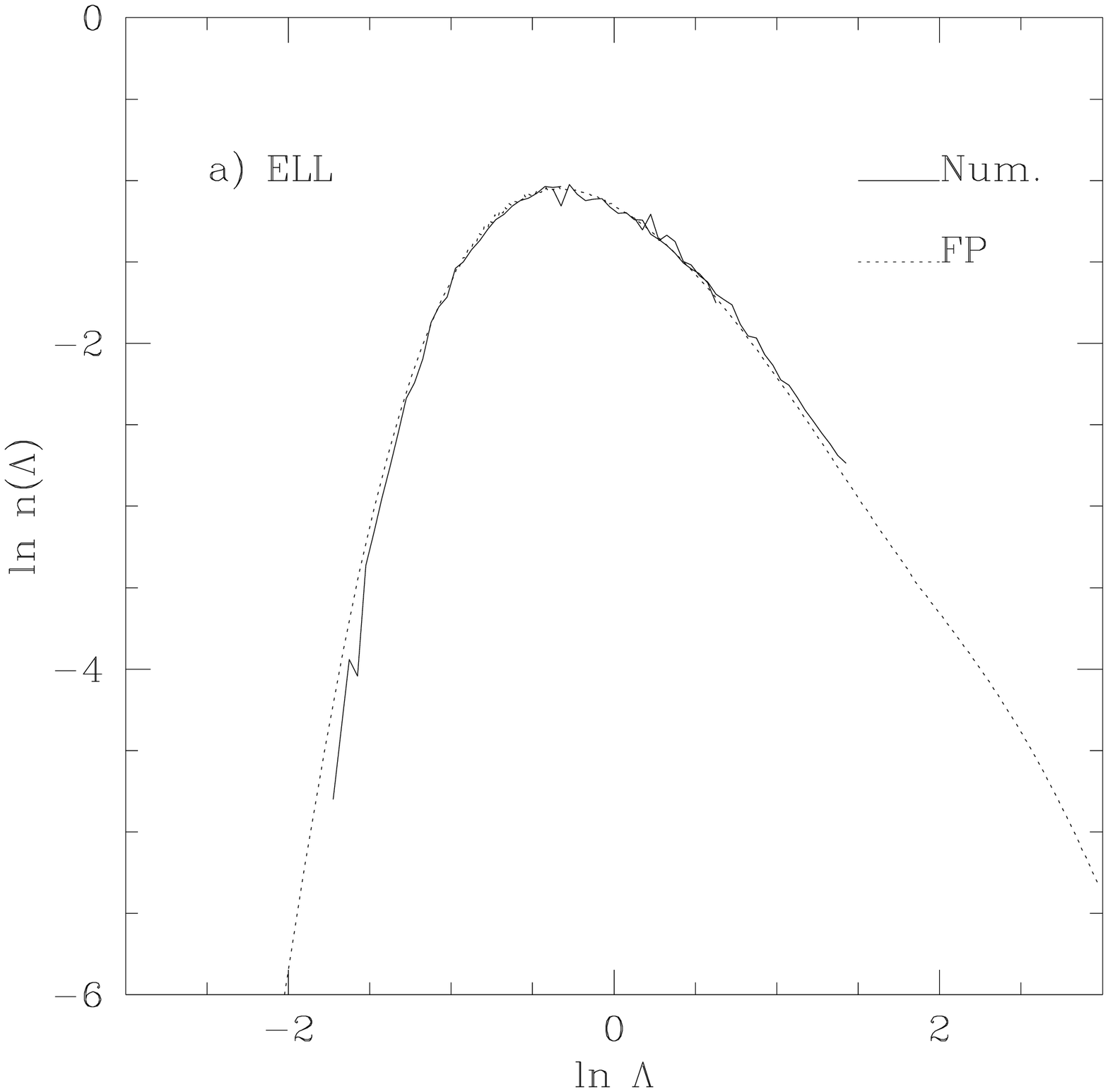,width=7cm}
      \epsfig{file=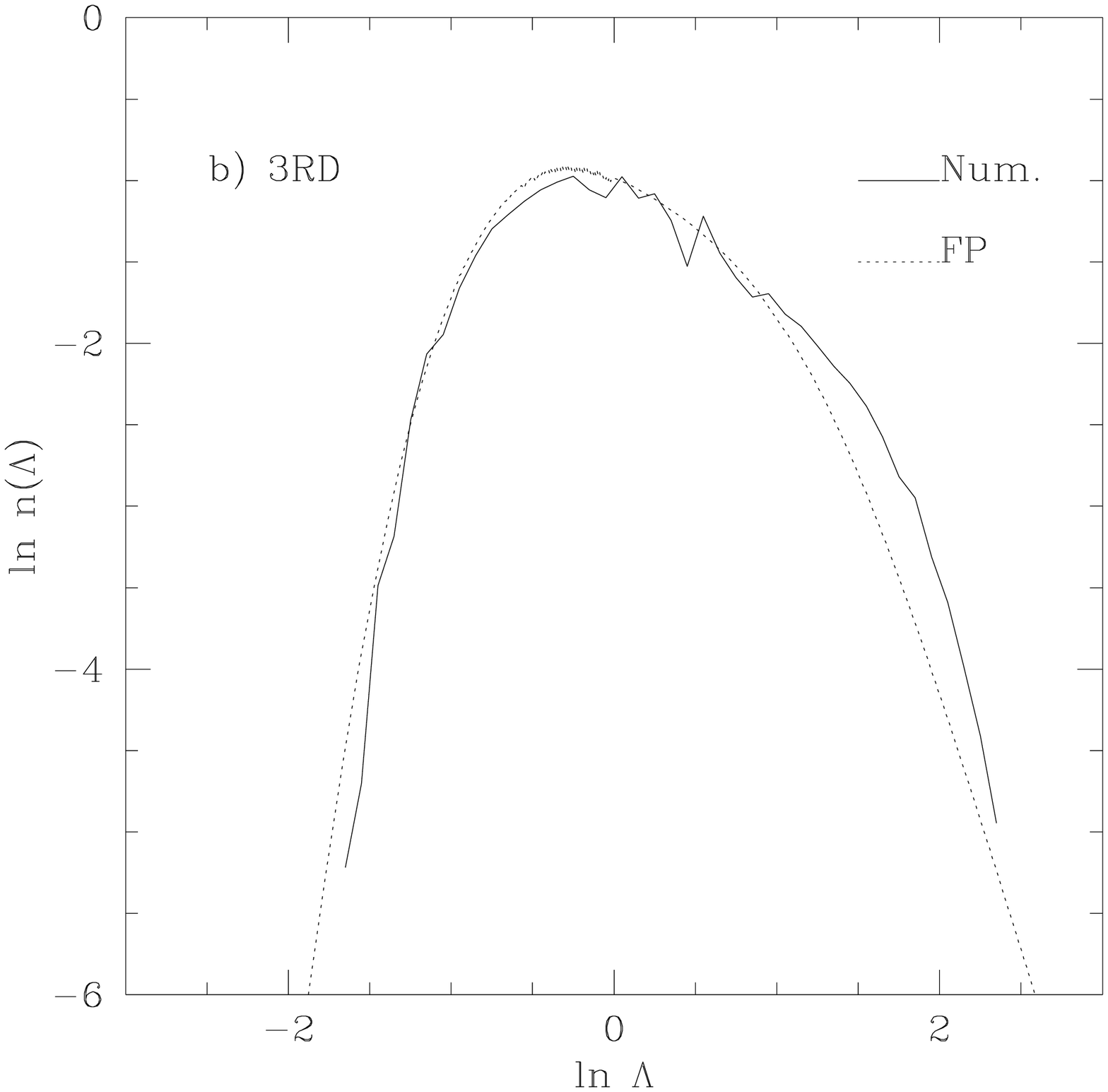,width=7cm}}
\caption{Numerical and FP $n(\mres)$ curves, SKS filtering.}
\end{center}
\end{figure}

Finally, it is necessary to verify whether the solution of the 1D
problem, obtained by assuming that $F(\mres)$ is a diffusion process,
gives the correct solution of the multidimensional problem described
in \S 4.2.1. To do this, the Monte Carlo simulations already described
in \S 3.3.2 have been used. Every realization has been smoothed by
means of a hierarchy of SKS filters, taking care to follow the spacing
of the modes of the cubic grid. To do this, 116 filterings of the
16$^3$ grids and 464 of the 32$^3$ grids have been performed.  For
every filtering, the collapse time has been calculated at every point
which had not collapsed at smaller resolution, thus determining the
first upcrossing rate as a function of resolution.  Given the limited
dynamical range of the grids, different samples of realizations with
different normalizations have been used. For ELL, three sets of 30
realizations 32$^3$, with total mass variances 0.8, 2 and 5 and
spectra with $n=-2$, have been used. The 3RD case is more delicate and
time consuming: the 16$^3$ realizations have been used in place of the
32$^3$ ones, and four sets of 60 realizations, with variances 0.8, 2,
5 and 12 have been performed. Fig. 4.10 shows the results of these
calculations, compared with the FP curves calculated before (with
complete barriers): the agreement is overall perfect for ELL, very
good for 3RD at large masses and good at intermediate masses (the
slight disagreement is too small to be really worrisome).  The
numerical 3RD curve appears significantly different from the FP one at
small masses, and this is presumably due to the poor fit of the
$F$-PDF given by Eq. (\ref{eq:Gauss_fit}). However, this disagreement
takes place in the range in which the MF theory is not considered
robust, so that an improved fitting of the PDF is considered
unnecessary.  Finally note how the numerical \dmfr\ curve for 3RD
accentuates the small-mass cutoff found in the other cases.

In conclusion, the solution of the 1D problem perfectly reproduces the
first upcrossing rate of the $F$ process in the SKS filtering case.
This provides the solution of the MF problem by extending the
diffusion formalism to the $F$ process, and gives precious information
on the possible Markov nature of $F$ itself, even though the explicit
demonstration of this remains an open problem.

\section{Gaussian smoothing}

The use of SKS smoothing is only motivated by the fact that it leads
to the elegant diffusion formalism, which allows us to find
(semi)analytical solutions even in the case of the non-Gaussian $F$
process. From a dynamical point of view, Gaussian smoothing is better
motivated than the SKS one, as it optimizes the dynamical predictions,
as commented in \S 3.5. But it is not possible to find exact solutions
for the absorbing-barrier problem in the Gaussian smoothing case, even
if $F$ is proportional to \dl.  In fact, when Gaussian smoothing is
used, the power spectrum is not sharply truncated at a given
resolution, so the $\varphi_i (\mres)$ processes contain information
about what is going to happen at larger resolutions.  As a
consequence, the $\varphi_i(\mres)$ processes lose their Markov
property, and the $F$ process can by no means be a diffusion
process. This implies that neither the {\pot}-PDF nor the $F$-PDF obey
a FP equation.

\begin{figure}
\begin{center}
\epsfig{file=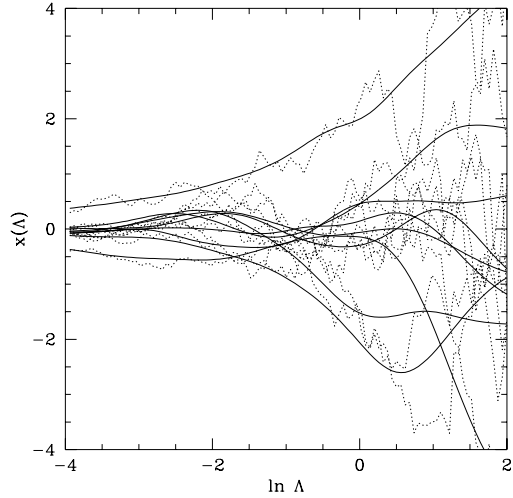,width=7cm}
\caption{SKS and Gaussian trajectories.}
\end{center}
\end{figure}

This fact can be better understood by looking at some SKS and
Gaussian-smoothed trajectories of the Wiener $x$ process, as shown in
Fig. 4.11. SKS trajectories are very noisy, while Gaussian
trajectories are much more stable. Their stability can be also shown
as follows; the normalized correlation function of Gaussian-smoothed
trajectories is approximately constant for small increments:

\be \frac{\langle x(\mres)x(\mres+\Delta\mres)\rangle}{\sqrt{\langle 
x(\mres)^2 \rangle \langle x(\mres+\Delta\mres)^2\rangle}} \simeq
\mres \left( 1 -\frac{1}{2}\left( \frac{\Delta\mres}{\mres_c}\right)^2
\right), \label{eq:normal_corfun} \ee

\noindent where $\mres_c$ is a coherence length equal to:

\be \mres_c = 2\mres\gamma(1-\gamma^2)^{-1/2}. \label{eq:lambda_c} \ee

\noindent
$\gamma$ is a standard spectral measure (see Bardeen et al.  1986),
equal to $((n+3)/(n+5))^{1/2}$ for scale-free power spectra, in which
case $\mres_c=\mres \sqrt{2(3+n)}$.  This coherence scale is quite
large, especially for large spectral indexes.  In the SKS case the
coherence scale vanishes, and the normalized correlation function
linearly decreases with $\Delta\mres$: trajectories are much less
correlated.

Note that such a stability, while being a problem from a mathematical
point of view, can be considered as positive feature from a dynamical
point of view, as a noisy trajectory corresponds to an unstable
dynamical prediction, which can change much when the resolution
changes.

Some consequences of the stability of Gaussian trajectories can
be appreciated immediately: 

\begin{enumerate}
\item 
Gaussian trajectories cannot easily downcross the barrier once they
have upcrossed it, as they need a resolution change of order $\mres_c$
to reverse their direction; as a consequence, the resulting MF will
reduce to the PS-like one at large masses, because no or a negligible
number of trajectories will have had time to upcross and then
downcross the barrier (note that this conclusion had already been
reached by Schaeffer \& Silk 1988a). This is at variance with the SKS
case, where an upcrossing trajectory had a large probability (one
half!) of downcrossing the barrier just after having upcrossed it, and
this causes a factor 2 more large-mass objects to be predicted to
form.
\item
The PS-like integral MF is a lower limit to the true integral MF (Bond
et al. 1991), as the PS-like approach counts the number of
trajectories which lie above the barrier at a given resolution, and
that have then been surely absorbed, but does not count al the
trajectories which have downcrossed the barrier.  As the Gaussian MF
reduces to the PS-like one at large masses, and as less than 10\% of
mass has to be redistributed to achieve the correct normalization, the
MF with Gaussian smoothing cannot be very different from the PS-like
one; this is at variance with what was found by Peacock \& Heavens
(1990) and Bond et al. (1991) in the canonical linear-theory case (the
PS-like and Gaussian curves are rather different), in which half of
the mass has to be redistributed.
\item
Since the correlation length is larger for larger spectral indexes, the
Gaussian \dmfr\ curve is expected to be more similar to the PS-like
one for larger spectral indexes.
\end{enumerate}

To rigorously solve the absorbing barrier problem in the
Gaussian-smoothing case, all the N-point correlation functions of the
process $F(\mres)$ at different resolutions have to be considered;
this makes the calculation prohibitive even in the spherical collapse
case.  However, some fairly motivated and successful analytical
approximations have been proposed both by Peacock \& Heavens (1990)
and by Bond et al. (1991); both the groups agree in stating that the
one proposed by the formers is slightly more successful than the
others.  The approximation proposed by Peacock and Heavens (hereafter
PH approximation) is obtained by assuming that Gaussian trajectories
are a random step process, constant over an interval equal to $\pi
\mres_c\ln 2$, whose transition probability can be written as:

\bea P(F,\mres;F',\mres') &=& \delta(F-F')\ \  {\rm if}\ \  
\mres/\mres'<\mres_c\\
                       &=& P(F,\mres)\ \ {\rm if}\ \  
\mres/\mres'\geq\mres_c\; ,
\label{eq:ph_trans_prob} \eea

The probability of never having upcrossed the barrier can be
calculated by multiplying over a discrete number of independent
probabilities:

\be \int_{-\infty}^{F_c} \mpdffu dF = \prod_i \int_{-\infty}^{F_c} 
P_F(F,\Lambda_i) dF. \label{eq:phapprox_uno} \ee

\noindent 
By taking the logarithm of the product on the right-hand-side, and by
changing it to an integral by means of a continuum limit, the
following integral MF is obtained (see PH, Bond et al. 1991, Monaco
1996b for details):

\be \mimfr = 1-\int_{-\infty}^{F_c} P_F(F,\mres) dF\label{eq:phapprox_due}\ee
$$ \times\exp \left(\int_0^\mres \ln \left(\int_{-\infty}^{F_c} P_F(F,\mres')
dF\right) \frac{d\mres'} {\pi\mres_c(\mres')\ln 2} \right). $$

\noindent 
Note that the PS-like expression is obtained if the exponential is set
equal to one.  The PH approximation was proposed to solve the
absorbing barrier problem in the $F\propto\delta_l$ case. In any case,
since the approximation is based on assuming that Gaussian
trajectories have a simple behavior, it can be readily applied, as it
stands, to the non-Gaussian $F$ processes considered here.  Expressing
the $F$-integrals in terms of the $x$ variable, the following
expression for the \dmfr\ curve is found:

\be n(\mres) = \left[\frac{\exp (-x_c^2/2\mres)}{\sqrt{2\pi\mres}} 
\left(\frac{x_c}{2\mres} - \frac{dx_c}{d\mres}
\right) - \int_{-\infty}^{x_c(\mres)} P_x(x,\mres) dx \right. \times 
\label{eq:phnofres}\ee
$$\left.\frac{1}{\pi\mres_c\ln 2}
\ln \left(\int_{-\infty}^{x_c(\mres)}P_x(x,\mres)dx \right)\right] 
\exp \left[\int_0^\mres \ln \left(\int_{-\infty}^{x_c(\mres')}
P_x(x,\mres')dx\right) \frac{d\mres'} {\pi\mres_c\ln 2} \right). $$

\begin{figure}
\begin{center}
\epsfig{file=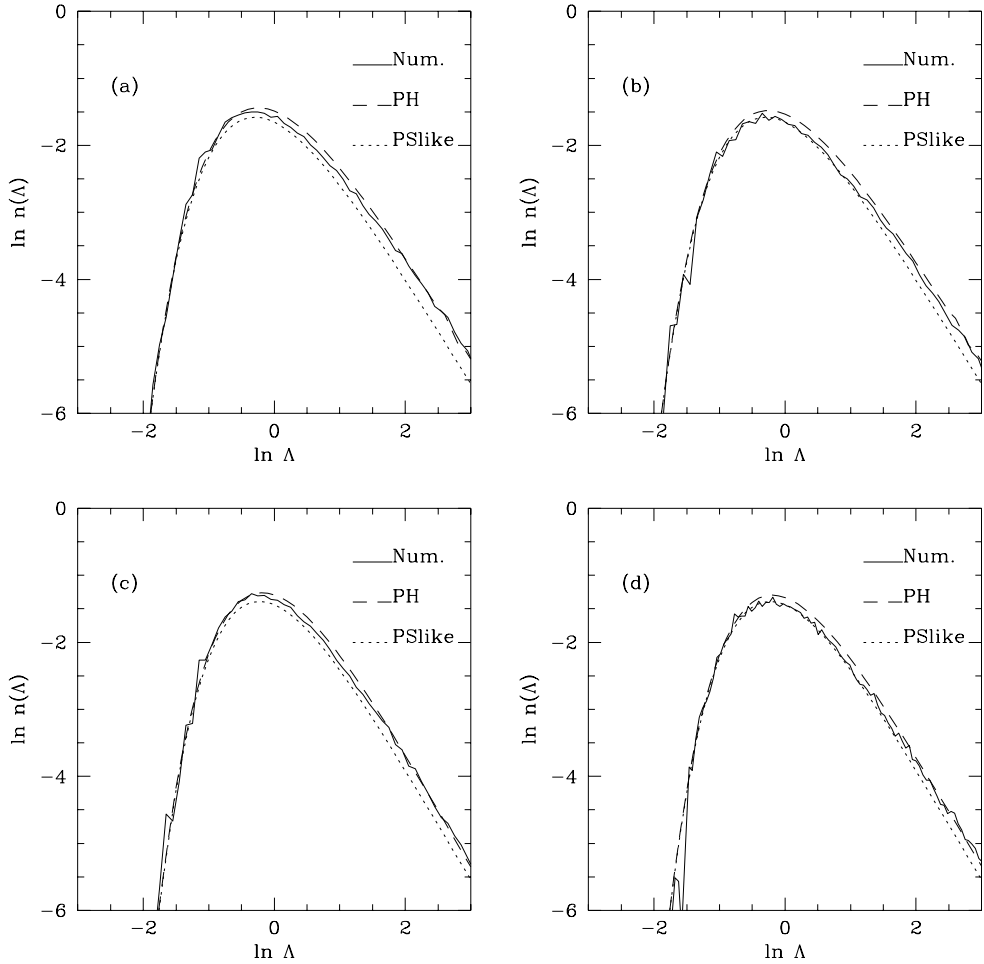,width=14cm}
\caption{Comparison between PH aproximation and Langevin simulations:
 (a): ELL, $n=-2$; (b): ELL, $n=-1$; (c): 3RD,  $n=-2$; (d): 3RD, $n=-1$.}  
\end{center}
\end{figure}

To test the goodness of the PH approximation in the case in which $F$
is given by ELL or 3RD, the absorbing-barrier problem with Gaussian
smoothing can be solved numerically, by simulating a large number of
Gaussian trajectories, and then counting the absorption rate as a
function of resolution.  Such simulations can be performed in two
ways. The first way is analogous to that used by Bond et al. (1991)
(see also Risken 1989): a large number of SKS $x$ trajectories are
simulated, then smoothed by means of a Gaussian filter and finally
absorbed by a moving $x_c(\mres)$ barrier. Such a procedure is valid
provided that true Gaussian trajectories, based on Gaussian smoothed
initial potentials, are equivalent to Gaussian-smoothed SKS
trajectories.  In other words, such calculations are valid provided
smoothing and dynamics commute:

\be {\cal F}[\varphi * W(\mres)] \equiv W(\mres) * {\cal F}[\varphi]
\label{eq:commute} \ee

\noindent
The $*$ operator denotes a convolution, and $W$ is the Gaussian window
function of width \res. This is strictly valid only in the case in
which the functional ${\cal F}$ is linear, i.e. in the linear theory
or spherical collapse case.

Bond et al. (1991) compared the results of such calculations to the PH
prediction, finding an overall good agreement, especially in the
asymptotic regimes, while the PH formula appeared to slightly
overestimate the height of the main peak.  Following them, 50000 SKS
trajectories (random walks) have been simulated, in a resolution range
from $\exp(-4)$ to $\exp(4)$, which has been divided into 2000
steps. Such trajectories have been smoothed for 100 or 150
\res\ values, the exact value depending on the spectral index; 
finer samplings are not worthwhile, as Gaussian trajectories are quite
stable. The calculations have been performed for $n=-2$ and $-1$. For
larger spectral indexes, the simulations become difficult, because of
the large coherence length $\mres_c$; moreover, the resulting PH
\dmfr\ curves are very similar to the PS-like ones (whose integral
gives a lower limit to the integral MF, as mentioned before), so it is
not very interesting to distinguish between the two.  Fig. 4.12 shows
the results for 3RD and ELL:as in Bond et al. (1991), the PH
approximation correctly reproduces the numerical curves, very slightly
overestimating them around \res=1; the asymptotic regimes are always
well reproduced. the numerical curves are always accurate enough to
prefer the PH curve to the PS-like one, especially at small masses.

\begin{figure}
\begin{center}
\epsfig{file=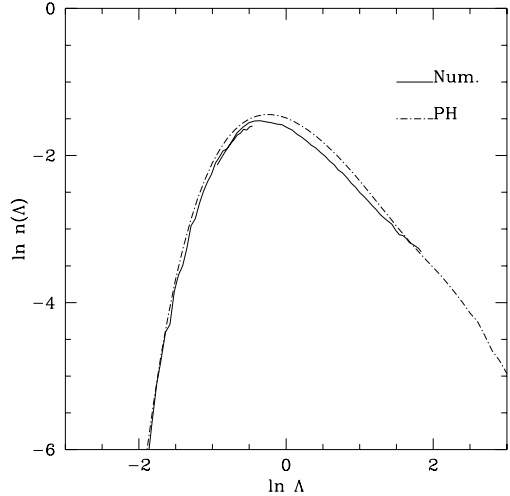,width=7cm}
\caption{Comparison between PH approximation and Monte Carlo simulations; 
ELL, $n=-2$.}
\end{center}
\end{figure}

The second way to generate Gaussian trajectories, analogous to that
used by Peacock \& Heavens (1990), is that already used in \S 4.2.2 in
the SKS case: the Monte Carlo realizations of the initial potential
are smoothed with a hierarchy of Gaussian filters, directly
calculating the collapse times for each point not collapsed at smaller
resolutions. This kind of calculation has the advantage of not
assuming the commutation between smoothing and dynamics, but is slow
and limited in resolution range; it can be used to verify that the
commutation hypothesis made above does not hamper the result found
above. Compared with the SKS calculations, the Gaussian smoothing case
has the advantage of not requiring a very fine sampling in
resolution, as trajectories are stable.  The ELL case with $n=-2$ has
been considered. Two sets of 30 realizations with different mass
variances have been used, to cover a significant range in resolution.
The calculated $F(\mres)$ trajectories have been absorbed, as usual,
by a barrier fixed at $F_c=1$. Fig. 4.13 shows the result: the
validity of the PH approximation is confirmed.

Figs. 4.7 and 4.8 show, for ELL and 3RD, the resulting PH \dmfr\
curves, for $n=-2$ and 1, compared to the PS-like and SKS ones; the
complete absorbing barriers have been used to calculate the Gaussian
PH curves. The qualitative conclusions given above about Gaussian
\dmfr\ curves are fully confirmed: they reduce to the PS-like ones at
large masses (then giving roughly a factor two less objects than the
SKS curves), and are overall very similar to them, especially for
$n=1$. It is then confirmed that a PS-like approach suffices in
determining the MF, if the Gaussian filter is preferred as more
``physical''.  Fig. 4.14 shows the Gaussian curves in comparison with
a PS one with \dc=1.5.  The following conclusions can be drawn:

\begin{figure}
\begin{center}
\epsfig{file=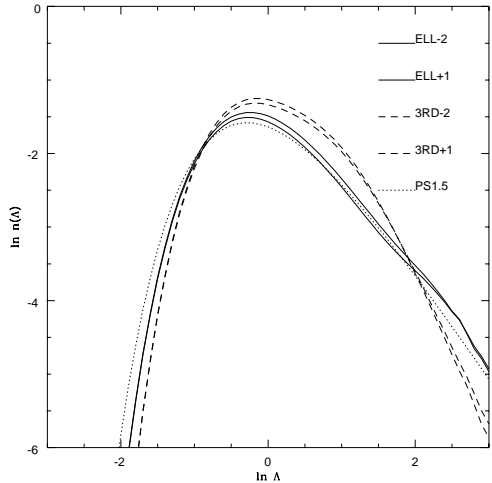,width=7cm}
\caption{$n(\mres)$ curves for ELL and 3RD with Gaussian smoothing, 
compared with a PS with \dc=1.5.}
\end{center}
\end{figure}

\begin{enumerate}
\item
Gaussian curves, just like PS-like ones, give more large-mass objects
than the canonical PS prediction (with \dc=1.69).
\item
Gaussian curves do not overproduce intermediate-mass objects, as it
happened with the SKS curves.
\item
The shape of the small-mass part crucially depends on the uncertain
particulars of the dynamical predictions at large resolutions.
\item
As in the SKS case, a drop in the number of small-mass objects is
observed when the non-linear barrier sets in.
\item
Gaussian curves, just like the PS-like ones, are generally similar to a
PS one with \dc=1.5; this is true especially for the ELL prediction.
\end{enumerate}

\section{From resolution to mass}

A general problem with excursion set-based approaches, including the
original PS one, is that the geometry of the collapsed regions (of the
excursion sets) in Lagrangian space is not properly taken into
account: the volume of the excursion sets as a function of resolution,
and then the mass of structures, is obtained by means of the
reasonable ``golden rule'' given in Eq.  (\ref{eq:ps_plain}), here
again reported:

\be M n(M) dM = \bar{\varrho} \left|\frac{d\Omega}{dM}\right| dM = 
\bar{\varrho} n(\mres) \left| \frac{d\Lambda}{dM}\right| dM. 
\label{eq:golden_rule} \ee

\noindent
This simple rule assumes that a single mass forms at every resolution:
$M=M(\mres)$.  Moreover, this mass is usually assumed to be that
included by the filter window (see, e.g., Bond et al. 1991; Lacey \&
Cole 1993), a concept which is completely clear only in the spherical
top-hat case, in which $M=4\pi \bar{\varrho} R^3/3$, where $R$ is the
radius of the top-hat filter. From this point of view, the choice made
by Lacey \& Cole (1994), of considering the exact proportionality
constant between $M$ and $R^3$ as a further free parameter, appears
quite reasonable. Moreover, the same authors (Lacey \& Cole 1993)
found some ``mild inconsistencies'', presumably due to the
oversimplified mass-resolution relation, in estimating the formation
time of halos.

In practice, such a golden rule is a reasonable zero-order
approximation, which is expected to give the correct order of
magnitude of the mean mass produced at a given resolution. To obtain a
more rigorous resolution-mass relation, the geometry of the excursion
sets ought to be properly taken into account. This raises an important
problem: with the absorbing-barrier procedure, it is possible to
determine the fraction of mass which is collapsed at a given
resolution, but no information is provided on how the collapsed medium
fragments into clumps. At small resolutions, the excursion sets of the
inverse collapse time $F$, in Lagrangian space, consist of isolated,
simply connected regions, every region containing a single peak (see,
e.g., Adler 1981). It is then quite reasonable to assume that each of
such regions will form a single structure. At moderate resolutions, of
order one, excursion sets start to be multiply connected, and their
topology becomes increasingly more complicated as the resolution
grows. In this case, a precise prescription for the fragmentation of
the collapsed medium has to be given in order to determine the number
of isolated structures; such a prescription would anyway be a new
element in the theory.

For instance, it could be assumed that structures form around the
peaks of the $F$ field (which is a different hypothesis with respect
to the canonical one in which structures are supposed to form in the
peaks of the linear density field).  Then, since the normalization of
the MF is fixed by the excursion sets approach, a procedure like that
proposed by Appel \& Jones (1990) and by Manrique {\&} Salvador-Sol\'e
(1995;1996) could be used to fragment the collapsed medium.  This
would be an interesting fusion of the excursion set and peak
approaches, but, given the complexity of the $F$ process, it would be
best addressed though the Monte Carlo simulations of the kind
presented in \S 3.3.2.

It appears that the mass-resolution relation is the place in which the
spatial correlations of the $F$ process, which were avoided by means
of the punctual interpretation of collapse times, come into play.
With the global interpretation proposed by Blanchard et al. (1992) and
Yano et al. (1996), spatial correlations are explicitly included in
the diffusion problem, with a consequent complication of the
formalism; however, the resolution-mass relation is in that case
``exact'', as all the points which collapse into the spherical region
are properly taken into account. This conclusion is true only provided
spherical collapse is true (and top-hat smoothing is used); in
practice such a procedure imposes spherical symmetry to the excursion
sets in Lagrangian space, which is a further approximation. Moreover,
Betancort-Rijo \& Lopez-Corredoira (1996), who explicitly assume the
sphericity of excursion sets, have concluded that the global
interpretation can be applied to peaks, but not to general field
points.

A more realistic resolution-mass relation would predict a whole
distribution of masses to form at a given resolution:

\be \mres \rightarrow p(M;\mres). \label{eq:mass_res} \ee

\noindent
The $p(M;\mres)$ function gives the probability that a mass $M$ is
formed at a resolution \res; its mean value will be of order:

\be \int_0^\infty M p(M;\mres) dM \sim \bar{\varrho}R(\mres)^3,
\label{eq:mean_mass} \ee

\noindent
as the smoothing length $R(\mres)$ is the relevant scale in this case;
the proportionality constant, of order one, will in general depend on
the shape of the filter, on the power spectrum and on the resolution.
The MF would then be given by:

\be M n(M)dM = \bar{\rho} \left( \int_0^\infty n(\mres)
p(M,\mres) d\mres \right)\, dM \; ,\label{eq:no_golden_rule} \ee 

\noindent 
i.e., by the \dmfr\ curve convolved with the distribution
$p(M,\mres)$.  According to the discussion given above, such a $p$
distribution is expected, at small resolutions, to be peaked at its
mean value, while its shape at large resolutions is expected be more
complex, probably influenced by the details of the prescription chosen
to fragment the collapsed medium. Then, with respect to using the
simple golden rule, the introduction of the $p$ distribution is
expected not to influence dramatically the large mass tail of the MF,
while it is likely to influence the small-mass part, which is
confirmed to be a not robust prediction of this kind of MF theories.
Finally, differences in the $p$ distributions for different filter
shapes could in principle at least attenuate the differences between
the MFs found with SKS and Gaussian filtering.

\begin{figure}
\begin{center}
\hbox{\epsfig{file=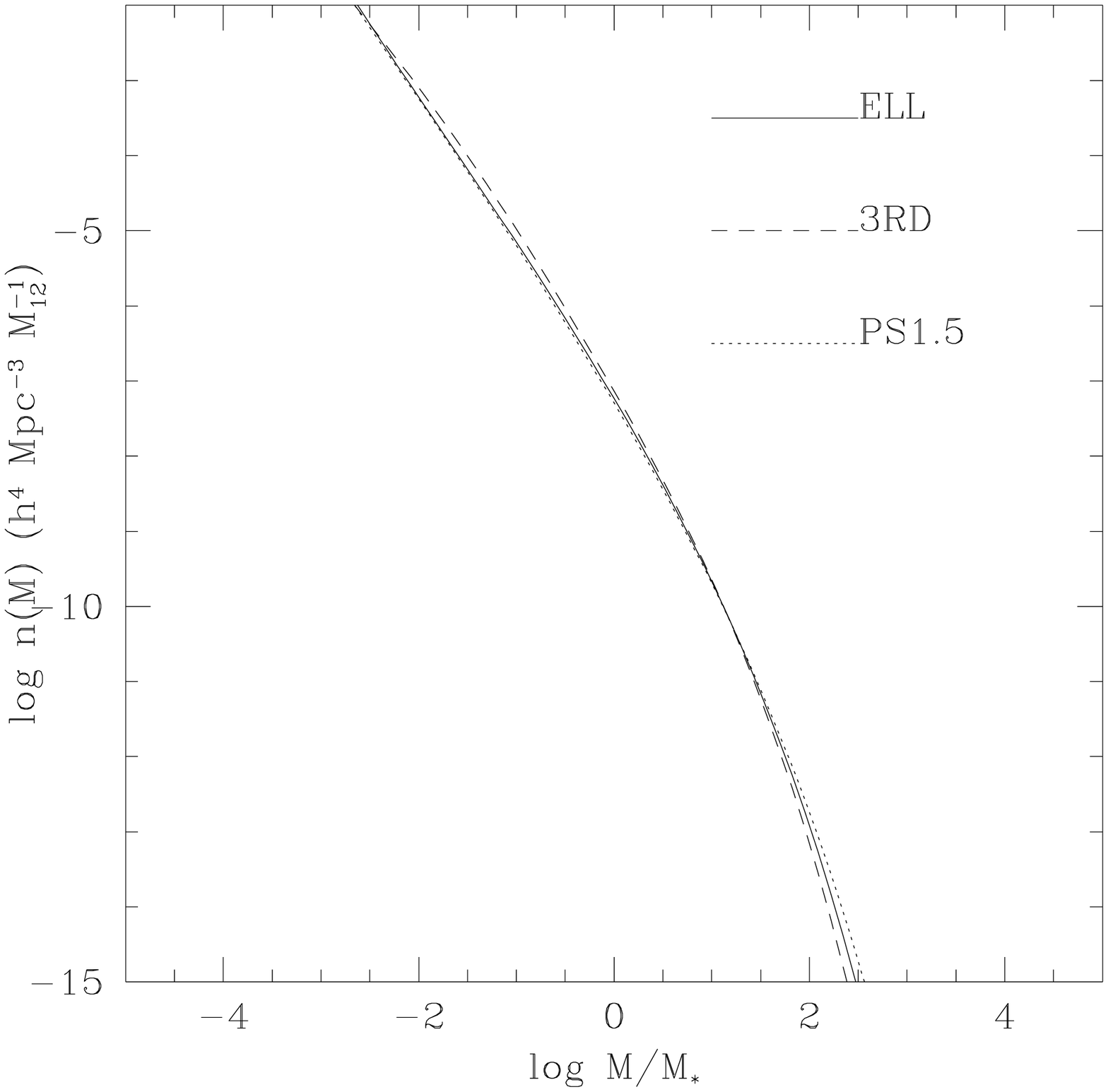,width=7cm}
      \epsfig{file=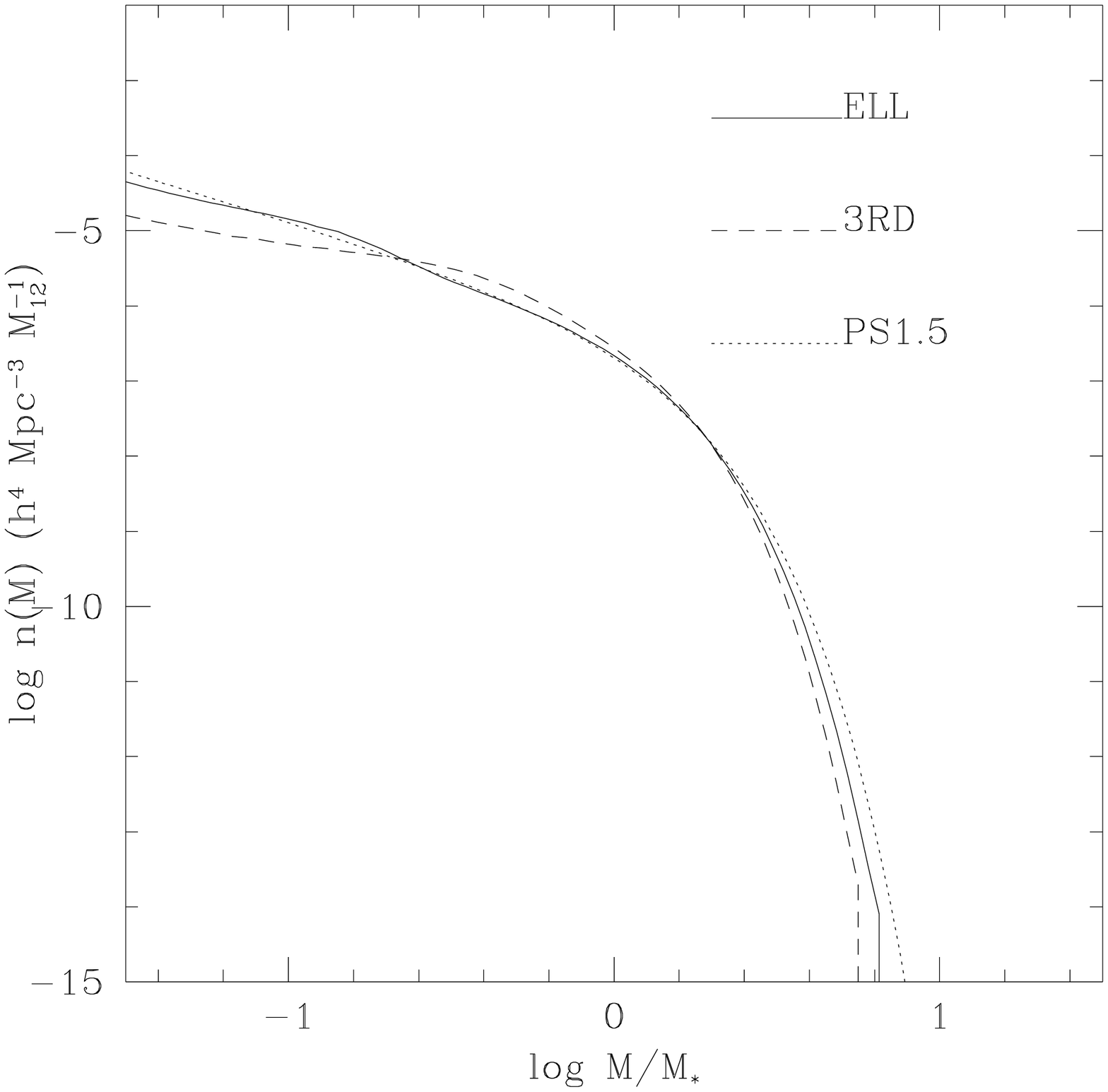,width=7cm}}
\caption{Mass functions: (a) $n=-2$, (b) $n=1$.}
\end{center}
\end{figure}

The MF is finally given by using the golden-rule, in the simple case
of scale-free spectra. In this case, the exact value of the
proportionality constant can be absorbed in the definition of a
typical mass $M_*$:

\be \mres = (M/M_*)^{-(3+n)/3}. \label{eq:res_mass_gr} \ee

\noindent 
$M_*$ is the mass corresponding to unit variance; note that this
definition is different from the usual PS one by a factor
$(\delta_c)^{3/(3+n)}$. Fig. 4.15 shows the resulting ELL and 3RD
Gaussian-smoothed MFs for the case $n=-2$ and 1 (the latter case is
cosmologically unrealistic, but of academic interest); the PS MF with
\dc=1.5 has been shown for comparison. Spectra have been normalized by
assuming unit variance over a top-hat sphere of radius 8 $h^{-1}$ Mpc.
It can be noted that, especially in the $n=-2$ case, the MF is
stretched along many orders of magnitude, so that the (small)
differences between the different curves are not very visible.  It is
confirmed that the PS MF with \dc=1.5 is, at large and intermediate
masses, a good approximation of the Gaussian-smoothing MFs presented
here. At small masses the dynamical MFs, especially the 3RD one, are
typically flatter than the PS prediction.

\section{Discussion}

In this Chapter it has been shown that the excursion set formalism can
be extended to the case in which the $F$ process, a non-linear and
non-local functional of the initial potential, is considered. A
PS-like procedure is enough to obtain the main features of the MF. IF
SKS filtering is used, it has been demonstrated that the diffusion
formalism can be extended to the non-Gaussian $F$ process, by
considering it as a diffusion process. The problem can then be
transformed to the diffusion of a Wiener process with a moving
absorbing barrier. For Gaussian smoothing, the simple approximation
proposed by Peacock \& Heavens (1991) has been shown very successful
in finding the MF.  The passage from the resolution to the mass
variable requires knowledge of how collapsed matter gathers in clumps,
an element which is missing in the excursion set approach; the MF has
been found by means of the usual resolution-mass relation (the
so-called golden rule, Eq. \ref{eq:golden_rule}), and the consequences
of this approximation have been discussed.

The main conclusions are the following:

\begin{enumerate}
\item 
A larger number of large-mass objects is expected to form with respect
to the canonical PS prediction with \dc=1.69.
\item
The large-mass part of the MF is considered robust with respect to the
dynamical prediction. The ELL prediction tends to give more objects
than the 3RD one in the large-mass tail; this is due to the fact that
3rd-order Lagrangian theory slightly underestimates spherical
collapse; thus the ELL prediction is considered more believable in
that range.
\item 
Reasonably accurate analytical solutions have been given for (at least
the large-mass parts of) the PS-like, SKS and Gaussian MFs.
\item
Both PS-like and Gaussian-smoothing MFs give fewer objects than the
SKS one, by roughly a factor of 2, in the large- and intermediate-mass
ranges.
\item
PS-like and Gaussian MFs very similar to the PS one with a $\delta_c
\simeq1.5$ parameter.
\end{enumerate}

The small-mass part of the MF is not considered a robust prediction of
the theory, for at least three reasons:

\begin{enumerate}
\item
The definition of collapse, which is based on the concept of orbit
crossing, is not expected to reproduce common small-mass structures
like virialized halos. OC regions rather represent those large-scale
collapsed environments in which the virialized halos are embedded.
\item
All the dynamical predictions used are considered good as long as the
inverse collapse time is not small. Thus, the small-mass part of the
MF is based on non-robust dynamical predictions.
\item
The $p$ distribution of the forming masses, at a given resolution, is
likely to significantly affect the small-mass part of the MF.
\end{enumerate}

Obviously, this criticism applies in the same way to all the MF
theories based on the excursion set approach, as well as on the peak
approach.

It has been noted that the dynamical predictions analyzed here,
especially the 3RD one, tend to cut off at small masses. This
corresponds to the onset of the non-linear barrier (and then to the
peak of the $F$-PDF curves, Fig. 3.12, whose actual existence is
considered rather certain), which somehow introduces a second
characteristic scale in the MF.

Gaussian smoothing is preferred for three reasons: 

\begin{enumerate}
\item it optimizes the dynamical predictions (as far as we know; see \S 3.5); 
\item it stabilizes the trajectories; 
\item it gives the right number of intermediate-mass objects.
\end{enumerate}

The prediction of more large-mass objects, caused by the improved
dynamical description of collapse, confirms some previous claims, for
example by Lucchin \& Matarrese (1988) and Porciani et al.  (1996),
who introduced non-Gaussianity of dynamical origin in the PS or
diffusion approaches (\S 2.3). This increase can be seen as the effect
of tides on the dynamics of the mass element (Bertschinger \& Jain
1994).  Besides, even spherical collapse, when the global
interpretation of collapse times is assumed (\S 2.3.4), leads to the
prediction of more large-mass objects (Blanchard et al. 1992; Yano et
al. 1996; Betancort-Rijo \& Lopez-Corredoira 1996).

The similarity of the (Gaussian-smoothed) dynamical MF with a PS one,
with \dc=1.5, makes the dynamical MF be consistent with many numerical
simulation (\S 2.2), even though this agreement is not a real proof of
validity as (i) OC regions are not the halos extracted from
simulations, and (ii) the resolution-mass relation is still treated in
a simplified way. However, the dynamical MF theory does not predict
the trend of lower \dc\ values at higher redshifts, observed in recent
N-body simulations (see references in \S 2.2). An explanation of this
behavior was proposed by Monaco (1995): the presence of small-scale
power can effectively slow down gravitational collapse, through
dynamical events of the kind of previrialization, proposed by Peebles
(1990; see also Lokas et al. 1996; Bouchet 1996), or through dynamical
friction with those possible particles which evaporate out of the
collapsed structures (Antonuccio \& Colafrancesco 1994). Such events
could be effective for power spectra with $n>-1$ (Lokas et al. 1996).
In CDM-like spectra, the effective spectral index at $M_*$ is smaller
than $-1$ at high redshifts, where \dc=1.5 (or even less) is found,
but becomes larger at lower redshifts, where \dc\ starts to increase to
1.7 or even larger values. However, such a behavior could also be due
to a dependence of the $p$ distribution on the power spectrum.

%
%

\chapter[The MF world]{The mass function world}

The MF theory is the starting point for the modeling of the
statistical behavior of astrophysical objects in hierarchical
cosmologies.  Originally, this theory was developed to describe galaxy
formation; Schechter (1976) used a simplified version of the
functional form of the PS MF to construct an analytical
fitting formula for the galaxy luminosity function:

\be \phi(L)dL = \phi_* (L/L_*)^\alpha \exp(-L/L_*)dL/L_*. 
\label{eq:schechter}\ee

\noindent
This formula has become standard for describing the observed
luminosity function of galaxies.

To model astrophysical objects, the knowledge of the MF is not enough:
gas heating and cooling, star formation, supernovae feedback and other
gas-dynamical events (see \S 1.3) have to be inserted in the
dark-matter halos, in order to get any predictions about observable
objects like galaxies, X-ray clusters, Lyman $\alpha$ clouds or
QSOs. For instance, the observed luminosity $L_*$ of the Schechter
distribution is not related to the critical mass $M_*$ of the MF,
which is today of the order of the mass of galaxy clusters, but is
related to the gas cooling time scale.

On the other hand, it is possible, through indirect methods, to probe
the depth of the potential well of astrophysical structures, and then
to determine their total mass. It is then possible to determine an
observational MF, which can be directly compared to the theory.

This chapter aims to give an overview of the MF world, which extends
to include many branches of cosmology.  The main results on
observational determinations of masses and on available observational
MFs will be reviewed, together with the cosmological constraints
obtained by comparing such data with the MF theory. The main
applications of the MF theory in modeling astrophysical objects, with
attention both to methodology and to the resulting cosmological
constraints, will be discussed. A special attention will be devoted to
the contribution that the dynamical MF theory, developed in Chapters 3
and 4, can give. \S 5.1 will deal with galaxies, \S 5.2 with galaxy
groups and clusters and \S 5.3 with high-redshift objects.

\section{Galaxies}

Galaxy formation has always played a special role in cosmology. In
fact, until the discovery of a hot gas component in galaxy clusters,
visible in X-rays, all the observable objects in the Universe (stars,
gas, AGNs etc.) were known to reside in galaxies, with the only
important exception of the CMB radiation. Galaxies are also the first
place in which a convincing proof of the existence of large amounts of
dark matter was found.

One of the most important (and easily measurable) properties of a
galaxy is its optical luminosity. The luminosity function of nearby
galaxies has been determined by many authors (see, e.g., Efstathiou,
Ellis \& Peterson 1988; Ellis 1997 and references therein).  Typically
the luminosity function is well fit by a Schechter function
(Eq. \ref{eq:schechter}), with $\alpha=\simeq-1$ and $L_*$
corresponding to an absolute magnitude of $\simeq-19.5-5\log h$.  In
the 70's and 80's, the theoretical MF was sometimes directly compared
to the observed luminosity function, assuming a constant stellar
mass/light (hereafter M/L) ratio (see the review by Lucchin
1989). This raised a problem: the luminosity function is significantly
flatter than the PS MF, whose low-mass slope is not much different
from $-2$.  Subsequently, it was shown that M/L systematically varies
with the luminosity (or the mass) of a galaxy; to perform a meaningful
comparison, the theoretical MF has to be compared to the distribution
of observed masses.

\subsection{The galactic MF}

The determination of galactic masses can profit from the fact that
galactic dynamics is traced by a large number of observable tracers,
namely stars and gas clouds. The exact procedure for determining
galaxy masses depends on the particular dynamical status of the
galaxy, i.e. on its morphology.

The masses of disk galaxies are probably easier to measure: disks are
known to perform a ``cold'' differential rotation, with rotational
velocities ($\sim$100--300 km/s) larger than velocity dispersions
($<$50 km/s). The rotation curve of galactic disks can easily be
reconstructed by means of the Doppler shift of spectral lines as they
are observed at different radii. Once the rotation curve is
determined, the mass profile can be estimated by solving the Newtonian
problem of a self-gravitating disk (plus a spheroidal bulge, which
influences only the inner parts of the galaxy). The mass profile so
determined can then be compared to the light distribution, to test
whether the luminous mass can be considered as responsible for the
observed rotation.

Rubin, Ford \& Thonnard (1980) were the first to discover that
luminous matter can not be held responsible for the outer parts of the
rotation curve, where the light fades down but the rotation curve
continues more or less flat, showing no Keplerian drop due to the lack
of matter (see also the review of Ashman 1992, or Persic \& Salucci
1996). It is now accepted that spiral galaxies are embedded in dark
matter halos which extends much further than their optical radius.

The most extended analysis of spiral rotation curves was given by
Persic \& Salucci (1995) and Persic, Salucci \& Stel (1995) (in
practice, because of selection biases, their galaxy sample does not
contain gas-poor lenticulars and Sa spiral galaxies). They showed that
rotation curves follow a universal relation, with a small scatter.
This relation, the universal rotation curve (first proposed by Rubin
et al. 1985), connects in a unique way the mass, luminosity, central
density and rotation-curve slope of any spiral galaxy. It is then
possible to find a total mass -- luminosity relation; such relation
has been shown, by the above mentioned authors, to be:

\be M \propto L^{0.5} \label{eq:m_su_l} \ee

\noindent
Then, faint galaxies contain more dark matter than luminous ones,
total masses not changing much with luminosity.

An open problem that remains is where the dark halo ends. Optical
rotation curves clearly indicate that the dark halo extends further
than the optical radius. Radio observations can push the observations
as far away as two or three optical radii, because neutral gas is
still present in the outer parts of the disk; in this case, the
rotation curve is observed to continue its trend up to there. The
extension of the dark halo can be probed even further by considering
the motion of satellite galaxies (Zaritsky et al. 1993), or the motion
of binary galaxy systems (Charlton \& Salpeter 1991); such
observations are consistent with halos which extend up to $\sim$5
optical radii.  Persic et al. (1995) give total mass values
extrapolated to the radius at which the mean density contrast reaches
200, a value inspired by spherical collapse.

Elliptical galaxies do not rotate and do not contain large amounts of
gas (except in some important cases). They can be considered as a gas
of stars in dynamical equilibrium; then, their mass profile can be
determined by means of the Jeans equation (see, e.g., Binney \&
Tremaine 1987; it is given in Eq. \ref{eq:jeans_gal}); however, the
kinematics of elliptical galaxies is not simple, as many hints of
triaxiality suggest.  The masses of elliptical galaxies can then be
determined by means of the luminosity density of stars and their
velocity dispersion, estimated by means of the width of galactic
absorption lines; in this way it is possible to probe only the inner
part of the potential well, in which there is not good evidence of
dark matter.  The external parts can be probed, for instance, by
exploiting the presence of gas in some ellipticals (see, e.g.,
Danziger 1996). According to recent studies (Bertola et al. 1993),
ellipticals do contain some amount of dark matter, and their total
mass -- luminosity relation is not very different from the one valid
for spiral galaxies, Eq. (\ref{eq:m_su_l}).

It is then possible to translate luminosity into mass, and then to
determine the galactic MF (Ashman, Salucci \& Persic 1993). At small
masses, it is a power-law with a slope slightly steeper than $\sim-1$,
but still different from the PS prediction. However, given the limited
mass range spanned by normal galaxies (not much more than an order of
magnitude), the exponential cutoff can lower the effective slope of
the observed MF to values not so different from $-2$; this can weaken
the problem, but probably does not completely solve it, as will be
shown later.

\subsection{Galactic formation and cosmology}

The topic of galaxy formation in cosmology has been reviewed by many
authors; one of the most recent reviews is given by White (1993),
while Matteucci (1996) has reviewed the closely connected topic of the
chemical evolution of galaxies.  The first ideas trace back to Binney
(1977), Rees \& Ostriker (1977), Silk (1977), White \& Rees (1978),
Blumenthal et al. (1984), Fall \& Rees (1986): if most matter in the
Universe is not baryonic, then dark-matter halos and gas follow
different evolutionary paths; dark-matter evolution is hierarchical
and nearly self-similar (if the spectrum is gently curved, as in the
CDM case), as described by the MF theories described in previous
Chapters, while gas dynamics is dominated by dissipation.  Gas falls
into dark-matter halos and, by dissipating part of its energy, settles
down into the core of the halo, fragmenting into clouds and giving
rise to star formation.  This process leads to a new characteristic
scale, the luminosity $L_*$ of the galaxy luminosity function; it
corresponds to the halo mass for which the gas cooling time equals the
dynamical time of the halo; gas is not able to effectively cool down
in larger halos.  After all gas has coalesced into halos, subsequent
hierarchical merging mainly involves dark matter: for instance, most
galaxies retain their identity inside groups or clusters. Note that
the problem of galactic formation is also connected to the
``Eulerian'' bias problem (see \S 1.1): once the kind of halos in
which galaxies form are determined, their spatial correlation
properties, as compared to those of dark matter, determine the bias
parameter.

To model the build-up of galaxies, dark-matter halos are usually
constructed by means of the Lacey \& Cole (1993) merging histories,
which fit the evolution histories of N-body halos (\S 2.3.3).  The
final properties of halos are obtained by means of the top-hat
spherical collapse model, or by assuming an isothermal profile for the
halo, or, more recently, by assuming that halos follow the simple
universal profile found in N-body simulations (Navarro, Frenk \& White
1996; Navarro 1996). Note that this is in surprising agreement with
the fact that spiral galaxies follow a universal rotation curve (\S
5.1.1), even though the two universal profiles do not agree in full
detail.

As the discussions in the previous Chapters have demonstrated,
dark-matter evolution is not understood in detail; nonetheless, the
modeling of virialized dark-matter halos is probably the safest step
in galaxy formation.  The gas fallen inside the halos consists of at
least three main phases, namely hot, X-ray emitting gas, cold neutral
gas and stars.  Shock heating, gas cooling, dissipation of angular
momentum, star formation, feedback from star formation (including
supernovae), further gas accretion, dissipative galaxy merging,
chemical evolution are the main events which have to be modeled.
Every step introduces poorly constrained parameters and oversimplified
assumptions; the main uncertainties probably come from the initial mass
function of newly formed stars, which determines the number of
supernovae, and the rate of dissipative galaxy mergers.  Such studies
have been performed by a number of authors, among whom are Frenk,
White \& Davis (1989), Peacock \& Heavens (1990), White \& Frenk
(1991), Cole (1991), Lacey \& Silk (1991), Blanchard, Valls-Gabaud \&
Mamon (1992), Kauffmann \& White (1993), Kauffmann, White \& Guideroni
(1993), Kauffmann, Guideroni \& White (1994), Kauffmann \& Charlot
(1994), Cole et al. (1994), Avila-Reese \& Firmani (1996).

Despite their complexity, such theories succeed in predicting the main
characteristics of galaxies, as the $L_*$ parameter, the fraction of
morphological types, the main colors, and some characteristics of
high-redshift galaxies, as number counts and color evolution.  Some
important points in such theories are: (i) gas can effectively cool in
small halos; as a consequence, all the available gas can very soon
cool in subgalactic clumps ({\it cooling catastrophe}), unless some
energetic events, like supernovae explosions, prevent it from
coalescing until galactic-size halos have assembled (White \& Rees
1978). (ii) As for the galactic MF, the predicted luminosity functions
are typically steeper than the observed one. (iii) Ellipticals are
usually assumed to form from mergers. (iv) Larger galaxies are
typically predicted to form later. These last two points can be tested
against chemical evolution models and observations of high-redshift
galaxies; however, they do not seem in clear agreement with these
observations.

On the other hand, galaxy formation can be studied by means of N-body
simulation with hydrodynamics, typically treated through the smoothed
particle hydrodynamics (SPH) technique. Such simulations have been
performed, for instance, by Katz (1992), Navarro, Frenk \& White
(1995a), Evrard, Summers \& Davis (1994), Steinmetz (1996), Katz,
Weinberg \& Hernquist (1996), Tormen, Bouchet \& White (1997), Gelato
\& Governato (1996).  The present state-of-the-art of hydrodynamic
simulations still does not reach firm conclusions about galaxy
formation, mainly because of the limited resolution, and because star
formation is still treated at a heuristic level; however, the
formation of gas disks is observed to take place inside dark-matter
halos, even though disks are usually too centrally concentrated.

Merging histories can be directly used to model astrophysical events
connected with interactions. For example, the aggregation formalism
proposed by Cavaliere, Colafrancesco \& Menci (1991) can be used as
follows. The kinetic Smoluchowsky aggregation equation
(Eq. \ref{eq:smoluchowsky}; see complete references in \S 2.5.3), can
in particular conditions give rise to a merging runaway, which
corresponds to the formation of a central massive objects with mass
comparable to the total mass of the system; such a merging runaway can
be taken as responsible for the formation of large cD galaxies in
protoclusters, or for the ``violent'' erasure of substructures in
galaxy clusters.  Kinetic aggregations can also take place in a more
quiet way, in suitable environments like large-scale galaxy filaments;
in this case they can cause a flattening of the luminosity function,
and can justify the increased number counts of blue galaxies at large
redshifts as an effect of merging-induced star formation.  Similarly,
Bower (1991) interpreted the increased number of blue galaxies in
high-redshift galaxy clusters (the so-called {\it Butcher-Oemler
effect}) as the effect of cluster formation, modeled by means of the
PS merging histories (\S 2.3.3). The time scales for halo formation
(\S 2.5.2) have been used by Blain \& Longair (1993a,b) to estimate
the radiation background in the millimetric band, caused by star
formation in galaxies.

In my opinion, galactic formation is more a problem of gas dynamics
than of dark matter. In fact, the numerical dark-matter halos, with
abundances predicted by the PS theory, probably suffice to model the
dark-matter side of galaxy formation in a satisfactory way, given that
the other steps, connected to the gas behavior, are very uncertain.
Moreover, any improvement of the modeling of merging histories, with
respect to the Lacey \& Cole ones, could be not very important, as
galaxy mergings follow a different path, in which dissipative events
are predominant. Besides, the necessity, expressed in the previous
chapters, of deepening the dynamical problem of halo formation,
remains.

With respect to the dynamical MF theory, today galaxies lie in the
small-mass, power-law part of the MF, which is considered as a not
robust prediction. On the other hand, if galaxies are typically found
in galaxy groups (or clusters), and if at least many galaxy groups
have already experienced their first collapse, then, from the MF point
of view, galaxies are not associated to isolated clumps, but are just
substructure of a larger structure. From this point of view, a direct
comparison of the galactic MF to the dynamical one does not make much
sense.  A more evolved dynamical theory, able to resolve the internal
structure of collapsed clumps, could be used to describe galaxies
inside larger-scale environments; the kinetic theory of \S 2.5.3
provides such an example. Then, we would be led to the existence of
{\it two} MFs, one for the large-scale environments, which could be
given by the dynamical theories presented above, and another one for
stable substructures, which would describe real galaxies.

On the other hand, high-redshift protogalaxies are thought to be
associated with dark-matter structures which were just collapsing at
the epoch of galaxy formation. At that stage, in which dynamical
transients may have had an important role, a dynamical MF theory can
apply (see \S 5.3). As a consequence, the dynamical MF theory can be
useful to understand those features which were imprint on galaxies at
the epoch of their formation.

\section{Galaxy groups and clusters}

Galaxy groups and clusters are observationally defined as local
density enhancements of the galaxy field, which satisfy some threshold
criterion. There is no crucial difference between the two classes of
objects: by definition (Abell 1958), rich galaxy clusters present at
least 50 galaxies, with apparent magnitude between $m_3$ and $m_3$+2
($m_3$ being the apparent magnitude of the third most luminous galaxy)
inside a sphere of radius 1.5 $h^{-1}$ Mpc, while galaxy groups are
usually sought for by means of a friends-of-friends percolation
algorithm (Hucra \& Geller 1982), or by means of a hierarchical
algorithm (Materne 1978).  Clusters can also be defined, in a
physically clearer way, by means of their X-ray emission (see, e.g.,
Sarazin 1986); on the other hand, it has been observed that many large
or compact groups show the same X-ray emission as clusters, although
with lower luminosities.

The first, pioneering comparison of the group (luminosity)
multiplicity function to the PS MF was made by Gott \& Turner (1977);
groups were identified as local overdensities in the number of
galaxies, and their multiplicity was estimated on the basis of the
number and luminosity of its component galaxies, which is equivalent
to assuming a constant M/L ratio.

Later, it became clear that such estimates, simply based on the
optical luminosity of structures (and then on the number of galaxies),
are unsatisfactory, at least when rather large mass ranges are
considered, because the M/L ratio is a function of the mass of the
structure, larger structures showing more dark matter (see, e.g.,
Bahcall 1988, even though, as stated in \S 5.1.1, galaxies show an
opposite trend). Nowadays, a number of different methods for
estimating the mass of galaxy structures have been developed, and mass
estimates for a discrete number of objects are available.

\subsection{Mass determination}

Up to now, only a few authors, whose works are described in the
following, have attempted to estimate the MF of galaxy systems; the
observational determination of this quantity is not definitive at all.
However, objects masses can be estimated by means of three independent
methods. Up to now, mass measures are affected by large errors, of
order at least of 50\%, and the different methods do not always agree,
but such mass estimates are expected to increase in precision with the
increase of the number and quality of observational data.

\subsubsection{Optical masses}

Optical determinations of cluster masses are obviously based on the
component galaxies.  Analogously to elliptical galaxies, a cluster can
be considered as a ``gas'' of galaxies; if this gas is in equilibrium,
and if the cluster is spherical, then the following Jeans equation
applies:

\be \frac{d\Phi(r)}{dr} = -\frac{1}{n(r)}\frac{d(n(r)\sigma_r^2(r))}{d r} -
\frac{2}{r}(\sigma^2_r(r)-\sigma^2_t(r)).  \label{eq:jeans_gal} \ee

\noindent
$\Phi$ is the gravitational potential of the cluster, $n(r)$ is the
galaxy number density, and $\sigma^2_r$, $\sigma^2_t$ are the radial
and transverse velocity dispersions. While the (projected) galaxy
number density is directly observed, only a combination of the two
velocity dispersions, namely the dispersion along the line of sight,
can be measured. It is possible to solve for this problem in a
self-consistent way (Dejonghe \& Merritt 1992), but such a refined
procedure requires a large number of galaxies, of order of a few
hundreds, to work in practice. On the other hand, by making some
further assumption on galaxy orbits, it is possible to determine mass
profiles with a lower number of galaxies (see, e.g., Merritt \&
Gebhardt 1994).

On the other hand, a ``cheaper'' mass determination can be obtained by
means of the virial theorem (see, e.g., Limber \& Mathews 1960;
Giuricin, Mardirossian \& Mezzetti 1982), according to which the mass
of a structure is related to its line-of-sight velocity dispersion
$\sigma^2_V$ and virial radius $R_v$ in the following way:

\be M\simeq\sigma^2_V R_v/G. \label{eq:virial} \ee

\noindent
The exact proportionality constant depends on the cluster profile; it
is usually assumed to be $3\pi/2$. The virial radius of a system of
$N$ masses $\{M_i\}$, with mutual projected distances $(r_{ij})_\bot$,
is defined as:

\be R_v = \frac{\pi}{2}\frac{\left(\sum_i M_i\right)}{\sum_{i<j}
\frac{M_iM_j}{(r_{ij})_\bot}} \simeq \frac{\pi}{2} \left( \sum_{i<j}
\frac{N^2}{(r_{ij})_\bot}\right)^{-1}. \label{eq:virial_radius} \ee

\noindent
The second expression, which is the one commonly used for estimating
$R_v$, assumes for simplicity that all galaxies have the same mass; it
was shown in \S 5.1.1 that the dependence of galactic mass on
luminosity is not very strong.  Virial masses are affected by rather
large errors, of order 50\%, but can be applied to clusters with at
least 30 measured redshifts (see Biviano et al. 1993).

Virial estimates are affected by a number of problems: first the
number of tracers (of galaxies) is limited in any case; this places a
lower limit on the precision which can be achieved by means of such
measures.  Second, foreground and background galaxies, not belonging
to the cluster, contaminate the determined quantities, especially the
velocity dispersion. This can be corrected for by developing
statistical tools to exclude such interlopers and obtain robust
estimates of velocity dispersions (Beers, Flynn \& Gebhardt 1990;
Girardi et al. 1992).  Third, the mass determination strongly relies
on the hypothesis that the cluster is virialized. Substructures in the
galaxy and X-ray distribution demonstrate that this is not the case in
a large number of clusters. This can be corrected for by eliminating
clusters with strong substructures.  On the other hand, a great
advantage of virial mass estimates is that they can be extended to the
outer parts of the cluster, at variance with X-ray estimates (see
below), and that they are more effective in detecting the case of two
clusters lying along the same line of sight.

To obtain an MF of clusters, it is necessary to estimate all the
masses of a complete catalog of galaxy clusters, or to estimate the
masses of an incomplete catalog, then correcting for incompleteness.
An MF of galaxy clusters, based on virial mass estimates, has been
given by Biviano et al. (1993) for a subsample of the Abell (1958) and
ACO (Abell, Corwin \& Olowin 1991) clusters. They found their MF to be
consistent with a power law of slope about $-2.5$; if the MF is
believed to be a power law of slope $-2$ at small masses, such a
steepening can be interpreted as an effect of the exponential cutoff.

Galaxy groups can be analyzed by means of the same methods, but the
number of galaxies which can be used is small, and the virialization
hypothesis is much more uncertain for such objects.  The effect of
incomplete virialization can be corrected for by following the recipe
of Giuricin et al. (1988), who modeled the groups as spherically
collapsing structures, and estimated the virialization state by means
of the crossing time, defined as the ratio between virial radius and
the square root of velocity dispersion; the difference between this
time scale and the Hubble time defines the dynamical state of the
structure, and then its displacement from virialization.  To find the
group MF, it is necessary to estimate the masses of an existing group
catalog, and then correct for the selective loss of small groups at
large distances.  This was done by Pisani et al. (1992), who
determined the MF of the galaxy groups of Tully (1987).  They found
their MF to be a power-law with slope about $-2$, in surprising
agreement with the typical prediction of MF theories.

\subsubsection{X-ray masses}

As mentioned above, galaxy clusters and large groups strongly emit in
the X-ray band; such emission is mainly due to thin bremsstrahlung
emission by the hot plasma which fills the gravitational well of the
cluster (see, e.g., Sarazin 1986; Cavaliere \& Colafrancesco 1990;
Jones \& Forman 1991).  This emission can be observed by means of
X-ray satellites, such as Exosat, Einstein, Rosat, Asca or the most
recent Beppo-Sax, or the future Axas.  Two main quantities are
observed, namely the luminosity and temperature of the gas; the first
quantity results of order $L_X\sim 10^{44}$ erg/s, the second of order
$T\sim 10^8$ K. By solving the Jeans equation of hydrostatic
equilibrium of the gas in spherical symmetry, it is possible to
estimate the total mass of the cluster through the temperature $T(r)$
and gas density $\rho_g$ profiles:

\be M(r) = -\frac{krT(r)}{G\mu m_p}\left(\frac{d\log\rho_g}{d\log r}+
\frac{d\log T}{d\log r}\right). \label{jeans_xray} \ee

\noindent
$k$ and $G$ are the Boltzmann and gravitational constants, $\mu$ is
the mean molecular weight and $m_p$ is the proton mass. The
temperature gradient is usually assumed to be small, as data seem to
indicate. The density profile is usually parameterized through the
$\beta$-model of Cavaliere \& Fusco Femiano (1976):

\be \rho_g(r)=\rho_0(1+(r/r_c)^2)^{-3\beta/2}. \label{eq:beta_model}\ee

\noindent
The $\beta$ and $r_c$ parameters are found by fitting the observed
X-ray profiles with the expected projected emission, proportional to
$\rho_g^2$.

Mass estimates based on this procedure require very good data, today
available only for a limited number of clusters.  On the other hand,
distributions of the X-ray luminosities (Edge et al. 1990) and
temperatures (Edge et al. 1990; Henry \& Arnaud 1991) of (X-ray)
complete samples of clusters are available in the literature. As will
be shown below, such quantities are known to tightly correlate with
the total mass of the cluster, so that it is often preferred to
compare theories and data in terms of luminosity or temperature
functions.

The definite advantage of X-ray masses is that they are based on gas,
whose physics is much simpler and well-understood than that of
galaxies, and which is not limited by limited number of tracers (gas
molecules are much more numerous than galaxies!). Moreover, only
galaxy groups and clusters show extended X-rays emission, so that
there is no interloper problem in this case, except the (not uncommon)
case in which more than one cluster lie in the same line of sight.
However, X-ray masses are still affected by large errors, mainly due
to the difficult temperature determination and to non-sphericity
effects; moreover, X-ray emission is limited only to the inner part
($\sim 0.5$ Mpc) of the cluster, so that the total mass of the cluster
has to be determined by means of some extrapolation.

The relation between optical and X-ray masses has not been well
settled as yet; some authors claim that, for some clusters, optical
masses tend to be larger than X-ray ones, but this conclusion could be
just due to the fact that both measures are still affected by large
errors of systematic nature.

Another determination of the cluster MF, which exploits both optical
and X-ray information, has been performed by Bahcall \& Cen (1993).
They determined the mass of cluster both by correlating it with the
number of core galaxies (which is equivalent to assuming a constant
M/L) and by using the X-ray temperature function of Henry \& Arnaud
(1991).  Both the relations used were calibrated on the Coma cluster.
A further measure at lower masses was added by using the group
luminosity function of Bahcall (1979).  Their resulting MF was fit by
means of a Schechter curve with a parameter $\alpha=-1$. Such MF is in
marginal disagreement with that given by Biviano et al. (1993).
Again, this disagreement could be due to systematics in the mass
determinations; the problem is regarded as still open.

\subsubsection{Lensing masses}

Another promising way of measuring cluster masses is by means of
gravitational lensing. The strong potential well of the cluster can
heavily or slightly distort the images of background galaxies; it is
possible, from the shape of the distorted images, to reconstruct the
depth of the potential well, and then the cluster mass. Of course, it
is necessary to determine whether such distortions are intrinsic of
the background object or caused by lensing. In some cases, galaxy
images are very heavily distorted into arc images; such events are
called giant arcs.  Making some hypotheses on cluster geometry, it is
possible to determine the projected mass inside the arc (see, e.g.,
Fort \& Mellier 1994). Unfortunately, such events are rare, and the
obtained masses deeply depend on the geometrical model of the cluster.
This method has recently been questioned by Bartlemann \& Steinmetz
(1996), who have found, using numerical simulations, that arcs form
predominantly in cluster which are experiencing a strong merging event
along the line of sight.

On the other hand, it is possible to reconstruct the potential of the
cluster by using the weak distortions of many background galaxies,
without making any hypothesis on cluster geometry; this is called weak
lensing method (see, e.g., Kaiser 1996). The small distortions of
galaxy images, caused by lensing from the cluster potential, can be
disentangled from the intrinsic ellipticity of foreground galaxies by
measuring the mean distortion of a large number of them, in the
hypothesis that intrinsic ellipticities do not correlate on the
relevant scales.  When the distortions are weak, they are linearly
related to the potential, but the signal is weak; on the other hand,
when the distortions are stronger, the relation becomes non-linear and
then more difficult to invert. The inversion techniques, necessary to
extract the signal in a bias-free way, have not been completely
developed as yet.

Mass estimates based on lensing are now available only for a limited
number of clusters. The main advantage of such mass determination is
that they directly probe the gravitational potential, so that no
equilibrium or virialization hypothesis is needed; in the weak lensing
case, no hypothesis on the geometry of the cluster is needed either.
They require high-quality imaging of the cluster field, but do not
require spectroscopy in general.  The reconstructed mass profiles are
in general compatible with optical and X-ray profiles; lens masses
tend to result larger than optical and X-ray masses (at the same
radius), by a factor of 2, but the opposite behavior can be noted in
some cases. Given the uncertainties in the present state-of-the-art,
it is not unrealistic to imagine that these discrepancies will be
reduced or removed with the refining of observations and analyses.

\subsection{Clusters and cosmology}

Galaxy clusters, being the largest collapsed structures of our
Universe, have long been recognized as objects of crucial importance
in the cosmological context.  In fact, their abundance and main
properties (velocity dispersion, X-ray luminosity and temperature,
substructure) can provide valuable constraints to be satisfied by
cosmological models. The PS MF, as an analytical fit of the results of
N-body simulations, has usually been used to determine the number of
clusters. However, comparisons with data have usually been performed
not in terms of mass but in terms of directly observed quantities,
such as velocity dispersions, X-ray luminosities and temperatures.

As a matter of fact, cluster masses have been shown to tightly
correlate with the above-mentioned quantities. Such correlations can
be theoretically found by using the hypotheses of spherical symmetry,
gas equilibrium and halo virialization, but their calibrations depend
on the assumed density profile, on the departures from sphericity and
on other particulars, so that they are often calibrated on numerical
simulations with hydrodynamics or real data.  Such correlations are:

\be \sigma^2_V \propto (1+z) M^{1/3} \label{eq:sigma_m_rel} \ee

\noindent 
(Evrard 1989), where $z$ is the redshift at which the cluster is
observed;

\be T \propto (1+z_c) M^{2/3} \Omega_0^{1/3} \rho_g^{1/3}
\label{eq:m_t_rel} \ee

\noindent
(see, e.g., Lilje 1992), where $z_c$ is the cluster redshift at
collapse and $\rho_g$ is the mean gas density;

\be L \propto (1+z)^{-3/5} M^{4/3} \rho_g^{7/6} K \label{eq:l_m_rel} \ee

\noindent 
(see, e.g., Colafrancesco \& Vittorio 1994), where $K$ is the standard
correction for the change of the spectral band as an effect of
redshift.

Using this formulas, the (PS) MF can be translated into a velocity
dispersion, X-ray temperature or X-ray luminosity function, and then
compared with data.  Observational X-ray temperature functions have
been given by Edge et al.  (1990), who also give the X-ray luminosity
function, and by Henry \& Arnaud (1991). Velocity dispersion functions
have been given by Zabludoff et al. (1993), Girardi et al. (1993),
Mazure et al. (1996), Fadda et al. (1996).  Notably, different
velocity dispersion functions, given by different authors, are
consistent among them within the limits of completeness of the sample;
the same holds true for the two temperature functions.

Many authors have compared the predictions of cosmological models with
the observed distributions. Typically, the PS formula with the
spherical \dc=1.69 value has been used, as galaxy clusters, being rare
structures, collapse quasi-spherically (as shown in Chapter 4, a \dc\
value of 1.5 would have been better; for instance, Klypin \& Rhee 1994
adopt this value).  Evrard (1989) and Peebles, Daly \& Juszkiewicz
(1989), and more recently Cavaliere, Menci \& Tozzi (1994), claimed
that the existence of a significant number of clusters at high
redshifts ($z>0.5$) and with large velocity dispersions ($\sigma^2_V>
700$ km/s) would be hard to reconcile with a CDM scenario. Such a
conflict would be solved by considering low-$\Omega_0$ universes (with
or without cosmological constant), in which structures have stopped
forming at small redshift, and then clusters have been assembled at
large redshifts. The consistency of a low-density Universe with data
was also advocated by Bahcall \& Cen (1992; 1993) and by Kofman,
Gnedin \& Bahcall (1993), by comparing the theoretical MF directly
with the observational one by Bahcall \& Cen (1993).  Another possible
explanation would be an effective slow-down of gravitational
clustering as the logarithmic slope of the power spectrum crosses
$-1$, with the corresponding effective lowering of the $\delta_c$
parameter at large redshifts, as commented at the end of \S 4.5
(Monaco 1995).

A comparison of PS-based predictions with X-ray data has been
performed by many authors, for instance Cavaliere \& Colafrancesco
(1988), Schaeffer \& Silk (1988b), Lilje (1990; 1992), Kaiser (1991),
Oukbir \& Blanchard (1992), Hanami (1993), Cavaliere, Colafrancesco \&
Menci (1993), Bartlett \& Silk (1993), Colafrancesco \& Vittorio
(1994), Cavaliere, Menci \& Tozzi (1994), Balland \& Blanchard (1995),
Kitayama \& Suto (1996a,b), Liddle et al. (1996), Viana \& Liddle
(1996), Eke, Cole \& Frenk (1996), Pen (1996).  A general conclusion
is that standard CDM fails to reproduce the observed abundance of
X-ray clusters, especially at large redshifts; again, a low-density
Universe, with or without cosmological constant, or a mixed C+HDM
Universe, is usually reported to give improved fits. These conclusions
have been confirmed by works based on N-body simulations, as that of
Ueda, Itoh \& Suto (1993) or of Klypin et al. (1993).

Cluster abundances can be used to constrain the mass variance at
scales of order $\sim$10 Mpc, as the position of the exponential
cutoff of the MF is more sensitive to the normalization than to the
shape of the power spectrum.  This has been done by White, Efstathiou
\& Frenk (1993), Bond \& Myers (1993c), and Eke, Cole \& Frenk (1996);
they all reported the mass variance on 8 $h^{-1}$ Mpc to be about 0.6
if $\Omega_0=1$, and about 1 if $\Omega_0\sim0.3$. Another important
observational quantity connected with galaxy clusters is their imprint
on the CMB, through the so called Sunyaev-Zel'dovich effect (Sunyaev
\& Zel'dovich 1972). This effect will be extensively measured by the
next generation of satellites dedicated to CMB observations. A number
of theoretical works have been devoted to such a topic (see, e.g.,
Schaeffer \& Silk 1988b; Cole \& Kaiser 1989; Cavaliere, Menci \&
Setti 1991; Makino \& Suto 1993; Bartlett \& Silk 1994; Barbosa et
al. 1996; Bond \& Myers 1996c). Finally, galaxy clusters have been
simulated by means of N-body simulations with hydrodynamics; see,
e.g., Evrard (1990), Thomas \& Couchman (1992), Navarro, Frenk \&
White (1995b), Ostriker \& Cen (1996), Lubin et al. (1996), Anninos \&
Norman (1996).

Interestingly, while cluster abundances and evolution seem to suggest
a low-density Universe (apart the possibility of a C+HDM one), other
observational evidences on clusters weight toward critical density.
Many clusters show strong or weak substructures (see, e.g., West
1994); even the core of the Coma cluster, when observed accurately
enough, shows some degree of subclustering. If subclustering is a
remnant of mergers, and if substructures survive much less than one
Hubble time inside the cluster, then the degree of subclustering would
indicate that they have recently been assembled, and then that the
cosmological density is nearly critical (see, e.g., Richstone, Loeb \&
Turner 1992; Evrard 1994; Crone. Evrard \& Richstone 1996; West, Jones
\& Forman 1995). However, a flat Universe with a cosmological constant
term could produce a significant amount of subclustering (Jing et
al. 1995). An analogous argument has also been given for compact
groups of galaxies: they can arise as the result of secondary infall
on a normal galaxy group, provided the density is critical; a Universe
with cosmological constant would be only marginally inconsistent with
the observational evidence (Governato, Tozzi \& Cavaliere 1996).

This brief review has been limited to a ``Lagrangian'' view of galaxy
clusters, i.e. biased toward the problems of mass distribution and
internal dynamics. The important ``Eulerian'' topics of the strongly
clustered space distributions of galaxy clusters, of their power
spectrum, their peculiar velocities, their dipole and so on are not of
immediate interest in the present context. However, it is interesting
to note that such topics have been analyzed, among the various
methods, by constructing mock cluster catalogs by means of large Monte
Carlo realizations, in which galaxy clusters are associated with the
peaks of the Zel'dovich-evolved density field (Borgani, Coles \&
Moscardini 1993); this procedure has an immediate interesting
connection with the dynamical MF theory.

\subsection{Discussion}

As mentioned before, clusters are usually assumed to evolve
spherically, as they are rare fluctuations (see, e.g., Bernardeau
1994). In Chapters 3 and 4 it was argued that spherical collapse does
not reproduce the statistics of collapse times of fast collapsing
points.  As a matter of fact, the two statements are not immediately
in contradiction, as the global collapse of an extended structure is a
different thing from the local collapse of a vanishing mass element
(see the example at the end of \S 1.3).  Anyway, it is clear from
theoretical arguments, N-body simulations and observations that
non-sphericity can play an important role in cluster formation.
However, all these arguments are only qualitative, as an important
point is missing.

The main point with cluster formation has been well expressed by
Cavaliere, Colafrancesco \& Menci (1991): clusters are {\it ill
defined} objects in space and time. While it is quite clear that
clusters exist (their X-ray emission is a crucial proof), it is not
clear where they end, and this indeterminacy hampers the same
definition of cluster total mass.  Moreover, clusters are usually
assumed to be spherical, while N-body simulations suggest that a
sheet-like or filamentary pattern dominates the outer parts of
clusters, and accretion of matter does not take place isotropically
but through filaments. 

The dynamical MF theory can give an important contribution to cluster
formation. The matter which dynamically belongs to the cluster can be
defined as the matter which has entered a multi-stream region; if the
Lagrangian excursion sets of the inverse collapse time $F$ are simply
connected, the cluster mass is well-defined, at least from the
theoretical point of view. A comparison with N-body simulation can
help to understand whether such matter is found in a single relaxed
clump or in a more complicated structure (in the Eulerian space).
Moreover, a dynamical description of the kind given in Chapter 3 takes
into account the important transient features of clusters. Finally, if
substructures are the remnants of recent mergers, a dynamical MF
theory can be used to connect the statistics of substructures to
cosmological models.

\section{High-redshift objects}

{\it Active galactic nuclei} (AGN) are the most luminous compact
objects which we can observe in our Universe.  They are believed to
consist of supermassive black holes (SBH) which accrete matter from a
host structure, which is believed to be a galaxy or a protogalaxy.
The geometry of the infalling matter determines the kind of emission
and its spectral features.  Very luminous AGNs are usually called
quasars (quasi-stellar radio sources) or, more generally, QSOs
(quasi-stellar objects), even though the hosting galaxy of many QSOs
have recently been observed. The SBH paradigm is the basis of the
so-called unified model, which aims to describe all the kinds of AGNs
as different realizations of a single kind of object. In practice, the
SBH paradigm has never been observationally demonstrated, but the
present constraints of high efficiency in the release of energy, high
compactness of the central engine and short variability time (which
can be less than a day) strongly hamper any model based on
non-gravitational events. The AGN problem has widely been studied, and
the relevant literature is very vast; the main topics have been
reviewed, for instance, by Rees (1984), Lawrence (1987) and Blanford
(1990).

The relevance of QSOs in cosmology is great, as such objects are
visible even at very high redshifts, up to $z=4$. They can then be
used to probe the early Universe. Moreover, it has been discovered
that a ``forest'' of Lyman $\alpha$ lines, with a variety of
redshifts, can be recognized in absorption in their spectra.  The
structures which originate these lines, called {\it Lyman $\alpha$
clouds} are probably gas clouds which lie in the same line of sight of
the QSO; the ones associated with the strongest lines, the {\it damped
Lyman $\alpha$ systems}, are probably associated with high-redshift
protogalactic disks; some of them have been observed with optical
telescopes. This class of objects is a further precious tool to
constrain cosmological models at high redshifts.

\subsection{QSO formation and cosmology}

The event of QSO formation couples what happens at very small scales,
of the size of the SBH ($\sim$0.01 pc) with very large, cosmological
scales, which at $z\sim 2$ are of the order of $\sim$1 Mpc. This makes
the problem very hard to address. It is essentially impossible to follow
the cosmological creation of a SBH with present-day numerical
simulations, and probably with that of the near future, as the range
of scales involved is too large, and the physical mechanisms are too
complex (see, e.g., Rees 1984).  Then, heuristic analytical arguments
are the only way to address the problem.

The major phases of QSO formation and activity are believed to be the
following (see, e.g., Haehnelt \& Rees 1993; Eisenstein \& Loeb
1995b):

\begin{enumerate}
\item 
The perturbation within which the SBH is going to form, which is
reasonably assumed to be a rather massive, rare fluctuation, acquires
angular momentum during its mildly non-linear evolution (Peebles 1969;
White 1984).  At this stage, gas follows dark matter.
\item 
The perturbation collapses and forms a virialized halo, which probably
constitutes the nucleus or the bulge of the forming galaxy. The gas is
shock heated, but at least a part of it can cool and form a
rotationally-supported disk.
\item 
The disk cools and fragments, forming stars. Supernovae explosions may
be able to sweep the gas away from the forming nucleus, if the potential
well is not deep enough.
\item 
On the other hand, the central part of the gas disk can lose angular
momentum by means of a number of viscous processes, among which are
gravitational instabilities, turbulence (eventually induced by
supernovae) or magnetic fields.
\item 
If the central part of the disk shrinks at small enough radii,
relativistic effects can provide enough viscosity to make the gas
collapse at even smaller radii.
\item 
If a runaway contraction sets in, the disk can evolve into a
relativistic disk or a supermassive star, depending on the amount of
angular momentum which is not dissipated. Such configurations are
highly unstable, and a SBH forms in a (cosmologically) short time.
\item 
Matter starts to accrete onto the SBH; radiation is emitted with a
high efficiency, of order of up to 10\% of $mc^2$. After some possible
but probably short transients, the SBH radiates at the Eddington limit.
\item 
The ``fuel'' stops accreting onto the SBH, whose luminosity suddenly
decreases much below the Eddington luminosity. The bursting phase
lasts less than the Hubble time, about 10$^7$ yr.
\item 
Sometimes new fuel becomes available for the SBH; this can be due to
interaction or merging events which are experienced by the host
object. This can give rise to new AGN activity (see Monaco et al. 1995
and references therein).
\end{enumerate}

Although this chain of events is quite schematic and contains many
questionable hypotheses, it gives an idea of the complexity of the QSO
phenomenon in connection to cosmology. 

The main observational constraints that QSO formation theories have to
fulfill are the luminosity function of QSOs (see, e.g., Pei 1995), as
function of redshift and in different spectral bands, and the upper
limits on the remnant SBH masses in normal galaxies (see, e.g.,
Cavaliere \& Padovani 1988; Salucci \& Szuszkiewicz 1996).  Luminous
QSOs show a comoving number density which steeply rises up to $z\simeq
2$, stays more or less constant up to $z\simeq 3$ and then falls
again, even though high-redshift measures are very uncertain.  The
redshift evolution of the luminosity function is compatible with a
pure luminosity evolution up to $z=2$ and a number antievolution at
$z>3.5$.

If SBHs radiate at the Eddington luminosity ($L_E= 1.3\times 10^{46}
(M/10^8 M_\odot)\ erg/s$; see, e.g., Cavaliere \& Padovani 1988), the
luminosity function can be transformed into a mass distribution of
remnant black-hole masses, with the conclusion that black-hole masses
decrease with their formation time. Besides, QSO activity has to last
much less than a Hubble time, otherwise it would imply too large
remnant black-hole masses, in contradiction with the observational
evidence (see, e.g., Cavaliere et al. 1983).

The problem of the QSO formation has been studied from a cosmological
point of view by a number of authors, among whom are Efstathiou \&
Rees (1988), Cole \& Kaiser (1989), Carlberg (1990), Nusser \& Silk
(1993), Kashlinsky (1993), Katz et al. (1994) Fugikita \& Kawasaki
(1994), Yi (1996). Such authors generally estimate the number of
quasars by assuming {\it a priori} in which kind of host object they
are going to form. The number of halos is generally estimated by means
of the PS theory, sometimes checked by a comparison to a N-body
simulation (e.g., the works of Efstathiou \& Rees 1988 and Katz et
al. 1994).  Haehnelt \& Rees (1993), Haehnelt (1993), Eisenstein \&
Loeb (1995a,b) and Cavaliere, Perri \& Vittorini (1997) have addressed
in more detail the problem of the cosmological formation of QSOs;
their works will be now outlined.

Haehnelt \& Rees (1993) argued that QSOs are associated with
newly-formed SBHs, and that SBHs form in common, already assembled
proto-bulges of galaxies. A lower limit on the depth of the potential
well of the host object can be set by imposing that supernovae are not
able to sweep all the gas away; halos with virial velocities lower
than $\sim$200 km/s can not host SBHs.  The efficiency of SBH
formation was assumed to depend on the central density of the halo,
which is larger at higher redshift (see \S 1.2.4), in order to obtain
smaller SBH masses at lower redshifts.  This efficiency was inserted
into the PS MF to give the number density of SBHs; the resulting curve
was convolved with a reasonable but speculative light curve in order
to obtain the luminosity function of QSOs. The resulting cosmological
constraints were examined by Haehnelt (1993). This procedure has the
merit of relying on the simplest and safest hypotheses about QSO
formation; however, it contains many free parameters, and its validity
can not be extrapolated to low redshifts, where the refueling of
existing SBHs is probably the most important mechanism in play.

Eisenstein \& Loeb (1995a,b) concentrated their attention on the
formation of SBH and on the role of angular momentum; this quantity is
of great importance, as it is necessary to dissipate it to form a
compact structure.  To estimate the angular momentum acquired by
protostructures, they modeled the structures as collapsing
ellipsoids.  In this way, they calculated the angular momentum
distribution of rare massive concentrations; a small but sufficient
number of massive clumps was then found to collapse with a final size
small enough that efficient viscosity can take place.  Then, the time
scale at which angular momentum is dissipated was compared to the
star-formation time scale (a free parameter): whenever the disk
shrank fast enough, a black hole was assumed to form, otherwise star
formation would have prevailed.  As a result, black holes can form at
high redshifts, $z>10$; they become visible when a galactic-size
structure supplies enough gas to give rise to luminous QSO
activity. These works provide an interesting attempt to follow in some
detail the physics of SBH formation, at the expenses of a strong
simplification of the problem. A general criticism could be that, in a
realistic situation, the angular momentum of the central gas clump
could be uncorrelated with that of the dark-matter halo. The idea that
SBHs form before the observable QSO activity had already been proposed
by Loeb (1993; see also Umemura, Loeb \& Turner 1993).

Cavaliere et al. (1996) proposed a double QSO and galaxy connection:
SBHs form in galaxies, and are recurrently reactivated by galaxy
interactions.  To avoid too large a number of free parameters, they
supposed a constant efficiency of black-hole formation. Using the
time-scale formalism of Cavaliere et al. (1991), they introduced a
formation time scale for SBHs and an interaction time scale for
galaxies; the first time scale dominates at high redshift, the second
one at small redshifts.  As a result, they succeeded in fitting the
luminosity function of QSOs with a limited number of free parameters.
However, in this procedure the formation of SBHs is still treated at a
heuristic level: it is not unreasonable to think that the efficiency
of SBH formation crucially depends on some important parameter.

The dynamical MF described in Chapters 3 and 4 can give a substantial
contribution to this problem, especially if the angular momentum of a
collapsing structure is important in determining whether a SBH is
going to form inside that structure. As shown by White (1994),
Lagrangian dynamics is suitable to follow the angular momentum
acquisition of a collapsing clump: in fact, angular momentum is
acquired during the mildly non-linear evolution.  More recently,
Catelan \& Theuns (1996a,b) have used Lagrangian perturbation theory
to determine the angular momentum of a collapsing ellipsoidal peak.
An intrinsic difficulty of these calculations resides in the
determination of the set of Lagrangian space which is going to end up
in the structure. Through the rigorous definition of collapse as OC,
the dynamical MF theory can provide such sets as the excursion sets of
the inverse collapse time $F$; these objects can be investigated
through Monte Carlo simulations.

Because of the wealth of dynamical information available, the
dynamical MF theory can be extended to give not only the value of the
angular momentum acquired by a structure, but also a joint merging --
angular momentum acquisition history. This could give some hints on
how angular momentum is redistributed inside the clump.  All this
information, obtained without assuming any special symmetry for the
collapsing structures, could be used to try to understand whether and
when suitable conditions for the formation of a SBH can be obtained.

\subsection{Lyman $\alpha$ clouds}

As mentioned above, QSO spectra show, at shorter wavelengths than
their Lyman $\alpha$\footnote{These far-ultraviolet lines are usually
very hard to detect, as they are absorbed by the neutral hydrogen of
our Galaxy.  However, they are so redshifted that fall in the optical
or near ultraviolet.} emission line, a ``forest'' of thin absorption
lines, of width $\sim$25 km/s. Lynds (1971) was the first to recognize
such lines as Lyman $\alpha$ absorption lines, associated to gas
clouds which lie in the same line of sight of the QSO.  These clouds
constitute a new class of high-redshift objects, whose properties can
be used to constrain cosmological models.

The observational state-of-the-art, relative to the search and
interpretation of these objects, can be found in the proceedings book
edited by Meylan (1995). The main observational evidences can be
listed as follows:

\begin{itemize}
\item
There are three broad classes of Lyman $\alpha$ clouds. The objects of
the first class are associated to lines with a high column density of
neutral hydrogen, $N_{HI}\geq 3 \times 10^{20}$ cm$^{-2}$\footnote{The
column density is estimated by means of the depth of the line, the
velocity dispersion by means of its width.} and small velocity
dispersion, $\sim$ 10 km/s. These lines are rare with respect to the
whole population, show strong damped wings, and metal lines can be
found at the same redshift. The structures associated to them are
called {\it damped Lyman $\alpha$ systems}, and are usually
interpreted as the inner parts of protogalactic disks.
\item
The clouds associated to lines with column density larger than
$N_{HI}\geq 3\ 10^{17}$ cm$^{-2}$ (these lines are saturated but do
not show damped wings) are called {\it Lyman limit systems}; they are
in general interpreted as the outer, less dense part of a
protogalactic disk.
\item
The objects of the third class are associated to lines with column
densities smaller than the Lyman limit, and constitute the so-called
{\it Lyman $\alpha$ forest}. They are thought to be associated to
clouds of low density contrast.
\item
Not much is known on the actual geometry of these structures.  In some
cases the galaxies corresponding to Lyman $\alpha$ lines have been
observed at intermediate or high redshifts (see, e.g., Lanzetta et
al. 1995, Steidel et al. 1997).  There is some observational hint,
based on observations of close QSO pairs or triplets, that Lyman
$\alpha$ clouds are quite spatially extended (see, e.g, Dinshaw et
al. 1995; Bajtlik 1995).
\end{itemize}

I am neglecting the important topics of element abundances in Lyman
$\alpha$ clouds and their interpretation, or of their space
correlations; these arguments are reviewed in Meylan (1995).

If damped Lyman $\alpha$ systems are generally thought to be
associated to protogalaxies, the interpretation of the Lyman $\alpha$
forest is more debated. Sargent et al. (1980) proposed a model in
which the cloud is confined by pressure exerted by the surrounding hot
intergalactic medium; however such a model cannot easily reproduce the
different column densities observed. Another model, due to Rees (1986)
and Ikeuchi, Murikama \& Rees (1988), associates the clouds with small
CDM halos (see also Bond, Szalay \& Silk 1988; Mo, Miralda-Escud\'e \&
Rees 1993; Gnedin \& Hui 1996; Hui, Gnedin \& Zhang 1997). This model
probably would not satisfy the constraint mentioned before of extended
Lyman $\alpha$ clouds. Recently it has been proposed that Lyman
$\alpha$ clouds are associated to ``pancake'' structures, which have
collapsed along one axis but are still expanding on the other axes
(Haehnelt 1995). Recent N-body+SPH numerical simulations show that
damped Lyman $\alpha$ and Lyman limit clouds can be associated to more
or less virialized halos or to gas filaments (see, e.g., Gardner et
al. 1997; Rauch, Haehnelt \& Steinmetz 1997; Weinberg 1996).

The abundance of damped Lyman $\alpha$ systems, interpreted as
protogalaxies, has now become a standard constraint to cosmological
models; it has been used, for instance, by Mo \& Miralda-Escud\'e
(1994), Klypin et al. (1995) and Liddle et al. (1996).  These authors
typically used the PS theory (sometimes tested against N-body
simulations) to find the number of structures which can host a Lyman
$\alpha$ cloud. Notably, some of them followed the suggestion of
Monaco (1995), using a \dc\ parameter of about 1.5: Klypin et al.
(1995) found \dc=1.5 also by comparing the PS with their simulations,
while Liddle et al. (1996) used the 1.5 value only for Lyman $\alpha$
clouds, interpreted as structures which have just experienced their
first collapse, while the spherical value 1.69 was reserved to more
relaxed structures such as galaxy clusters.  The cosmological use of
the Lyman $\alpha$ forest is still debated; see, for instance,
Haehnelt (1996), Cen et al. (1994), Cen (1997); a general criticism to
these approaches is that they are still rather model-dependent.

The dynamical MF theory can manifestly give a contribution to this
topic. Dynamical pancake-like and filamentary transients, which
probably dominate the dynamics of Lyman $\alpha$ clouds, are treated
in a realistic way, and then structures of not large density contrast
can be described. Moreover, the large amount of dynamical information
available can be used to model gas dynamics in a more realistic way.

As a final comment, the dynamical MF theory can also give an important
contribution to the problem, deeply connected to QSO and Lyman
$\alpha$ formation and to the late reionization of primordial
hydrogen, of the formation of the first luminous objects (Tegmark et
al. 1997; Rees 1996), which could be connected to the collapse of the
first structures, again possibly dominated by gravitational
transients.

%
%

\chapter{Prospects}

The previous Chapters have shown that the cosmological MF is a lively
and promising field, both from the theoretical and from the
observational point of view: the firm results are just a few, the open
problems are many, and many procedures have to be developed in detail.

From the theoretical point of view, the main problems are the
following:

\begin{itemize}
\item
It is necessary to understand in detail what is the total mass of a
structure, and to find a rigorous, well-posed and easily interpretable
definition for it, suitable to help in the interpretation of
observations.  In searching for this definition, it is not opportune
to suppose {\it a priori} the presence of some symmetry, or a certain
shape for the density profile of structures; this definition has to
take into account that already collapsed and still infalling matter
can coexist in the external regions of structures.
\item
It is necessary to understand the role of filtering in the MF problem,
as different kinds of filters (sharp $k$-space, Gaussian) lead to
different and not equivalent formulations of the problem.
\item
It is necessary to take in full account the geometry of collapsed
regions in Lagrangian space, going beyond the usual golden rule
described in \S 4.4. In the realistic case in which the stochastic
process on which the MF is based is non Gaussian, this investigation
can be performed through Monte Carlo simulations of initial fields.
\item
It is necessary to add more physics to the dynamical MF theory, to
describe events such as the aggregation and fragmentation of already
collapsed structures. In this way, the dynamics inside structures
could be resolved; this is necessary to describe objects such as
galaxies inside larger structures, as groups or clusters. The kinetic
theories presented in \S 2.5.3 go in this direction.
\end{itemize}

The state-of-the-art of the observational MF is quite promising,
especially for galaxy clusters, for which three independent procedures
for mass estimates are available. The main problem is again, in my
opinion, the definition of the total mass of clusters: without a deep
understanding of what a cluster is, any comparison between theory and
observations will have a certain degree of uncertainty; on the other
hand, available observational mass estimates are so uncertain that
this indetermination now has a modest impact, but it can become
important in the near future. Another problem with mass estimates is
that they often rely on some restrictive hypothesis: virial estimate
are obviously based on the hypothesis that the cluster is virialized
and on other hypotheses, X-ray estimates are based on spherical
symmetry and gas equilibrium, giant arc estimates are based on
assuming a certain geometry for the cluster, while weak lensing
estimates still present technical problems. All these techniques will
be improved, relaxing some of the restrictive hypotheses, when
available data will increase in number and improve in quality.
Presently, comparisons between theoretical predictions and data are
usually performed in terms of direct observables, as galaxy velocity
dispersions, X-ray luminosities or X-ray temperatures.

The dynamical MF theory is a promising starting point for a further
deepening of the MF problem and its applications. The main ideas that
can be developed are the followings:

\begin{itemize}
\item
{\bf Comparison to simulations:} the first necessary step is to test
the validity of the dynamical MF theory against N-body simulations.
This is already an active project, and will be briefly described in
the next Section.
\item
{\bf From resolution to mass:} this important passage has been dealt
with the simple golden rule; the $p$ distribution of forming masses,
introduced in \S 4.4, has to be investigated before giving definite
predictions on the MF. This is also necessary to understand in detail
merging histories.
\item
{\bf Merging histories:} these are one of the most interesting products
of MF theories. To obtain them from the dynamical MF theory, it is
necessary to understand the role of smoothing: Gaussian smoothing
allows us to distinguish (maybe artificially) between merging and
accretion, while the rather special SKS filter predicts only mergings
with small or large halos.
\item
{\bf Joint mass-angular momentum statistics:} the wealth of dynamical
information on every collapsed structure allows us to describe in
detail the behavior of structures; in particular, it is possible to
obtain the joint merging -- angular momentum histories, which can be a
very important tool for galaxy formation theories.
\item
{\bf QSO activity:} the joint mass-angular momentum statistics can
provide useful information for understanding cosmological QSO
activity, in the hypothesis that the initial angular momentum of the
halo is important.
\item
{\bf Lyman $\alpha$ clouds:} a subclass of these structures is
probably observed during a not much evolved phase, maybe after their
collapse along one axis (or even just before). Then, the dynamical MF
theory, which takes into account the relevant dynamical transients and
the geometry of collapse, can be used to describe these structures in
a satisfactory way.
\item
{\bf Galaxy groups:} the dynamical state of many open groups of
galaxies is not clear: they could be real structures which are
relaxing, or could be experiencing their first collapse, or could even
be structures without a precise dynamical meaning. The dynamical MF
theory can help to understand whether such groups are stable,
well-defined structures or pancake-like transients, and if they can be
used to obtain useful cosmological constraints.
\item
{\bf Galaxy clusters:} again, the wealth of dynamical information
given by the dynamical MF theory can lead to a more detailed
description of galaxy clusters, especially in their external parts,
whose dynamical role can be clarified. In particular, as suggested
before, it is important to understand what is the total mass of a
galaxy cluster, a thing which can not be addressed properly if
dynamics is not described in a realistic way.
\item
{\bf Substructures in galaxy clusters:} if such substructures are a
remnant of recent merging events of structures of comparable size,
merging histories can be used to predict the expected number of
substructures, a quantity which is known to depend strongly on the
cosmological density parameter. Given the transient nature of these
objects, the dynamical MF theory appears suitable to address this
problem.
\end{itemize}

\section{Comparison to simulations}

In comparing the Press \& Schechter theory with numerical simulations,
the halos are typically extracted by means of an overdensity
criterion, as in the DENMAX algorithm, or through the famous
friends-of-friends algorithm, based on percolation (see \S 2.2). These
algorithms are parametric, in the sense that their result depends on a
free parameter, such as a threshold overdensity or a percolation
parameter.  The values of such parameters can be weakly constrained
through heuristic arguments based on spherical collapse.

In the case of the dynamical MF theory, collapse is defined in a
precise way as the OC event. It is then not opportune to compare the
dynamical MF with the abundances of friends-of-friends or DENMAX
halos; it is instead opportune to construct an algorithm able to
reproduce those structures which are predicted to form by the theory.
The collapse definition used in the theory is based on orbit crossing,
which corresponds to the vanishing of the Jacobian determinant
of the transformation from Lagrangian to Eulerian space (Eq.
\ref{eq:jacobian}); large densities are just a consequence of it.
Orbit crossing is very easy to understand: mass elements coming from
far away get very near. However, OC is defined on a smoothed field,
while small-scale structure is always present; the most important
hypothesis of the dynamical MF theory is that small-scale structure
does not dramatically influence larger-scale dynamics. The likely
effect of small-scale structure is to spread collapsed regions, thus
lowering the actual density (and avoiding the formation of real
caustics). It is then possible to clarify why an overdensity
criterion, which would anyway be parametric, can be not suitable to
find OC regions: such a criterion would preferentially select caustics
with more compact (Eulerian) geometry, which, when spread by
small-scale structure, would tend to remain at high density, while
filamentary or pancake-like caustics would be spread to lower
effective densities.

In Monaco (1996a) the following algorithm was proposed as a natural
implementation of OC in realistic situations, as, e.g., an N-body
simulation:

\be \exists \mq_1, \mq_2\, :\ \ |\mq_1-\mq_2|\geq L, \ \ |\mx(\mq_1,t)-
\mx(\mq_2,t)|<\varepsilon,\ \ \ \ \varepsilon \ll L. \label{eq:map} \ee

\noindent
In other words, mass elements in a neighborhood of size $\varepsilon$
of the Eulerian point \x\ belong to a multi-stream region, on scale
$\geq L$, if at least two mass elements $\mq_1$ and $\mq_2$ come from
a distance greater than $L$. This algorithm appears simple to
implement in a numerical simulation, but its success does not appear
guarenteed. For this definition to make sense, it is necessary to find
an interval of $\varepsilon$ values, connected with the typical
dimension of a structure on a certain scale, for which the
determination of the collapsed regions is stable; in other words,
$\varepsilon$ does not have to be a free parameter. The quantity $L$
highlights that this definition is naturally scale-dependent; it has
to be put in relation with the smoothing radius, and then with the
resolution \res.  The exact relation has to be chosen so as to
optimize the dynamical predictions; this optimization also has to
allow for the shape of the filter: the Gaussian filter is expected to
be prefered.  This procedure is analogous to what has been done for
the truncated Zel'dovich approximation (see \S 1.2.6).

This comparison between theory and simulations will allow us to
clarify whether and to what extent the total mass of a structure is a
well defined quantity, and what a structure exactly is; indeed, the
particles defined as belonging to a structure are not only those in
its densest, central part, but all the particles which have got in
``dynamical contact'' with the structure by getting into its
multi-stream region.

\bigskip
\bigskip

\section*{Acknowledgments}

I wish to thank my PhD supervisor, Fabio Mardirossian, for his
constant guidance and encouragement, Dennis Sciama and Luigi Danese
for having hosted me in the stimulating SISSA environment for the
whole PhD course, Alfonso Cavaliere and Sabino Matarrese for many
enlightening discussions on the mass function problem, Francesca
Matteucci for her suggestions, and Steve Maddox for his help in
improving the text.  I have profited much from my discussions with
Sergei Shandarin, Giuliano Giuricin, Marino Mezzetti, Stefano Borgani,
Marco Bruni, Thomas Buchert, Paolo Catelan, Nicola Menci, Massimo
Persic, Manolis Plionis, Cristiano Porciani, Paolo Salucci, Paolo
Tozzi and Riccardo Valdarnini. This work has been partially supported
by the Italian Research Council (CNR-GNA) and by the Ministry of
University and of Scientific and Technological Research (MURST).

%
%

\chapter{Bibliography}

\refe Abell G.O., 1958, ApJS, 3, 211
\refe Abell G.O., Corwin H.G., Olowin R.P, 1989, ApJS, 70, 1
\refe Adler R.J., 1981, The Geometry of Random Fields. Wiley, New York
\refe Anninos P., Norman M.L., 1996, ApJ, 459, 12
\refe Antonuccio-Delogu, V., Colafrancesco, S., 1994, ApJ, 427, 72 
\refe Appel L., Jones B.J.T., 1990, MNRAS, 245, 522
\refe Arnold L., 1973, Stochastic Differential Equations. Wiley, New York
\refe Ashman, K.M., 1992, PASP, 104, 1109
\refe Ashman, K.M., Salucci, P., Persic, M. 1993, MNRAS 260, 610
\refe Audit E., Alimi J.M., 1996, A\&A, 315, 11
\refe Avila-Reese V., Firmani C., 1996, in Persic M., Salucci P. eds., 
Dark Matter 1996. Pasp Conf. Ser., in press 

\refe Bagla, J.S., Padmanabhan, T. 1994, MNRAS 266, 227
\refe Bahcall N.A., 1988, ARA\&A, 26, 631
\refe Bahcall N.A., Cen, R. 1992, ApJ, 398, L81
\refe Bahcall N.A., Cen, R. 1993, ApJ, 407, L49
\refe Bajtlik S., 1995, in Meylan G. ed., QSO Absorption Lines. Springer, 
Berlin
\refe Balland C., Blanchard A., 1995, A\&A, 298, 323
\refe Barbosa D., Bartlett J.G., Blanchard A., Oukbir J., 1996, A\&A, 314, 13
\refe Bardeen J.M., Bond J.R., Kaiser N., Szalay A.S., 1986, ApJ, 304, 15
\refe Barrow J.D., Silk J., 1981, ApJ, 250, 432
\refe Bartlemann M., Ehlers J., Shneider P. 1993, A\&A 280,351
\refe Bartlemann M., Steinmetz M., 1996, MNRAS, 283, 431
\refe Bartlett J.G., Silk J., 1993, ApJ, 407, L45
\refe Bartlett J.G., Silk J., 1994, ApJ, 423, 12
\refe Beers T.C., Flynn K., Gebhardt K., 1990, AJ, 100, 32
\refe Bernardeau F., 1994a, A\&A, 291, 697
\refe Bernardeau F., 1994b, ApJ, 427, 51 
\refe Bernardeau F., Kofman L., 1995, ApJ, 443, 479
\refe Bernardeau F., Schaeffer R., 1991, A\&A, 250, 23
\refe Bertola F., Pizzella A., Persic M., Salucci P., 1993, ApJ, 416, L45
\refe Bertschinger, E., Jain, B. 1994, ApJ, 431, 486
\refe Betancort-Rijo J., Lopez-Corredoira M., 1996, ApJ, 470, 674
\refe Binney J., 1977, ApJ, 215, 483
\refe Binney J., Tremaine S., 1987, Galactic Dynamics. Princeton 
University Press, Princeton
\refe Biviano A., Girardi M., Giuricin G., Mardirossian F., Mezzetti M., 
1993, ApJ, 411, L13
\refe Blanford R.D., 1990, in Courvoisier T.J.L., Mayor M. eds., Active 
Galactic Nuclei. Springer, Berlin
\refe Blain A.W., Longair M.S., 1993a, MNRAS, 264, 509
\refe Blain A.W., Longair M.S., 1993b, MNRAS, 265, L21
\refe Blanchard A., Valls-Gabaud D., Mamon G.A., 1992, A\&A, 264, 365
\refe Bond J.R., 1989, in Large-Scale Motions in the Universe, eds. V. Rubin,
G. Coyne (Princeton: Princeton University Press)
\refe Bond J.R., Cole S., Efstathiou G., Kaiser N., 1991, ApJ, 379, 440
\refe Bond J.R., Myers S.T., 1996a, ApJS, 103, 1
\refe Bond J.R., Myers S.T., 1996b, ApJS, 103, 41
\refe Bond J.R., Myers S.T., 1996c, ApJS, 103, 63
\refe Bond J.R., Szalay A.S., Silk J., 1988, ApJ, 324, 627
\refe Borgani S., Coles P., Moscardini L., 1994, MNRAS, 271, 223
\refe Bouchet F.R., 1996, in Bonometto S., Primack J., Provenzale A., eds.,
Dark Matter in the Universe. In press (preprint astro-ph/9603013)
\refe Bouchet F.R., Colombi S., Hivon E., Juszkiewicz R., 1995, A\&A, 296, 
575
\refe Bouchet F.R., Juszkiewicz R., Colombi S., Pellat R., 1992, ApJ, 394, 
L5
\refe Bower R.G., 1991, MNRAS, 248, 332
\refe Brainerd T.G., Villumsen J.V., 1992, ApJ, 394, 409
\refe Brainerd T.G., Sherrer R., Villumsen J.V., 1993, ApJ, 418, 570
\refe Buchert T., 1989, A\&A, 223, 9
\refe Buchert T., 1992, MNRAS, 254, 729
\refe Buchert T., 1994, MNRAS, 267, 811
\refe Buchert T., 1996, in Bonometto S., Primack J., Provenzale A. eds., 
Dark Matter in the Universe. In press (preprint astro-ph/9603013)
\refe Buchert T., Ehlers J., 1993, MNRAS, 264, 375
\refe Buchert T., Melott A.L., Wei\ss\  A.G., 1994, A\&A, 288, 349

\refe Carlberg R.G., 1990, ApJ, 350, 505
\refe Carlberg R.G., Couchman H.M.P., 1989, ApJ, 340, 47
\refe Catelan P., 1995, MNRAS, 276, 115
\refe Catelan P., Lucchin F., Matarrese S., 1988, Phys.Rev.Lett., 61, 267
\refe Catelan P., Theuns T., 1996a, MNRAS, 282, 436
\refe Catelan P., Theuns T., 1996b, MNRAS, 282, 455
\refe Cavaliere A., Colafrancesco S., 1988 ApJ, 331, 660 
\refe Cavaliere A., Colafrancesco S., 1990, in Oegerle W.R., Fitchett M.J.,
Danly L. eds., Clusters of galaxies. Cambridge University Press, Cambridge
\refe Cavaliere A., Colafrancesco S., Menci N., 1991a, ApJ, 376, L37 
\refe Cavaliere A., Colafrancesco S., Menci N., 1991b, in Fabian A.C. ed., 
Clusters and Superclusters of galaxies.Kluwer Ac. Pub., Dordrecht
\refe Cavaliere A., Colafrancesco S., Menci N., 1992, ApJ, 392, 41 
\refe Cavaliere A., Colafrancesco S., Menci N., 1993, ApJ, 415, 50 
\refe Cavaliere A., Colafrancesco S., Scaramella R., 1991, ApJ, 380, 15 
\refe Cavaliere A., Fusco Femiano R., 1976, A\&A, 49, 137
\refe Cavaliere A., Giallongo E., Vagnetti F., Messina A., 1983, ApJ, 269, 57
et al. 1983, 
\refe Cavaliere A., Menci N., Setti G., 1991, A\&A, 245, L21
\refe Cavaliere A., Menci N., 1993, ApJ, 407, L9 
\refe Cavaliere A., Menci N., 1994, ApJ, 435, 528 
\refe Cavaliere A., Menci N., Tozzi P., 1994, in Seitter W.C., ed., 
Cosmological Aspects of X-ray Clusters of Galaxies.  Kluwer Ac. Pub., 
Dordrecht 
\refe Cavaliere A., Menci N., Tozzi P., 1996, ApJ, 464, 44 
\refe Cavaliere A., Padovani P., 1988, ApJ, 333, L33
\refe Cavaliere A., Perri F., Vittorini V., 1997, MemSAIt, 68, 27
\refe Cen R., 1997, ApJ, 479. 285
\refe Cen R., Miralda-Escud\'e  J., Ostriker J.P., Rauch M., 1994, ApJ, 437, L9
\refe Chandrasekhar S., 1943, Rev. Mod. Phys., 15, 2
\refe Charlton J.C., Salpeter E.E., 1991, ApJ, 375, 517
\refe Colafrancesco S., Lucchin F., Matarrese S., 1989, ApJ, 345, 3
\refe Colafrancesco S., Vittorio N., 1994, ApJ, 422, 443
\refe Cole S., 1991, ApJ, 367, 45
\refe Cole S., Kaiser N., 1988, MNRAS, 233, 637
\refe Cole S., Kaiser N., 1989, MNRAS, 237, 1127
\refe Cole S., Aragon-Salamanca A., Frenk C.S., Navarro J.F., Zepf S.E., 1994,
MNRAS, 271, 781
\refe Coles P., Jones B., 1991, MNRAS, 248, 1
\refe Coles P., Lucchin F., 1995, Cosmology. Wiley, New York
\refe Coles P., Melott A.L., Shandarin S.F., 1993, MNRAS, 260, 765
\refe Crone M.M., Evrard A.E., Richstone D.O., 1996, ApJ, 467, 489

\refe Danziger I.J., 1996, in Persic M., Salucci P. eds., Dark Matter 1996. 
Pasp Conf. Ser., in press 
\refe Davis M., 1996, in Turok N. ed., Critical Dialogues in Cosmology. In press  
\refe Davis M., Peebles P.J.E., 1983, ApJ, 267, 465
\refe Dejonghe H., Merritt D., 1992, ApJ, 391, 531
\refe Dekel A., Rees M., 1987, Nature, 326, 455
\refe Dinshaw N., Foltz C.B., Impey C.D., Weymann R.J., Morris S.L., 1995, 
in Meylan G. ed., QSO Absorption Lines. Springer, Berlin
\refe Doroshkevich A.G., 1967, Astrofizika 3, 175
\refe Doroshkevich A.G., 1970, Astrofizika 6, 581 (transl.: 1973, 
Astrophysics 6, 320)
\refe Doroshkevich A.G., Kotok T.V., 1990, MNRAS, 246, 10

\refe Edge A.C., Stewart G.C., Fabian A.C., Arnaud K.A., 1990, MNRAS, 245, 559
\refe Efstathiou G., 1989, in Peacock J.A., Heavens A.F., Davies A.T. eds., 
Physics of the Early Universe. Edinburgh University Press, Edinburgh
\refe Efstathiou G., 1995, MNRAS, 272, L25
\refe Efstathiou G., Ellis R.S., Peterson B.A., 1988, MNRAS, 232, 431
\refe Efstathiou G., Fall S.M., Hogan C., 1979, ApJ, 189, 203
\refe Efstathiou G., Frenk C.S., White S.D.M., Davis M., 1988, MNRAS, 235, 
715
\refe Efstathiou G., Rees M.J., 1988, MNRAS, 230, 5P
\refe Ehlers J., Buchert T., 1997, GRG, in press (astro-ph/9609036)
\refe Eisenstein D.J., Loeb A., 1995a, ApJ, 439, 520
\refe Eisenstein D.J., Loeb A., 1995b, ApJ, 443, 11
\refe Eke V.R., Cole S., Frenk C.S., 1996, MNRAS, 282, 263
\refe Ellis, G.F.R. 1971, in General Relativity and Cosmology, ed. 
R.K.Sachs (New York: Academic Press)
\refe Ellis R.S., 1997, ARA\&A, in press
\refe Ellis R.S., Colless M., Broadhurst T., Heyl J., Glazebrook K., 
1996, MNRAS, 280, 235
\refe Epstein R.I., 1983, MNRAS, 205, 207
\refe Epstein R.I., 1984, ApJ, 281, 545
\refe Ernst M.H., 1986, in Fractals in Physics, eds. L. Pietronero, E. 
Tosatti (Elsevier Science Publisher)
\refe Evrard A.E., 1989, ApJ, 341, L71
\refe Evrard A.E., 1990, ApJ, 363, 349 
\refe Evrard A.E., 1994, in Durret F., Mazure A., Tran Thanh Van J. eds., 
Clusters of Galaxies. Editions Frontieres, Gif-sur-Yvette
\refe Evrard A.E., Summers F.J., Davis M., 1994, ApJ, 422, 11 

\refe Fadda D., Girardi M., Giuricin G., Mardirossian F., Mezzetti M., 1996, 
ApJ, 473, 670
\refe Fort B., Mellier Y., 1994, A\&A Rev., 5, 239
\refe Fugikita M., Kawasaki M., 1994, MNRAS, 269, 563

\refe Gardner J.P., Katz N., Hernquist L., Weinberg D.H., 1997, ApJ, 484, 31
\refe Gelato S., Governato F., 1996, in Persic M., Salucci P. eds., Dark 
Matter 1996. Pasp Conf. Ser., in press 
\refe Gelb J.M., Bertschinger E., 1994, ApJ, 436, 467
\refe Girardi M., Biviano A., Giuricin G., Mardirossian F., Mezzetti M., 
1993, ApJ, 404, 38
\refe Giuricin G., Mardirossian F., Mezzetti M., 1982, ApJ, 255, 361
\refe Giuricin G., Mardirossian F., Mezzetti M., Monaco P., 1993, ApJ, 407, 22
\refe Giuricin G., Gondolo P., Mardirossian F., Mezzetti M., Ramella M., 
1988, A\&A, 199, 85
\refe Gnedin N.Y., Hui L., 1996, ApJ, 472, L73
\refe Gott J.R. III, Turner E.L., 1977, ApJ, 216, 357
\refe Governato F., Tozzi P., Cavaliere A., 1996, ApJ, 458, 18
\refe Gunn J.E., Gott J.R., 1972, ApJ, 176, 1
\refe Gurbatov S.N., Saichev A.I., Shandarin S.F., 1989, MNRAS, 236, 385

\refe Haehnelt M.G., 1993, MNRAS, 265, 727
\refe Haehnelt M.G., 1995, in Cold Gas at High Redshift, eds. Bremer et al.,
in press (preprint astro-ph/9512024)
\refe Haehnelt M.G., Rees M.J., 1993, MNRAS, 263, 168
\refe Hanami H., 1993, ApJ, 415, 42
\refe Harrison E.R., 1970, Phys. Rev. D, 1, 2726
\refe Henriksen R.N., Lachi\`eze-Rey M., 1990, MNRAS, 245, 255
\refe Henry J.P., Arnaud K.A., 1991, ApJ, 372, 410
\refe Hoffman Y., 1986, ApJ, 308, 493
\refe Hoffman Y., 1988, ApJ, 329, 8
\refe Hucra J.P., Geller M.J., 1982, ApJ, 257, 423
\refe Hui L., Gnedin N.Y., Zhang Y., 1997, ApJ, 486, 599

\refe Ikeuchi S., Murikama I., Rees M.J., 1988, MNRAS, 236, 21P

\refe Jain B., Bertschinger E., 1994, ApJ, 431, 495
\refe Jedamzik K., 1995, ApJ, 448, 1
\refe Jing Y.P., Mo H.J., Borner G., Fang L.Z., 1995, MNRAS, 276, 417
\refe Jones C., Forman W., 1991, in Fabian A.C. ed., Clusters and 
Superclusters of galaxies.Kluwer Ac. Pub., Dordrecht

\refe Kaiser N., 1984, ApJ, 284, L9
\refe Kaiser N., 1991, ApJ, 383, 104
\refe Kaiser N., 1996, in Schramm D. ed., Generation of Large-Scale 
Cosmological Structures. In press
\refe Kashlinsky A., 1987, ApJ, 317, 19
\refe Kashlinsky A., 1993, ApJ, 406, L1
\refe Katz N., 1992, ApJ, 391, 502
\refe Katz N., Quinn T., Bertschinger E., Gelb J.M., 1994, MNRAS, 270, L71
\refe Katz N., Quinn T., Gelb J.M., 1993, MNRAS, 265, 689 
\refe Katz N., Weinberg D.H., Hernquist L., 1996, ApJS, 105, 19
\refe Kauffmann G., Guideroni B., White S.D.M., 1994, MNRAS, 267, 981
\refe Kauffmann G., White S.D.M., 1993, MNRAS, 261, 921
\refe Kauffmann G., White S.D.M., Guideroni B., 1993, MNRAS, 264, 201
\refe Kitayama T., Suto Y., 1996a, MNRAS, 280, 638
\refe Kitayama T., Suto Y., 1996b, ApJ, 469, 480
\refe Klypin A., Holtzman J., Primack J., Reg\"os E., 1993, ApJ, 416, 1
\refe Klypin A., Borgani S., Holtzman J., Primack J., 1995, ApJ, 444, 1
\refe Klypin A., Rhee G., 1994, ApJ, 428, 399
\refe Kofman L.A., Gnedin N.Y., Bahcall N.A., 1993, ApJ, 413, 1
\refe Kofman L., Pogosyan D., 1995, ApJ, 442, 30
\refe Kofman L., Pogosyan D., Shandarin S., Melott A., 1992, ApJ, 393, 437
\refe Kolb E.W., Turner M.S., 1990, The Early Universe. Addison-Wesley, 
New York
\refe Kontorovich V.M., Kats A.V., Krivitsky D.S., 1992, Sov. Phys. JETP 
Lett., 55, 1

\refe Lacey C., Cole S., 1993, MNRAS, 262, 627
\refe Lacey C., Cole S., 1994, MNRAS, 271, 676
\refe Lacey C., Silk J., 1991, ApJ, 381, 14
\refe Lachi\`eze-Rey M., 1993a, ApJ, 407, 1
\refe Lachi\`eze-Rey M., 1993b, ApJ, 408, 403
\refe Lahav O., Lilje P.B., Primack J.R., Rees M.J., 1991, MNRAS, 251, 128
\refe Lanzetta K.M., Bowen D.B., Tytler D., Webb J.K., 1995, ApJ, 442, 538
\refe Lawrence A., 1987, PASP, 99, 309
\refe Lilje P.B., 1990, ApJ, 351, 1 
\refe Lilje P.B., 1992, ApJ, 386, L33 
\refe Limber D.N., Mathews W.G., 1960, ApJ, 132, 286
\refe Liddle A.R., Lyth D.H., Schaefer R.K., Shafi Q., Viana P.T.P., 1995,
MNRAS, in press (preprint astro-ph/9511057)
\refe Loeb A., 1993, ApJ, 403, 542
\refe Lokas E.L., Juszkiewicz R., Bouchet F.R., Hivon E., 1996, ApJ, 467, L1
\refe Lubin L.M., Cen R., Bahcall N.A., Ostriker, J.P., 1996, ApJ, 460, 10
\refe Lynds R., 1971, ApJ, 168, L87
\refe Lucchin F., 1989, in Flin P., Duerbeck H.W., eds., Morphological 
Cosmology. Springer Verlag, Berlin 
\refe Lucchin F., Matarrese S., 1988, ApJ, 330, 535
\refe Lynden-Bell D., 1967, MNRAS, 136, 101 

\refe Ma C., Bertschinger E., 1994, ApJ, 434, L5
\refe Makino N., Suto S., 1993, ApJ, 405, 1
\refe Manrique A., Salvador-Sol\'e E., 1995, ApJ, 453, 6
\refe Manrique A., Salvador-Sol\'e E., 1996, ApJ, 467, 504
\refe Mart\'{\i}nez-Gonz\'alez E., Sanz J.L., 1988a, ApJ, 324, 653
\refe Mart\'{\i}nez-Gonz\'alez E., Sanz J.L., 1988b, ApJ, 332, 89
\refe Matarrese S., 1996, in Bonometto S., Primack J., Provenzale A., eds.,
Dark Matter in the Universe. In press (preprint astro-ph/9601172)
\refe Matarrese S., Lucchin F., Moscardini L., Saez D., 1992, MNRAS, 259, 
437
\refe Matarrese S., Pantano O., Saez D., 1993, Phys. Rev. D 47, 1311
\refe Matarrese S., Pantano O., Saez D., 1994, Phys. Rev. Lett. 72, 320
\refe Materne J., 1978, A\&A, 63, 401
\refe Matteucci F., 1996, Fund. Cosm. Phys., 17, 283
\refe Mazure A., et al., 1996, A\&A, 310, 31
\refe Melott A.L., Buchert T., Wei\ss\ A.G., 1995, A\&A, 294, 345
\refe Melott A.L., Pellman T., Shandarin S.F., 1994, MNRAS, 269, 626
\refe Menci N., Colafrancesco S., Biferale L., 1993, J. Phys., 3, 1105
\refe Menci N., Valdarnini R., 1995, ApJ, 445, 1019
\refe Merritt D., Gebhardt K., 1994, in Durret F., Mazure A., Tran Thanh 
Van J. eds., Clusters of Galaxies. Editions Frontieres, Gif-sur-Yvette
\refe Meylan G. (ed.), 1995, QSO Absorption Lines. Springer, Berlin
\refe Mo H.J., Miralda-Escud\'e J., 1994, ApJ, 430, L25
\refe Mo H.J., Miralda-Escud\'e J., Rees M.J., 1993, MNRAS, 264, 705
\refe Monaco P., 1994, in Durret F., Mazure A., Tran Thanh Van J. eds., 
Clusters of Galaxies. Editions Frontieres, Gif-sur-Yvette
\refe Monaco P., 1995, ApJ, 447, 23
\refe Monaco P., 1996a, MNRAS, in press (preprint astro-ph/9606027)
\refe Monaco P., 1996b, MNRAS, submitted (preprint astro-ph/9606029)
\refe Monaco P., 1996c, Astrophys. Lett. Comm., 33, 79
\refe Monaco P., 1996d, Astrophys. Lett. Comm., in press
\refe Monaco P., 1996e, in Persic M., Salucci P. eds., Dark Matter 1996. Pasp
Conf. Ser., in press
\refe Monaco P., 1996f, in Schramm D. ed., Generation of Large-Scale 
Cosmological Structures. In press
\refe Monaco P., Giuricin G., Mardirossian F., Mezzetti M., 1994, ApJ, 436, 
576 
\refe Moutarde F., Alimi J.M., Bouchet F.R., Pellat R., Ramani A., 1991, 
ApJ, 382, 377 

\refe Narayan R., White S.D.M., 1988, MNRAS, 231, 97P
\refe Navarro J.F., 1996, in Persic M., Salucci P. eds., Dark Matter 
1996. Pasp Conf. Ser., in press  
\refe Navarro J.F., Frenk C.S., White S.D.M., 1995a, MNRAS, 275, 56 
\refe Navarro J.F., Frenk C.S., White S.D.M., 1995b, MNRAS, 275, 720 
\refe Navarro J.F., Frenk C.S., White S.D.M., 1996, ApJ, 462, 563 
\refe Newman W.I., Wasserman I., 1990, ApJ, 354, 411
\refe Nusser A., Dekel A., 1990, ApJ, 362, 14
\refe Nusser A., Silk J., 1993, ApJ, 411, L1

\refe Occhionero F., Scaramella R., 1988, A\&A, 204, 3
\refe Ostriker J.P., Cen R., 1996, ApJ, 464, 27
\refe Oukbir J., Blanchard A., 1992, A\&A, 262, L21

\refe Padmanabhan T., 1993, Structure Formation in the Universe. Cambridge
University Press, Cambridge
\refe Peacock J.A., Heavens A.F., 1985, MNRAS, 217, 805
\refe Peacock J.A., Heavens A.F., 1990, MNRAS, 243, 133
\refe Peebles P.J.E., 1969, ApJ, 155, 393
\refe Peebles, P.J.E. 1980, The Large Scale Structure of the Universe 
(Princeton: Princeton Univ. Press)
\refe Peebles P.J.E., 1990, ApJ, 365, 27
\refe Peebles P.J.E., 1993, Principles of Physical Cosmology. Princeton 
University Press, Princeton
\refe Peebles P.J.E., Daly R.A., Juszkiewicz R., 1989, ApJ, 347, 563
\refe Pei Y.C., 1995, ApJ, 438, 623
\refe Pen U., 1996, preprint astro-ph/9610147
\refe Persic M., Salucci P., 1995, ApJS, 99, 501
\refe Persic M., Salucci P., 1996, in Persic M., Salucci P. eds., Dark Matter 
1996. Pasp Conf. Ser., in press
\refe Persic M., Salucci P., Stel F., 1996, MNRAS, 281, 27
\refe Pietronero L., Montuori M., Sylos-Labini F., 1996, in Turok N. ed., 
Critical Dialogues in Cosmology. In press (preprint astro-ph/9611197)
\refe Pisani A., Giuricin G., Mardirossian F., Mezzetti M., 1992, ApJ, 389, 68
\refe Porciani C., Ferrini F., Lucchin F., Matarrese S., 1996, MNRAS, 281, 311
\refe Press W.H., Flannery B.P., Teukolsky S.A., Vetterling W.T., 1992, 
Numerical Recipes in Fortran. Cambridge University Press, Cambridge
\refe Press W.H., Schechter P., 1974, ApJ, 187, 425
\refe Press W.H., Teukolsky S.A., 1990, Computers in Physics Jan/\-Feb, 
1990, 92

\refe Rauch M., Haehnelt M.G., Steinmetz M., 1997, ApJ, 481, 601
\refe Rees M.J., 1984, ARA\&A, 22, 471
\refe Rees M.J., 1986, MNRAS, 218, 25P
\refe Rees M.J., 1996, preprint astro-ph/9608196
\refe Rees M.J., Ostriker J.P., 1977, MNRAS, 179, 541
\refe Richstone D., Loeb A., Turner E.L., 1992, ApJ, 393, 477
\refe Risken H., 1989, The Fokker-Planck Equation. Springer Verlag, Berlin  
\refe Rodrigues D.D.C., Thomas P.A., 1996, MNRAS, 282, 631
\refe Rubin V.C., Ford W.K.Jr, Thonnard N., 1980, ApJ, 238, 471
\refe Rubin V.C., Burstein D., Ford W.K.Jr, Thonnard N., 1985, ApJ, 289, 81
\refe Ryden B.S., 1988, ApJ, 333, 78

\refe Sahni V., Coles P., 1996, Phys. Rep., in press (preprint astro-ph/9505005)
\refe Sahni V., Shandarin S.F., 1996. MNRAS, 282, 641
\refe Salucci P., Szuskiewicz E., 1996, in preparation
\refe Sarazin C.L., 1986, Rev. Mod. Phys., 58, 1
\refe Sargent W.L.W., Young P.J., Boksenberg A., Tytler D., 1980, ApJS, 42, 41
\refe Sasaki S., 1994, PASJ, 46, 427
\refe Schaeffer R., Silk J., 1985, ApJ, 292, 319
\refe Schaeffer R., Silk J., 1988a, ApJ, 332, 1 
\refe Schaeffer R., Silk J., 1988b, ApJ, 333, 509
\refe Shandarin S.F., Zel'dovich Ya.B., 1989, Rev. Mod. Phys., 61, 185
\refe Shandarin S.F., Doroshkevich A.G., Zel'dovich Ya.B., 1983, Sov. 
Phys. Usp., 26, 46
\refe Shaviv N.J., Shaviv G., 1993, ApJ, 412, L25
\refe Shaviv N.J., Shaviv G., 1995, ApJ, 448, 514
\refe Schechter P.L., 1976, ApJ, 203, 297
\refe Sheth R.K., 1995, MNRAS, 276, 796
\refe Sheth R.K., 1996, MNRAS, 261, 1277
\refe Silk J., 1977, ApJ, 211, 638
\refe Silk J., 1978, ApJ, 220, 390
\refe Silk J., White S.D., 1978, ApJ, 223, L59
\refe Smoluchowski M., 1916, Phys. Z., 17, 557
\refe Steidel C., Dickinson M., Meyer D., Adelberger K., Sembach K., 1997,
ApJ, 480, 568
\refe Steinmetz M., 1996, in Bonometto S., Primack J., Provenzale A., eds.,
Dark Matter in the Universe. In press (preprint astro-ph/9512013)
\refe Sunyaev R.A., Zel'dovich Ya.B., 1970, Astrophys. Space Sci., 7, 3

\refe Tegmark M., Silk J., Rees M.J., Blanchard A., Abel T., Palla F., 1997,
ApJ, 474, 1
\refe Thomas P.A., Couchman H.M.P., 1992, MNRAS, 257, 11
\refe Tormen G., Bouchet F.R., White S.D.M., 1997, MNRAS, 286, 865
\refe Tully R.B., 1987, ApJ, 321, 280

\refe Ueda H., Itoh M., Suto Y., 1993, ApJ, 408, 3
\refe Umemura M., Loeb A., Turner E.L., 1993, ApJ, 419, 459

\refe van de Weygaert R., Babul A., 1994, ApJ, 425, L59
\refe Vergassola M., Dubrulle B., Frisch U., Noullez A., 1994, A\&A, 289, 325
\refe Viana P.T.P., Liddle A.R., 1996, MNRAS, 281, 323

\refe Weinberg S., 1972, Gravitation and Cosmology. Wiley, New York
\refe Wei\ss\ A.G., Gottl\"ober S., Buchert T., 1996, MNRAS, 278, 953
\refe West M.J., 1994, in Durret F., Mazure A., Tran Thanh Van J. eds., 
Clusters of Galaxies. Editions Frontieres, Gif-sur-Yvette
\refe West M.J., Jones C., Forman W., 1995, ApJ, 451, L5
\refe White S.D.M., 1984, ApJ, 286, 38
\refe White S.D.M., in Schaeffer R. ed., Les Houches 1993. In press
\refe White S.D.M., Frenk C.S., 1991, ApJ, 379, 52
\refe White S.D.M., Efstathiou G., Frenk C.S., 1993, MNRAS, 262, 1023
\refe White S.D.M., Rees M.J., 1978, MNRAS, 183, 341
\refe White S.D.M., Silk J., 1979, ApJ 231, 1
\refe Williams B.G., Heavens A.F., Peacock J.A., Shandarin S.F., 1991, MNRAS,
250, 458

\refe Yano T., Nagashima M., Gouda N., 1996, ApJ, 466, 1
\refe Yi I., 1996, ApJ, 473, 645

\refe Zabludoff A.I., Geller M.J., Hucra J.P., Vogeley M.S., 1993, 
AJ, 106, 1273
\refe Zaritsky D., Smith R., Frenk C.S., White S.D.M., 1993, ApJ, 405, 464
\refe Zel'dovich, Ya.B. 1970, Astrofizika 6, 319 (transl.: 1973, 
Astrophysics 6, 164)
\refe Zel'dovich Ya.B., 1972, MNRAS, 160, 1P
\refe Zhan Y., 1990, ApJ, 355, 387
\end{document}